\documentclass[aps,floats,floatfix,showpacs,amssymb,prd,onecolumn,superscriptaddress,nofootinbib,nolongbibliography,10pt]{revtex4-2}

\usepackage{caption}
\usepackage{multirow}
\usepackage{amssymb,amsmath,verbatim,mathtools,needspace,enumitem,etoolbox,graphicx,physics,microtype,afterpage,bigints,textcomp,gensymb,tabularx,bm,booktabs,multirow,ragged2e,nicefrac,colortbl,scalerel,comment}
\usepackage[dvipsnames]{xcolor}
\definecolor{linkcolor}{rgb}{0.0,0.3,0.5}
\definecolor{lightblue}{RGB}{200,220,240}
\definecolor{dodgerblue}{HTML}{1E90FF}
\usepackage[unicode, colorlinks=true, linkcolor=linkcolor, citecolor=linkcolor, filecolor=linkcolor,urlcolor=linkcolor, pdfusetitle]{hyperref}
\usepackage[all]{hypcap}
\usepackage[T1]{fontenc}
\usepackage[utf8]{inputenc}
\usepackage{orcidlink}
\usepackage{newunicodechar}
\newunicodechar{−}{\ensuremath{-}}
\usepackage{array}
\usepackage[normalem]{ulem}
\usepackage{wrapfig2}
\usepackage[breakable]{tcolorbox}
\definecolor{ocre}{RGB}{52,177,201} %
\usepackage{ragged2e}
\usepackage[capitalise]{cleveref}

\usepackage{tabularray}
\UseTblrLibrary{booktabs} %

\makeatletter
\newif\ifaffil@first \affil@firsttrue
\def\frontmatter@above@affiliation@script{%
  \skip@\@flushglue
  \@flushglue\z@ plus.3\hsize\relax
  \centering
  \@flushglue\skip@
  \parskip 2\p@\relax
  \ifaffil@first
    \addvspace{9\p@}%
    \global\affil@firstfalse
  \else
    \addvspace{2\p@}%
  \fi
}
\def\frontmatter@affiliationfont{\footnotesize\itshape}
\makeatother

\definecolor{outrageousorange}{rgb}{1.0, 0.43, 0.29}

\makeatletter
\newcommand*{\balancecolsandclearpage}{\close@column@grid \cleardoublepage \twocolumngrid}
\makeatother

\makeatletter
\def\frontmatter@title@format{\LARGE\bfseries\centering\parskip\z@skip}
\def\frontmatter@title@above{\addvspace{12\p@}}
\def\frontmatter@title@below{\addvspace{6\p@}}
\makeatother

\usepackage{lipsum}

\newcommand{\jhu}{\affiliation{William H. Miller III Department of Physics and Astronomy,\\ Johns Hopkins University, 3400 North Charles Street, Baltimore, Maryland, 21218, USA}}
\newcommand{\UU}{\affiliation{Institute for Gravitational and Subatomic Physics (GRASP), Utrecht University, \\Princetonplein 1, 3584 CC Utrecht, The Netherlands}}
\newcommand{\Nikhef}{\affiliation{Nikhef -- National Institute for Subatomic Physics, Science Park 105, 1098 XG Amsterdam, The Netherlands}}
\newcommand{\mbi}{\affiliation{Marietta Blau Institute - Austrian Academy of Sciences, 1010 Vienna, Austria}}
\newcommand{\AEI}{\affiliation{Max Planck Institute for Gravitational Physics (Albert Einstein Institute), Am M\"uhlenberg 1, Potsdam 14476, Germany}}
\newcommand{\UMD}{\affiliation{Department of Physics, University of Maryland, College Park, MD 20742, USA}}
\newcommand{\GSSI}{\affiliation{Gran Sasso Science Institute (GSSI), Via Michele Iacobucci 2, I-67100 L'Aquila, Italy}}
\newcommand{\LNGS}{\affiliation{INFN, Laboratori Nazionali del Gran Sasso, Via Giovanni Acitelli 22, I-67100 Assergi, Italy}}
\newcommand{\UPotsdam}{\affiliation{Institut f\"ur Physik und Astronomie, Universit\"at Potsdam, Haus 28, Karl-Liebknecht-Str.\ 24/25, 14476 Potsdam, Germany}}
\newcommand{\UGhent}{\affiliation{Department of Physics and Astronomy, Universiteit Ghent, B-9000 Ghent, Belgium}}
\newcommand{\PSU}{\affiliation{Institute for Gravitation and the Cosmos, Department of Physics, Pennsylvania State University, University Park, PA 16802, USA}}
\newcommand{\Berkeley}{\affiliation{Department of Physics, University of California, Berkeley, CA 94720, USA}}
\newcommand{\Northwestern}{\affiliation{Department of Physics and Astronomy, Northwestern University, 2145 Sheridan Road, Evanston, IL 60208, USA}}
\newcommand{\CIERA}{\affiliation{Center for Interdisciplinary Exploration and Research in Astrophysics (CIERA), Northwestern University, 1800 Sherman Ave, Evanston, IL 60201, USA}}
\newcommand{\UAntwerpen}{\affiliation{Universiteit Antwerpen, Prinsstraat 13, B-2000 Antwerpen, Belgium}}
\newcommand{\Minnesota}{\affiliation{School of Physics and Astronomy, University of Minnesota, Minneapolis, MN 55455, USA}}

\begin{document}

\title{The Science Chapter of the Einstein Telescope \\ Geometry Comparison Report}

\author{Alessandra Buonanno}\AEI\UMD
\author{Federico De Lillo}\UAntwerpen
\author{Arnab Dhani}\AEI
\author{Tim Dietrich}\UPotsdam\AEI
\author{Ulyana Dupletsa}\mbi
\author{Archisman Ghosh}\UGhent
\author{Ish Gupta}\Berkeley\Northwestern\CIERA
\author{Jan Harms}\GSSI\LNGS
\author{Francesco Iacovelli}\jhu
\author{Vuk Mandic}\Minnesota
\author{Harsh Narola}\UU\Nikhef
\author{B.\,S. Sathyaprakash}\PSU
\author{Chris Van den Broeck}\UU\Nikhef
\author{Haowen Zhong}\Minnesota\jhu

\date{\rule{0pt}{16pt}\today}

\begin{abstract}
	
	This is the Scientific Added Value Chapter of the Einstein Telescope (ET) Geometry Comparison Report, commissioned from the ET Organization by the ET Board of Governmental Representatives.\footnote{The Geometry Comparison Report and Executive Summary are publicly  available~\cite{ET-GCR,ET-GCR-ExecSummary}.} We compare three detector configurations: a single-site 10-km triangle ($\Delta$), and two 15-km L-shaped detectors (2L) at widely separated sites, either nearly aligned or misaligned. To assess their performance, we use a broad set of metrics representative of the ET science case: compact-binary populations, dense-matter physics, cosmography, stochastic backgrounds, multimessenger astronomy, fundamental physics and tests of general relativity, and continuous and burst gravitational-wave sources. All three configurations successfully deliver the broad ET science case, but they perform best in different areas. For most sensitivity-driven science, the 2L configurations outperform the $\Delta$. The misaligned 2L is best for localizing and characterizing individual sources, multimessenger astronomy, and bright-siren cosmology. The nearly aligned 2L is best for stochastic backgrounds. The $\Delta$ localizes sources better than the aligned 2L, and performs comparably to both 2L configurations for continuous and burst sources and at kilohertz frequencies. An intermediate 2L orientation of $30^\circ$ offers a good compromise between stochastic-background sensitivity and source localization. We systematically validate the Fisher information matrix (FIM) forecasts against full Bayesian inference for $\sim550$ binary black hole and neutron star-black hole signals. Despite substantial differences, the FIM shows no systematic preference for either configuration in terms of intrinsic parameters. For distance and sky localization, the inability of the FIM to capture multimodality and long-range parameter correlations can lead to larger discrepancies. Nevertheless, these discrepancies do not affect the relative ranking of the detector configurations. Beyond the scope of the report, we repeat the comparison with Cosmic Explorer added to the network. This improves the absolute reach of all three ET configurations, most notably in volume localization and characterization of high-redshift sources. The differences between geometries reduce, but the ranking is largely unchanged.
	
\end{abstract}

\raggedbottom 
\maketitle

\section*{Executive Summary}

\noindent
\begin{minipage}{0.45\textwidth}
The Einstein Telescope (ET) is Europe's proposed next-generation
gravitational-wave (GW) observatory. Building on the discoveries of the 
LIGO-Virgo-KAGRA Collaboration, ET is designed not merely to detect GWs, but to turn GW astronomy into a precision observational science. Its substantially greater sensitivity and access to frequencies from a few hertz to several kilohertz will allow it to observe black-hole and neutron-star mergers across much of the history of the Universe, unravel the origin of the supermassive black holes at the centers of galaxies and their role in galaxy formation and evolution, and search for new populations of compact objects. It will study the nature of gravity and black holes in regimes inaccessible by other means, and probe matter at densities beyond those achievable in terrestrial laboratories. It will also measure the expansion history of the Universe and potentially detect GW backgrounds from processes in the very early Universe, providing a glimpse of the initial conditions that shaped the Universe soon after the Big Bang. A summary of the science capabilities of ET is given in Fig.\,\ref{fig:science wheel}.
\vskip5pt
Realizing this scientific potential requires a major European research infrastructure that will operate for decades. An important decision is therefore not only \emph{where} ET should be built, but \emph{what geometry it should have}. 
\end{minipage}
\hfill
\begin{minipage}{0.48\textwidth}
\begin{tcolorbox}[standard jigsaw,colframe=gray,colback=gray!10!white,opacityback=0.0,coltext=black, title=\small\sc Science Targets of the Einstein Telescope]
\centering\includegraphics[width=1.02\textwidth]{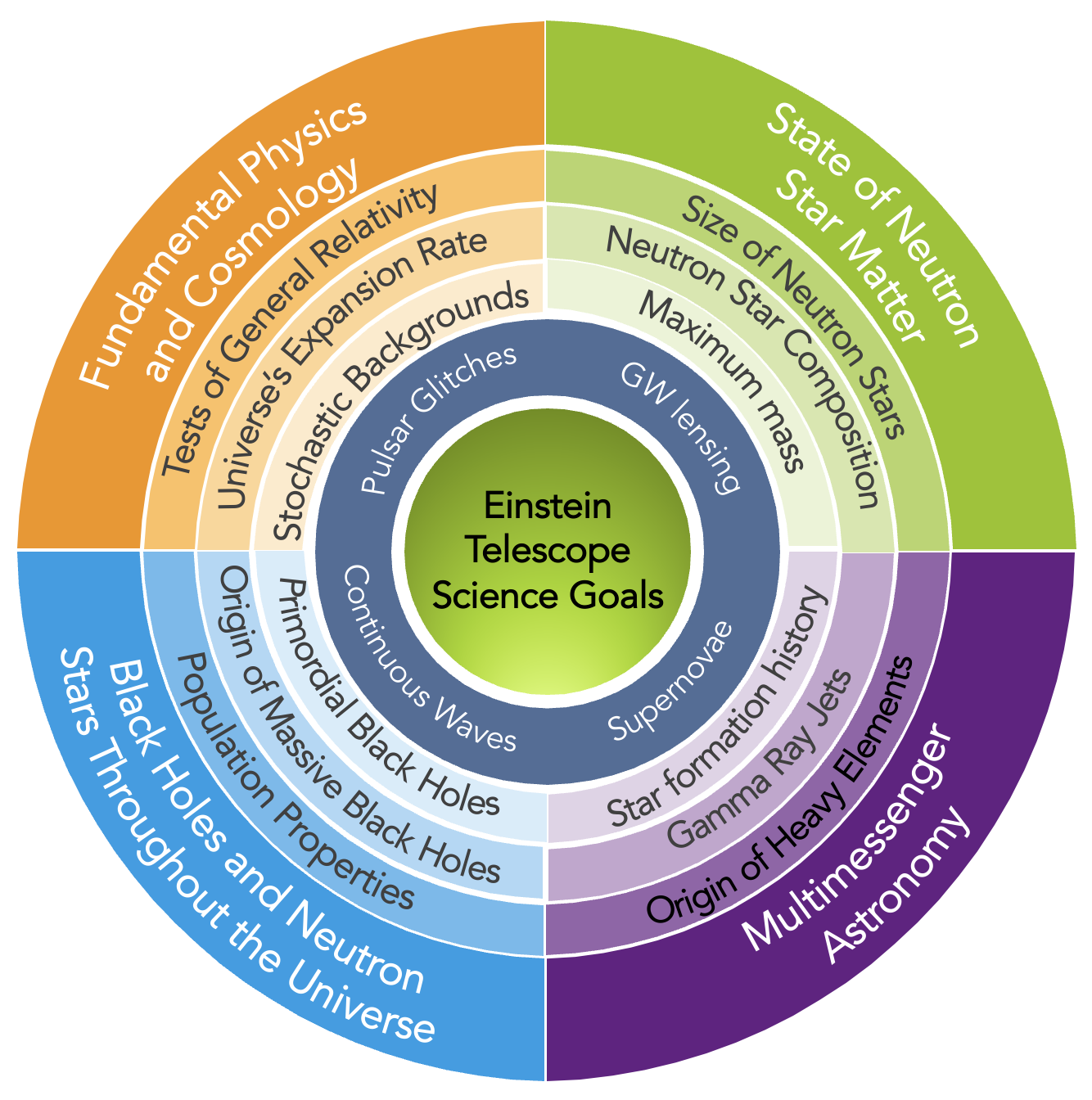}
\captionof{figure}{The illustration shows the main science goals of ET some of which were used in this study and form the core of the geometry-comparison report.}
\vskip-5pt
\label{fig:science wheel}
\end{tcolorbox}
\end{minipage}

\vskip5pt
\noindent The original ET design consists of three
nested 10-km interferometers arranged as an underground triangle at a single
site. An alternative now under consideration consists of two 15-km
L-shaped observatories at sites separated by more than 600 km. For the
latter, the two detectors can be oriented nearly parallel to one another
or rotated relative to each other. This summary compares the scientific
performance of the 10-km triangle with two representative orientations
of the 15-km two-site configuration: a nearly aligned geometry at $-2.5^\circ$, and one 
misaligned by $+42.5^\circ$.

\begin{tcolorbox}[colback=teal!5!white,colframe=ocre!75!black,title=\sc Naming of Detector Geometry Configurations Used Here]
\begin{itemize}[leftmargin=*,itemsep=1pt,topsep=2pt]
\item \textbf{Triangle} $(\bm{\Delta})$:  An equilateral triangular configuration with 10\,km arms.
\item \textbf{2L} $\bm{(-2.5^\circ)}$: Two distant detectors with 15\,km arms rotated by $-2.5^\circ$ relative to each other.
\item \textbf{2L} $\bm{(42.5^\circ)}$: Two distant detectors with 15\,km arms rotated by $42.5^\circ$ relative to each other.
\end{itemize}
\end{tcolorbox}

We find that much of the apparent complexity of the geometry question follows from \emph{two basic physical principles}.

First, \emph{longer arms provide greater sensitivity}. The 15-km arms
of the two-site configurations generally produce stronger signals than
the 10-km arms of the triangle. This raises the number of detectable
sources, sharply increases the number of exceptionally loud events,
and improves every measurement whose precision is set by
signal strength. These include measurements of black-hole masses and
spins, neutron-star tidal properties, and tests of Einstein's theory of general
relativity. The advantage is not universal: at frequencies above about 1\,kHz, the long-wavelength approximation does not hold anymore, and accurate modeling shows that the detector response to GWs is generally reduced for detectors with longer arms. Science relying on the
highest-frequency signals, including much of neutron-star
post-merger physics, therefore gains less from the longer arms. 

Second, \emph{the relative orientation of the detectors determines what
information can be extracted from a gravitational wave}. Gravitational
waves have two polarizations. Detectors oriented differently measure
different combinations of them, making it possible to disentangle source
position, distance and orientation. Two widely separated and substantially
misaligned detectors therefore provide powerful triangulation and
polarization information. Nearly aligned detectors, by contrast, measure
almost the same combination of the two polarizations. This sacrifices
some ability to locate and characterize individual sources, particularly
at large distances, but greatly enhances sensitivity to a stochastic GW
background of astrophysical and cosmological origin through cross-correlation of the two detectors.

\begin{figure*}
\begin{tcolorbox}[standard jigsaw,colframe=gray,colback=gray!10!white,opacityback=0.0,coltext=black, title=\small\sc Polar Histograms Comparing the Science Reach of the Two Geometries]
    \includegraphics[width=0.43\textwidth]{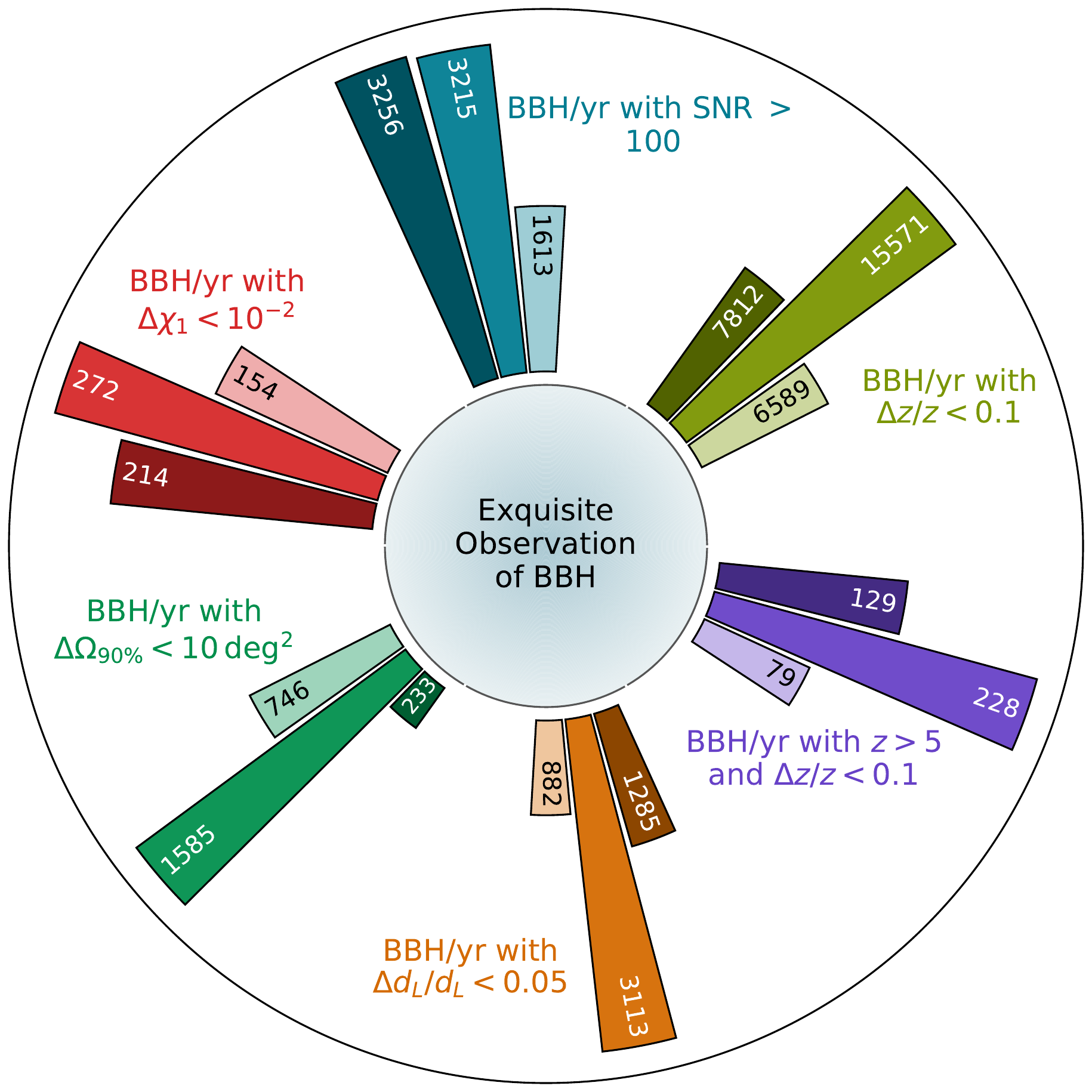} \hspace{0.5cm}
    \includegraphics[width=0.43\textwidth]{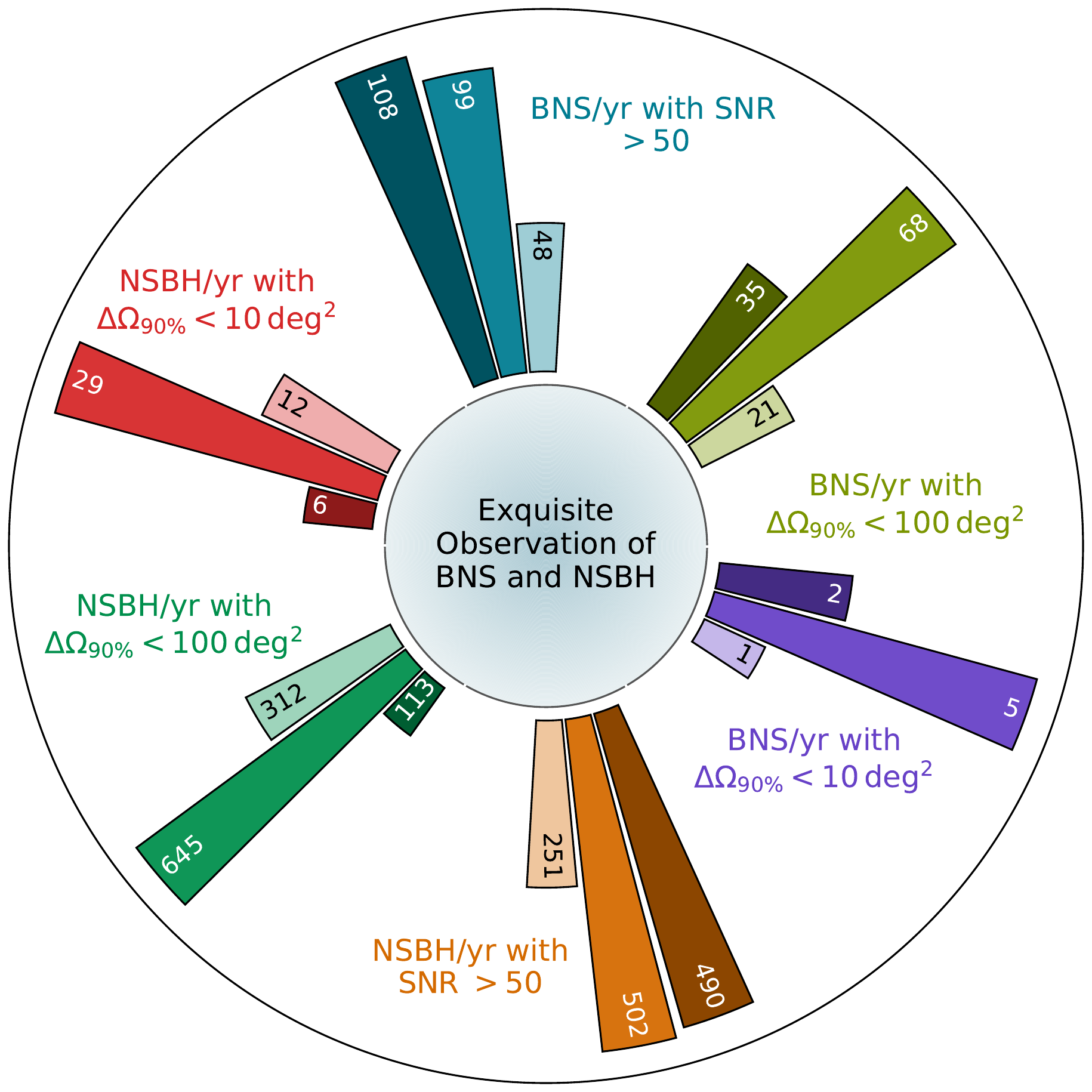}\\[8pt]
    \vskip-10pt
    \includegraphics[width=0.40\textwidth]{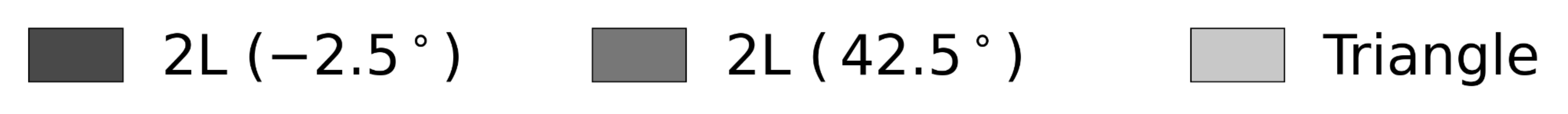}\\[2pt]
    \vskip-7pt
    \includegraphics[width=0.43\textwidth]{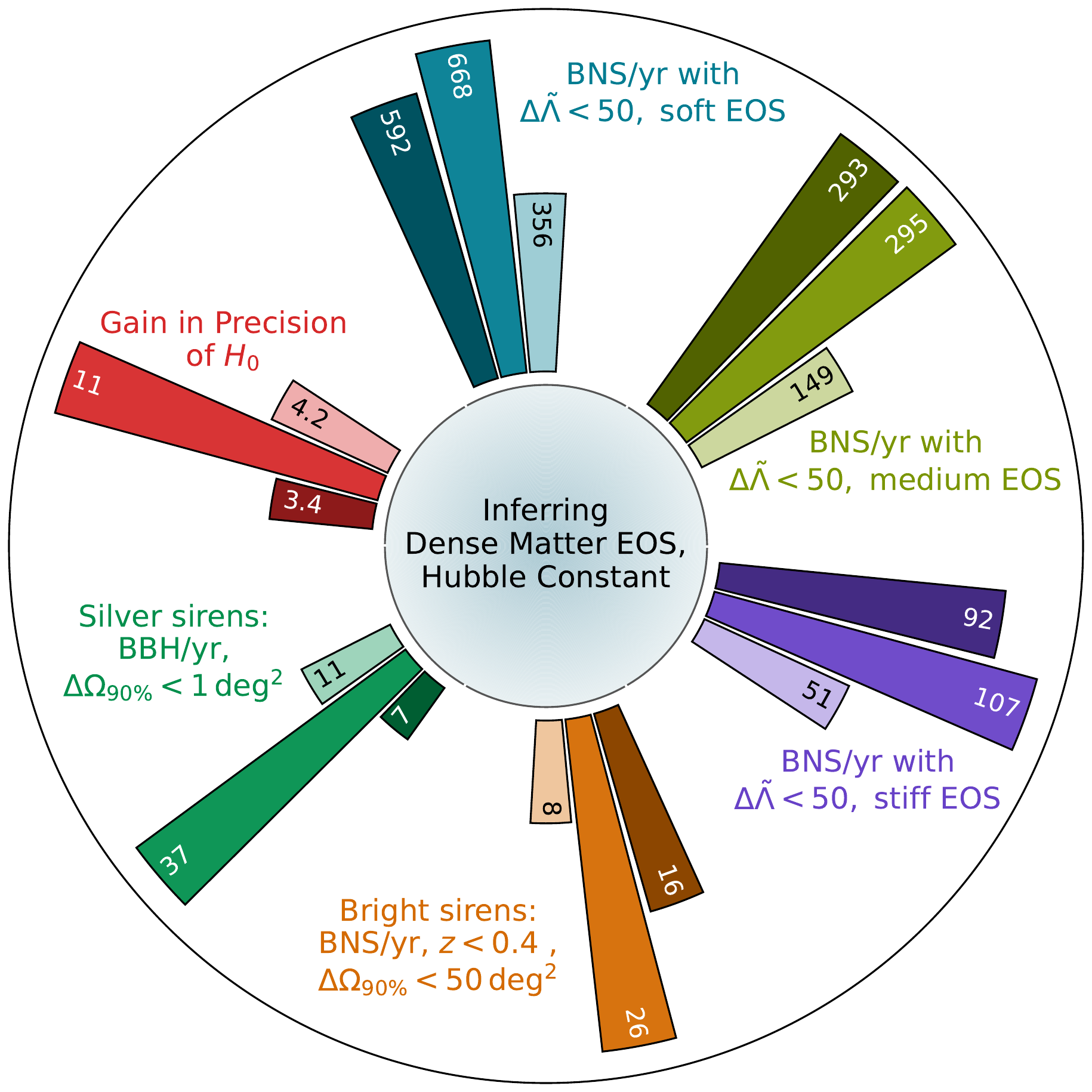} \hspace{0.5cm}
    \includegraphics[width=0.43\textwidth]{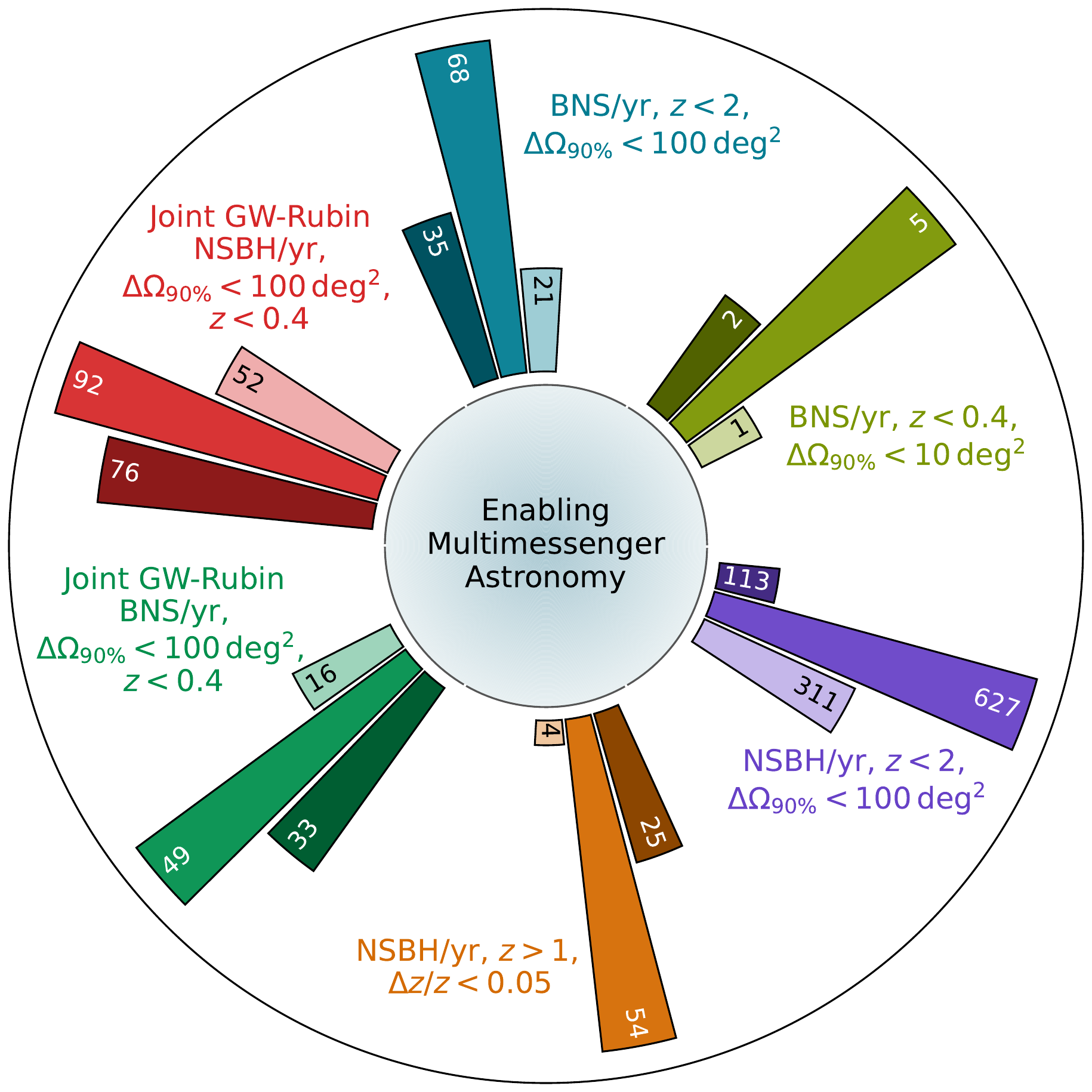}
    \caption{\justifying \textbf{Comparison of selected science metrics for the ET geometries.} 
    Within each wedge the three bars
are, from dark to light, the nearly aligned 2L ($-2.5^\circ$), the
misaligned 2L ($42.5^\circ$) and the triangle. Bar lengths are
normalized within each wedge, so they compare the geometries
with one another and not one metric against another; the number within on each bar is the absolute result.
The four polar plots show what each geometry could deliver for binary black
holes (BBH, top left), binary neutron stars (BNS) and neutron-star--black-hole
(NSBH) binaries (top right), the equation of state (EOS) of dense matter and
the Hubble constant $H_0$, which sets the expansion rate of the Universe
(bottom left), and multimessenger astronomy (bottom right). For $H_0$, bar lengths 
and numbers show the improvement factor in precision relative to the current $\sim10\%$ GW measurement. 
Throughout, ``X/yr'' is the number of events detected per year, and a
quantity written $\Delta X/X$ is the fractional uncertainty on $X$; the symbols
are the signal-to-noise ratio SNR, which measures how strong a signal is; the
redshift $z$, which measures how distant and how far back in time a source lies;
the luminosity distance $d_L$; the sky area $\Delta\Omega_{90\%}$, in square
degrees, within which the source lies with $90\%$ probability; the dimensionless
spin $\smash[b]{\chi_{1}}$ of the primary object; and the mass-weighted tidal deformability
$\tilde\Lambda$, which measures how far a neutron star is deformed by its
companion and is quoted for soft, medium and stiff equations of state. Bright
sirens are mergers with an identified electromagnetic counterpart, which supplies
the redshift; silver sirens are dark sirens (no counterpart) localized within $1 \deg^2$ 
with $\Delta d_L/d_L < 0.1$, for which a statistical host-galaxy association is possible; 
``Joint GW-Rubin'' counts mergers at $z < 0.4$ localized within $100 \deg^2$, 
the approximate reach of the Vera C. Rubin Observatory for kilonovae.}
\label{fig:polar1}
\end{tcolorbox}
\end{figure*}

\begin{figure*}
\begin{tcolorbox}[standard jigsaw,colframe=gray,colback=gray!10!white,opacityback=0.0,coltext=black, title=\small\sc Polar Histograms Comparing the Science Reach of the Two Geometries]
\centering
    \includegraphics[width=0.43\textwidth]{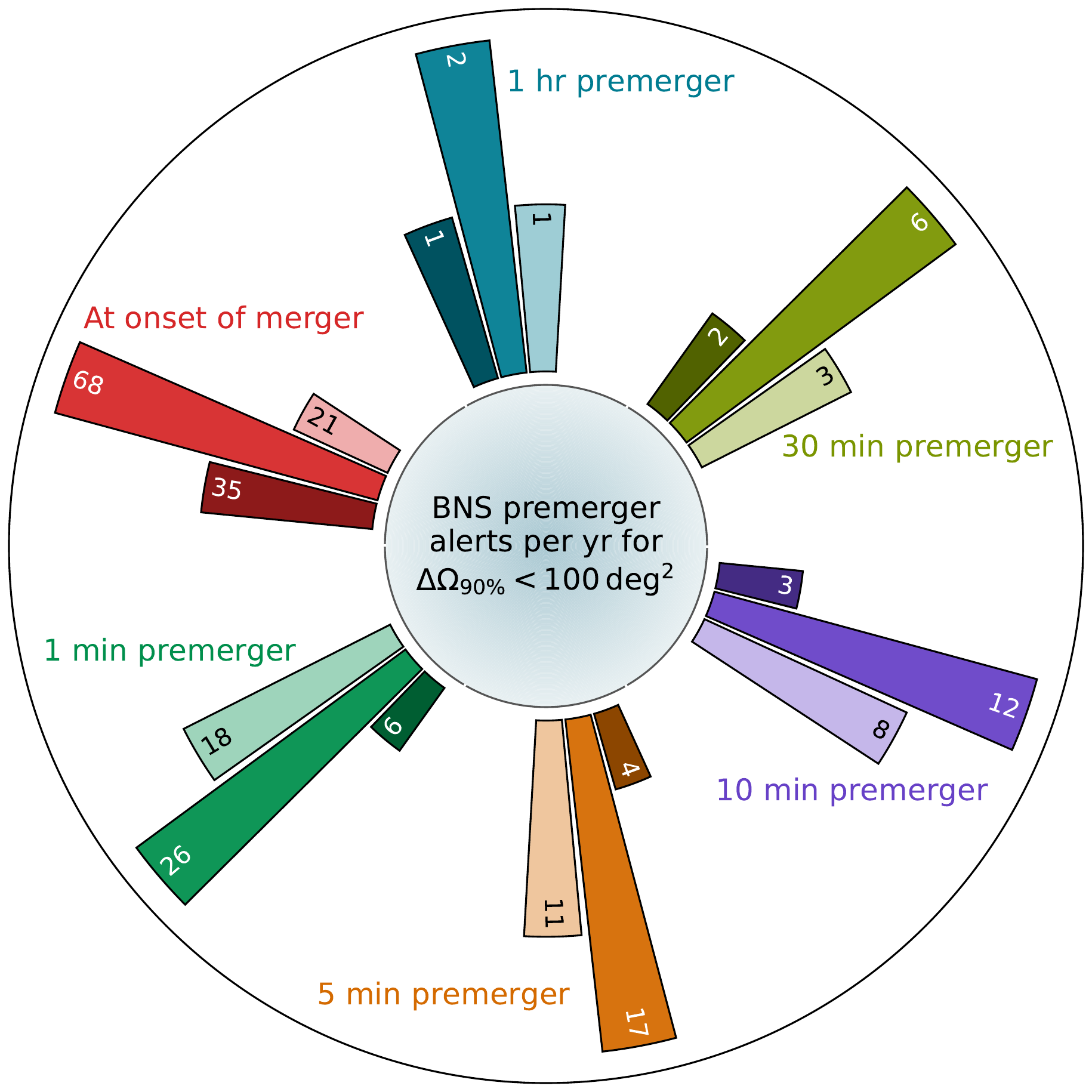} \hspace{0.5cm}
    \includegraphics[width=0.43\textwidth]{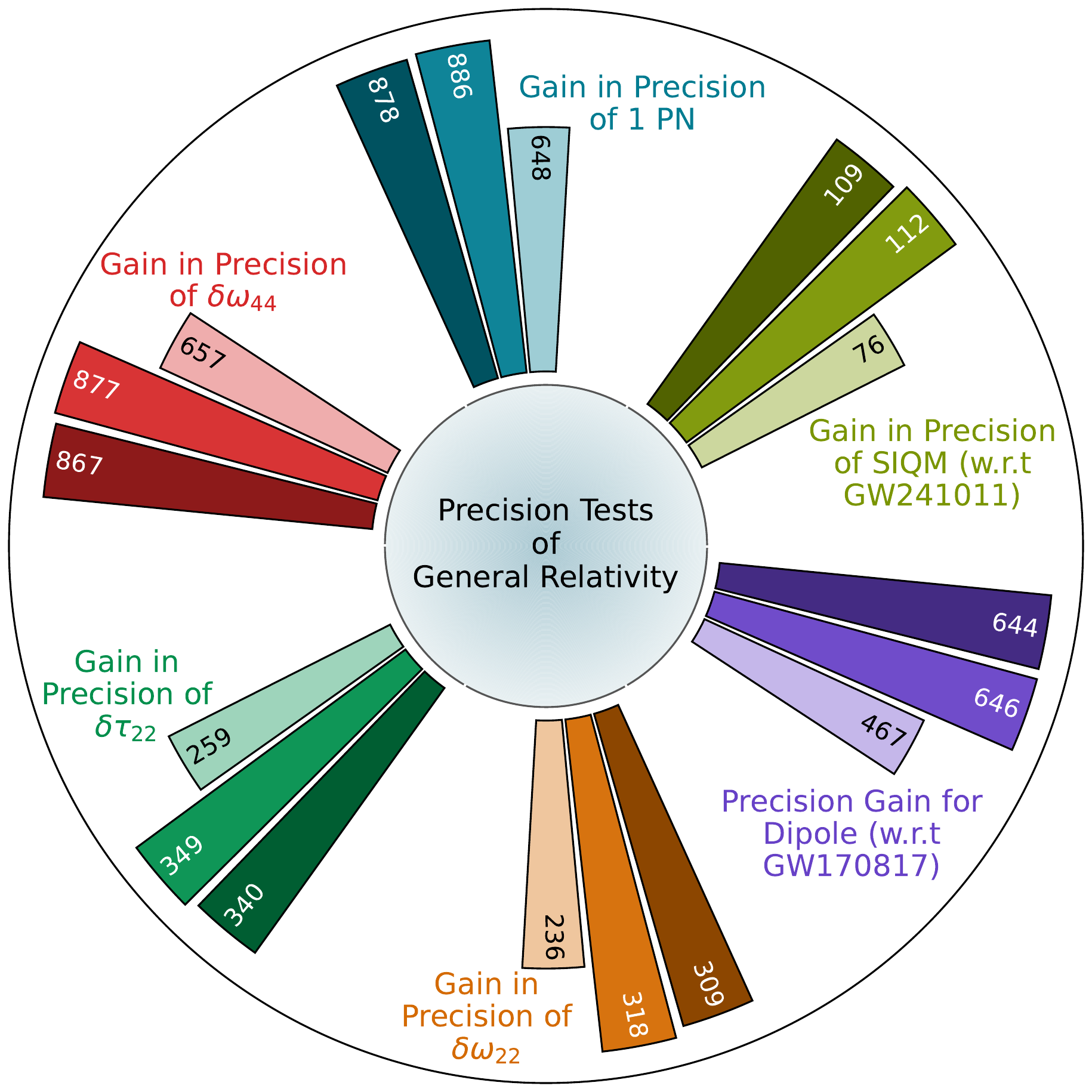}\\[8pt]
    \vskip-7pt
    \includegraphics[width=0.40\textwidth]{src/legend.pdf}\\[8pt]
    \vskip-7pt
    \includegraphics[width=0.43\textwidth]{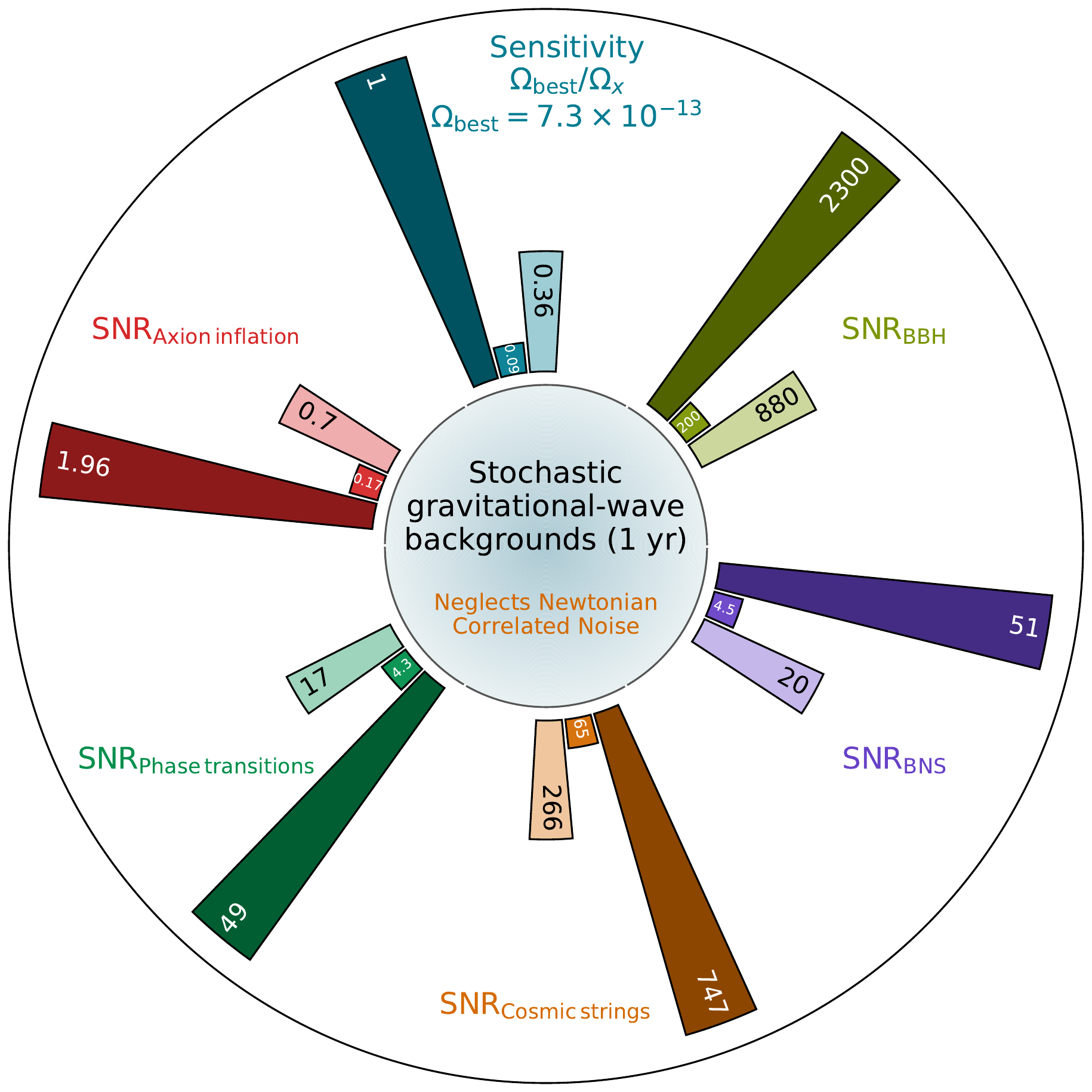} 
\caption{\justifying\textbf{Comparison of further science metrics for the ET geometries.}
Same as Fig.\,\ref{fig:polar1}, but for a different set of science metrics.
\textit{Top left: BNS pre-merger alerts.}
Expected annual numbers of BNS mergers localized to
$\Delta\Omega_{90\%}<100\,{\rm deg}^2$ at selected times before merger,
from 1~h to 1~min, together with the number localized by the onset of
merger. These metrics characterize both the rate and advance warning
time of alerts suitable for electromagnetic follow-up.
\textit{Top right: Precision tests of general relativity.}
Improvement in the precision with which representative deviations from
general relativity (GR) can be constrained including the 
first post-Newtonian (1PN) inspiral coefficient, the
ringdown frequencies of the 22- and 44-modes, $\delta\omega_{22}$ and 
$\delta\omega_{44},$ and the damping time of the 22-mode,
$\delta\tau_{22}$, the spin-induced quadrupole moment (SIQM), and 
dipole radiation. In all cases the quoted gains are relative to the BBH event 
GW250114, except for the dipole and SIQM tests, which are quoted relative to GW170817
and GW241011, respectively;
larger values correspond to greater improvement in precision.
\textit{Bottom left: Stochastic gravitational-wave backgrounds.}
One-year sensitivity to an isotropic stochastic background, expressed
through the ratio $\Omega_{\rm best}/\Omega_x$ between the best minimum detectable
amplitude among the geometries, $\Omega_{\rm best}(= 7.3 \times 10^{-13}$ at SNR=5), and 
that of each geometry, $\Omega_\mathrm{x}$ and SNRs for
representative astrophysical and cosmological backgrounds from BBHs, BNSs, cosmic strings, phase
transitions, and axion inflation; correlated (e.g., Newtonian) noise is neglected. 
Smaller $\Omega_{\rm best}$ indicates better sensitivity, 
whereas larger SNR indicates better detectability.}
\label{fig:polar2}
\end{tcolorbox}
\end{figure*}

These two principles lead to a clear overall picture, summarized in the text box titled
{\sc Geometry Comparison Summary}, and quantified in Figs.\,\ref{fig:polar1} and \ref{fig:polar2} for some of the science cases.
\emph{The two-site 15-km configurations generally outperform the 10-km triangle for science
driven primarily by sensitivity}, while their relative orientation determines 
which measurements benefit most. The misaligned
two-site configuration provides the strongest overall performance for
observing and characterizing individual sources. Its combination of
longer arms, geographic separation and complementary polarization
response gives it a particularly strong advantage for sky localization,
luminosity-distance measurements, multimessenger astronomy, early
warning of neutron-star mergers and cosmological measurements using
electromagnetic counterparts. We find that differences between geometries are 
typically at the tens-of-percent level for the precision of individual measurements 
and combined constraints, except when these are driven by localization. The differences 
grow to factors of a few in the number of sources measured to a given precision 
and become larger still for sources localized well enough for astronomical follow-up.

The nearly aligned two-site configuration strikes a different balance. It
retains the sensitivity and reach provided by 15-km arms but gives up
much of the polarization information obtained from differently oriented
detectors. Its principal advantage is correspondingly in searches for a
\emph{stochastic gravitational-wave background}---a persistent background
that could contain signals from unresolved astrophysical populations or
from physical processes in the early Universe. Here, alignment is an asset
rather than a liability, because the stochastic signal is extracted by
correlating the outputs of the two detectors. A relative orientation
close to $45^\circ$, although highly effective for localization of
individual sources, performs particularly poorly for this purpose. The
comparison therefore exposes a genuine design tension: \emph{the
orientation that is best for characterizing individual sources is not
the orientation that is best for stochastic-background science}.

The triangular configuration remains scientifically powerful and would
deliver strong results across every major source class considered here. Its three interferometers at one site provide complementary
responses and allow construction of a null stream---a combination of
detector outputs that contains no GW signal and can therefore be valuable
for diagnosing some of the instrumental disturbances. The present comparison, though, does
not quantify that potential advantage under realistic non-Gaussian
detector noise. 
With respect to stochastic-background searches, the triangle is superior to the two-site configuration above 100\,Hz; however, below 40\,Hz --— the intended prime-performance band for ET —-- its strong exposure to correlated environmental noise prevents it from reaching competitive sensitivity.
Its shorter arms
generally give it lower sensitivity than the 15-km configurations for measurements dominated by signal strength, while having all interferometers at a single site limits the geographical baseline available for source localization. 
Nevertheless, because its three differently oriented interferometers 
resolve both polarizations, the triangle localizes sources better than the
nearly aligned two-site configuration. 

\begin{tcolorbox}[float=t, colback=teal!5!white,colframe=OliveGreen!75!black,
                  title=\sc Geometry Comparison Summary]
All three configurations deliver the broad ET science case. The principal
differences arise from arm length and relative orientation.
\begin{itemize}[leftmargin=*,itemsep=2pt,topsep=3pt]

  \item \emph{Longer arms:} The 15-km 2L configurations provide higher signal-to-noise ratio and better precision than the 10-km triangle for most sources and measurements. This advantage largely disappears at the highest
        frequencies.

  \item \emph{Misaligned 2L:} The misaligned configuration is strongest
        overall for localization and characterization of individual sources,
        with particular advantages for distance measurements, multimessenger
        astronomy and bright-siren cosmology.

  \item \emph{Nearly aligned 2L:} The nearly aligned configuration retains
        the sensitivity and reach of 15-km arms while providing the best
        sensitivity to a stochastic background and the early Universe. 
        This comes at the expense of polarization information, and hence of localization accuracy.

  \item \emph{Triangle:} The 10-km triangle remains scientifically powerful across
        the full ET science program, but is generally limited by its shorter
        arms and single site. Its three interferometers uniquely enable the construction of a null stream, but they are also the most exposed to correlated environmental noise. It nevertheless localizes sources better than the nearly aligned 2L.

\item \emph{Some results are geometry independent.} Several important measurements depend only weakly on geometry, including binary population inference ( studied here for binary black holes), spectral-siren cosmology and neutron-star post-merger physics. For these, all three configurations give broadly comparable results.

\item \emph{An intermediate 2L orientation.} For the 2L configuration, an orientation around $30^\circ$ may provide an attractive compromise, retaining much of the localization capability of the misaligned geometry while preserving substantially better stochastic-background sensitivity.

\end{itemize}
\end{tcolorbox}

Importantly, \emph{not every part of the ET science case strongly
discriminates among the geometries}. All three configurations produce
enormous catalogs of compact-binary mergers and reach source populations
far beyond those accessible to today's observatories. Population-level inference on compact binaries (studied here for binary black holes), cosmological measurements that do not depend strongly on
localization, neutron-star post-merger studies, continuous-wave searches
and observations of Galactic supernovae show relatively modest geometry
dependence compared with either statistical or astrophysical
uncertainties. The geometry decision is therefore driven less by whether
ET can accomplish its broad scientific mission---all three configurations
can---than by \emph{how well it performs the most demanding measurements
and which unique discoveries it is optimized to make}.

Some considerations go beyond the idealized source calculations. ET will operate underground at unprecedented sensitivity,
where environmental disturbances can produce correlated noise. Two
geographically separated sites naturally have largely independent local
seismic environments, whereas the interferometers of the triangle share
the same environment. This distinction is particularly relevant to
searches for weak stochastic backgrounds of cosmological and astrophysical origin. 
The magnitude of the resulting two-site advantage depends on the eventual sites and on noise-mitigation
technologies that have not yet been demonstrated at the required level,
and it should therefore be considered alongside, rather than folded into,
the source-based performance comparison.

The comparison also points toward a possible compromise rather than a binary 
choice between nearly aligned and strongly misaligned two-site configurations. An \emph{intermediate relative orientation} can
retain much of the localization and polarization capability of a
misaligned network while preserving substantially better
stochastic-background sensitivity. The calculations presented here
indicate that an orientation around $30^\circ$ merits particular
attention as the design is optimized across the full ET science program.

Taken together, the results show that \emph{there is no single geometry
that maximizes every element of the Einstein Telescope science case}.
The 15-km two-site concept generally provides greater reach and precision
than the 10-km triangle, but its scientific performance depends critically
on the relative orientation of the two observatories. A substantially
misaligned network is favored for localization, multimessenger astronomy
and precision characterization of individual sources; a nearly aligned
network is favored for stochastic-background searches; and intermediate
orientations offer attractive compromises. The choice of
geometry is therefore not merely an engineering decision. \emph{It
determines the balance of scientific capabilities that Europe will
possess in gravitational-wave astronomy for decades.}

These conclusions should also be interpreted in the context of the
future global GW network. The quantitative comparison in this report
deliberately considers ET operating on its own, because it is not yet
known which present or planned observatories will be operating alongside
ET. In practice, ET may form part of a worldwide network that includes
LIGO-India, Cosmic Explorer in the United States and other facilities. Such a network would provide additional baselines and polarization information and could therefore alter some of the trade-offs---particularly those involving
localization---without removing the fundamental dependence of ET's own
capabilities on arm length and geometry.

Finally, we compared our forecasts of the statistical uncertainties, obtained with the Fisher information matrix, with full Bayesian parameter estimation for a subset of simulated signals. For individual events, especially at low signal-to-noise ratio, the forecasts can differ from the full analysis, most notably for sky localization. However, both methods rank the detector geometries in the same way, which supports the main conclusions of this report.

\begin{tcolorbox}[colback=teal!5!white, colframe=BurntOrange!80!black,
title=\sc What This Comparison Does Not Yet Include]
The results above assume ET operating alone. We do not include the impact of transient, non-Gaussian detector noise on any geometry. In particular, we have not applied the null stream---the combination of the triangle's three outputs that is insensitive to GWs and
can, in principle, identify non-environmental instrumental artifacts---nor compared it with glitch-mitigation strategies available to the 2L configuration.
\vskip5pt

We have also not explored the possibility of compensating for calibration-system failures in one of the three interferometers of the triangle configuration through astrophysical calibration with loud binary black-hole signals and the null stream.
Moreover, the standard time-shift method of estimating the noise background in template searches, in which triggers from different detectors are displaced by more than their light-travel time, may not be applicable when ET data are dominated by astrophysical signals. 
\vskip5pt

Finally, we do not include other observatories operating concurrently with
ET, which would improve sky coverage and further break localization and
polarization degeneracies.
\end{tcolorbox}

\newpage 

\clearpage

\flushbottom

\tableofcontents

\section*{List of acronyms}

\begin{center}
\begin{tabular}{@{}ll@{}}
  BBH     & Binary black hole \\
  BH      & Black hole \\
  BNS     & Binary neutron star \\
  CBC     & Compact binary coalescence \\
  CCSN    & Core-collapse supernova \\
 CMB & Cosmic microwave background \\
 EEZ & Exclusive Economic Zone\\
  EM      & Electromagnetic \\
  EOS     & Equation of state \\
  ET      & Einstein Telescope \\
  FIM     & Fisher information matrix \\
  FOPT    & First-order phase transition \\
  GR      & General relativity \\
  GRB     & Gamma-ray burst \\
  GW      & Gravitational wave \\
  GWTC    & Gravitational-Wave Transient Catalog \\
  IMBH    & Intermediate-mass black hole \\
  LIGO    & Laser Interferometer Gravitational-Wave Observatory \\
  LISA    & Laser Interferometer Space Antenna \\
  LVK     & LIGO--Virgo--KAGRA (Collaboration) \\
  $\Lambda$CDM & Lambda cold dark matter\\ 
  NN      & Newtonian noise \\
  NR      & Numerical relativity \\
  NS      & Neutron star \\
  NSBH    & Neutron-star--black-hole binary \\
  PBH     & Primordial black hole \\
  PI      & Power-law integrated (sensitivity curve) \\
  PN      & Post-Newtonian \\
  Pop. III & Population III \\
  PSR     & Pulsar \\
  QCD     & Quantum chromodynamics \\
  QNM     & Quasinormal mode \\
  SGWB    & Stochastic gravitational-wave background \\
  SIQM    & Spin-induced quadrupole moment \\
  SNR     & Signal-to-noise ratio \\
\end{tabular}
\end{center}

\newpage

\section{Preamble}

The Einstein Telescope (ET) science case, presented by the ET Collaboration in Ref.~\cite{ET:2025xjr}, builds on a decade of GW discoveries. Since
the first detection of GWs from a binary black hole (BBH) merger in
2015~\cite{LIGOScientific:2016aoc}, 
the LIGO-Virgo-KAGRA (LVK) Collaboration has recorded several hundred compact-binary detections through its most recent observing run~\cite{LIGOScientific:2026wfs,KAGRA:2021vkt}, including binary neutron stars
(BNSs)~\cite{LIGOScientific:2017vwq} and neutron-star--black-hole (NSBH) systems~\cite{LIGOScientific:2021qlt}. The
multimessenger event GW170817 identified BNS mergers as progenitors of kilonovae and short gamma-ray bursts (GRBs)~\cite{LIGOScientific:2017ync}, while population analyses have delineated initial mass and spin distributions of stellar-mass BHs and NSs~\cite{LIGOScientific:2026ctl},
produced the first standard-siren measurement of the Hubble--Lema\^itre
parameter~\cite{LIGOScientific:2017adf}, and constrained the NS tidal deformability~\cite{LIGOScientific:2018cki}. With typical signal-to-noise ratios (SNRs) of ten to a few tens (and $\sim 80$ for the loudest event to date), however, the current detector network remains sensitivity limited: the BBH
horizon reaches only $z\lesssim 1$ and the BNS horizon $z\lesssim
0.08$, restricting most events to discovery rather
than precision characterization.

Around 2035, the Laser Interferometer Space Antenna (LISA)~\cite{LISA:2024hlh}, funded by ESA with NASA support, will open the millihertz band, $10^{-4}$--$10^{-1}\,$Hz, inaccessible from the ground. LISA will be complementary to ground-based detectors. It will observe mergers of $10^{4}$--$10^{7}\,M_\odot$ BHs
across most of cosmic history, extreme
mass-ratio inspirals, resolvable Galactic compact binaries, and
cosmological and astrophysical GW stochastic backgrounds. A subset of stellar-mass BBHs first detected by LISA years before coalescence will sweep into the ET
band and merge there, providing multi-band sources whose joint
analysis will sharpen tests of general relativity and the
identification of binary formation channels.

With a factor $\sim 10$ (or more, depending on the frequency) improvement in strain sensitivity and broadband coverage from a few Hz to several kHz, ET will yield $\sim
10^{4}$--$10^{5}$ BBH and BNS detections per year, extend the BBH
horizon to $z\gtrsim 10$ (and to $z\sim 100$ for optimally oriented sources of favorable mass)---reaching remnants of Population III (Pop. III) stars,
intermediate-mass BHs (IMBHs), and primordial BHs (PBHs), if they exist---detect BNS mergers out to $z\sim2$ and enable tight constraints on the NS equation of state (EOS) from the loudest of them, and probe stochastic backgrounds well below the compact-binary
foreground. Pre-merger sky localization of BNS and NSBH events, minutes to hours before coalescence, will enable prompt electromagnetic and neutrino follow-up across cosmological distances, turning multimessenger astronomy from a rare opportunity
into a routine observational mode.

\begin{figure}
\begin{tcolorbox}[standard jigsaw,colframe=gray,colback=gray!10!white,opacityback=0.0,coltext=black, title=\small\sc Sensitivity target of the Einstein Telescope]
\center{
\includegraphics[width=0.4\textwidth]{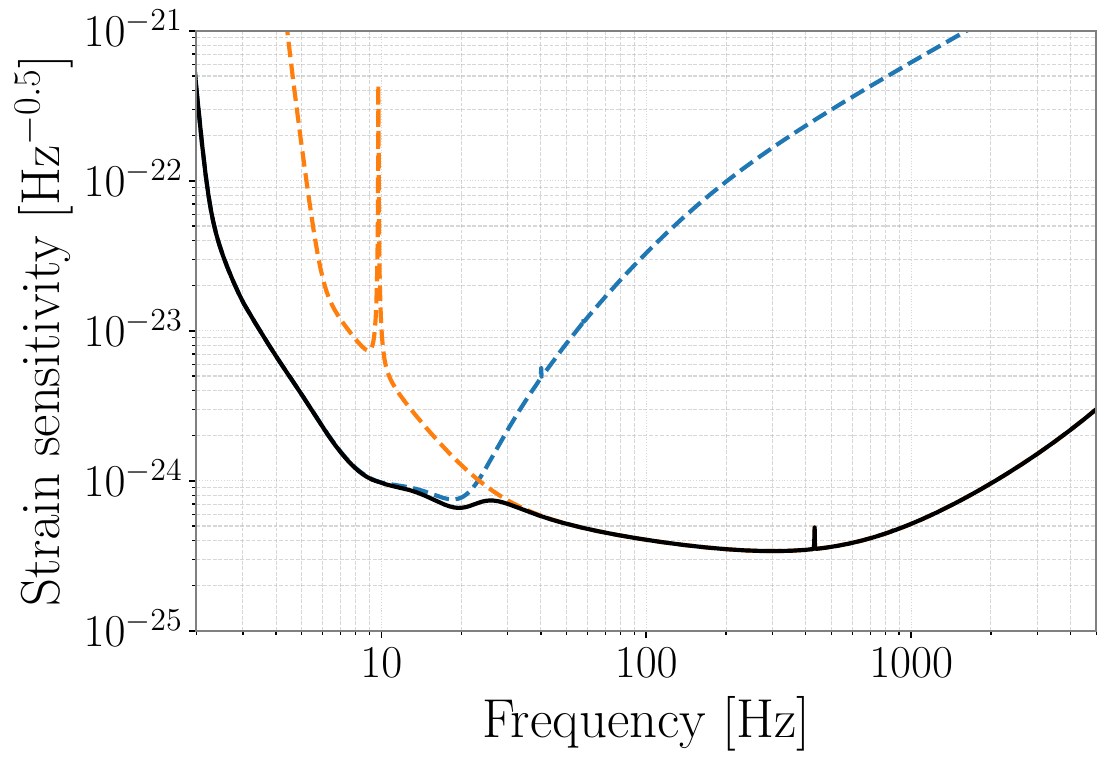}
\includegraphics[width=0.4\textwidth]{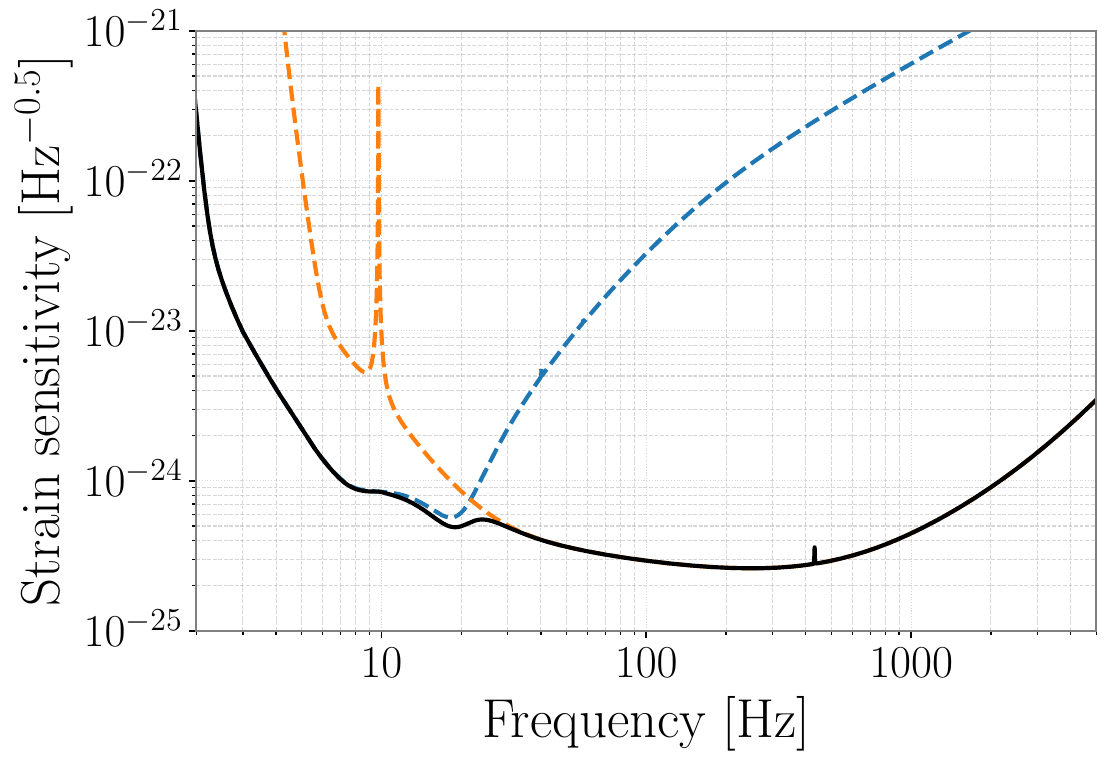}
}
\caption{\justifying Differential strain noise of a single Einstein Telescope detector, i.e., the combination of its low-frequency  and high-frequency interferometers. Left: 10-km triangle. Right: 15-km L-shaped detector.}
\label{fig:sensitivity}
\end{tcolorbox}
\vskip-15pt
\end{figure}

The ET Instrument Science Board has recently published updated sensitivity models for interferometers of the 10-km triangle, and 15-km 2L configurations \footnote{\url{https://gitlab.et-gw.eu/pub/et-sensitivity-curves}}, which are used in this study. The sensitivity models are shown in Fig.~\ref{fig:sensitivity}. The curves represent the so-called differential strain sensitivity, which does not depend on the opening angle between the two arms of an interferometer. The angle is taken into account when the sensitivity curves are processed by the simulation tools. Each ET detector is composed of two interferometers: one optimized for low frequencies (ET-LF; blue dashed curve) and one for high frequencies (ET-HF; orange dashed curve). The sensitivities are combined to obtain the full detector sensitivity (black curve). A longer arm length generally gives better sensitivity  \cite{Evans:2016mbw}. Above about 1\,kHz, the long-wavelength approximation does not hold anymore, and accurate modeling shows that the detector response to GWs is generally reduced for detectors with longer arms \cite{Virtuoso:2024kyp}. As a consequence, the individual interferometers of the 15-km 2L configuration have a sensitivity advantage over the interferometer of the 10-km triangle configuration except for signals in the kilohertz band. 

While the sensitivity comparison at the level of individual detectors is straightforward, the network sensitivity also depends on the number of detectors, i.e., three for the triangle vs two for the 2L. The sensitivity comparison of the 2L with the triangle configuration is facilitated by noting that the triangle is equivalent to two collocated L-shaped detectors with $45^\circ$ relative orientation \cite{Freise2009,Harms2026}. The resulting comparison shows that the advantage of the individual 2L detectors seen in Fig.~\ref{fig:sensitivity} is similar to that of the 2L network over the triangle. The larger number of detectors of the triangle does not lead to better network sensitivity relative to the 2L because of its 60$^\circ$ opening angle between interferometer arms.

\section{Science themes and results of the geometry-comparison study}
\label{sec:req}

We organize the ET geometry comparison by science theme: compact-binary populations, dense-matter physics, cosmography, astrophysical and cosmological stochastic backgrounds, multimessenger astronomy with GWs, fundamental physics and tests of general relativity, and continuous and burst GW sources. For each theme we first discuss the scientific motivation and the targets identified in several ET-science studies~\cite{ET:2025xjr,Branchesi:2023mws}, translate them into quantitative observational criteria and present the corresponding results. The latter are also summarized in Table~\ref{tab:metrics} of Sec.~\ref{summary-table}. 

The scientific criteria and results are evaluated for the three ET configurations (two geometries) under study: a single-site triangular detector with $10\,$km arms (henceforth denoted $\Delta$), and two L-shaped detectors with $15\,$km arms with a relative orientation of either $42.5^\circ$  or $-2.5^\circ$ (henceforth denoted 2L ($42.5^\circ$) and 2L ($-2.5^\circ$), respectively); an intermediate orientation, 2L ($30^\circ$), is explored in Appendix~\ref{app:2L30}.
To remain as site-agnostic as possible while preserving European latitudes, we place both the 2L and the $\Delta$ configurations in the Atlantic Ocean, outside any Exclusive Economic Zone (EEZ), with the two L-shaped detectors $1200\,$km apart and the triangle at their midpoint, as shown in the right panel of Fig.~\ref{fig:2LT_sites} in Appendix~\ref{detectors}. We have verified that the results  are unchanged when the detectors are instead placed on land at the same latitudes but a different longitude --- for example, along the $21.7^\circ$E meridian through North Macedonia (see Fig.~\ref{fig:2LT_sites_macedonia} in Appendix~\ref{detectors}).

\subsection{Compact-binary populations}
\label{sec:population}

\begin{figure*}
\begin{tcolorbox}[standard jigsaw,colframe=gray,colback=gray!10!white,opacityback=0.0,coltext=black, title=\small\sc Anatomy of a compact-binary gravitational-wave signal]
\center{
\includegraphics[width=0.9\textwidth]{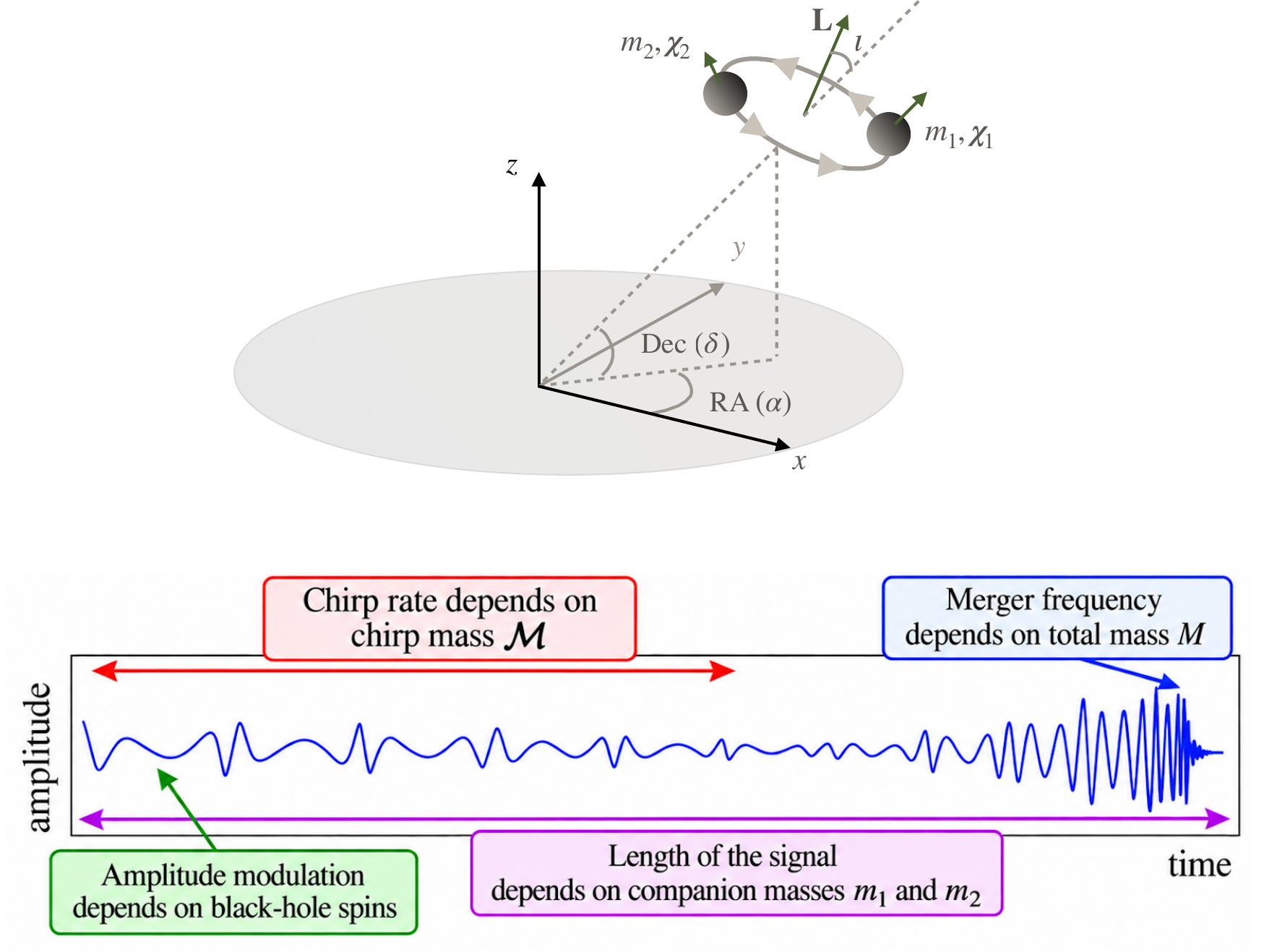}
}
\caption{\textbf{Parameters and waveform anatomy of a compact binary.}
\justifying
In the above diagram a gravitational-wave detector is assumed to be
at the origin of the coordinate system.
The binary is characterized by component masses $(m_1,m_2)$ and spins
$(\boldsymbol{\chi}_1,\boldsymbol{\chi}_2)$, inclination $\iota$, sky
position $(\alpha,\delta)$, luminosity distance $d_L$, polarization
$\psi$, coalescence time $t_c$, and phase $\phi_c$. The waveform
illustrates how the masses and spins determine its duration, chirp,
merger frequency, and modulations, while the extrinsic parameters
determine its observed amplitude, phase, and detector response.}
\label{fig:anatomy of a compact binary}
\end{tcolorbox}
\vskip-15pt
\end{figure*}

Compact binary coalescences (CBCs) are the only class of sources detected so far by LVK detectors. The waveform emitted by an 
inspiraling and merging binary carries the signature of the source
parameters as described below.

As depicted in Fig.~\ref{fig:anatomy of a compact binary}, a 
generic compact binary is described by the component masses
$(m_1,m_2)$ and the two spin vectors $\boldsymbol{\chi}_1$ and
$\boldsymbol{\chi}_2$, giving eight intrinsic parameters. Its orientation
relative to the observer is specified by the inclination $\iota$, while
its position and distance are given by the right ascension $\alpha$,
declination $\delta$, and luminosity distance $d_L$. The remaining
extrinsic parameters are the polarization angle $\psi$, coalescence time
$t_c$, and phase $\phi_c$ at coalescence.

Different features of the gravitational waveform encode different
combinations of these parameters. The chirp rate is governed primarily
by the chirp mass $\mathcal{M}$, while the characteristic frequency near
merger depends principally on the total mass $M=m_1+m_2$ and the component
masses largely determine the duration of the signal. The spins affect the
phase evolution and, when misaligned with the orbital angular momentum,
produce precessional amplitude and phase modulations. The observed
amplitude scales inversely with $d_L$ and depends on the inclination
$\iota$, polarization $\psi$, and sky position $(\alpha,\delta)$ through
the detector antenna response. The sky position also determines the
relative arrival times at geographically separated detectors, while
$t_c$ and $\phi_c$ set the time and phase origins of the signal.

These dependencies lead to significant correlations among the source
parameters. In particular, a single L-shaped interferometer measures only one linear
combination of the two GW polarizations. For short-lived
signals, a network of at least three non-collocated detectors is therefore
needed to exploit arrival-time differences, reconstruct the two
polarizations, localize the source, and disentangle the intrinsic and
extrinsic parameters.

\begin{tcolorbox}[colback=teal!5!white,colframe=yellow!75!black,title=\sc Science Target]
{\em Determine the demography of compact binaries in the Universe, shed light on their formation pathways, and understand their evolution across cosmic time.}
\end{tcolorbox}

GW observations of CBCs provide a direct way to study populations of stellar remnants across cosmic time. From the measured signals of merging BBHs, BNSs, and NSBH systems, we can infer source properties such as component masses, spins, luminosity distance, sky position, and orbital inclination. When combined over many detections, these measurements allow us to reconstruct the demographic properties of compact-binary populations, including their merger rates, mass and spin distributions, and redshift evolution; see e.g. Refs.~\cite{Sathyaprakash:2009xs,LIGOScientific:2026ctl} and references therein.

\begin{tcolorbox}[standard jigsaw,colframe=ocre,colback=ocre!10!white,opacityback=0.0,coltext=black,title=\sc Key Science Goals]
\small
Observing and characterizing compact-binary mergers across redshifts is one of the flagship science cases of ET. Individual loud and/or distant sources as well as population properties traced precisely thanks to sheer numbers enable stringent tests of astrophysical models and provide access to regimes and phenomena that are inaccessible to current detectors.

\begin{itemize}[leftmargin=*]
\item {\bf Detection range of compact binaries:} How far out can we detect BBH, BNS, and NSBH sources? How loud are they expected to be? The relevant metrics are detection horizons and signal-to-noise ratio (SNR) distributions.
\item {\bf Positive detection of a high-redshift source:} To what extent can we confirm the discovery of a high-redshift compact binary? This could be a major discovery by ET and is governed by the inference horizon instead.
\item {\bf Parameter inference of compact binaries:} How well do we expect to measure their masses and spins? How well do we expect to measure their distance and sky location?
\item {\bf Population-level inference:} By carrying out a population-level inference on all observed compact binaries, how well can we disentangle different formation channels, and uncover the physical processes governing their evolution?
\end{itemize}
\end{tcolorbox}

These population-level measurements are closely connected to questions about the formation and evolution of compact binaries. For example, the masses and spins of BBHs can encode information about stellar evolution, supernova physics, mass loss, natal kicks, and the environments in which the binaries formed~\cite{2002RvMP...74.1015W,Belczynski:2016obo}. Spin magnitudes and orientations can help distinguish between binaries formed through isolated binary evolution, where spin directions may be preferentially aligned with the orbital angular momentum, and dynamical formation channels in dense stellar environments, where spin orientations may be more isotropic~\cite{Rodriguez:2016vmx,Gerosa:2021mno}. The redshift dependence of the merger rate traces the history of star formation, metallicity evolution, and the time delays between binary formation and merger~\cite{Chruslinska:2022ovf,LIGOScientific:2026ctl}.

\begin{figure*}
\begin{tcolorbox}[standard jigsaw,colframe=gray,colback=gray!10!white,opacityback=0.0,coltext=black, title=\small\sc Observation of compact binaries across cosmic time]
\includegraphics[height=7cm]{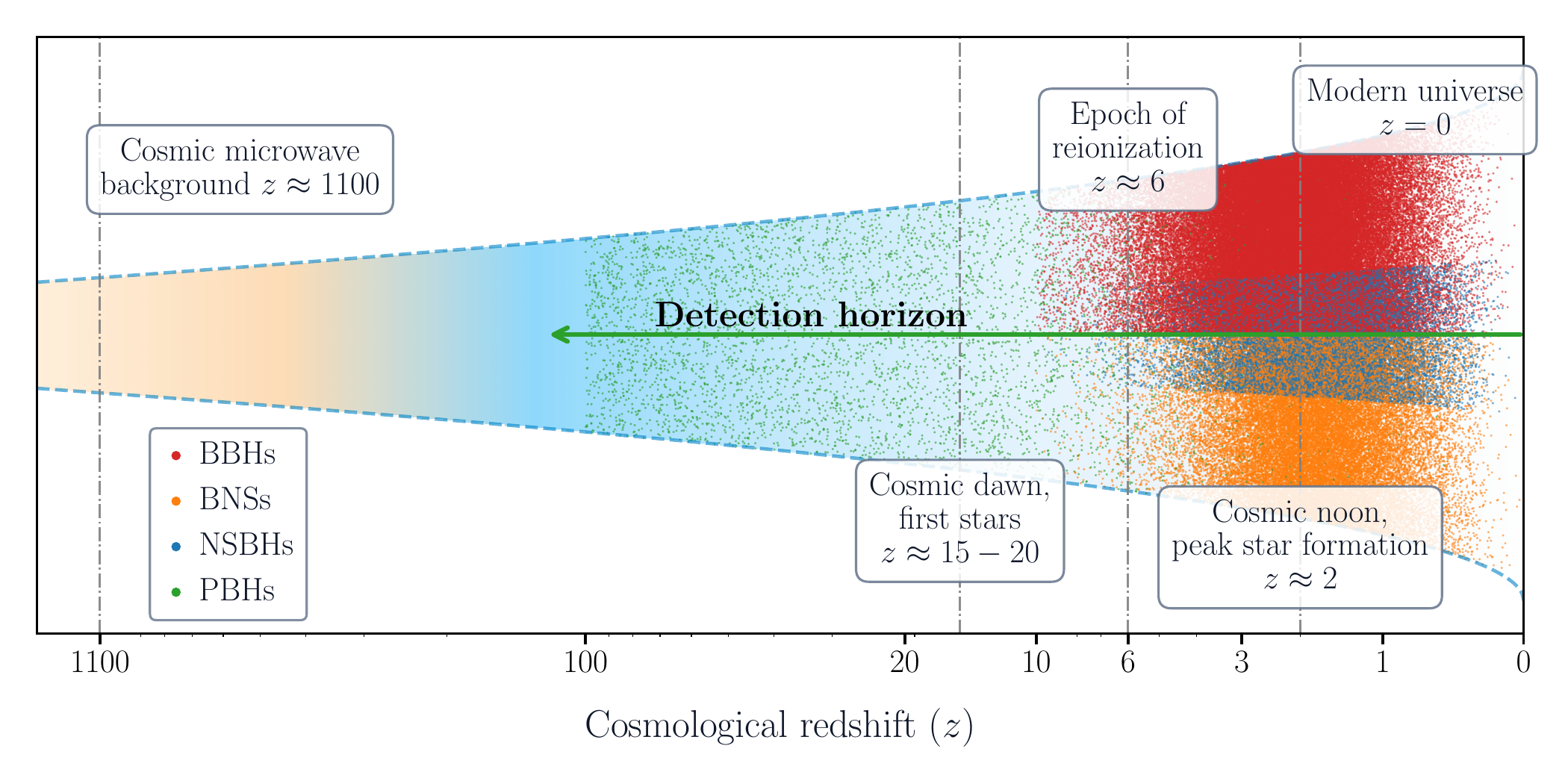}
\caption{\small\justifying 
The Einstein Telescope enables observations of compact-binary coalescences in the distant Universe, giving us access to the ``dark ages'' before the first stars formed. The figure shows a schematic timeline of the Universe (measured in terms of cosmological redshift $z$). The red, orange, and blue dots indicate, respectively, the BBHs, BNSs, and NSBHs employed in our simulations, given our current understanding of astrophysical compact-binary populations. In addition a simulated speculative population of PBHs is indicated by green dots. The approximate ET detection horizon for compact binaries is also marked.}
\label{fig:cosmic_bbh}
\end{tcolorbox}
\vskip-15pt
\end{figure*}

The scientific return of a GW detector configuration is therefore determined not only by whether it can detect compact-binary signals, but also by how many sources it can observe and how accurately their properties can be reconstructed. ET, with its high sensitivity across a wide frequency band, will allow us to access a larger volume of the Universe than current detectors~\cite{Maggiore:2019uih,ET:2025xjr}. This will increase the expected number of detections and extend population studies to higher redshifts. The improved sensitivity will also enable more precise measurements of source parameters, which can be used to better constrain the underlying astrophysical population models. As an example, thanks to its low-frequency sensitivity, ET will be able to observe signals for a longer duration, allowing for more accurate measurements of the spins during the inspiral phase.

\subsubsection{Detection and inference horizons for compact binaries}

Comparing GW detector configurations requires evaluating both detection yield and measurement performance. The number of observable compact binaries determines the statistical power available for population inference, while the accuracy of reconstructed source parameters determines how well individual systems can be classified and how effectively their properties can be combined hierarchically. 

\begin{figure*}
\begin{tcolorbox}[standard jigsaw,colframe=gray,colback=gray!10!white,opacityback=0.0,coltext=black, title=\small\sc Detection and inference horizons ]
\center{\includegraphics[width=\textwidth]{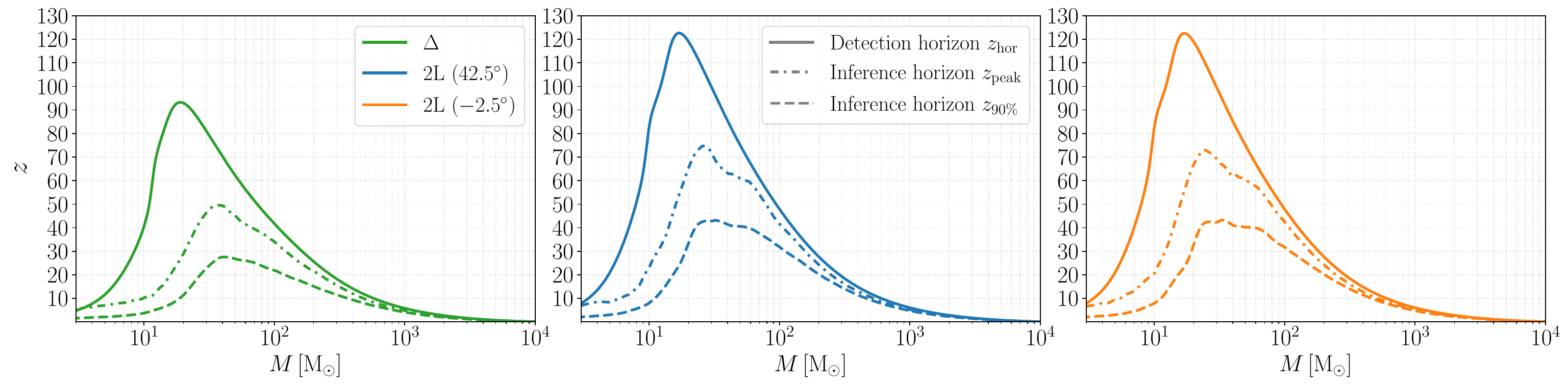}}
\captionof{figure}{\justifying Detection and inference horizons for equal-mass non-spinning binaries. The former indicates the farthest events detectable by each configuration, while the latter indicates how
confidently high-redshift events can be placed at large distances.}
\label{fig:horizons}
\end{tcolorbox}
\vskip-15pt
\end{figure*}

Before moving to population-specific metrics, in Fig.~\ref{fig:horizons} we report the redshift reach of the three detector configurations considered for equal-mass, non-spinning compact binaries as a function of total source-frame mass. For each configuration, the solid curve shows the detection horizon, $z_{\rm hor}$, defined as the maximum redshift at which an optimally oriented and located binary of a given total mass can be detected with ${\rm SNR}\geq8$. This threshold differs from the one adopted for BBHs and NSBHs in the rest of the document, and is chosen for consistency and ease of comparison with previous literature. This quantity therefore describes the maximum accessible volume for sources with those intrinsic parameters.

In addition to the detection horizon, the figure shows an ``inference horizon,'' which quantifies not only whether a source can be detected, but whether its redshift can be constrained with useful accuracy~\cite{Mancarella:2023ehn}. The dotted curve, $z_{\rm peak}$, indicates the redshift at which a source of a given total mass gives the largest lower edge of the 90\% credible interval on redshift. The dashed curve, $z_{90\%}$, gives this lower edge of the 90\% credible interval. Thus, $z_{\rm peak}$ identifies the true source redshift that maximizes the inferred high-redshift reach, while $z_{90\%}$ indicates the largest redshift beyond which a source can be confidently placed at 90\% credibility.

For example, for a binary with total mass $M=20\,{\rm M}_\odot$, the 10-km $\Delta$ configuration has a detection horizon of approximately $z_{\rm hor}\simeq 93$. However, the source redshift that maximizes the lower edge of the 90\% credible interval is lower, $z_{\rm peak}\simeq 29$. At this redshift, the 90\% credible interval on the redshift has a lower edge at approximately $z \approx11$, i.e., the source can be confidently placed beyond this value. This illustrates the distinction between detectability and parameter inference: although sources may be detectable out to very large redshifts, the redshift at which the measurement provides the strongest inference reach can occur well inside the formal detection horizon.

The same qualitative behavior is seen for all three configurations. The redshift reach is largest for intermediate total masses, where the signal is best matched to the most sensitive frequency band of the detectors. At low masses, the intrinsic amplitude of the emitted GW signal is smaller, so that the SNR drops below threshold at lower redshift. At very large masses, the merger and ringdown are shifted to lower frequencies, eventually moving out of the most sensitive part of the band. As a result, both the detection and inference horizons peak at total masses of order tens of solar masses and decline toward both lower and higher masses.

Comparing the three panels shows that the 15-km 2L configurations extend the reach relative to the 10-km $\Delta$ configuration. In particular, both 15-km designs achieve larger maximum detection horizons, reaching $z\sim 120$ for favorable masses, compared with a peak of $z\simeq 93$ for the 10-km $\Delta$ configuration. The inference horizons follow the same trend, with the longer-arm configurations not only detecting sources to greater distances but also allowing redshift constraints to extend to higher redshifts. Notably, the 2L configurations provide a broad mass window, approximately $M\sim 20$–$100\,{\rm M}_\odot$, in which detected events can be placed at redshifts higher than $z\simeq 30$ within the 90\% credible interval. 

This high-redshift inference capability is particularly important for distinguishing between different formation scenarios. Astrophysical compact binaries produced by stellar evolution are expected to trace the cosmic history of star formation and metal enrichment, and should become increasingly rare at sufficiently high redshift, before the formation of the first stars. In contrast, a putative population of primordial black holes could exist and merge at much earlier cosmic times. Therefore, the ability to infer that a compact-binary merger lies at very high redshift would provide a powerful discriminant between primordial and astrophysical origins. 

\subsubsection{Individual compact binaries}

Before presenting the results, we briefly describe the source populations used in the simulations. Our baseline catalogs are based on analytical population models consistent with the latest LVK analyses. For BBHs we adopt the \textsc{Broken Power-Law + 2 Peaks} mass model, together with the Gaussian spin component, including spin precession~\cite{LIGOScientific:2025pvj,LIGOScientific:2026ctl}. For BNSs we use a flat source-frame mass distribution and small aligned spins, while for NSBH binaries we use the \textsc{NSBH-pop} model~\cite{Biscoveanu:2022iue} calibrated to the latest LVK data release~\cite{LIGOScientific:2024elc}, allowing the BH mass distribution to extend naturally toward higher masses. The catalogs correspond to one year of mergers in the Universe for BBHs, and five years for BNSs and NSBHs, so that each catalog contains more than $10^5$ sources.

For all source classes, the merger-rate evolution is modeled using a Madau-Dickinson form~\cite{Madau:2014bja}. For BNSs and NSBHs we additionally include a power-law delay-time distribution with a minimum delay time of 40 Myr. For systems containing NSs, we consider three representative EOSs, spanning relatively soft to stiff models, in order to capture the impact of the NS matter properties on the signals. In particular, we use the APR4 EOS as the reference soft model, SFHo as the medium and H4 as the stiff model. 

In addition to these standard compact-binary populations, we include three putative BBH subpopulations: binaries originating from Population III (Pop. III) stars (using the catalog from Ref.~\cite{Mestichelli:2024djn}), IMBHs (obtained from \textsc{B-POP}~\cite{Sedda:2021vjh}), and PBHs (using the catalog from Ref.~\cite{Franciolini:2023opt}). These components are particularly relevant for ET because of its improved low-frequency sensitivity. Signals from high-mass systems, such as IMBH binaries, and from very high-redshift sources, such as Pop. III or primordial BBHs, are shifted toward lower observed frequencies. ET's sensitivity at low frequency therefore opens a unique observational window onto these populations, and including them allows us to assess not only the performance of different configurations for established source classes, but also their discovery potential for rare compact binaries.

\begin{figure*}
\begin{tcolorbox}[standard jigsaw,colframe=gray,colback=gray!10!white,opacityback=0.0,coltext=black, title=\small\sc SNR distributions for the different source classes ]
\center{\includegraphics[width=0.325\textwidth]{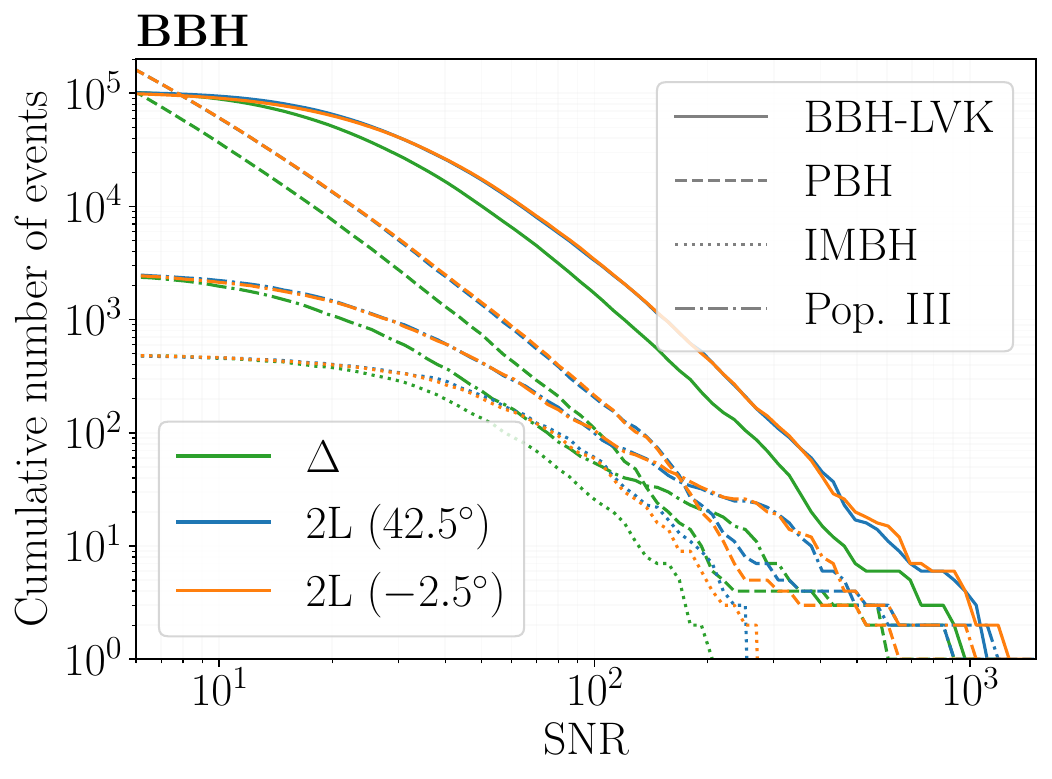} \includegraphics[width=0.325\textwidth]{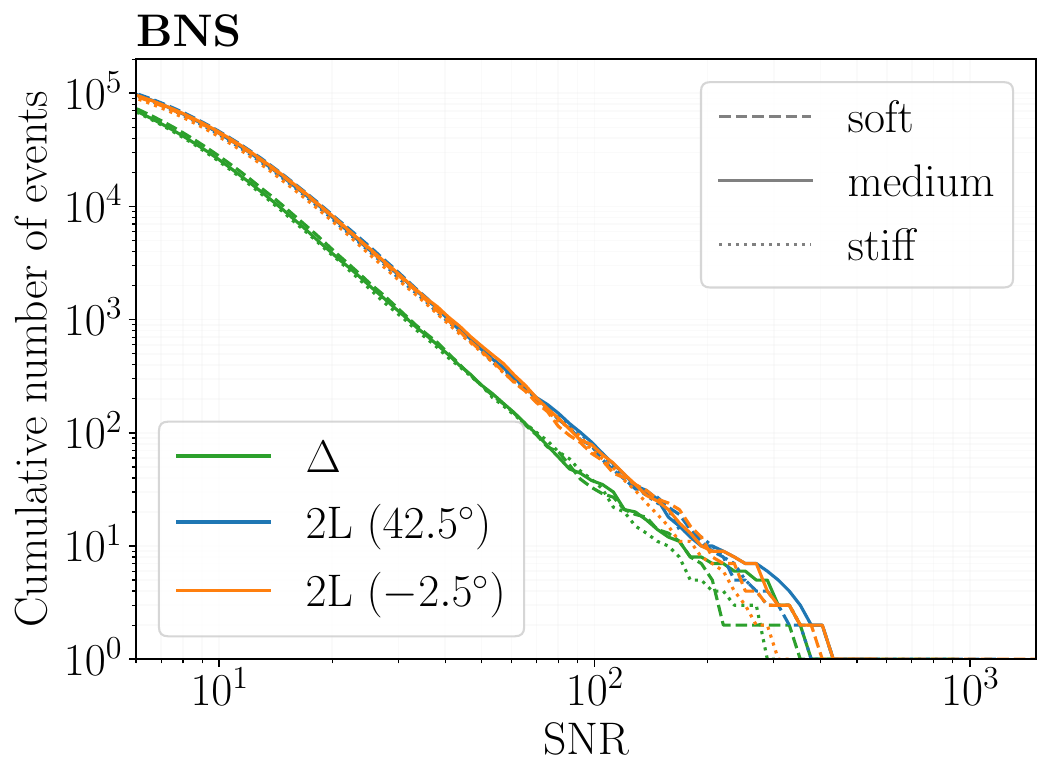} \includegraphics[width=0.325\textwidth]{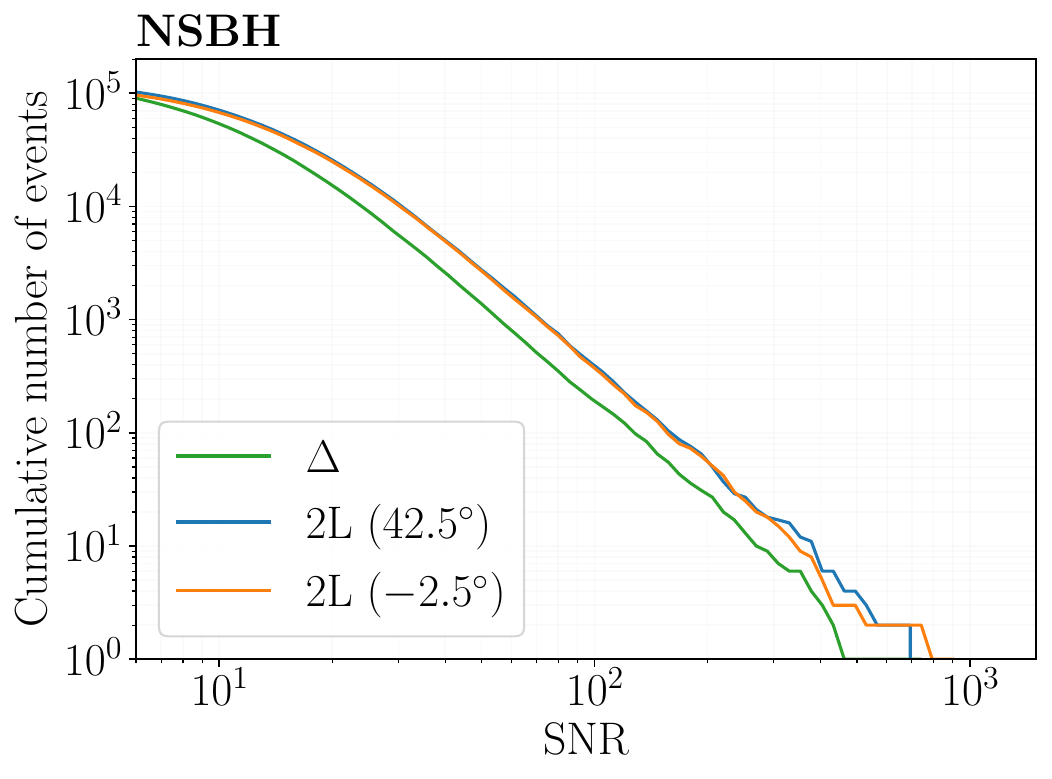} }
\captionof{figure}{\justifying Inverse cumulative distributions of the SNR of the events in each simulated population.}
\label{fig:snrs_ds1}
\end{tcolorbox}
\vskip-15pt
\end{figure*}

We first compare the SNR distributions obtained for the different source populations and detector configurations. The cumulative distributions show that all configurations would detect very large numbers of compact-binary mergers, implying an excellent scientific return in all cases. For the BBH-LVK population generated from the analysis results of the fifth Gravitational-Wave Transient Catalog (GWTC-5.0), the number of detections with ${\rm SNR}>12$ is $\sim 7.9\times 10^4$ for the $\Delta$ configuration, increasing to $\sim 8.4$–$8.8\times 10^4$ for the 2L configurations; see Fig.~\ref{fig:snrs_ds1}. The same trend is visible for the additional BBH subpopulations: ET would detect hundreds of IMBH and Pop. III binaries, and tens of thousands of PBH candidates in the adopted model, although the latter correspond to only a small fraction of the full simulated PBH catalog.

The 2L configurations generally yield a larger detected fraction, mainly due to the longer arm length and the resulting sensitivity improvement. This advantage becomes particularly clear at high SNR. For example, the number of BBH events with ${\rm SNR}>100$ roughly doubles, from $\sim 1.5\times 10^3$ for the $\Delta$ configuration to $\sim 3.1\times 10^3$ for the 2L configurations. Similar gains are found for IMBHs, Pop. III binaries, PBHs, BNSs and NSBHs. These high-SNR events are especially valuable, since they will dominate precision measurements.

For BNSs, using a detection threshold ${\rm SNR}>8$, the $\Delta$ configuration detects about $3.6$–$4.2\times 10^4$ events in 5 years of observations depending on the assumed EOS, while the 2L configurations detect about $5.6$–$6.5\times 10^4$ in the same amount of time. The dependence on the neutron-star EOS is modest compared to the difference between detector configurations. For NSBHs, the number of detections with ${\rm SNR}>12$ in 5 years increases from $\sim 3.9\times 10^4$ for the $\Delta$ configuration to $\sim 5.3$–$5.6\times 10^4$ for the 2L configurations.

Overall, these results show that all proposed configurations enable large compact-binary catalogs, including standard stellar-origin systems and more exotic high-mass or high-redshift populations. The 2L layouts provide a systematic improvement in detection yield and, more importantly, in the number of loud events.

\begin{figure*}
\begin{tcolorbox}[standard jigsaw,colframe=gray,colback=gray!10!white,opacityback=0.0,coltext=black, title=\small\sc Distributions for the statistical uncertainties on the masses and spins for BBHs ]
\center{\includegraphics[width=0.65\textwidth]{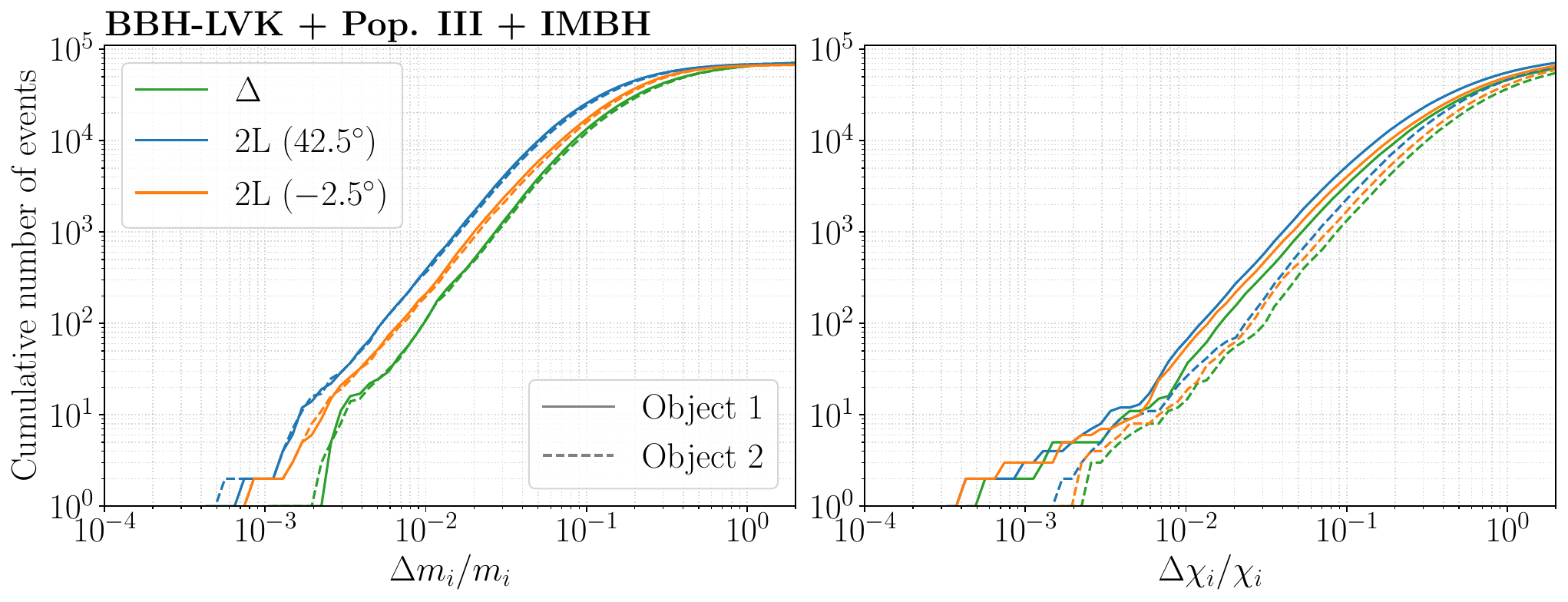} \includegraphics[width=0.33\textwidth]{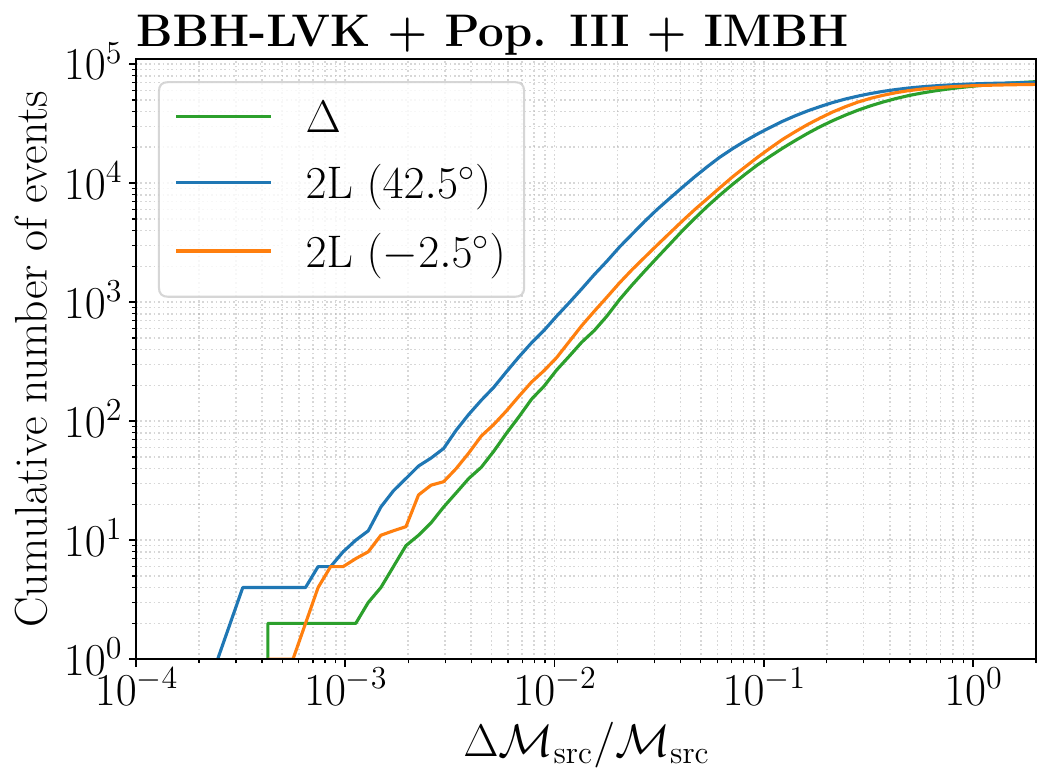}}
\captionof{figure}{\justifying Cumulative distributions of the relative uncertainty on the individual source-frame masses, spin magnitudes, and source-frame chirp masses of the astrophysical BBH populations used in this work.}
\label{fig:errs_ds1_bbh_intr}
\end{tcolorbox}
\vskip-15pt
\end{figure*}

We next consider the statistical uncertainties on the BBH parameters. In this case, we merge the results for the populations of stellar-origin BBHs generated from the LVK results and the simulated IMBH and Pop. III binaries. The cumulative distributions show a clear hierarchy among the configurations: the 2L layouts generally improve parameter reconstruction with respect to the $\Delta$ configuration, with the 2L $(42.5^\circ)$ configuration providing the best overall performance; see Fig.~\ref{fig:errs_ds1_bbh_intr}.

\noindent
\begin{minipage}{0.27\textwidth}
For intrinsic parameters, the 2L ($42.5^\circ$) configuration gives the largest number of systems with accurate measurements of the source parameters. For example, the number of BBHs with source-frame chirp-mass uncertainty below $1\%$ increases from $\sim 220$ for the $\Delta$ configuration to $\sim 620$ for the 2L ($42.5^\circ$) one, while the 2L ($-2.5^\circ$) configuration gives $\sim 290$. A similar improvement is seen for the individual component masses: the number of events with $\Delta m_1/m_1<10\%$ rises from $1.2\times 10^4$ to $2.4\times 10^4$, and for $m_2$ from $1.1\times 10^4$ to $2.2\times 10^4$. Spin measurements are more challenging, but the same trend is visible, especially for the primary spin.
\end{minipage}
\hfill
\begin{minipage}{0.70\textwidth}
\begin{tcolorbox}[standard jigsaw,colframe=gray,colback=gray!10!white,opacityback=0.0,coltext=black, title=\small\sc Distributions for the statistical uncertainties on the luminosity distance and sky localization for BBHs ]
\center{
\includegraphics[width=0.48\textwidth]{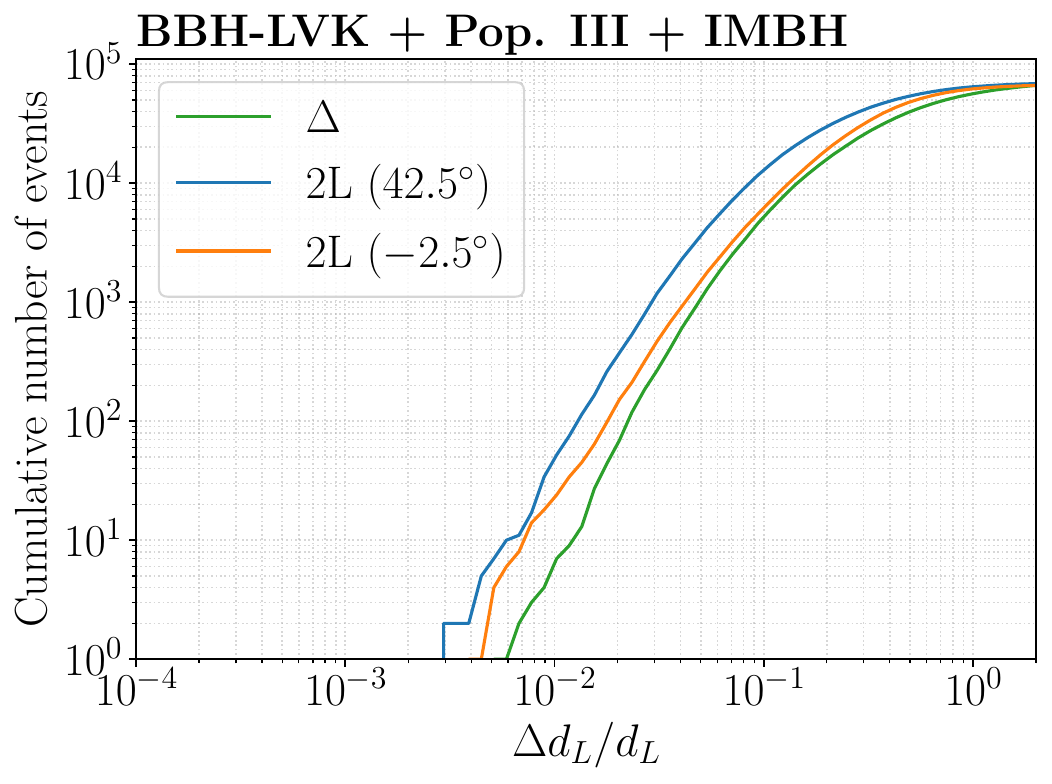} 
\includegraphics[width=0.48\textwidth]{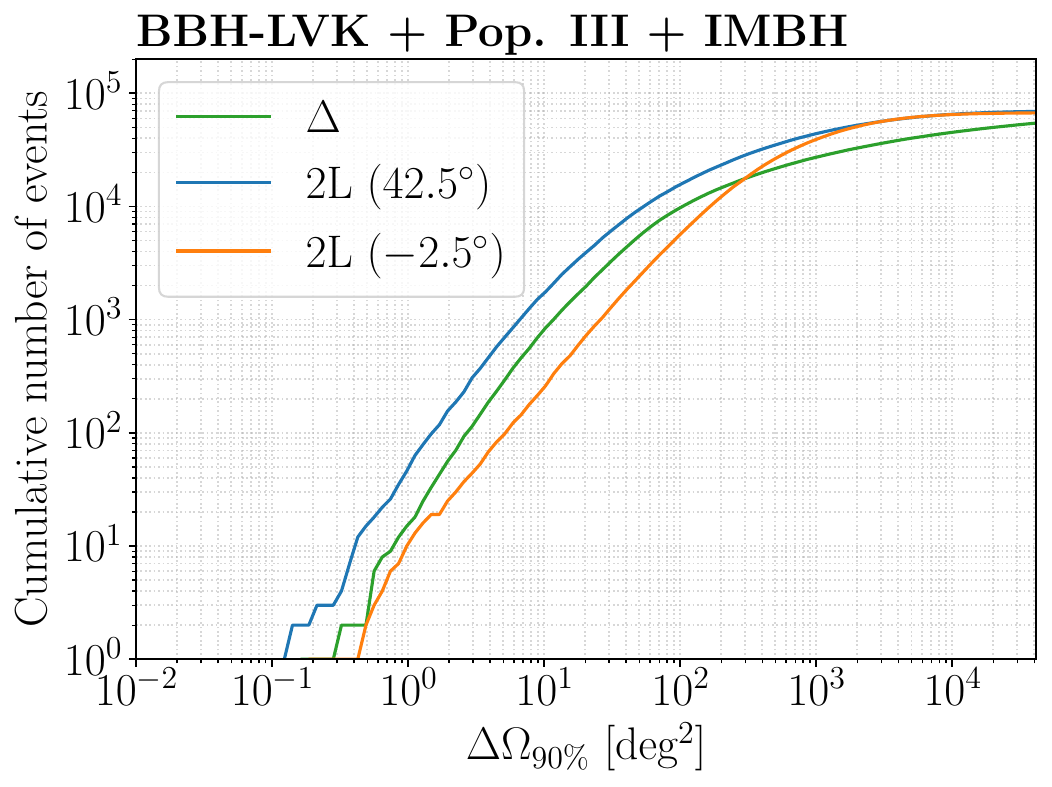}}
\captionof{figure}{\justifying Cumulative distributions of the relative uncertainty on the luminosity distance and 90\% sky localization of the astrophysical BBH populations used in this work.}
\label{fig:errs_ds1_bbh_extr}
\end{tcolorbox}
\end{minipage}

The improvement is even more pronounced for extrinsic parameters. For luminosity distance, the number of events with $\Delta d_L/d_L<5\%$ increases from $\sim 900$ for the $\Delta$ configuration to $\sim 3000$ for the 2L ($42.5^\circ$) one, compared to $\sim 1300$ for the 2L ($-2.5^\circ$) configuration; see Fig.~\ref{fig:errs_ds1_bbh_extr}. The same configuration also performs best for high-redshift distance inference: among sources at $z\geq 5$, $\sim 230$ events have $\Delta z/z<10\%$ for the 2L ($42.5^\circ$) configuration, compared to 79 for the $\Delta$ one and $\sim 130$ for the 2L ($-2.5^\circ$) configuration.

Sky localization highlights the importance of detector geometry. The 2L ($42.5^\circ$) configuration yields $\sim 1600$ BBHs localized within $10\,{\rm deg}^2$ at 90\% credibility, compared to $\sim 750$ for the $\Delta$ configuration and only $\sim 230$ for the 2L ($-2.5^\circ$) case. This reflects the improved polarization information and angular resolution provided by the relative orientation of the two interferometers, and explains the superior performance of the $\Delta$ configuration compared to the 2L ($-2.5^\circ$) one despite the lack of a long baseline.

Overall, while the longer arms of the 2L configurations improve sensitivity and hence parameter estimation, the misaligned geometry provides an additional advantage for breaking degeneracies among distance, inclination, polarization, and sky position. This makes the 2L ($42.5^\circ$) configuration the most effective for BBH parameter reconstruction across both intrinsic and extrinsic source properties.

\begin{figure*}
\begin{tcolorbox}[standard jigsaw,colframe=gray,colback=gray!10!white,opacityback=0.0,coltext=black, title=\small\sc Distributions for the statistical uncertainties on the source parameters for BNSs ]
\center{\includegraphics[width=0.32\textwidth]{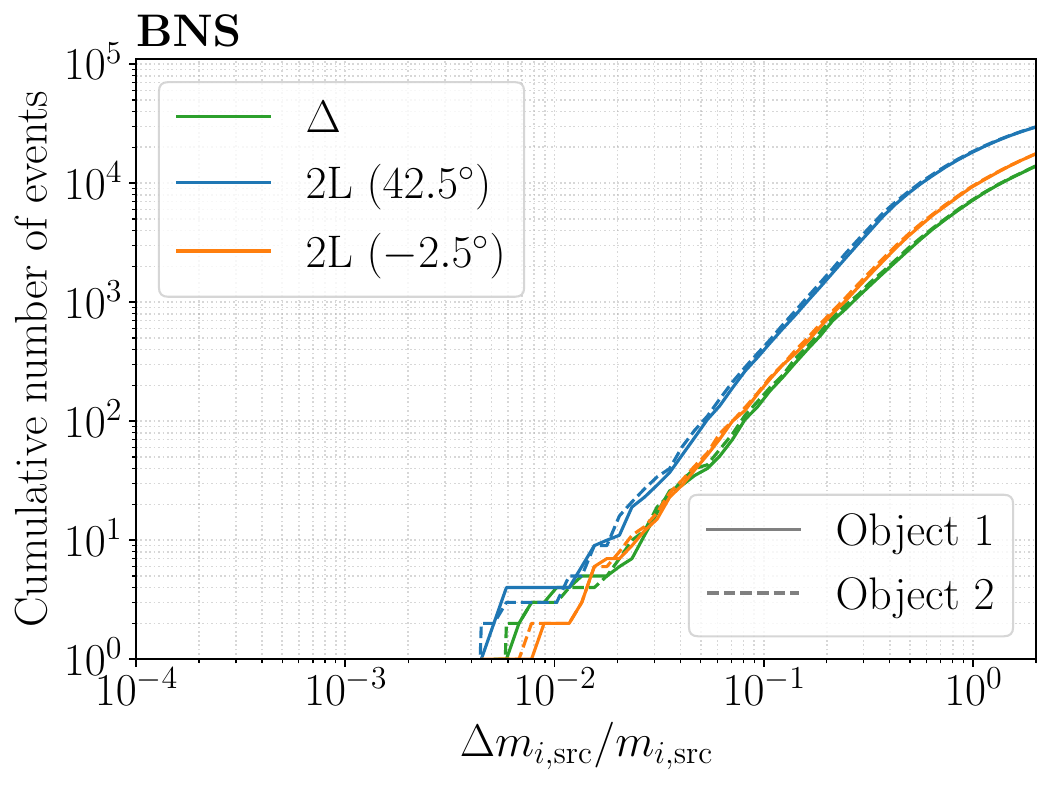}\hfill\includegraphics[width=0.32\textwidth]{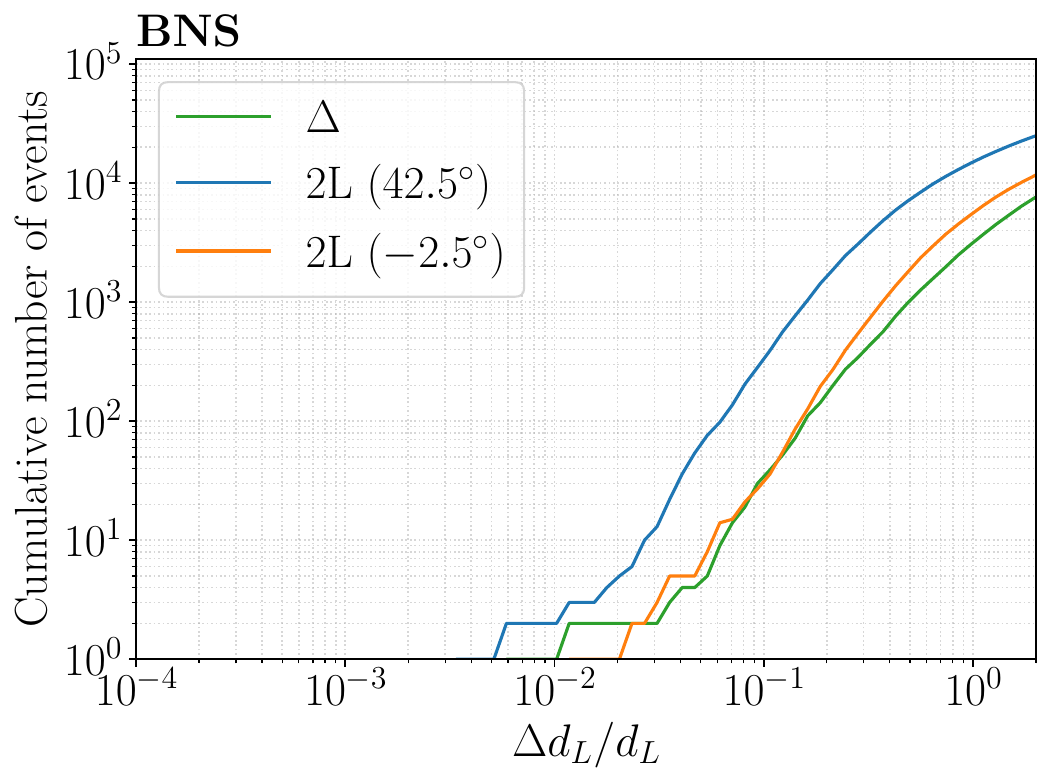}\hfill\includegraphics[width=0.32\textwidth]{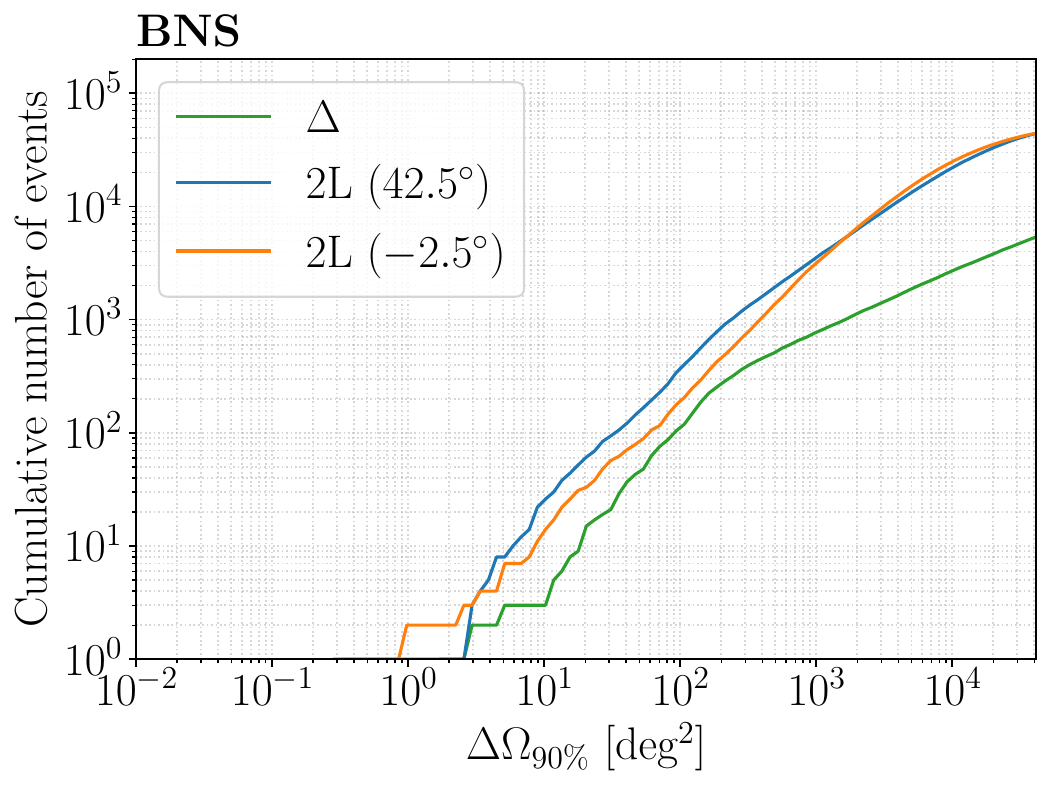}}
\captionof{figure}{\justifying Cumulative distributions of the relative uncertainty on the individual source-frame masses, luminosity distance, and 90\% sky localization of the BNS population used in this work. We only show the results for the medium EOS.}
\label{fig:errs_ds1_bns}
\end{tcolorbox}
\vskip-15pt
\end{figure*}

We perform the same comparison for BNS systems, focusing on the mass parameters, luminosity distance and sky localization; see Fig.~\ref{fig:errs_ds1_bns}. All BNS and NSBH counts quoted in the following refer to the five-year catalogs, whereas
Table~\ref{summary-table} reports numbers per year. In the figures we show only the results  for the medium EOS, since the results obtained with the soft and stiff ones are very similar and do not change the qualitative conclusions.

The 2L ($42.5^\circ$) configuration again gives the smallest overall statistical uncertainties. The number of events with source-frame chirp-mass uncertainty below $1\%$ increases from 30 for the $\Delta$ configuration to 102 for the 2L ($42.5^\circ$) one, while the 2L ($-2.5^\circ$) configuration gives 24. The improvement is also visible for the individual masses: the number of events with $\Delta m_1/m_1<10\%$ rises from $\sim 130$ to $\sim 340$, and for $m_2$ from $\sim 150$ to $\sim 370$. Although these numbers remain small compared to the full detected BNS sample, they still correspond to a sizeable set of well-measured systems.

The largest gain is found for the luminosity distance. For the medium EOS, $\sim 140$ BNSs have $\Delta d_L/d_L<20\%$ with the $\Delta$ configuration, compared to $\sim 1400$ with the 2L ($42.5^\circ$) configuration and $\sim 200$ with the 2L ($-2.5^\circ$) one. This demonstrates the importance of the misaligned 2L geometry in breaking the distance--inclination degeneracy. Consistently, only a small number of BNSs at $z\geq1$ have redshift measured to better than $10\%$, but this number increases from zero for the $\Delta$ and 2L ($-2.5^\circ$) configurations to 13 for the 2L ($42.5^\circ$) one.

Sky localization also benefits from the 2L configurations. For the medium EOS, the number of events localized within $100\,{\rm deg}^2$ at 90\% credibility increases from $\sim 100$ for the $\Delta$ configuration to $\sim 340$ for the 2L ($42.5^\circ$), with $\sim 180$ for the 2L ($-2.5^\circ$). For tighter localizations below $10\,{\rm deg}^2$, the corresponding numbers are 3, 24 and 12. In the BNS case, localization is aided by the long signal duration: many systems remain in band long enough for the rotation of the Earth to modulate the detector response, improving angular resolution even for a single-site detector.

Overall, the BNS results confirm the trends observed for BBHs. The 2L configurations, and especially the 2L ($42.5^\circ$) layout, provide better distance, mass and sky-position reconstruction, while the dependence on the neutron-star EOS is subdominant for the quantities considered here.

\begin{figure*}
\begin{tcolorbox}[standard jigsaw,colframe=gray,colback=gray!10!white,opacityback=0.0,coltext=black, title=\small\sc Distributions for the statistical uncertainties on the source parameters for NSBHs ]
\center{\includegraphics[width=0.32\textwidth]{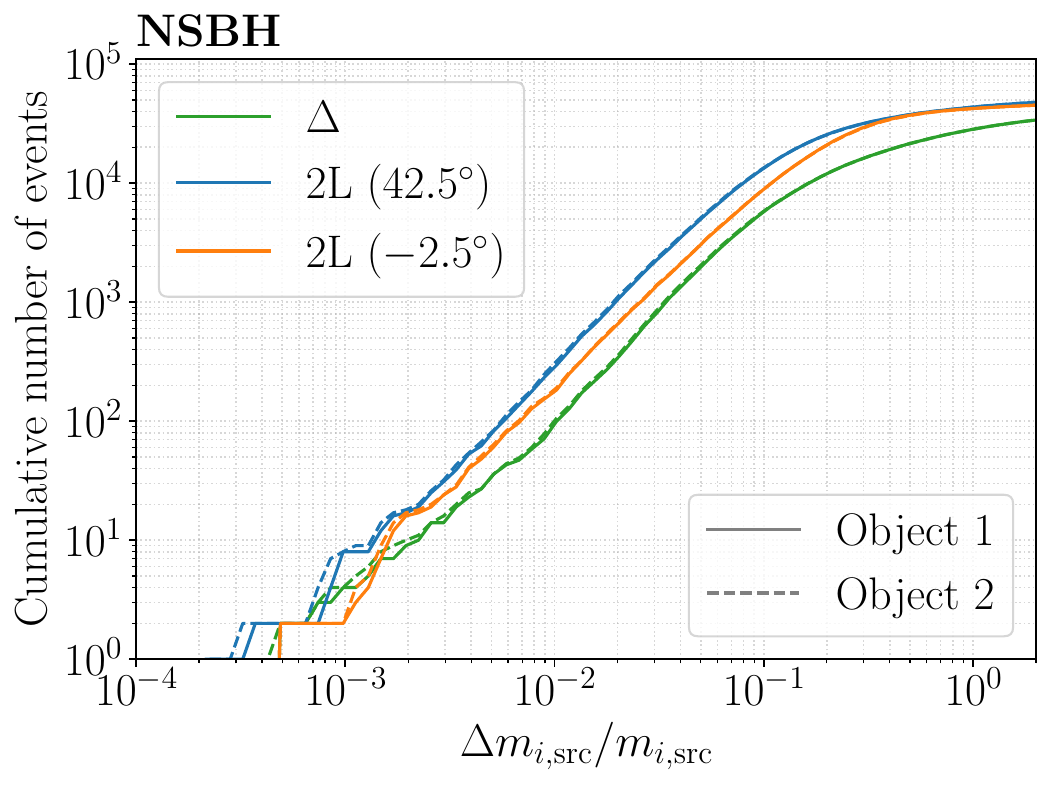}\hfill\includegraphics[width=0.32\textwidth]{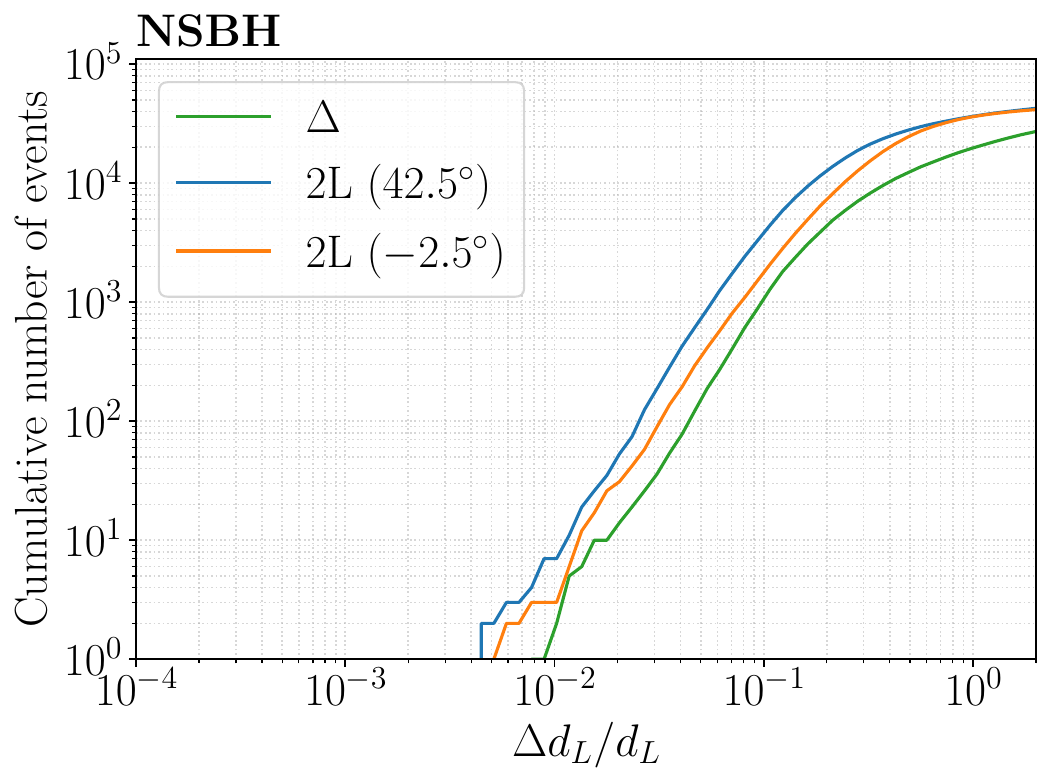}\hfill\includegraphics[width=0.32\textwidth]{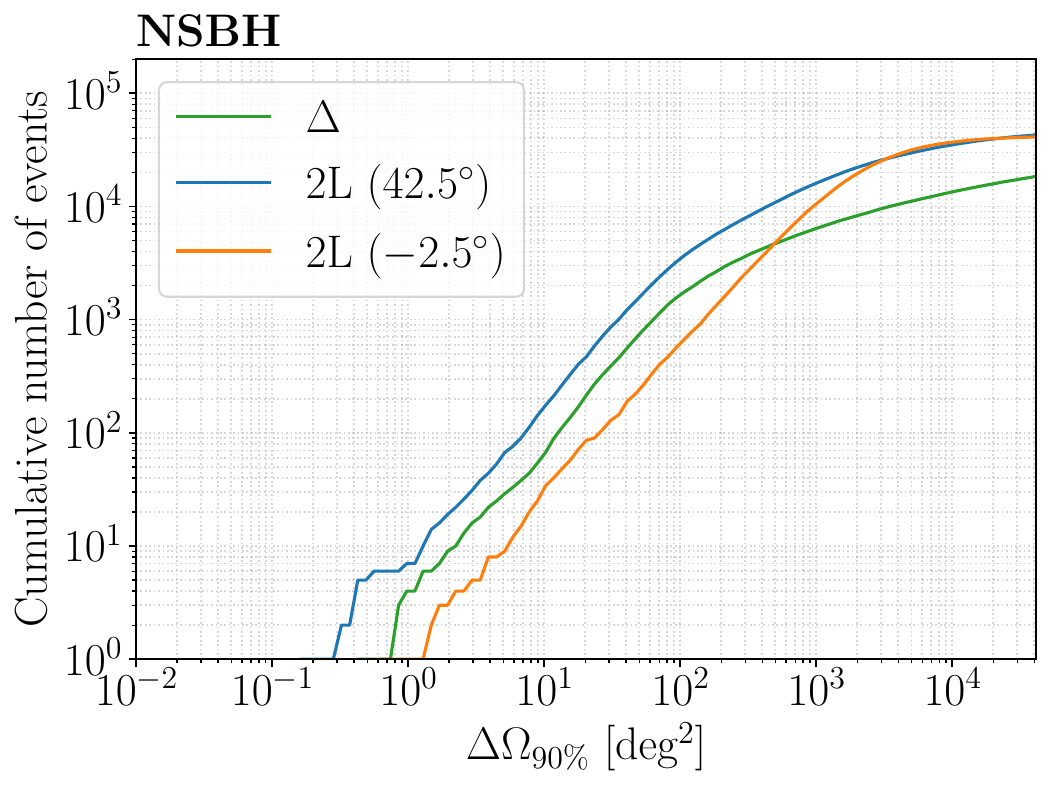}}
\captionof{figure}{\justifying Cumulative distributions of the relative uncertainty on the individual source-frame masses, luminosity distance, and 90\% sky localization of the NSBH population used in this work. We show the results obtained when employing a waveform including higher-order modes and no tidal effects.}
\label{fig:errs_ds1_nsbh}
\end{tcolorbox}
\vskip-15pt
\end{figure*}

Finally, we consider the NSBH population. As expected, the statistical uncertainties are generally intermediate between the BBH and BNS cases: NSBHs are lighter and remain in band longer than typical BBHs, but are still louder and shorter than BNSs for much of the population. The same hierarchy among configurations is found. The two 15-km 2L layouts improve the number of well-measured systems relative to the 10-km $\Delta$ configuration, with the misaligned configuration giving the best overall performance, especially for distance, redshift, and sky localization; see Fig.~\ref{fig:errs_ds1_nsbh}.

The number of NSBH detections with chirp-mass precision better than $1\%$ is $90$ for the $\Delta$ configuration, $\sim 300$ for the 2L ($42.5^\circ$), and $\sim 180$ for the 2L ($-2.5^\circ$). For luminosity distance measured to better than $20\%$, the corresponding numbers are $\sim 3.9 \times10^3$, $\sim 1.2 \times10^4$, and $\sim 6.5 \times10^3$. Component masses are also substantially better recovered in the 2L ($42.5^\circ$) layout: requiring $\Delta m_{1}/m_{1}<0.1$ gives $\sim 5.2 \times10^3$, $\sim 1.2 \times10^4$, and $\sim 8.1 \times10^3$ events, and similarly for the secondary mass. For sources at $z\geq 1$, the number with $\Delta z/z<0.1$ increases from $\sim 530$ in the $\Delta$ configuration to $\sim 2.6 \times10^3$ for the 2L ($42.5^\circ$) one, compared with $\sim 1000$ in the 2L ($-2.5^\circ$) case.

The localization results again show most clearly the benefit of the 2L ($42.5^\circ$) configuration. The number of NSBHs localized within $100\,{\rm deg}^2$ is $\sim 1.6 \times10^3$ for the $\Delta$ configuration, $\sim 3.2 \times10^3$ for the 2L ($42.5^\circ$) configuration, and $\sim 560$ for the 2L ($-2.5^\circ$) one; the corresponding numbers within $10\,{\rm deg}^2$ are $58$, $145$, and $30$. NSBHs also benefit, although less dramatically than BNSs, from Earth-rotation modulation during their longer in-band evolution. Overall, the NSBH results reinforce the trends seen for BBHs and BNSs: all configurations deliver large samples with strong population-science potential, the 2L designs increase the number of high-quality measurements, and the misaligned configuration is generally the most effective.

\subsubsection{Population-level inference}

\noindent
\begin{minipage}{0.27\textwidth}
Finally, we also assess the impact of the different geometries on BBH population-level inference. We perform this analysis with the same population model used to generate the injections for the BBH-LVK population and restrict the analysis to those systems. We estimate the expected constraints through a population-level Fisher matrix. 

The three detector geometries give essentially indistinguishable results at this level. In both the $m_1$ and $z$ distributions, the median reconstructed population and the corresponding $90\%$ credible intervals overlap almost perfectly for the three considered configurations; see Fig.~\ref{fig:popfisher_bbh_ds1}.

\end{minipage}
\hfill
\begin{minipage}{0.70\textwidth}
\begin{tcolorbox}[standard jigsaw,colframe=gray,colback=gray!10!white,opacityback=0.0,coltext=black, title=\small\sc Recovered primary-mass and
redshift distributions of the BBH-LVK population]
\center{\includegraphics[width=\textwidth]{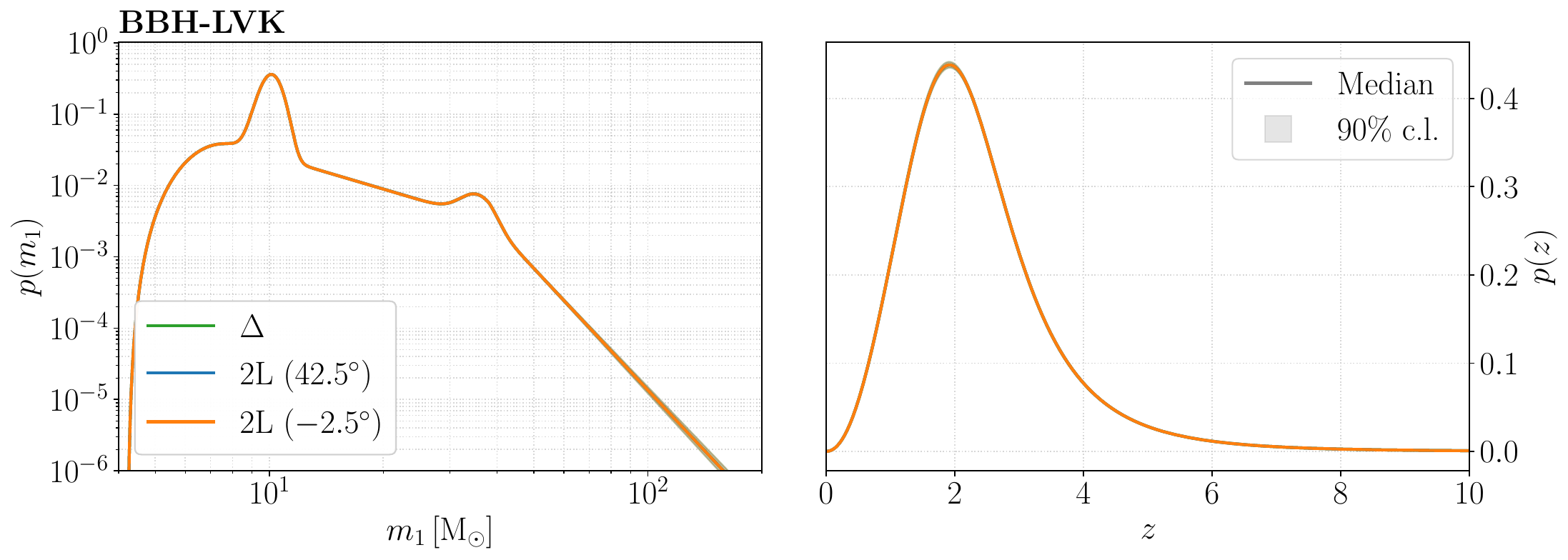}}
\captionof{figure}{\justifying Reconstructed source-frame primary mass and redshift distributions for the BBH-LVK population. The solid lines show the median and the bands the 90\% credible region. The bounds obtained from the different detector configurations almost perfectly overlap.}
\label{fig:popfisher_bbh_ds1}
\end{tcolorbox}
\end{minipage}

This reflects the very large BBH samples delivered by all configurations: once the population is constrained by tens of thousands of detections, the statistical uncertainty on the hyperparameters becomes extremely small, and the residual differences between geometries have little visible impact on the recovered population distributions.

These results underline the outstanding population-science potential of ET. Irrespective of the specific geometry, ET will measure the BBH mass spectrum and merger-rate evolution with very high precision, enabling stringent tests of compact-binary formation models and of the redshift dependence of the BBH population.

\begin{tcolorbox}[colback=teal!5!white,colframe=red!75!black,title=\sc Geometry Comparison]
{
ET alone will be able to deliver an impressive number of CBC detections, with tens to hundreds of thousands of events per year. It will also allow the detection of candidates at high redshift. Moreover, ET will make it possible to reconstruct source properties with unprecedented accuracy, and tightly constrain the population properties of the observed systems. 
\begin{itemize}
    \item  {\em 2Ls perform better for most metrics.} The 2L geometries yield better results for most metrics across different source classes. This is driven by the longer arm length, which leads to higher SNRs. The differences become even more significant for distance and sky location reconstruction.
    \item  {\em Sky localization is best for 2L ($42.5^\circ$) and suboptimal for 2L ($-2.5^\circ$) configurations.} Precision in sky localization measurement comes from the ability to disentangle the two polarizations. This is best enabled by well-separated detectors that are misaligned. Aligned detectors are suboptimal for probing different polarizations.
    \item  {\em Comparable results at population level.} The three geometries considered yield roughly identical results for population-level inference of BBHs. In this case, the result is driven by the large number of compact-binary sources observed by ET and the high accuracy in reconstructing their properties. 
\end{itemize}
}
\end{tcolorbox}

\subsection{Dense-matter physics}

\vskip10pt
\begin{tcolorbox}[colback=teal!5!white,colframe=yellow!75!black,title=\sc Science Target]
{\em Understanding and constraining the properties of matter at extreme densities to reveal the state of matter under extreme conditions, discover potential phase transitions and the emergence of quarks.}
\end{tcolorbox}

\begin{tcolorbox}[standard jigsaw,colframe=ocre,colback=ocre!10!white,opacityback=0.0,coltext=black,title=\sc Key Science Goals]
\small
Multimessenger observations of numerous luminous events in the Universe involving dense matter in extreme environments will find answers to several key puzzles in fundamental physics:
\begin{itemize}[leftmargin=*]
\item {\bf The nature of matter at supranuclear densities.}
What are the fundamental properties of the densest matter in the cosmos? How do quarks and gluons manifest in the cores of the most massive NSs and in merger remnants? 
\item {\bf Production sites of heavy elements.} What elements are produced in NS mergers and how? Are they able to explain the abundances of elements heavier than iron in the Solar System and in stars?
\item {\bf Formation and evolution of compact binaries.} How do NS binaries form and evolve? What are their merger rates, as well as mass and spin distributions through cosmic time?
\end{itemize}
\end{tcolorbox}

\subsubsection{Understanding the nature of matter at the highest densities} 
\label{sec:ns}

NSs enable us to probe matter under unique conditions, including supranuclear densities, i.e., densities exceeding even those inside atomic nuclei ($\sim 2.7 \times 10^{14}\,\mathrm{g/cm^3}$), extreme magnetic fields, and, in the case of merger remnants, high temperatures of about $10^{12}\,\mathrm{K}$, which is approximately a hundred thousand times hotter than the center of our Sun~\cite{Lattimer:2004pg}. While in the following we focus on the potential to constrain the EOS governing the interior of NSs using GW signals alone, it is important to emphasize the complementary role of multimessenger observations of BNS and NSBH mergers, as also discussed in Sec.~\ref{sec:mma}. Combined GW and electromagnetic observations generally lead to tighter and more robust constraints on the properties of matter~\cite{Dietrich:2020efo}. ET will therefore not only provide direct constraints on matter at extreme densities, but will also facilitate synergies with large-scale astronomical observatories and terrestrial particle colliders, thereby contributing to a more comprehensive understanding of the properties of matter under extreme conditions.

\begin{figure*}
\begin{tcolorbox}[standard jigsaw,colframe=gray,colback=gray!10!white,opacityback=0.0,coltext=black, title=\small\sc NS structure and QCD phase diagram]
\begin{center}
\includegraphics[height=5.7cm]{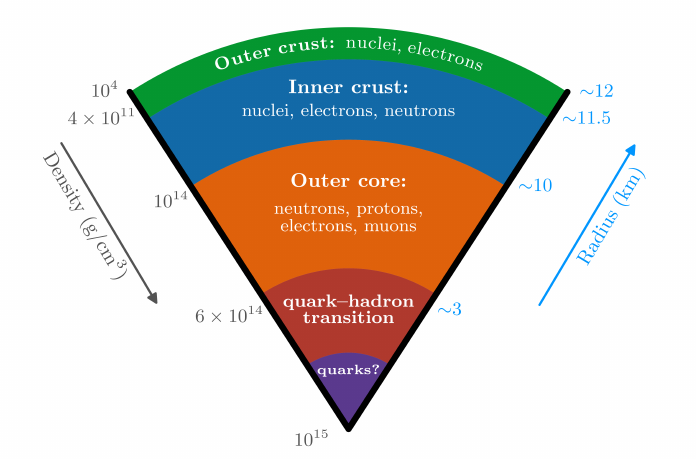} \hfill
\includegraphics[height=5.7cm]{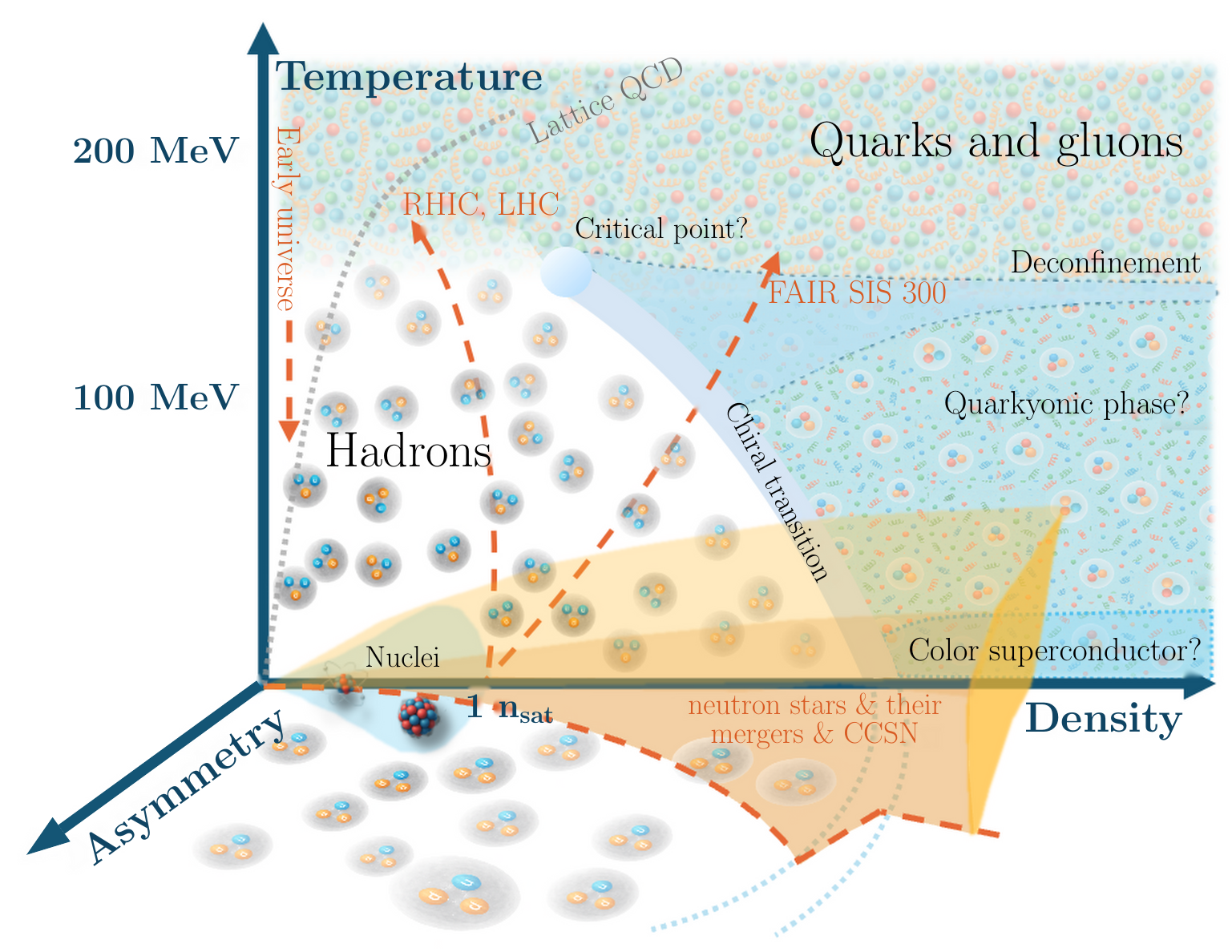}
\end{center}
\caption{\small\justifying 
Left: Composition of matter in the interior of an NS. Nuclei in the outer part of the star form a liquid of neutrons, protons, electrons and muons. At even higher densities, moving further inside the star, the quarks that make up neutrons and protons may no longer remain confined inside them. Neither the density at which this happens nor the character of the change is known: it may occur gradually, or as a sharp phase transition. Figure adapted from~\cite{Adhikari:2022sve}. \newline
Right: Phase diagram of quantum chromodynamics (QCD), the theory of the strong nuclear force, drawn as a function of temperature, density, and isospin asymmetry, i.e., the imbalance between the number of neutrons and protons. Highlighted are parts of the diagram probed by various (planned) terrestrial facilities, as well as the range spanned by NSs, NS mergers, and core-collapse supernovae. The figure is adapted from \cite{Dietrich:2025dde}.}
\label{fig:BNS_intro}
\end{tcolorbox}
\vskip-15pt
\end{figure*}

Our current understanding of the NS interior is illustrated in the left panel of Fig.~\ref{fig:BNS_intro}. 
The properties of matter up to densities characteristic of terrestrial nuclei are relatively well understood and determine the properties of the star's outer layers. 
In the outer core, nuclei dissolve into a liquid composed predominantly of neutrons, with a small admixture of other particles, including protons, electrons, and muons.

\begin{figure*}
\begin{tcolorbox}[standard jigsaw,colframe=gray,colback=gray!10!white,opacityback=0.0,coltext=black, title=\small\sc Probing the BNS coalescence with ET]
\begin{center}
\includegraphics[width=0.9\textwidth]{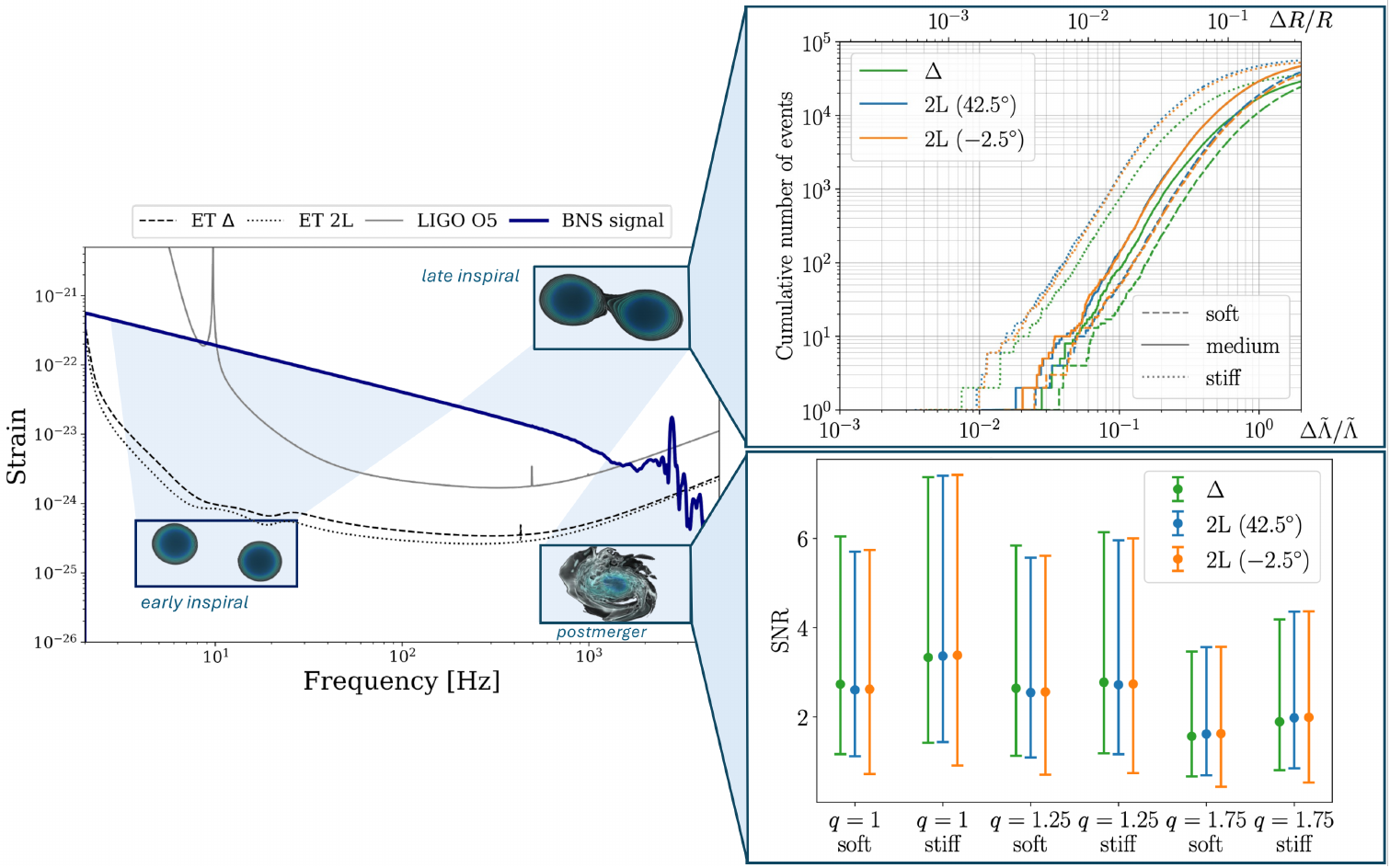}
\end{center}
\caption{\small\justifying 
GW strain of a BNS signal with a source at $100\,\mathrm{Mpc}$ distance. In contrast to current detectors, ET will enable us to observe the whole spectrum of the coalescence (left panel), incorporating the early inspiral, in which potential resonance effects might provide information about the NS interior, the late inspiral, enabling tidal-deformability measurements, and the post-merger phase. This part of the figure is adapted from~\cite{Dietrich:2025dde}. Focusing on the late inspiral (top right panel), we present the cumulative number of BNS mergers with their respective relative uncertainty in the mass-weighted tidal deformability $\tilde{\Lambda}$. Given the influence of the EOS, we have used three different EOSs to describe the NS interior, namely APR4, labeled as soft, SFHo, labeled as medium, and H4, labeled as stiff. These EOSs cover a reasonable range between soft and stiff EOSs and provide us with an understanding of the respective uncertainty. Rough estimates of the relative uncertainty on the NS radius are indicated on the top axis of the same plot. \newline
To determine the post-merger properties, we compute the post-merger SNR for a set of numerical-relativity waveforms (bottom right panel) assuming the source is located at a distance of $100\, \rm{Mpc}$. These simulations are the same as those used in~\cite{Dupletsa:2026jzs}. We note that the employed EOSs do not correspond one to one to those employed in the upper panel, but likewise span a large range of potential scenarios. The presented uncertainty results from variations of the extrinsic parameters such as the sky location and orientation of the system.}
\label{fig:BNS_results}
\end{tcolorbox}
\vskip-15pt
\end{figure*}

In NSs with sufficiently large masses, or in the remnants of BNS mergers, the density in the core may become high enough for exotic states of matter to emerge. Under these conditions, the distance between nucleons becomes comparable to their size, and their quark substructure is expected to become relevant. At such densities, phase transitions to new states of matter containing deconfined quarks may occur. The corresponding regions of the phase diagram of quantum chromodynamics (QCD) are not directly accessible with terrestrial experiments. Although heavy-ion collisions do reach comparable densities, they create hot, short-lived matter with roughly equal numbers of neutrons and protons~\cite{Danielewicz:2002pu}, whereas NSs and their mergers probe strongly neutron-rich material~\cite{Huth:2021bsp}. NSs and their mergers are therefore unique laboratories for studying the properties of strongly interacting matter at high baryon density and, in the case of merger remnants, finite temperature. Extracting information about their internal structure relies on measuring the properties of the stars before the merger as well as the properties of the merger remnant, all of which are encoded in the GW signal. It is important to highlight that, in contrast to existing GW detectors, the increased sensitivity and broader frequency coverage of ET will allow us to probe the full BNS coalescence, from the early inspiral up to the merger and post-merger phase (Fig.~\ref{fig:BNS_results}). Each stage of the signal contains important information about the properties of NS matter.

\subsubsection{Constraints through measurements of the tidal deformability} 

In contrast to BHs, NSs have a finite size and are deformed during the inspiral by the gravitational pull of their companion. This mechanism is similar to the one by which the Moon raises tides in Earth's oceans. Here, the companion pulls harder on the near side of the star than on the far side, and the star is stretched along the line joining them. 
This tidal deformation depends on the internal structure of the stars and therefore provides a direct probe of the EOS. It is characterized by the tidal deformability $\Lambda$ of each star. The GW signal is sensitive to a particular mass-weighted combination of the two tidal deformabilities, denoted by $\tilde{\Lambda}$. The parameter $\tilde{\Lambda}$ enters directly into the description of the GW phase~\cite{Flanagan:2007ix} and can therefore be extracted from the observed data.
It is worth highlighting that the leading-order tidal deformability is only one component of the rich tidal physics encoded in the GW signal. Among others, higher-order tidal effects, dynamical tides, and resonant excitations may provide additional information about the NS structure and composition. In particular, the observation of tidal resonances~\cite{Gittins:2026ntx}, potentially during the early inspiral, could offer complementary probes and reveal features of the stellar oscillation spectrum that cannot be captured with the current generation of GW detectors~\cite{Pratten:2019sed}.

Generally, the number of detected BNS mergers will also be critical (Fig.~\ref{fig:snrs_ds1}). While the majority of individual events will provide only limited information on the tidal deformability owing to parameter degeneracies and finite SNRs, a large population of detections will allow these uncertainties to be reduced by combining information from multiple events. This will enable population-level constraints on the NS EOS. The sensitivity and expected event rate of ET will therefore be essential not only for improving the precision of tidal measurements in individual high-SNR systems, but also for exploiting a large population of NS mergers. Considering the accuracy with which the tidal deformability can be measured (top right panel of Fig.~\ref{fig:BNS_results}), we find slight differences between the employed detector configurations. Given the lack of knowledge about the true nature of the EOS, we select a set of three EOSs to cover the range from soft to stiff, where generally softer EOSs lead to NSs with smaller radii and, consequently, smaller tidal deformabilities. 
This is also in agreement with our findings, where the stiff EOS leads to larger tidal deformabilities and therefore to more pronounced tidal effects in the GW signal. Consequently, for the stiff EOS the absolute uncertainty $\Delta \tilde{\Lambda}$ is larger while the relative uncertainty $\Delta \tilde{\Lambda}/\tilde{\Lambda}$ is smaller (Fig.~\ref{fig:BNS_results}), and the number of systems measured to a given relative precision is therefore lower for the softer EOSs. Comparing different detector configurations, we find that the 2L configurations generally lead to better measurements of the tidal deformability, which is directly connected to the higher SNR obtained for these detector configurations, as outlined in Sec.~\ref{sec:population}. The improved measurements of the tidal deformability can in turn be translated into constraints on NS radii. We show an estimate of the measured relative uncertainty on the NS radius based on individual detections in Fig.~\ref{fig:BNS_results} using the top axis of the top-right plot. This estimate is derived following the very approximate scaling relation of $\Lambda \propto R^6$~\cite{Branchesi:2023mws}. Using this relation, the fractional error in the radius measurement can be estimated according to $\Delta R/R \approx (1/6) \Delta \tilde{\Lambda}/\tilde{\Lambda}$. We emphasize that the measurement accuracy of the source masses of the stars also plays a critical role in determining the stellar radius. Here, we also point out differences between the two 2L configurations as seen in Fig.~\ref{fig:errs_ds1_bns}, where the misaligned 2L configuration generally outperforms the other setups. 

\begin{figure*}
\begin{tcolorbox}[standard jigsaw,colframe=gray,colback=gray!10!white,opacityback=0.0,coltext=black, title=\small\sc NS radius constraints]
\begin{center}
\includegraphics[width=0.95\textwidth]{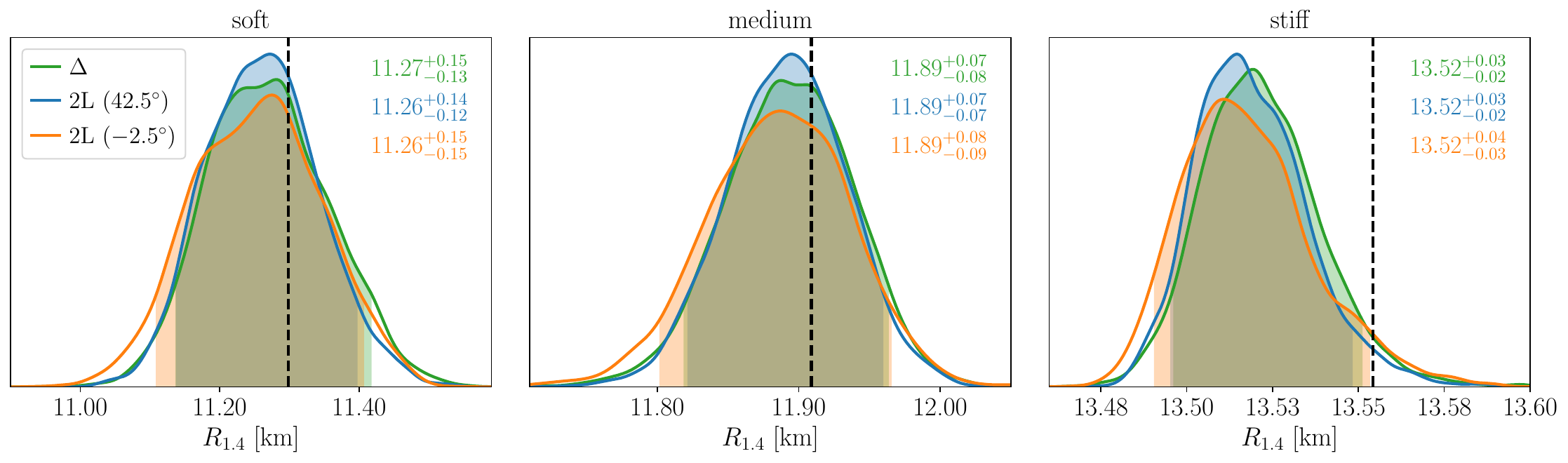}
\end{center}
\caption{\small\justifying Estimated NS radius for a star of about $1.4\,M_\odot$. Employing the framework of~\cite{Wouters:2025zju}, we present a radius estimate by combining the loudest events (SNR above 100). The dashed, black vertical line represents the radius for the injected EOSs, i.e., those EOSs used to create our mock data. Small biases are visible in all configurations and highlight the need to further improve our theoretical description of dense matter. The 90\% credible intervals of the obtained constraints are shaded and the results are reported in the top right of each panel.}
\label{fig:NS_radius}
\end{tcolorbox}
\vskip-15pt
\end{figure*}

Going beyond the rough estimate of radius measurement from individual events, we can combine information from the loudest events, i.e., here we restrict ourselves to an SNR above 100, to present a population-based estimate for the radius measurement (Fig.~\ref{fig:NS_radius}). We find that the accuracy with which the radius can be measured is very similar for all different configurations, with the misaligned 2L (42.5$^\circ$) configuration performing slightly better. This is a direct consequence of the higher SNR of the detected signals due to the longer arm length and the better source-frame mass estimate for the misaligned detectors. 

We add that, although generally weaker than constraints derived from BNSs, NSBH systems can also provide constraints on the EOS. These constraints are again determined mainly by the number of sources and the measurement accuracy, which is dominated by the SNR of the detected signals (Sec.~\ref{sec:population}). We find that the 2L configurations generally lead to larger detected SNRs and will consequently provide tighter constraints than the $\Delta$ configuration.

\subsubsection{Constraints from post-merger detection} 

The exact outcome of a BNS merger is highly uncertain and depends on numerous factors. In particular, whether the remnant survives as a massive NS or collapses to a BH is primarily determined by its mass relative to the maximum mass that the EOS can support against gravitational collapse. Depending on the system parameters and the properties of the EOS, collapse may occur promptly or after a short-lived post-merger phase. The time between merger and the possible collapse therefore provides information about the EOS and the maximum NS mass, a quantity of great interest not only for nuclear physics but also for understanding the NS and BH populations. If the remnant survives, its extreme densities, temperatures, and compositions offer a unique opportunity to probe regions of the QCD phase diagram that are inaccessible to terrestrial experiments.

One particularly interesting possibility is that the merger remnant undergoes a transition from hadronic matter to a state containing deconfined quarks. Such a transition could leave characteristic imprints on the post-merger GW spectrum, providing a new way to probe the QCD phase diagram (right panel of Fig.~\ref{fig:BNS_intro}). While detailed studies are needed to establish how specific EOS properties are reflected in the measured post-merger signal, the amount of information that can ultimately be extracted will depend strongly on the SNR of the post-merger emission. In contrast to the inspiral, where the long duration of the signal allows substantial SNR to accumulate, the post-merger emission is typically weaker and concentrated at frequencies of a few kHz (Fig.~\ref{fig:BNS_results}). Resolving the characteristic post-merger spectral peaks, as well as possible shifts or additional features associated with phase transitions, therefore requires sufficiently high SNR~\cite{Bauswein:2018bma}. 
An important difference between the inspiral and post-merger regimes arises from the frequency of the signal; most notably, the high post-merger frequency leads to a breakdown of the long-wavelength approximation. As a result, the detector response becomes frequency-dependent across the post-merger band, and the improvement in sensitivity expected from the longer arm lengths of the 2L configurations does not translate directly into a corresponding increase in post-merger SNR. This behavior is illustrated in the bottom right panel of Fig.~\ref{fig:BNS_results}, where post-merger SNRs are computed for the same set of numerical-relativity simulations used in the previous design study~\cite{Dupletsa:2026jzs}. Based on our findings, none of the configurations shows a clear advantage for inferring properties from the post-merger signal. 

\begin{tcolorbox}[colback=teal!5!white,colframe=red!75!black,title=\sc Geometry Comparison]
{
The unique capability of ET will enable us to test the nature of dense matter with great sensitivity and to probe matter under conditions unreachable in terrestrial facilities.
\begin{itemize}
    \item  {\em An order-of-magnitude greater sensitivity than existing GW detectors:} Independent of the exact design, ET will enable observations of the NS binary population and the full GW spectrum of the BNS coalescence. 
    \item {\em Premerger information:} Due to slightly larger SNRs of the detected signals for the 2L configurations, and consequently the better measurement of the tidal deformability, the derived radius constraint marginally differs between the configurations. The misaligned 2L (42.5$^\circ$) configuration provides slightly tighter constraints due to the large number of detected high-SNR signals and the most accurate measurement of the source-frame mass.  
    \item {\em Post-merger detection:} For the loudest detections, ET will enable us to probe matter at extreme densities and temperatures. Regarding the ability to measure the post-merger signal, none of the configurations shows a clear benefit over the others. 
\end{itemize}
}
\end{tcolorbox}

\subsection{Cosmography} \label{sec:cosmo}

\begin{tcolorbox}[colback=teal!5!white,colframe=yellow!75!black,title=\sc Science Target]
{\em Map the late-time expansion history of the Universe with gravitational-wave sirens and constrain the Hubble constant, matter density, and the properties of dark energy.}
\end{tcolorbox}

\begin{figure*}
\begin{tcolorbox}[standard jigsaw,colframe=gray,colback=gray!10!white,opacityback=0.0,coltext=black, title=\small\sc Gravitational-wave sirens across cosmic history]
\center{\includegraphics[width=1\textwidth]{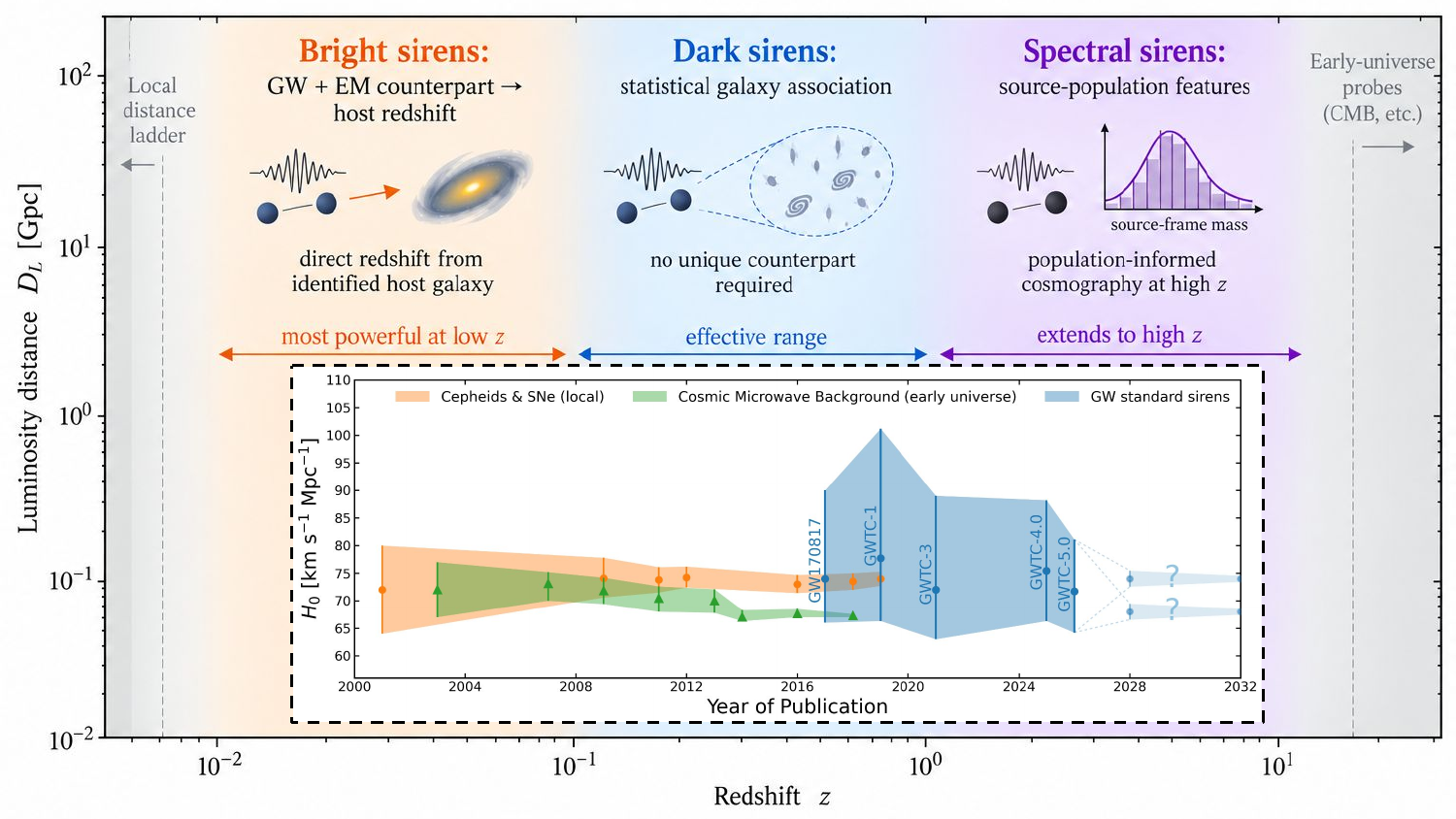}}
\caption{\justifying Compact binary mergers provide a direct measurement of luminosity distance from the GW signal, while the source redshift can be obtained in different ways: from an electromagnetic counterpart and identified host galaxy for bright sirens, statistically from possible host galaxies for dark sirens, or from features in the compact-binary population itself for spectral sirens. These methods probe complementary regimes of cosmic history: nearby bright sirens are particularly sensitive to the present-day expansion rate, characterized by the Hubble constant $H_0$, while the much larger populations of more distant dark and spectral sirens provide increasing sensitivity to the matter density $\Omega_m$ and to the properties and evolution of dark energy. The inset places this opportunity in the context of existing measurements of $H_0$: current GW siren constraints are considerably broader than those obtained from the local distance ladder or the early Universe, but future GW observations can become substantially more precise and provide an independent measurement with different physical assumptions and systematic uncertainties. Together, GW sirens offer a route from the local expansion rate to the expansion history of the Universe, complementary to both Cepheid--supernova and CMB-based cosmological measurements.} \label{fig:cosmo_schematic}
\end{tcolorbox}
\vskip-15pt
\end{figure*}

Determining the expansion history of the Universe is central to modern cosmology. The Hubble constant, $H_0$, sets the present-day expansion rate, while the matter density ($\Omega_m$) and the dark energy equation of state determine how that expansion has evolved over cosmic time. Independent measurements of these quantities are particularly important because existing cosmological probes rely on different physical assumptions and are affected by different systematic uncertainties. Most notably, the persistent discrepancy between measurements of $H_0$ based on the local distance ladder and those inferred from the early Universe motivates new methods that are independent of both~\cite{DiValentino:2021izs}. Extending such measurements over a broad redshift range also provides a direct test of the standard Lambda cold dark matter ($\Lambda$CDM) model and of whether cosmic acceleration is consistent with a cosmological constant or instead points to evolving dark energy. Gravitational-wave sources offer such an independent route to cosmography, allowing the present-day expansion rate, the late-time expansion history, and the properties of dark energy to be constrained using observables and systematics distinct from conventional electromagnetic probes.

\begin{tcolorbox}[standard jigsaw,colframe=ocre,colback=ocre!10!white,opacityback=0.0,coltext=black,title=\sc Key Science Goals]
\small
Gravitational-wave cosmography provides an independent measurement of the late-time Universe, complementary to electromagnetic probes and subject to different observational systematics.
\begin{itemize}[leftmargin=*]
\item {\bf Determine the expansion rate:} Measure the Hubble constant, $H_0$, independently of the cosmic distance ladder and early-Universe inference, providing a new route to address the current Hubble tension.
\item {\bf Constrain the matter content of the Universe:} Measure the matter density, $\Omega_m$, independently with GW sirens and test the consistency of the standard flat $\Lambda$CDM cosmology.
\item {\bf Probe the nature of dark energy:} Constrain the dark energy equation of state and its possible redshift evolution to test whether cosmic acceleration is consistent with a cosmological constant.
\end{itemize}
\end{tcolorbox}

\begin{figure*}
\begin{tcolorbox}[standard jigsaw,colframe=gray,colback=gray!10!white,opacityback=0.0,coltext=black, title=\small\sc Distance precision and sky localization of compact-binary sirens]
\center{\includegraphics[width=0.9\textwidth]{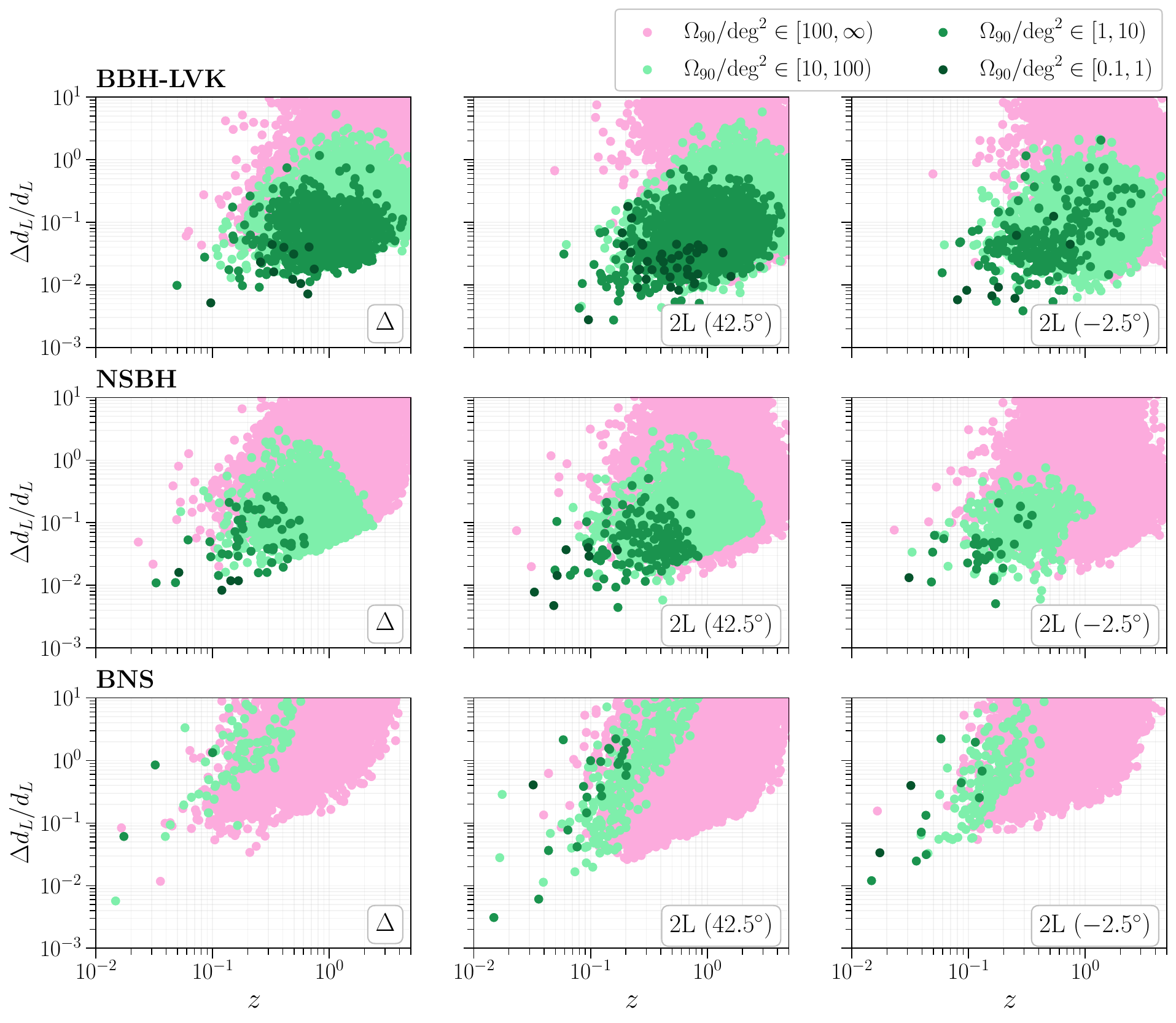}}
\caption{\justifying Fractional luminosity distance uncertainty, $\Delta d_L/d_L$, as a function of redshift for detected BBH (top), NSBH (middle), and BNS (bottom) mergers. The dots correspond to qualifying detections from a 1-year BBH population and 5-year NSBH and BNS populations. The columns show the triangular, misaligned $\mathrm{2L}\ (42.5^{\circ})$, and nearly aligned $\mathrm{2L}\ (-2.5^{\circ})$ configurations, while the colors indicate the corresponding 90\% credible sky-localization area. The figure highlights the two measurements most relevant for bright and dark siren cosmography: an accurate GW luminosity distance and a sufficiently small localization region to identify, or statistically associate, the source with galaxies.} \label{fig:cosmo_all_cbc_scatter}
\end{tcolorbox}
\vskip-15pt
\end{figure*}

Compact-binary mergers can be used as standard sirens because the amplitude and phase evolution of their GW signals provide a direct measurement of the luminosity distance, $d_L$, without calibration through the cosmic distance ladder. To turn this distance into a measurement of cosmology, however, it must be connected to a redshift. The most direct route is provided by \emph{bright sirens}, for which an electromagnetic counterpart identifies the source and its host galaxy, yielding an independent redshift measurement~\cite{LIGOScientific:2017adf}. This is the case for NS mergers accompanied by kilonovae, gamma-ray bursts (GRBs), and their afterglows. When no electromagnetic counterpart is available, the event can instead be used as a \emph{dark siren}, with potential host galaxies within the GW localization volume assigned statistically using a galaxy catalog~\cite{Schutz:1986gp}, a procedure that requires careful modeling of catalog incompleteness. Exceptionally well-localized dark sirens can have only one plausible host galaxy and effectively provide an individual redshift measurement~\cite{Borhanian:2020vyr}, although whether a source reaches this ``golden'' regime depends on its full three-dimensional localization and the galaxy density at that redshift, rather than on sky area alone~\cite{Dang:2025vqx}. 

The cosmological value of an individual siren is determined by more than its detectability. A precise luminosity distance directly reduces the uncertainty on the inferred distance--redshift relation, while a small sky area facilitates counterpart or host identification and reduces the number of candidate galaxies in a dark siren analysis. These requirements also depend on the cosmological parameter of interest. Nearby sources are particularly effective for measuring $H_0$, since the low-redshift distance--redshift relation is dominated by the present-day expansion rate. Measurements of $\Omega_m$ and especially of the dark energy equation of state require observations at larger redshifts, making well-measured sirens at cosmological distances increasingly valuable.

Figure~\ref{fig:cosmo_all_cbc_scatter} illustrates these considerations for the BBH population calibrated to LVK observations
(BBH-LVK; see Sec.~\ref{sec:population}), the NSBH, and the BNS populations used here. In this section, we present numbers for qualifying detections from a 1-year BBH-LVK population and 5-year BNS and NSBH populations. For these simulated populations, BBHs generally produce the largest SNRs, followed by NSBHs and BNSs, reflecting their typical mass hierarchy. While luminosity distance uncertainties scale approximately as the inverse SNR, they are also strongly affected by degeneracies with the inclination angle. This is especially relevant for NSBHs: higher-order GW multipoles become important due to their asymmetric masses and carry additional inclination information that helps break the distance-inclination degeneracy. Consequently, NSBH systems can achieve substantially better distance measurements than would be expected from SNR alone, particularly when higher-order modes are included~\cite{Gupta:2022fwd}.

The detector geometry then determines how often a precise distance measurement is accompanied by useful localization. At the stringent threshold of $\Delta\Omega_{90\%}<1\,{\rm deg}^2$ with $\Delta d_L/d_L \leq 0.1$, the detections are localized to a small volume. While not enough to uniquely identify the host, such detections can still have a precise redshift measurement. Hence, these are called silver sirens~\cite{Dang:2025vqx}. With these criteria, the triangular, 2L $(42.5^\circ)$, and 2L $(-2.5^\circ)$ configurations localize 11, 37, and 7 BBHs, respectively; 2, 4, and 5 of these also have exceptionally precise luminosity distance measurements with $\Delta d_L/d_L \leq 0.01$. NSBHs reach similarly precise localizations less frequently, with 4, 7, and 1 events within $1\,{\rm deg}^2$ and $\Delta d_L/d_L \leq 0.1$ for the triangle, 2L $(42.5^\circ)$, and 2L $(-2.5^\circ)$ configurations, respectively, of which 1, 2, and 0 also satisfy $\Delta d_L/d_L\leq0.01$. Relaxing the localization requirement to $10\,{\rm deg}^2$ and removing the distance-measurement restriction yields much larger NSBH samples: 58, 145, and 30 events for the three main configurations, of which 1, 6, and 1, respectively, reach percent-level distance precision. Note that in this section, whenever the number of events for BNSs or NSBHs are reported, they refer to the 5 years of observations, unless otherwise stated. Furthermore, we use the medium EOS for BNS events.

BNSs are more challenging, with only 3, 24, and 12 systems localized within $10\,{\rm deg}^2$ for the triangle, 2L $(42.5^\circ)$, and 2L $(-2.5^\circ)$ configurations, respectively. Percent-level $d_L$ precision is only reached with the 2L $(42.5^\circ)$ configuration, that too for just 2 systems. Thus, the misaligned 2L $(42.5^\circ)$ configuration generally produces the largest well-localized samples across all three source classes. The nearly aligned 2L $(-2.5^\circ)$ configuration can still produce individual events with excellent distance precision, particularly for BBHs, but does so for a smaller well-localized population.

None of the systems considered here is localized within $0.1\,{\rm deg}^2$ and qualifies as a golden dark siren, illustrating the difficulty of identifying a unique host from ET alone without an electromagnetic counterpart. Of course, this target can be achieved by ET together with a network of detectors, leading to a sub-percent $H_0$ measurement within a year of observation~\cite{Gupta:2023lga}. We emphasize, however, that $0.1\,{\rm deg}^2$ should not be regarded as a universal threshold for a golden dark siren: unique host identification depends on the localization volume, redshift, and local galaxy density. Nevertheless, the absence of extremely small sky areas in these catalogs indicates that conventional host-based dark siren cosmography with ET alone will generally remain statistical rather than relying on large samples of uniquely identified hosts. However, as part of a network, ET is expected to achieve a sub-percent $H_0$ measurement using the statistical route~\cite{Muttoni:2023prw}.

A complementary route avoids individual host identification altogether. In \emph{spectral siren} cosmography, features in the source-frame compact-binary mass distribution provide a statistical reference scale. Because GW observations measure redshifted masses, changing the cosmological parameters alters the mapping between the observed detector-frame mass spectrum, source-frame masses, and luminosity distance~\cite{Ezquiaga:2022zkx}. A sufficiently structured and well-measured population can constrain cosmology without assigning a host galaxy to each event. BBHs are particularly promising for this purpose: the loudest systems in Fig.~\ref{fig:cosmo_all_cbc_scatter} combine precise $d_L$ measurements with accurately measured masses, and the detectable population extends to redshifts well beyond those accessible to electromagnetic counterparts. As we show below, this makes the BBH population an important counterpart-independent probe of the expansion history and, in particular, of dark energy parameters.

We begin with the most direct standard siren measurement: bright sirens from NS mergers with identified kilonova counterparts, for which the GW luminosity distance can be paired with an electromagnetic host redshift.

\subsubsection{Determining the expansion rate and matter content of the Universe}

\begin{figure*}
\begin{tcolorbox}[standard jigsaw,colframe=gray,colback=gray!10!white,opacityback=0.0,coltext=black, title=\small\sc Bright siren constraints on flat $\Lambda$CDM]
\center{\includegraphics[width=0.4\textwidth]{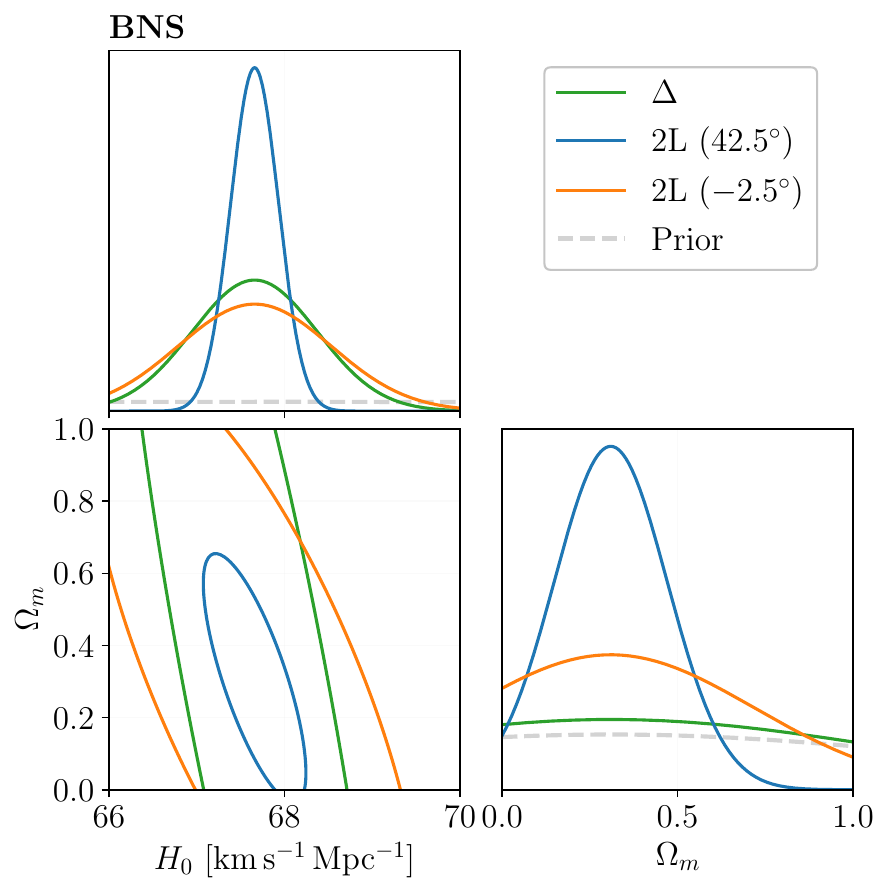}
\includegraphics[width=0.4\textwidth]{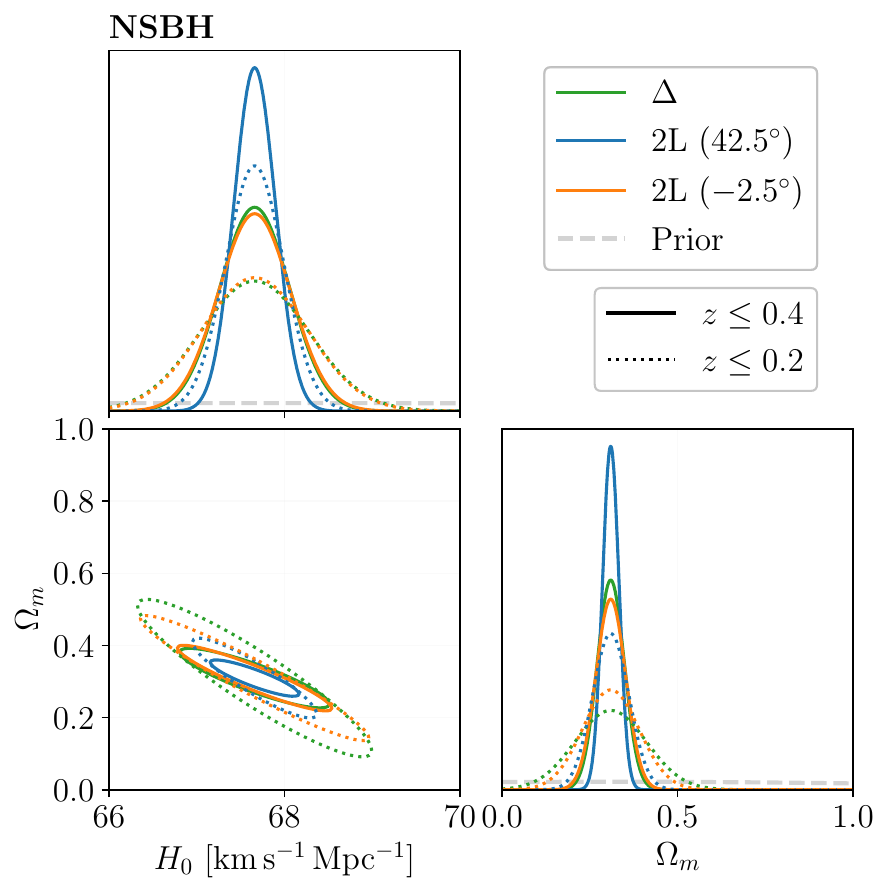}}
\caption{\justifying Forecast constraints on the flat $\Lambda$CDM parameters $H_0$ and $\Omega_m$ from bright sirens associated with BNS mergers (left) and NSBH mergers (right). The BNS results use the medium EOS, while the NSBH results use the higher-mode waveform adopted throughout this work. Contours at $90\%$ credibility are shown for the triangular, misaligned 2L $(42.5^\circ)$, and nearly aligned 2L $(-2.5^\circ)$ configurations, with the prior shown as dashed curves for comparison. For NSBHs, the additional contours illustrate the effect of restricting the bright siren events to $z\leq0.2$, compared with the baseline $z\leq0.4$ sample.} \label{fig:cosmo_bns_nsbh_corner_flcdm}
\end{tcolorbox}
\vskip-15pt
\end{figure*}

Bright sirens provide the most direct route from a GW observation to cosmology. For NS mergers, a kilonova can identify the electromagnetic counterpart and its host galaxy, supplying the redshift that is otherwise absent from the GW signal, while the GW measurement provides the luminosity distance. The resulting $(d_L,z)$ pair can then be used to constrain the cosmological expansion. We refer to Sec.~\ref{sec:mma} for a detailed discussion of kilonova emission and counterpart detectability. Here, we adopt a more conservative localization requirement than in the multimessenger analysis, restricting the bright siren events to $\Delta\Omega_{90\%}\leq50\,{\rm deg}^2$, rather than $100\,{\rm deg}^2$, to increase the likelihood of successful counterpart and host identification. For BNSs, we consider sources out to $z\leq0.4$. For NSBHs, we show both this baseline selection and a more conservative $z\leq0.2$ sample, to bracket the possibility that their kilonovae have a smaller practical detection horizon.

Figure~\ref{fig:cosmo_bns_nsbh_corner_flcdm} shows the resulting constraints on $H_0$ and $\Omega_m$ in a flat $\Lambda$CDM cosmology. The BNS and NSBH catalogs used here correspond to 5 years of mergers. We also impose broad Gaussian priors with $\sigma_{H_0}=10\,{\rm km\,s^{-1}\,Mpc^{-1}}$ and $\sigma_{\Omega_m}=1$. For a fixed detector configuration and source selection, the statistical uncertainty from an ensemble of comparable sirens is expected to scale approximately as $1/\sqrt{N}$. Thus, we also quote the approximate 1-year $H_0$ precision obtained by rescaling the 5-year uncertainties by $\sqrt{5}$.

For BNS bright sirens, the triangular configuration yields 40 events and constrains $H_0$ to $1.06\%$ (at 68.3\% credible level unless otherwise stated), while $\Omega_m$ remains effectively unconstrained. The 2L $(42.5^\circ)$ configuration yields 128 events and improves these measurements to $0.40\%$ in $H_0$ and $52\%$ in $\Omega_m$. The 2L $(-2.5^\circ)$ configuration yields 78 events and gives weaker constraints of $1.29\%$ and $132\%$. These samples correspond to approximately 8, 26, and 16 bright sirens per year, giving 1-year statistical precisions on $H_0$ of approximately $2.4\%$, $0.9\%$, and $2.9\%$ for the triangle, misaligned $2{\rm L} (42.5^\circ)$, and nearly aligned $2{\rm L} (-2.5^\circ)$ configurations, respectively. 

The comparison again illustrates that the number of detections alone does not determine the cosmological performance. The 2L $(-2.5^\circ)$ configuration contains almost twice as many BNS bright sirens as the triangle, yet gives only a modestly different $H_0$ constraint. Its nearly parallel detector responses are less effective at breaking the distance--inclination and polarization degeneracies, so the luminosity distances of the individual events remain less precise. The misaligned 2L $(42.5^\circ)$ configuration performs substantially better, benefiting from the sensitivity of the longer arms while retaining the geometric information needed to constrain the source orientation and distance, and gives the strongest BNS constraints.

NSBH systems perform at least as well as, and in several respects substantially better than the BNS sample. For the $z\leq0.4$ selection, the triangular configuration yields 191 events and constrains $H_0$ to $0.59\%$ and $\Omega_m$ to $12.4\%$. The 2L $(42.5^\circ)$ configuration yields 313 events and reaches $0.35\%$ and $7.6\%$, respectively. The 2L $(-2.5^\circ)$ configuration yields 191 events and gives corresponding uncertainties of $0.61\%$ and $13.7\%$. These samples correspond to approximately 38, 63, and 38 events per year, implying 1-year statistical $H_0$ precisions of about $1.31\%$, $0.78\%$, and $1.35\%$.

Restricting the NSBH sample to the more conservative $z\leq0.2$ range reduces both the number of sirens and the cosmological impact. The triangular configuration retains 58 events and constrains $H_0$ to $0.92\%$ and $\Omega_m$ to $32.9\%$. The 2L $(42.5^\circ)$ configuration retains 91 events and achieves $0.49\%$ and $16.7\%$. The 2L $(-2.5^\circ)$ configuration retains 88 events and gives $0.90\%$ and $26.1\%$. The corresponding 1-year $H_0$ precisions are approximately $2.1\%$, $1.09\%$, and $2.0\%$. The degradation in $\Omega_m$ when the redshift range is reduced from $z\leq0.4$ to $z\leq0.2$ is expected, reflecting the limited sensitivity of very nearby sirens to parameters governing the shape of the expansion history.

The stronger NSBH constraints arise from two advantages acting together. More NSBH systems satisfy the bright siren selection, and their typically larger SNRs and asymmetric masses allow their luminosity distances to be measured more precisely than for BNSs. The latter is particularly important because higher-order modes in asymmetric binaries provide additional inclination information and help break the distance--inclination degeneracy. Consequently, the NSBH population can compensate for the greater astrophysical uncertainty surrounding its electromagnetic counterparts with substantially more informative GW distance measurements.

As a more conservative test, we also restrict the NSBH population to systems with $q\equiv m_{\rm NS}/m_{\rm BH}\geq1/4$, for which tidal disruption is generally more probable. The smaller number of systems naturally weakens the constraints. At $z\leq0.4$, the triangle, 2L $(42.5^\circ)$, and 2L $(-2.5^\circ)$ configurations retain 49, 86, and 47 events, respectively, with $H_0$ uncertainties of $3.24\%$, $1.58\%$, and $1.35\%$. The corresponding $\Omega_m$ uncertainties are approximately $61\%$, $32\%$, and $36\%$. Applying the additional $z\leq0.2$ restriction leaves only 19, 29, and 27 events, degrading the $H_0$ precision to $5.22\%$, $2.39\%$, and $1.55\%$, respectively, with $\Omega_m$ uncertainties of approximately $157\%$, $76\%$, and $54\%$. Note that the cosmological forecasts shown in Fig.~\ref{fig:cosmo_bns_nsbh_corner_flcdm} use the full NSBH population within the stated redshift cuts. The condition $q\geq1/4$ is only a simple proxy for disruption: sufficiently rapidly spinning BHs can tidally disrupt the NS even at smaller mass ratios. A hard mass-ratio cut would remove potentially viable electromagnetic counterparts and is used here only to illustrate how the forecasts change under a more restrictive source selection.

We also checked the dependence of the BNS bright siren constraints on the assumed NS EOS. Although the number of selected events remains broadly similar across the soft, medium, and stiff EOSs, the resulting cosmological uncertainties show noticeable scatter. For the misaligned 2L $(42.5^\circ)$ configuration, the soft, medium, and stiff populations contain 117, 128, and 131 bright sirens in the 5-year population and constrain $H_0$ to $0.63\%$, $0.40\%$, and $1.05\%$, respectively. Larger variations occur for the triangle, where 37, 40, and 46 events give $H_0$ uncertainties of $1.91\%$, $1.06\%$, and $8.63\%$. Since the three EOSs were already found to produce very similar overall SNR and localization distributions, this variation is unlikely to reflect a strong intrinsic EOS dependence of the cosmological measurement. Rather, the independently sampled populations contain different masses, orientations, and distances, and the combined constraint can be dominated by a relatively small number of exceptionally well-measured sirens. Thus, the exact forecast carries realization-to-realization scatter. Importantly, the misaligned 2L $(42.5^\circ)$ configuration consistently gives the strongest BNS bright siren constraints.

Overall, bright sirens can provide an independent sub-percent statistical measurement of $H_0$ with ET alone. For BNSs, the 2L $(42.5^\circ)$ configuration reaches sub-percent precision already after approximately 1 year. For NSBH bright sirens extending to $z\leq0.4$, the 2L $(42.5^\circ)$ configuration reaches sub-percent precision after one year, with projected uncertainties of approximately $0.8\%$. As expected, these forecasts worsen when only mass-symmetric NSBHs are considered. The results also show that extending precisely measured NSBH bright sirens to larger redshift substantially improves sensitivity to $\Omega_m$. This need for higher-redshift observations becomes even more important when moving beyond $\Lambda$CDM and attempting to constrain the equation of state and evolution of dark energy, which we consider next.

\subsubsection{Probing the nature of dark energy}

\begin{figure*}
\begin{tcolorbox}[standard jigsaw,colframe=gray,colback=gray!10!white,opacityback=0.0,coltext=black, title=\small\sc Luminosity distance precision across cosmic time]
{\centering\includegraphics[width=1.0\textwidth]{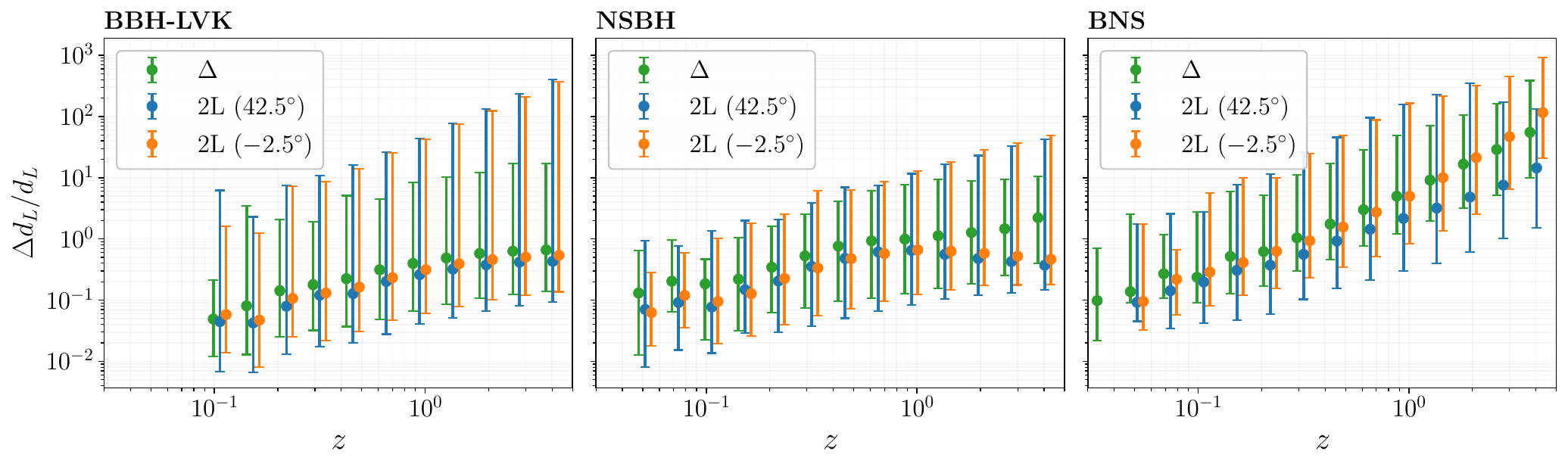}}
\caption{\justifying Median fractional luminosity distance uncertainty, $\Delta d_L/d_L$, as a function of redshift for BBH (left), NSBH (middle), and BNS (right) mergers detected with the three ET configurations. Markers show the median uncertainty in each redshift bin, with error bars indicating the spread of the recovered distance uncertainties within that bin. The progressive degradation of distance precision with redshift is strongest for BNSs, while the typically louder BBH and NSBH signals retain more informative distance measurements to larger cosmological distances.} \label{fig:cosmo_med_dl_errs}
\end{tcolorbox}
\vskip-15pt
\end{figure*}

Going beyond $\Lambda$CDM requires allowing the dark energy equation of state to differ from a cosmological constant and to evolve with redshift. We adopt the commonly used $w_0$--$w_a$ parametrization, where $w_0$ describes the present-day equation of state and $w_a$ captures its evolution with cosmic time~\cite{Chevallier:2000qy,Linder:2002et}. Our fiducial cosmology corresponds to a cosmological constant, with $w_0=-1$ and $w_a=0$. Constraining these additional degrees of freedom is substantially more demanding than measuring $H_0$, since their imprint on the expansion history becomes appreciable only over sufficiently large redshifts.

This is already evident from the nearby bright siren sample considered in the previous subsection. Even for the comparatively well-measured NSBH population extending to $z\leq0.4$, the triangle, 2L $(42.5^\circ)$, and 2L $(-2.5^\circ)$ configurations contain 191, 313, and 191 events, respectively, but constrain $w_0$ only to approximately $88\%$, $77\%$, and $87\%$. The corresponding $\Omega_m$ uncertainties are larger than $170\%$, and, together with $w_a$, remain effectively unconstrained. BNS bright sirens within $z\leq0.4$ provide essentially no useful constraint on either dark energy parameter. Thus, although nearby sirens can measure $H_0$ very precisely, probing dark energy requires extending the measurement to substantially larger redshifts.

Figure~\ref{fig:cosmo_med_dl_errs} illustrates the difficulty of doing this with bright sirens. BBHs retain significant luminosity distance precision to high redshift, with median fractional uncertainties in their highest populated bins of approximately $0.51$, $0.59$, and $0.76$ for the misaligned 2L $(42.5^\circ)$, 2L $(-2.5^\circ)$, and triangular configurations, respectively. NSBHs show a stronger geometry dependence: the triangular configuration eventually reaches median uncertainties above unity, whereas both $2{\rm L}$ configurations retain a high-redshift subset with typical uncertainties of order $35$--$40\%$. BNSs fare considerably worse, with the median distance uncertainty rapidly exceeding unity as redshift increases. The 2L $(42.5^\circ)$ configuration improves these constraints, but even its high-redshift BNS distances become too poorly measured to provide useful cosmological information.

Short GRBs offer a possible way to supply electromagnetic redshifts beyond the kilonova horizon. For BNSs, however, the gain in redshift is overwhelmed by the worse luminosity distance precision. We explicitly tested the BNS--GRB population and find that the resulting dark energy parameters remain effectively unconstrained. Thus, high-redshift BNS detections with ET alone are not sufficient for useful dark energy cosmography.

NSBHs are more promising because their higher SNRs and asymmetric masses allow substantially better $d_L$ measurements at the same redshift. We consider a putative NSBH--GRB bright siren population, additionally requiring $q\geq1/4$ to favor systems for which tidal disruption and electromagnetic emission are more plausible. The short GRBs are modeled using a Gaussian structured-jet profile, automatically including a line-of-sight dependence~\cite{Dhani:2022ulg}. Further, we assume a sky coverage of $60\%$, a host-association probability of $80\%$~\cite{OConnor:2022lpb,
Fong:2022mkv}, and  $\Delta\Omega_{90\%} \leq 10^4\,{\rm deg}^2$. Using this framework, the triangle, 2L $(42.5^\circ)$, and 2L $(-2.5^\circ)$ configurations yield 323, 661, and 574 usable GW--GRB associations, respectively. These constrain $w_0$ to approximately $60\%$, $49\%$, and $48\%$, while $\Omega_m$ is measured to roughly $80\%$, $43\%$, and $44\%$. The evolution parameter $w_a$ remains largely unconstrained in all three cases. The two $2{\rm L}$ configurations perform comparably for this high-redshift sample, and both are substantially better than the triangle. Alternatively, in a model-agnostic approach, simply applying a viewing angle cut of $15^{\circ}$, imposing $z\leq 2$ and $\Delta\Omega_{90\%} \leq 10^4\,{\rm deg}^2$ changes the number of detections to 369, 602, and 513 for the triangle, 2L $(42.5^\circ)$, and 2L $(-2.5^\circ)$ configurations, respectively. These constrain $w_0$ to approximately $66\%$, $52\%$, and $51\%$, while $\Omega_m$ is measured to roughly $80\%$, $58\%$, and $63\%$.

Note that these NSBH--GRB forecasts should be interpreted cautiously. No NSBH merger has yet been securely associated with a GRB, and the conditions under which an NSBH launches an observable relativistic jet remain uncertain. While they demonstrate the potential of a high-redshift bright siren population, they do not offer a guaranteed route to dark energy measurements. This motivates a complementary approach that can exploit the large high-redshift BBH population without requiring an electromagnetic counterpart.

\begin{figure*}
\begin{tcolorbox}[standard jigsaw,colframe=gray,colback=gray!10!white,opacityback=0.0,coltext=black, title=\small\sc Cosmological constraints from BBH spectral sirens]
\center{\includegraphics[width=0.6\textwidth]{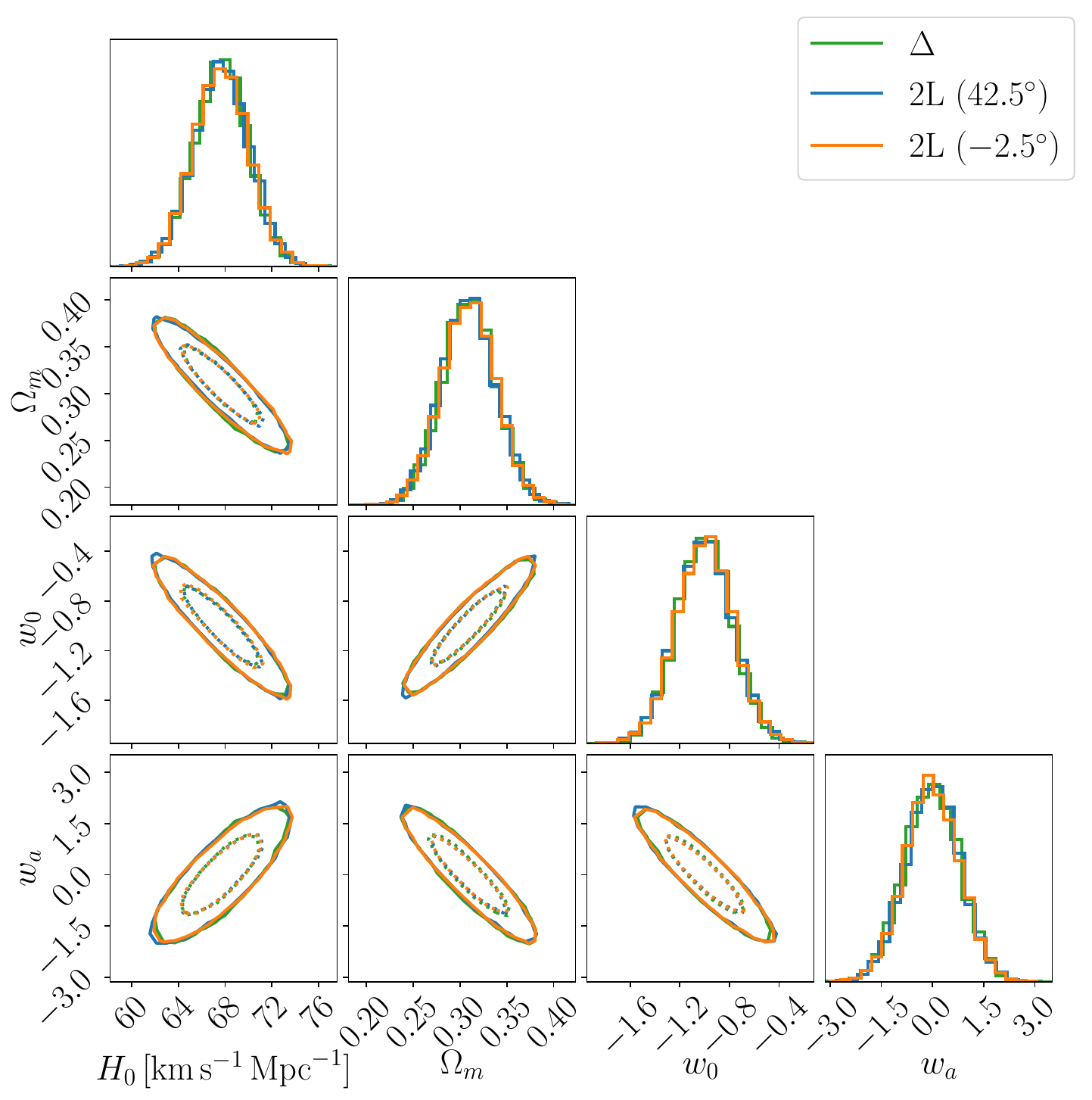}}
\caption{\justifying Forecast constraints on $H_0$, $\Omega_m$, $w_0$, and $w_a$ from the BBH--LVK population using the spectral siren method. The diagonal panels show the marginalized posterior distributions, while the lower-triangle panels show the corresponding joint constraints for the triangular, misaligned $2{\rm L}\ (42.5^{\circ})$, and nearly aligned $2{\rm L}$ configurations. The near-complete overlap between the three results shows that the spectral siren constraints are largely insensitive to the ET geometry.} \label{fig:cosmo_spectral}
\end{tcolorbox}
\vskip-15pt
\end{figure*}

Features in the compact-binary mass distribution can act as statistical reference scales for redshift. GW observations measure redshifted masses and luminosity distances, so the cosmological model determines how the observed population is mapped back to its source-frame mass spectrum. If that spectrum contains sufficiently well-measured structure, the population and cosmology can be inferred jointly, allowing the mass distribution itself to provide information about the distance--redshift relation. BBHs are particularly well suited to this approach. As seen in Fig.~\ref{fig:cosmo_med_dl_errs} and Sec.~\ref{sec:population}, ET detects BBHs with significant SNR out to high redshifts, while their intrinsic mass parameters remain well measured over a substantial fraction of this range. 

Figure~\ref{fig:cosmo_spectral} shows the resulting joint constraints on $H_0$, $\Omega_m$, $w_0$, and $w_a$, using the spectral siren method and assuming that the mass spectrum does not evolve with redshift. The three detector configurations give almost indistinguishable posterior distributions. At $90\%$ credibility, the triangular configuration constrains $H_0$, $\Omega_m$, and $w_0$ to approximately $5.6\%$, $15.4\%$, and $37.0\%$, respectively, with an absolute uncertainty of $1.33$ on $w_a$. The corresponding constraints are $5.8\%$, $15.4\%$, $37.7\%$, and $1.32$ for the $2{\rm L}\ (42.5^{\circ})$ configuration, and $5.8\%$, $15.5\%$, $37.8\%$, and $1.32$ for the $2{\rm L}\ (-2.5^{\circ})$ configuration. Thus, the differences between the geometries are negligible compared to the statistical uncertainties of the measurement.

The weak dependence on geometry contrasts with the bright siren results. Bright siren cosmography benefits directly from precise sky localization and from breaking the distance--inclination degeneracy, giving the $2{\rm L}\ (42.5^{\circ})$ configuration a clear advantage. Spectral sirens instead draw their constraining power from the enormous BBH catalog and the accurately measured population structure across a broad redshift range. Once tens of thousands of BBHs contribute to the hierarchical inference, the differences in individual-event localization and distance precision among the geometries have little impact on the final cosmological constraints. Thus, spectral sirens provide a promising counterpart-independent route to probing dark energy with ET, while also illustrating an important complementarity between cosmographic methods: detector geometry is critical when cosmology relies on identifying individual counterparts or hosts, but becomes considerably less important when the information is extracted statistically from the full compact-binary population.

Taken together, these results show that ET alone can deliver precision cosmological measurements through several complementary siren channels. Bright sirens from nearby neutron star mergers provide the cleanest route to the present-day expansion rate and, for the $2{\rm L}\ (42.5^{\circ})$ configuration, can reach sub-percent statistical precision on $H_0$ after approximately one year of observations. Extending bright siren cosmography to dark energy parameters is more difficult because the gain via high-redshift GRB detections is accompanied by rapidly worsening luminosity-distance measurements, particularly for BNSs. NSBHs remain more promising, although their association with GRBs is still uncertain. Spectral sirens offer a complementary route by exploiting the large, high-SNR BBH population without requiring electromagnetic counterparts, and can constrain $H_0$, $\Omega_m$, $w_0$, and $w_a$ with little dependence on detector geometry. The geometry choice matters most for counterpart-based cosmography, where the $2{\rm L}\ (42.5^{\circ})$ configuration consistently provides the best combination of localization and distance precision, while population-based spectral siren measurements are largely insensitive to the ET geometry. Table~\ref{tab:h0_summary} summarizes the results discussed in this section.

\begin{table}
\caption{\justifying
Summary of uncertainties on the measurement of cosmological parameters considering 5 years of observations for BNS and NSBH mergers, and 1 year of observation for BBH mergers. Entries in each triple follow the configuration order: $2\rm{L}\,(-2.5^\circ)$, $2\rm{L}\,(42.5^\circ)$,
$\Delta$. The ``/'' symbol means that the parameter was not part of the analysis, whereas ``--'' means that it is largely unconstrained. All the reported values refer to the $68.3\%$ credible interval.}\label{tab:h0_summary}
\centering
\resizebox{\textwidth}{!}{
\renewcommand{\arraystretch}{1.1} 
\begin{tabular}{llcccccr}
\toprule
&Analysis
 & Number of events
 & $\Delta H_0/H_0$
 & $\Delta\Omega_m/\Omega_m$
 & $\Delta w_0/w_0$
 & $\Delta w_a$ \\
\midrule

\multirow{12.5}{*}{\rotatebox{45}{Bright}}
 &BNSs ($\Delta\Omega_{90\%} < 50$\,deg$^2$
 &\multirow{2}{*}{$78,\ 128,\ 40$}
 &\multirow{2}{*}{$1.29\%,\ 0.40\%,\ 1.06\%\ $}
 &\multirow{2}{*}{$132\%,\ 52\%,\ -$}
 &\multirow{2}{*}{/} 
 &\multirow{2}{*}{/} &\multirow{9.5}{*}{\rotatebox{-90}{$\Lambda$CDM}} \\
 
 &and $z \leq 0.4$) & & & & & & \\

 \cline{2-7}

 &NSBHs ($\Delta\Omega_{90\%} < 50$\,deg$^2$
 &\multirow{2}{*}{$191,\ 313,\ 191$}
 &\multirow{2}{*}{$0.61\%,\ 0.35\%,\ 0.59\%\ $}
 &\multirow{2}{*}{$13.7\%,\ 7.6\%,\ 12.4\%$}
 &\multirow{2}{*}{/} 
 &\multirow{2}{*}{/} & \\
 & and $z \leq 0.4$) & & & & & & \\

 \cline{2-7}

 &NSBHs ($\Delta\Omega_{90\%} < 50$\,deg$^2$           
 &\multirow{2}{*}{$88,\ 91,\ 58$}
 &\multirow{2}{*}{$0.90\%,\ 0.49\%,\ 0.92\%\ $}
 &\multirow{2}{*}{$26.1\%,\ 16.7\%,\ 32.9\%$}
 &\multirow{2}{*}{/}
 &\multirow{2}{*}{/} & \\
 & and $z \leq 0.2$) & & & & & &\\

 \cline{2-7}

 &NSBHs ($\Delta\Omega_{90\%} < 50$\,deg$^2$             
 &\multirow{2}{*}{$47,\ 86,\ 49$}
 &\multirow{2}{*}{$1.35\%,\ 1.58\%,\ 3.24\%\ $}
 &\multirow{2}{*}{$36\%,\ 32\%,\ 61\%$}
 &\multirow{2}{*}{/}
 &\multirow{2}{*}{/} & \\
 &$z \leq 0.4$, and $q \geq 1/4$) & & & & & &\\

\cline{2-7}

 &NSBHs ($\Delta\Omega_{90\%} < 50$\,deg$^2$             
 &\multirow{2}{*}{$27,\ 29,\ 19$}
 &\multirow{2}{*}{$1.55\%,\ 2.39\%,\ 5.22\%\ $}
 &\multirow{2}{*}{$54\%,\ 76\%,\ 157\%$}
 &\multirow{2}{*}{/}
 &\multirow{2}{*}{/} & \\
 &$z \leq 0.2$, and $q \geq 1/4$) & & & & & &\\
 
\cline{2-8}

 &NSBHs + GRBs (structured-jet,                     
 &\multirow{2}{*}{$574,\ 661,\ 323$}
 &\multirow{2}{*}{$ 5.0\%,\  5.6\%,\ 5.6\% $}
 &\multirow{2}{*}{$44\%,\ 43\%,\ 80\%$}
 &\multirow{2}{*}{$48\%,\ 49\%,\ 60\%$}
 &\multirow{2}{*}{$-,\ -,\ -$} & \\
 &$q \geq 1/4$, $\Delta\Omega_{90\%} < 10^4$\,deg$^2$) & & & & & &\multirow{3}{*}{\rotatebox{-90}{$w$CDM}}\\

 \cline{2-7}

 &NSBHs + GRBs ($z\leq2$, $q \geq 1/4$,                     
 &\multirow{2}{*}{$513,\ 602,\ 369$}
 &\multirow{2}{*}{$ 5.70\%,\  6.20\%,\ 6.92\% $}
 &\multirow{2}{*}{$63\%,\ 58\%,\ 80\%$}
 &\multirow{2}{*}{$51\%,\ 52\%,\ 66\%$}
 &\multirow{2}{*}{$-,\ -,\ -$} & \\
 &$\Delta\Omega_{90\%} < 10^4$\,deg$^2$, $\Theta<\pi/12$) & & & & & & \\

\cline{1-7}
\vspace{1pt}
\multirow{3.5}{*}{\rotatebox{45}{Spectral}} 
&\multirow{3}{*}{BBH--LVK} 
&\multirow{3}{*}{$ 86577,\ 90602,\ 81459$}
&\multirow{3}{*}{$3.48\%,\ 3.50\%,\ 3.52\%$}
&\multirow{3}{*}{$9.34\%,\ 9.34\%,\ 9.35\%$}
&\multirow{3}{*}{$22.7\%,\ 22.8\%,\ 23.0\%$}
&\multirow{3}{*}{$0.798,\ 0.796,\ 0.801$} &\\\\\\
\bottomrule
\end{tabular}
}
\end{table}

\begin{tcolorbox}[colback=teal!5!white,colframe=red!75!black,title=\sc Geometry Comparison]
{
ET alone can deliver precision cosmology, including sub-percent statistical measurements of $H_0$ on approximately one-year timescales. The relative importance of detector geometry depends on how the source redshift is obtained: it is critical for bright sirens, but much less important when cosmology is inferred statistically from the BBH population.
\begin{itemize}
    \item {\em Misaligned $2{\rm L}\ (42.5^{\circ})$ is optimal for bright sirens:} It provides the largest samples with both precise luminosity distances and useful sky localizations, enabling the strongest constraints on $H_0$ and $\Omega_m$.
    \item {\em Aligned $2{\rm L}\ (-2.5^{\circ})$ loses cosmological information through localization degeneracies:} Its greater sensitivity does not fully compensate for poorer polarization resolution, weakening bright siren constraints relative to the misaligned configuration and, in some cases, the triangle.
    \item {\em Spectral siren cosmography is largely geometry independent:} The enormous high-redshift BBH population yields nearly identical constraints on $H_0$, $\Omega_m$, $w_0$, and $w_a$ for all three ET configurations.
\end{itemize}
}
\end{tcolorbox}

\subsection{Astrophysical and cosmological stochastic backgrounds}
\label{sec:backgrounds}

\begin{tcolorbox}[colback=teal!5!white,colframe=yellow!75!black,title=\sc Science Target]
{\em %

Determine the energy budget of the gravitational-wave background, understand the properties and the evolution of the population of compact binaries, and probe the physics of the very high energy scales that drove the first minute of the evolution of the Universe.

}
\end{tcolorbox}

\begin{tcolorbox}[standard jigsaw,colframe=ocre,colback=ocre!10!white,opacityback=0.0,coltext=black,title=\sc Key Science Goals]
\small
\begin{itemize}[leftmargin=*]
\item {\bf Cosmological backgrounds:} Explore the physics of the highest energy scales by measuring properties of the GW background generated during the first minute in the history of the Universe.
\item {\bf Compact-binary foreground:} Develop a detailed understanding of the population of compact binaries, including their distribution and evolution in mass, distance, and spin, by measuring the astrophysical GW foreground generated by these systems.
\item {\bf Other astrophysical foregrounds:} Search for astrophysical GW foregrounds due to other astrophysical sources, including PBHs, Pop. III BHs, and IMBHs.
\end{itemize}
\end{tcolorbox}

\begin{figure*}
\begin{tcolorbox}[standard jigsaw,colframe=gray,colback=gray!10!white,opacityback=0.0,coltext=black, title=\small\sc Cosmology and astrophysics with stochastic GW backgrounds]
\center{
\includegraphics[width = \textwidth]{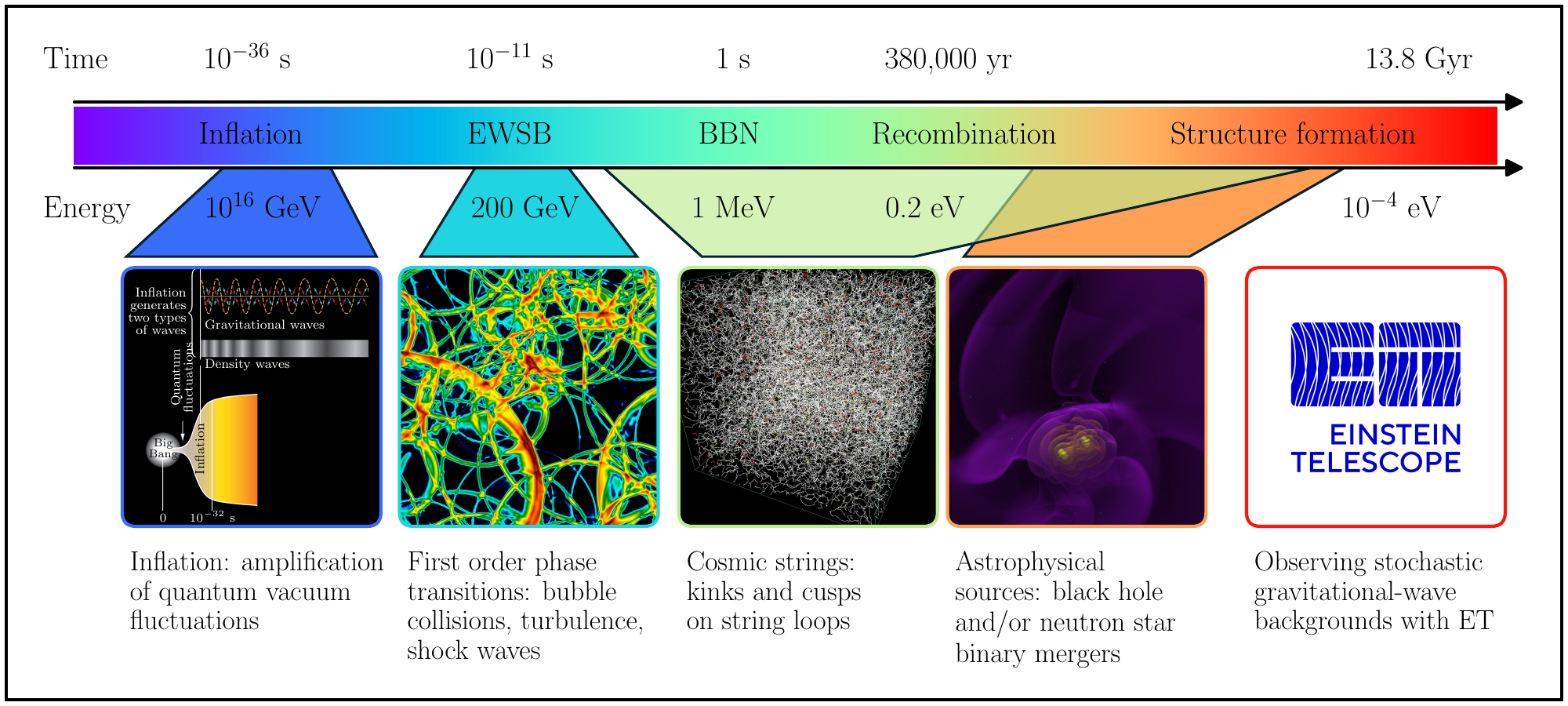}
}
\caption{\justifying The stochastic gravitational-wave background (SGWB) receives contributions from a broad variety of cosmological and astrophysical sources. Processes associated with inflation and phase transitions during the first minute after the Big Bang could leave observable SGWB signatures, potentially revealing the physics of very high energy scales not accessible in laboratories. Cosmic string networks produced in early Universe phase transitions could continue to emit GWs until late times. Astrophysical mechanisms such as mergers of BBH, BNS, and NSBH systems create foregrounds that carry information about the formation of structure in the late Universe. ET will enable observation of these cosmological and astrophysical sources by measuring the frequency content, spatial anisotropy, polarization, and temporal variability of the SGWB. Image credits, left to right: NSF/BICEP II Collaboration~\cite{NSF:2014HistoryOfUniverse};  D.~Weir~\cite{Weir:2016PTsim,Hindmarsh:2015qta,Hindmarsh:2017gnf};  C.~Ringeval~\cite{Ringeval:2018stringmovie,Ringeval:2005kr};  NASA/B.~J.~Kelly, C.~Henze and T.~Sandstrom~\cite{NASA:SVS13197}.}
\label{fig:SGWBintro}
\end{tcolorbox}
\vskip-15pt
\end{figure*}

Stochastic GW backgrounds (SGWBs) arise as a superposition of GW signals generated by many uncorrelated sources of either astrophysical or cosmological origin. 
Cosmological backgrounds may be generated by processes such as the amplification of quantum-vacuum fluctuations during inflation, phase transitions due to spontaneous symmetry breaking in the early Universe. Such processes would have taken place during the first minute after the Big Bang, an epoch that is not accessible to traditional messengers such as light. They are also associated with energy scales that are not achievable in colliders, making this kind of science unique for the field of GW observations. Detection of a cosmological background is, therefore, arguably the most fundamentally impactful observation that GW observatories could make. 

On the other hand, a myriad of individual astrophysical sources, such as compact binary mergers, could also generate a stochastic GW {\it foreground}. Such a foreground is of significant interest in its own right, as it contains information about the properties of the population of sources that generated it---distribution and evolution of the masses, distances, and spins of black holes and/or neutron stars---which, in turn, carry signatures
of the evolution of structure in the relatively late Universe.  Furthermore, the astrophysical foreground may mask the cosmological background over much of the accessible spectrum. Understanding the astrophysical GW foregrounds is, therefore, imperative for detecting and characterizing cosmological GW backgrounds.

GW backgrounds and foregrounds are typically described in terms of their normalized energy density $\Omega_{\rm gw}(f) = \frac{f}{\rho_c} \; \frac{d\rho_{\rm GW}}{df}$, %
where $f$ is the GW frequency, $\rho_{\rm GW}$ is the energy density carried by GWs, and $\rho_c$ is the critical energy density of the Universe. Different contributors to the stochastic background typically produce different frequency spectra, as well as different anisotropy (angular spectra), polarization, and potentially temporal variability. These properties are used in statistical analyses of GW data to disentangle different contributions and estimate the energy budget of the stochastic background.

\subsubsection{Cosmological backgrounds}

\begin{figure*}
\begin{tcolorbox}[standard jigsaw,colframe=gray,colback=gray!10!white,opacityback=0.0,coltext=black, title=\small\sc Sensitivity for stochastic gravitational-wave backgrounds]
\center{%
\includegraphics[width = \textwidth]{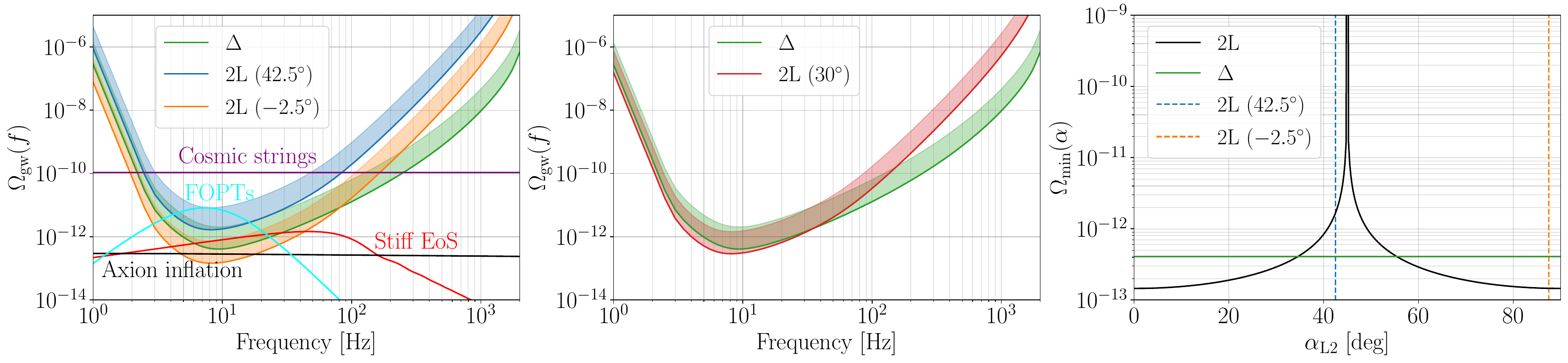}
}
\caption{\justifying Left panel: Comparison of the sensitivity to stochastic gravitational-wave backgrounds (SGWBs) for the three ET configurations. The green, blue, and orange curves correspond to the power-law integrated (PI) sensitivity curves for the triangle, the 2L ($42.5^{\circ}$), and the 2L ($-2.5^{\circ}$) configurations, respectively. The curves assume one year of continuous observation and SNR = 1; the shaded regions extend to SNR = 5, the nominal detection threshold. The lower the PI curve, the greater the sensitivity to SGWBs. The other solid curves show representative cosmological SGWBs, namely first-order phase transitions \cite{Hasegawa:2019amx}, cosmic strings \cite{Blanco-Pillado:2013qja}, axion inflation \cite{LIGOScientific:2025kry}, and inflation with a stiff equation of state \cite{Duval:2024jsg}. 
Central panel: Comparison of the PI sensitivity curve of the intermediate 2L ($30^\circ$) configuration (red) with that of the triangle configuration (green). 
Right panel: Sensitivity of the 2L configuration as a function of the relative angle between the detector arms. The black curve shows the minimum value of the PI sensitivity curve (SNR = 1) at each relative angle, while the green line shows the corresponding minimum for the triangle configuration. The dashed orange and blue vertical lines mark the 2L ($-2.5^\circ$) and 2L ($42.5^\circ$) configurations, respectively.}

\label{fig:PI_curves}
\end{tcolorbox}
\vskip-15pt
\end{figure*}

A cosmological SGWB can be generated by a variety of early-Universe processes. Its observation would provide a novel probe of high-energy physics at energy scales well beyond those accessible to the Large Hadron Collider, while also delving into the physics of the earliest phases of cosmic history \cite{Caldwell:2022qsj, Maggiore:1999vm, Caprini:2018mtu}. 
Extensions of the Standard Model of particle physics and cosmology predict a multitude of phenomena that may leave observable imprints in the form of an SGWB. In the following, we briefly discuss three particularly relevant sources, namely inflation, first-order phase transitions (FOPTs), and topological defects. We then illustrate examples of the resulting SGWB spectra in Fig. \ref{fig:PI_curves} and compare them with the sensitivity of different geometries for the ET detector network. %

\textit{The inflation} paradigm predicts an early phase of accelerated expansion of the Universe \cite{Starobinsky:1979ty}, where quantum fluctuations of massless fields grow from sub-Hubble to super-Hubble scales. Tensor perturbations that re-enter the Hubble radius after inflation turn into an irreducible SGWB \cite{Grishchuk:1974ny, Caprini:2018mtu}. The resulting energy density spectrum from the standard slow-roll inflation is likely to be too faint to be detected by ET. However, the physics of inflation is currently unknown and a variety of mechanisms that may occur during or after inflation can enhance the inflationary SGWB amplitude, potentially within the reach of ET. Examples include models of inflation driven by axion inflatons \cite{LIGOScientific:2025kry} and an additional phase with a stiff equation of state in a post-inflationary epoch \cite{PhysRevD.78.043531,Duval:2024jsg}. The corresponding SGWB spectra are illustrated in Fig. \ref{fig:PI_curves}.

\textit{Cosmological FOPTs} are a prediction of multiple extensions of the Standard Model of particle physics exhibiting spontaneous symmetry breaking. Such transitions may have occurred within the first trillionth of a second after the Big Bang, at energy scales even above the electroweak scale %
\cite{Witten:1984rs, Caprini:2024ofd}. As the Universe cools down, it can transition to a new energy minimum, the true vacuum, that is separated from the previous metastable minimum, the false vacuum, by a potential energy barrier. 
This happens through the nucleation of bubbles of true vacuum at different locations throughout the Universe. The bubbles expand, then collide, and eventually fill the entire space, generating GWs via bubble collisions, %
sound waves propagating through the early-Universe plasma,
and magnetohydrodynamic turbulence \cite{Caprini:2018mtu}. %
The resulting SGWB may be accessible to ET, as illustrated in Fig. \ref{fig:PI_curves}. Such phase transitions could be associated with the electroweak symmetry breaking (EWSB) and supersymmetry breaking, among others. In all cases, detection of this background would be a unique opportunity to probe regions of particle-physics parameter space that are otherwise inaccessible to, or complementary to, current and future planned colliders, possibly shedding light on outstanding questions such as the nature of dark matter and the origin of the matter--antimatter asymmetry of the Universe~\cite{SNOWMASS_White_Paper}.

\textit{Topological defects} are extended structures that may form as the aftermath of a phase transition in the early Universe. 
As an example, cosmic strings \cite{Kibble:1976sj, Jeannerot:2003qv} are one-dimensional topological defects that produce a dynamically evolving network characterized by intersection and intercommutation of strings and the production of closed string loops. The loops decay by emitting GWs, with additional GW bursts from the cusps, kinks, and kink-kink collisions %
\cite{Blanco-Pillado:2013qja, Lorenz:2010sm}. The amplitude of the resulting SGWB depends on the string tension, which encodes the energy scale associated with the underlying symmetry-breaking phase transition. Fig. \ref{fig:PI_curves} shows that the SGWB from cosmic strings is a promising cosmological SGWB that ET can target. A detection would provide a direct probe of the physics of the phase transition.

\textit{Sensitivity} -- Fig. \ref{fig:PI_curves} shows the \textit{power-law integrated} (PI) sensitivity curves \cite{Thrane:2013oya} for isotropic SGWBs for different geometries of the ET network. 
The best sensitivity in the $1$--$100\,\mathrm{Hz}$ frequency band is achieved by the 2L ($-2.5^\circ$) configuration, whereas the triangular configuration provides the best sensitivity at frequencies between approximately $100$ and $2000\,\mathrm{Hz}$. The 2L configuration with a relative angular displacement of $42.5^\circ$ is instead one or more orders of magnitude less sensitive than the 2L $(-2.5^\circ)$ configuration. %
Assuming a signal-to-noise ratio (SNR) of 5 as a detection threshold and one year of observation, the best sensitivities are approximately $2\times10^{-12}$ for the triangular configuration, $7.3\times10^{-13}$ for the 2L $(-2.5^\circ)$ configuration, and $8.3\times10^{-12}$ for the 2L $(42.5^\circ)$ configuration.

The strong dependence of the SGWB sensitivity on the relative orientation of the arms of the two detectors in the 2L configuration is highlighted in the right panel of Fig. \ref{fig:PI_curves}. We stress the presence of a critical angle around $45^\circ$,\footnote{This angle differs from that reported in \cite{Branchesi:2023mws} because the two L detectors considered here lie on the same meridian. In this configuration, the critical angle is expected to occur exactly at $45^\circ$.} near which the sensitivity to an isotropic SGWB deteriorates by several orders of magnitude, becoming worse than that of the current ground-based GW-detector network. Hence, choosing a near-$45^\circ$ relative orientation of the two L-shaped detectors severely compromises the SGWB science potential of the ET network and is unsuitable for pursuing this type of science. We also note the possibility of an intermediate 2L orientation choice that may optimize the science potential of the ET network \cite{Bruno:2026rwf}: 
a relative orientation of $30^\circ$ between the two L-shaped detectors would degrade the SGWB sensitivity by only a factor of $\sim 2$ relative to the optimal 2L configuration, while still enabling other science goals discussed in this document. Further, the central panel of Fig. \ref{fig:PI_curves} compares the sensitivity in the $30^\circ$-orientation configuration with that of the triangle geometry, showing that the two configurations perform similarly in the most sensitive $10$-$100$ Hz band. 

\textit{Correlated noise} -- %
The above analysis does not account for the possible correlated noise between ET detectors \cite{Branchesi:2023mws}, which can arise due to instrumental effects or environmental contamination (seismic noise, magnetic noise, Newtonian noise, etc.), as discussed in Appendix \ref{sec:NoiseModels}. As in the existing GW detector network, the geographic separation between two L-shaped detectors will effectively suppress most of these noise sources. The notable exception are the Schumann magnetic resonances in the Earth's ionosphere \cite{1952ZNatA...7..250S, 1952ZNatA...7..149S}, but these can be effectively monitored with sensitive magnetometers \cite{Coughlin:2018tjc}, and hence partially subtracted from the data, although residual correlated magnetic noise may still degrade the sensitivity of both geometries below $\sim 10\, \mathrm{Hz}$ as shown in Fig. \ref{fig:excess_omega}. Restricting the SGWB search to frequencies above 10 Hz would degrade the SGWB sensitivity by $\sim 2\times$ for each geometry and for the frequency-independent energy density. The situation is much more complex in the case of the triangle ET configuration, where local seismic and Newtonian noise can generate strong additional correlations between different detectors in the triangle, in the most sensitive frequency band. Mitigating such correlations would require a sophisticated network of environment monitors (accelerometers, tiltmeters etc), as well as modern (likely AI-based) methods for using these monitors to filter the GW data channels. Fig. \ref{fig:excess_omega} projects the effectiveness of such mitigation techniques with a large uncertainty, indicating that correlated noise will likely not be fully mitigated at frequencies below $\sim 40$ Hz. Restricting the SGWB search to frequencies above 40 Hz would degrade the SGWB sensitivity by a factor of $\sim 25$ for the triangle configuration for the frequency-independent energy density \cite{Branchesi:2023mws}.

Although the cosmological SGWB is one of the prime science targets of GW observations, its detection with ET is complicated by the presence of (likely more intense) astrophysical stochastic foregrounds. Characterization and subtraction of astrophysical foregrounds will, therefore, be critical in detecting the cosmological SGWB, as discussed next.

\subsubsection{Compact-binary coalescence foregrounds}
\begin{figure*}[!htbp]
\begin{tcolorbox}[standard jigsaw,colframe=gray,colback=gray!10!white,opacityback=0.0,coltext=black, title=\small\sc $\Omega_\mathrm{gw}(f)$ for three types of CBCs]
\center{\includegraphics[width=0.32\textwidth]{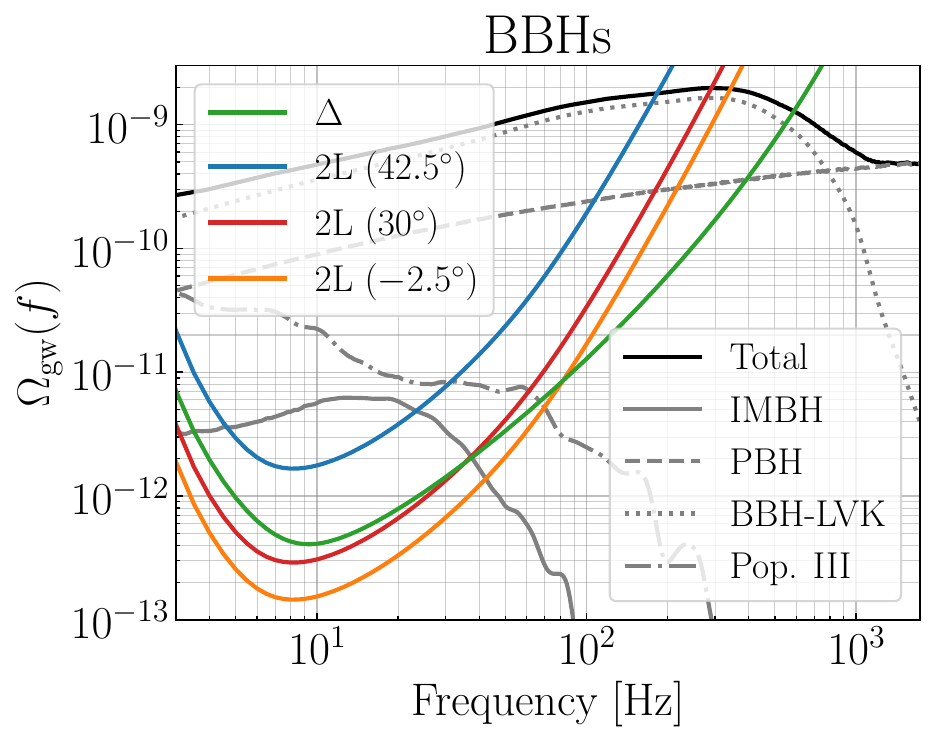}
\includegraphics[width=0.32\textwidth]{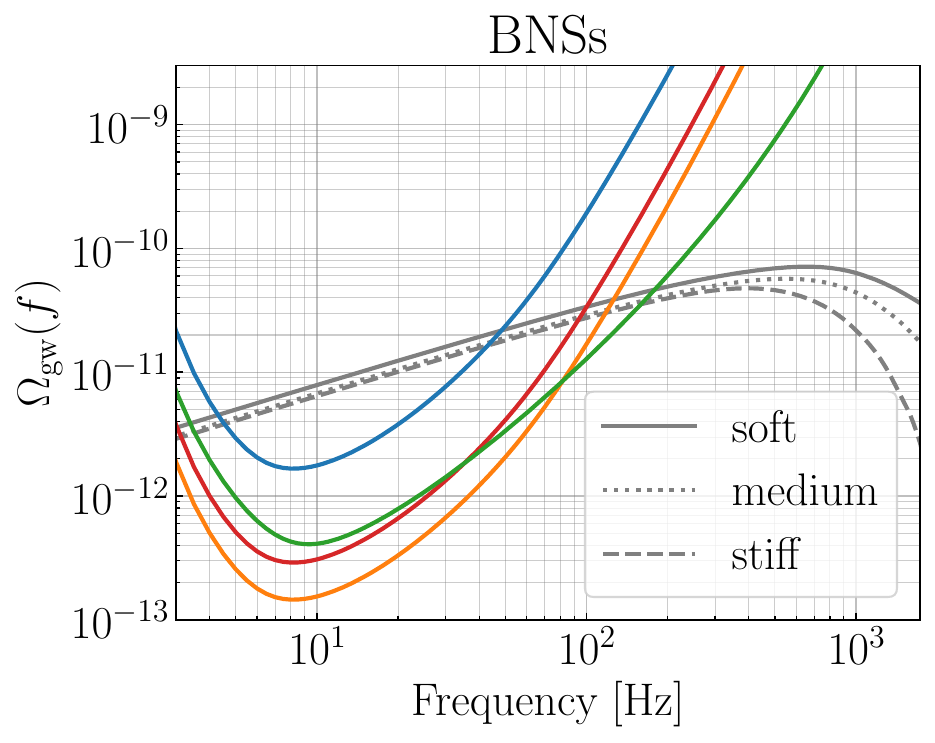}
\includegraphics[width=0.32\textwidth]{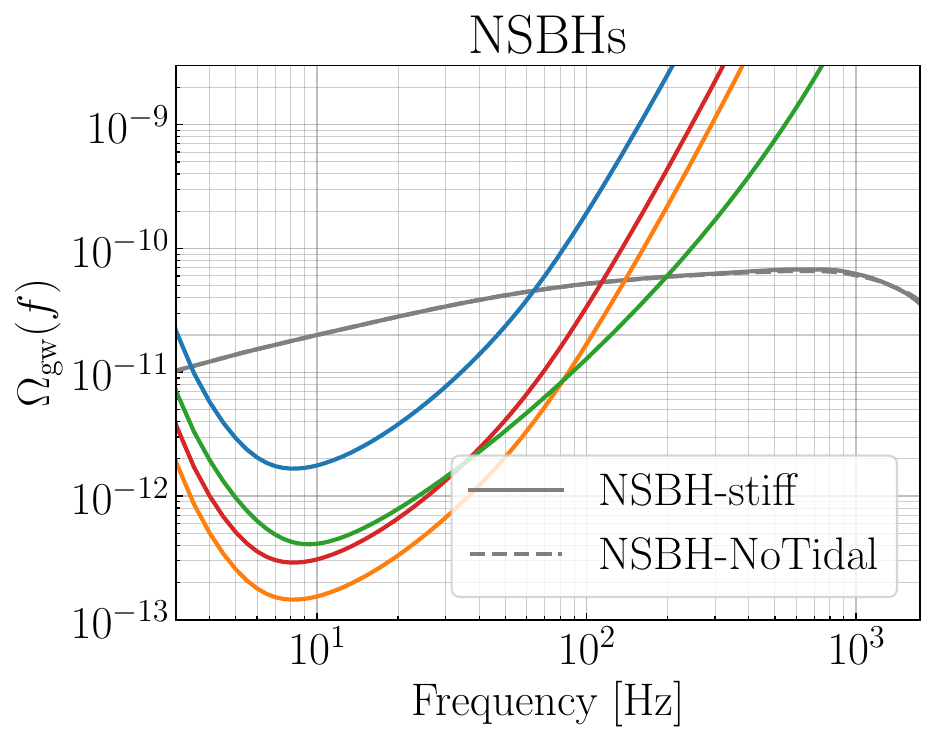}}
\caption{\justifying Forecast of the dimensionless energy density spectra of the astrophysical foreground from all three classes of CBC systems. The left, middle, and right panels correspond to the BBH, BNS, and NSBH foregrounds, respectively. In each panel, the colored curves show the SNR = 1 power-law integrated sensitivity curves for different ET detector geometries, with orange, red, blue, and green curves corresponding to the 2L $(-2.5^\circ)$, 2L $(30^\circ)$, 2L $(42.5^\circ)$, and triangular configurations, respectively.}

\label{fig:Omega_before_subtraction}
\end{tcolorbox}
\vskip-15pt
\end{figure*}
Astrophysical foregrounds can also be generated by a variety of sources, including unresolvable continuous waves~\cite{Ferrari:1998jf}, supernova bursts~\cite{SN_SGWB}, and other unmodeled long-lasting transient signals of astrophysical origin~\cite{Thrane:2010ri}. Most of these are expected to be too weak to be observed by ET---the notable exception is the superposition of GWs emitted by CBC events, which is one of the primary science targets of both current and next-generation SGWB searches~\cite{LIGOScientific:2026mjf}.
In this section, we exclusively focus on the astrophysical GW foreground emitted by compact binary coalescences.

Given the CBC populations described in Sec.~\ref{sec:population}, we compute the corresponding dimensionless energy density spectrum $\Omega_\mathrm{gw}(f)$ for all three types of CBC systems, shown in Fig.~\ref{fig:Omega_before_subtraction}. This result is independent of the detector geometry, depending only on the details of the population. 
The left panel of Fig.~\ref{fig:Omega_before_subtraction} shows the BBH foreground broken down by subpopulations, with the black solid curve giving the total obtained by summing $\Omega_\mathrm{gw}(f)$ over all subpopulations. Notably, BBHs-LVK dominate the BBH foreground across the band from $5$ to $800\, \mathrm{Hz}$, while the PBH foreground takes over above $800\,\mathrm{Hz}$. The IMBH and Pop. III foregrounds are orders of magnitude weaker than the other two components across the entire frequency range.
The middle panel presents the BNS foreground for three EOS models. The three curves share the same overall shape: $\Omega_\mathrm{gw}(f)\propto f^{2/3}$ at low frequencies, corresponding to the inspiral phase of the coalescence, followed by a turnover at high frequencies. The soft EOS model yields the largest BNS foreground, roughly 23\% greater than the other two models at low frequencies. The turnover frequencies differ between models: the solid curve (soft EOS model) turns over at the highest frequency, around $650\,\mathrm{Hz}$, whereas the dotted (medium EOS model) and dashed curves (stiff EOS model) turn over around $550$ and $400\,\mathrm{Hz}$, respectively.
The right panel shows the NSBH foreground. The solid curve corresponds to the case where the tidal effects are modeled, and the dashed curve to the case where they are neglected. 
The absolute difference between the two curves lies orders of magnitude below the foreground itself, indicating that tidal effects have only a minimal impact on $\Omega_\mathrm{gw}(f)$ for NSBH mergers. 

\begin{minipage}{0.67\textwidth}
\begin{tcolorbox}[standard jigsaw,colframe=gray,colback=gray!10!white,opacityback=0.0,coltext=black, title=\small\sc Residual background from LVK-confirmed CBCs for four configurations]
\center{\includegraphics[width=0.49\textwidth]{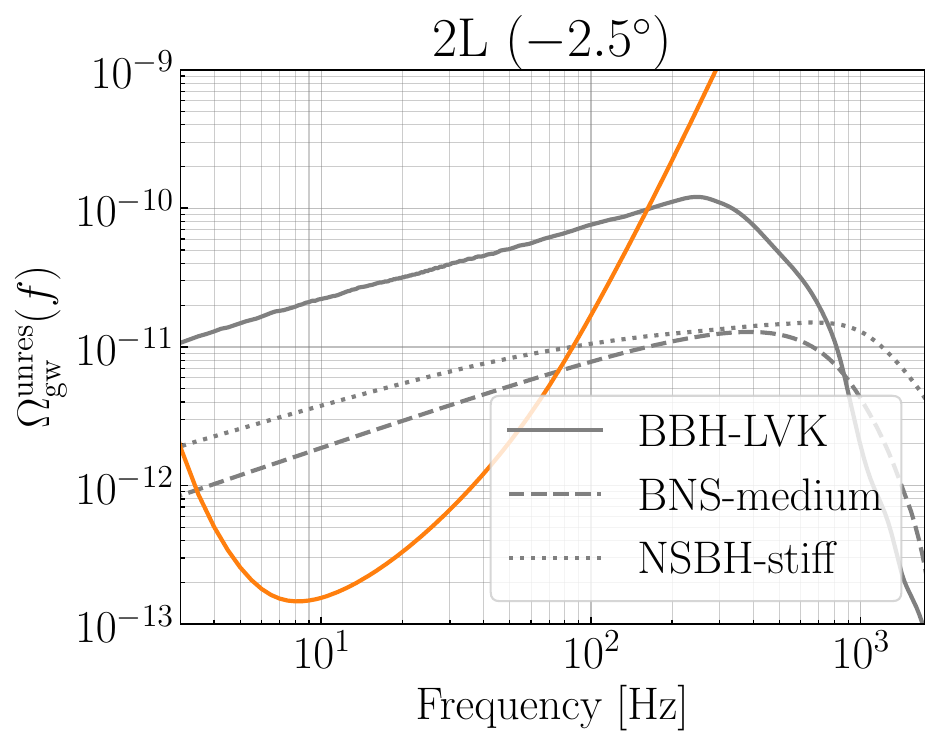}
\includegraphics[width=0.49\textwidth]{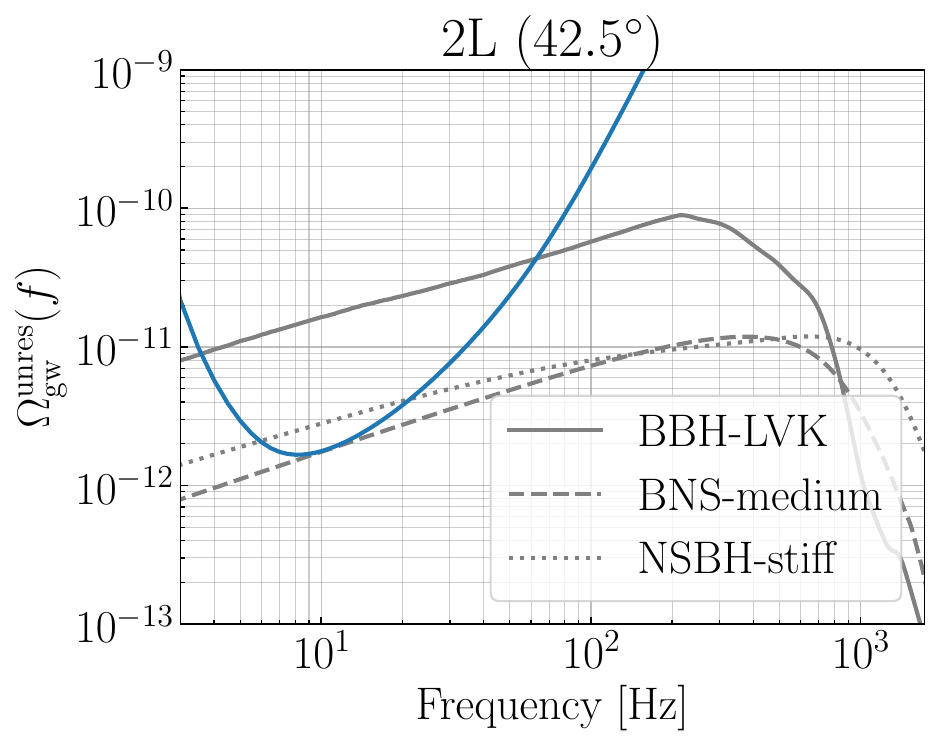}\\
\includegraphics[width=0.49\textwidth]{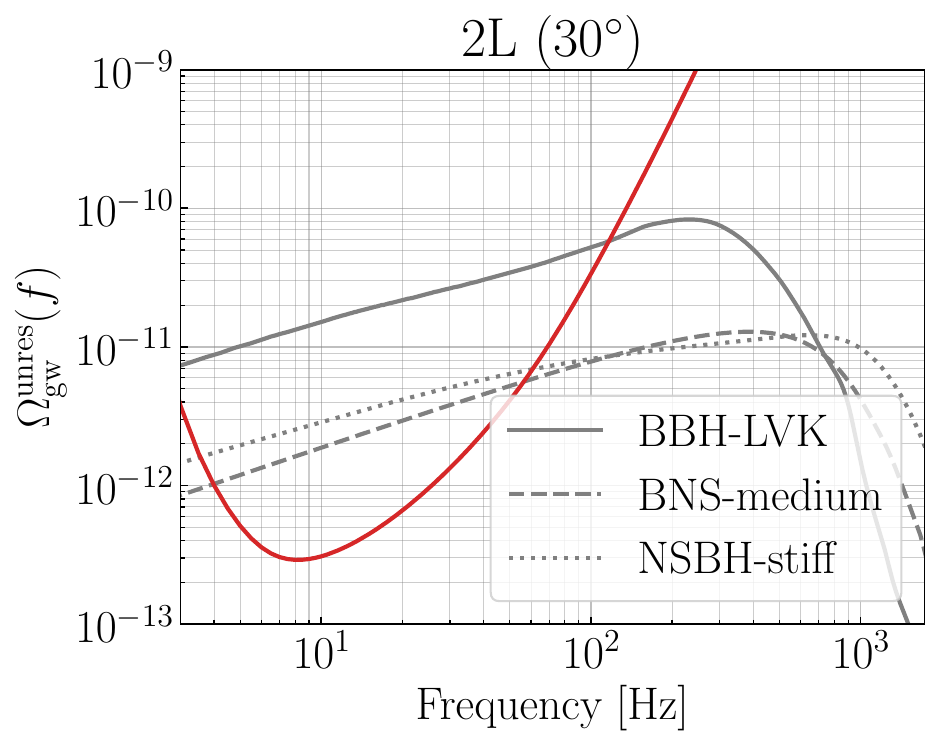}
\includegraphics[width=0.49\textwidth]{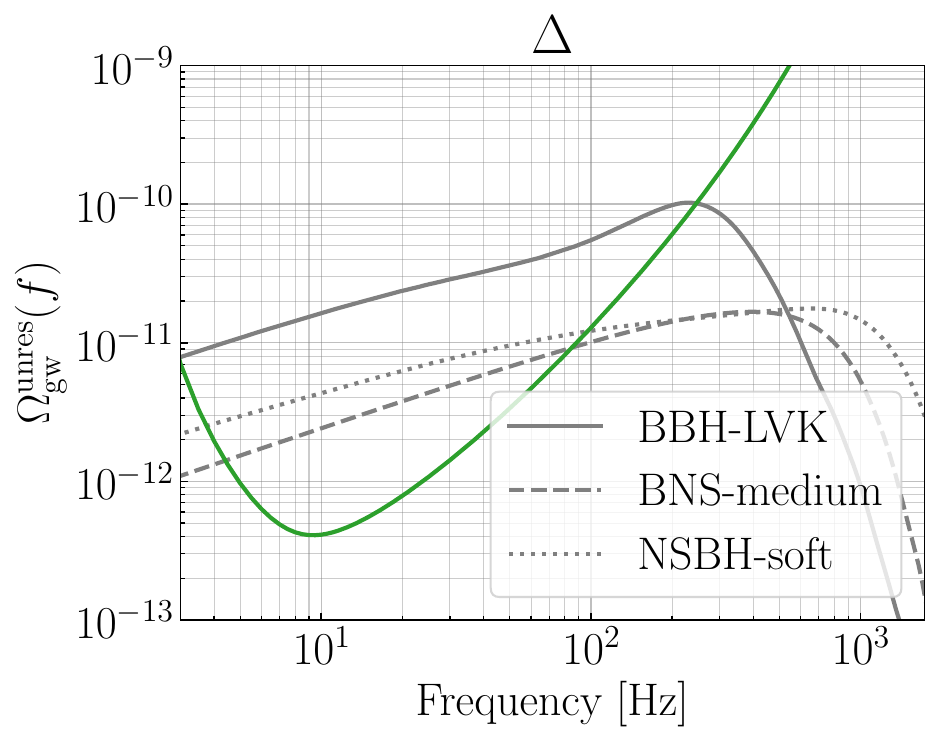}}
\captionof{figure}{\justifying Forecast of the residual background from LVK-confirmed CBCs for four configurations. The four panels correspond to the 2L $(-2.5^\circ)$, 2L $(42.5^\circ)$, 2L $(30^\circ)$,  and triangular configurations, respectively. The colored solid curves show the SNR = 1 power-law integrated sensitivity curves for each case.}

\label{fig:Omega_after_subtraction_regular}
\end{tcolorbox}
\end{minipage}
\hfill
\begin{minipage}{0.3\textwidth}

Comparing the three panels, the BBH foreground exceeds the BNS and NSBH foregrounds by at least an order of magnitude across the band, and is therefore the dominant contribution to the total CBC foreground.
In all three panels we also overlay the PI curves for the four geometrical configurations, where 2L $(-2.5^\circ)$, 2L  $(30^\circ)$, 2L $(42.5^\circ)$, and triangular configurations are shown in orange, red, blue, and green, respectively.
Every subpopulation curve rises above the PI curves in at least a part of the frequency band, indicating that the astrophysical GW foreground from every (sub)population lies within reach of all four geometrical configurations. We also compute the cross-correlation SNR assuming one-year of observation of BBH, BNS (soft), and NSBH foregrounds for 2L $(-2.5^\circ)$, 2L $(42.5^\circ)$, and $\Delta$ configurations. For the BBH foreground, we obtain 2300 for 2L $(-2.5^\circ)$, 200 for 2L $(42.5^\circ)$, and 880 for $\Delta$; for the BNS (soft) foreground, we obtain 51, 4.5, and 20, respectively; for the NSBH foreground, we obtain 129, 11, and 49, respectively.
\end{minipage}

The above results indicate a promising outlook for observing the GW foreground of astrophysical origin, but they also raise another challenge: the astrophysical GW foreground becomes a new contaminant in the measurement of the cosmological GW background. 
In practice, a fraction of CBC events are individually resolvable, i.e. they have a matched-filtering SNR above a chosen threshold, and these can be removed through the notching procedure~\cite{Zhong:2022ylh}, leaving the remaining data for further analysis. To provide a forecast of the sensitivity of the four ET configurations to the SGWB after removing resolvable CBC events, we make the strong assumption that notching is perfect so that all resolvable events are removed without introducing any error or residual. Under this assumption, we compute the residual background for all populations and all four geometrical configurations. In this subsection, we focus on the BNS, NSBH, and the BBH-LVK populations, which are better understood than those of the Pop. III, IMBH, and PBH components; the latter are discussed in the following subsection.

We show the results in Fig.~\ref{fig:Omega_after_subtraction_regular}, where the four panels correspond to the 2L $(-2.5^\circ)$, 2L $(42.5^\circ)$, 2L $(30^\circ)$, and triangular configurations, respectively. The magnitude of the residual backgrounds is nearly identical across the four cases, whereas the detector sensitivities indicated by the PI curves differ significantly. Except for the 2L $(42.5^\circ)$ configuration, all residual SGWB components remain detectable; disentangling them and searching for a cosmological SGWB would then require the joint analysis framework proposed in~\cite{Zhong:2025qno}.
To perform the joint search analysis, one combines the individual detections of CBC events and the cross-correlation SGWB search data to simultaneously constrain the CBC population hyperparameters and the parameters that control the cosmological SGWB model. In Ref.~\cite{Zhong:2025qno}, the overall sensitivity of the joint analysis framework to the SGWB of cosmological origin was found to be only $\sim 2\times$ weaker than the idealized case in which the sensitivity is limited purely by detector noise (i.e no astrophysical foreground present). We expect a comparable retention of sensitivity in our case when performing the joint analysis.
In Table~\ref{summary-table}, we assume a factor-of-$2$ degradation in sensitivity to the primordial SGWB from the joint analysis, relative to the ideal case in which no foreground is present. A more precise evaluation remains for future investigation.
The situation is qualitatively different for the 2L $(42.5^\circ)$ case, where the PI curve only marginally touches the dashed (BNS) curve, so that detector noise, rather than the astrophysical foreground, becomes the limiting factor in searching for a cosmological SGWB.

\subsubsection{Other astrophysical foregrounds}
\hfill
\begin{minipage}{0.67\textwidth}
\begin{tcolorbox}[standard jigsaw,colframe=gray,colback=gray!10!white,opacityback=0.0,coltext=black, title=\small\sc Residual background from three putative subpopulations of BBHs for four configurations]
\center{\includegraphics[width=0.49\textwidth]{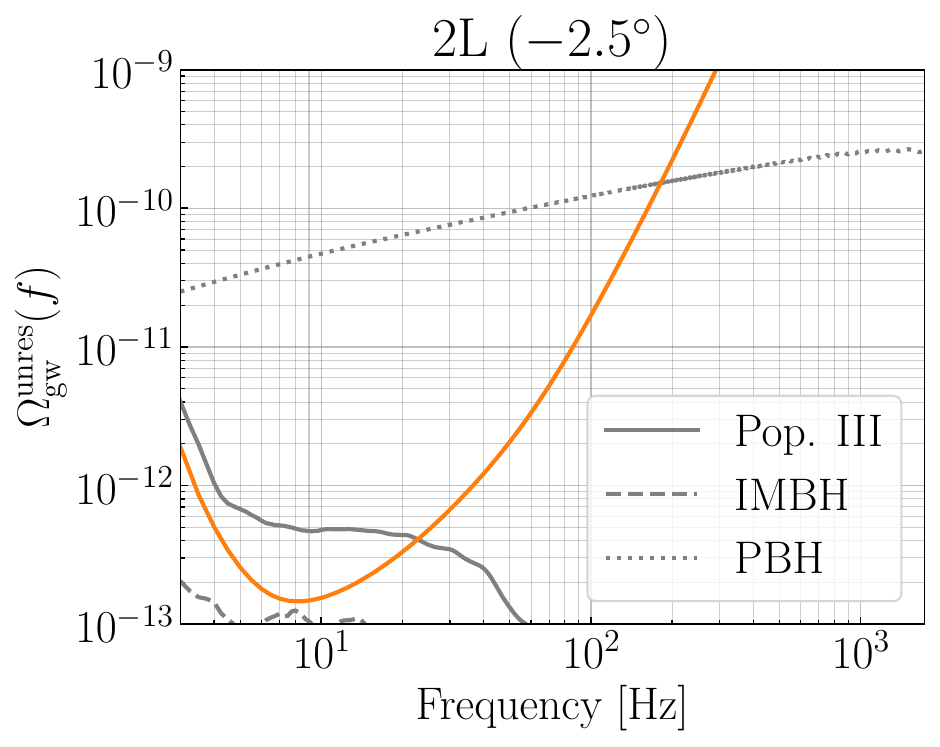}
\includegraphics[width=0.49\textwidth]{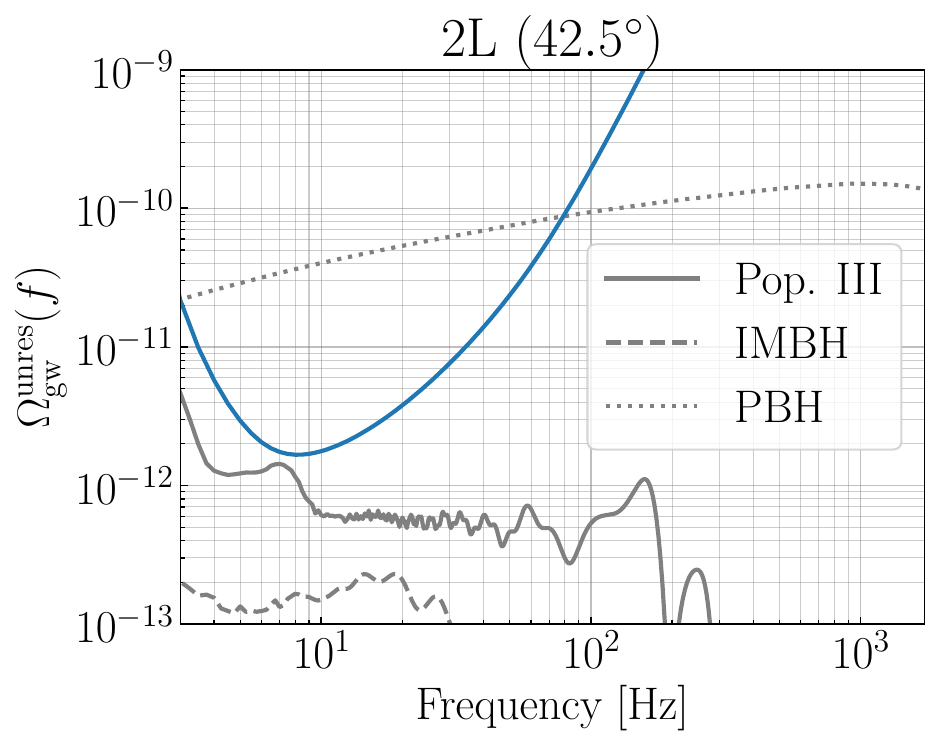}
\includegraphics[width=0.49\textwidth]{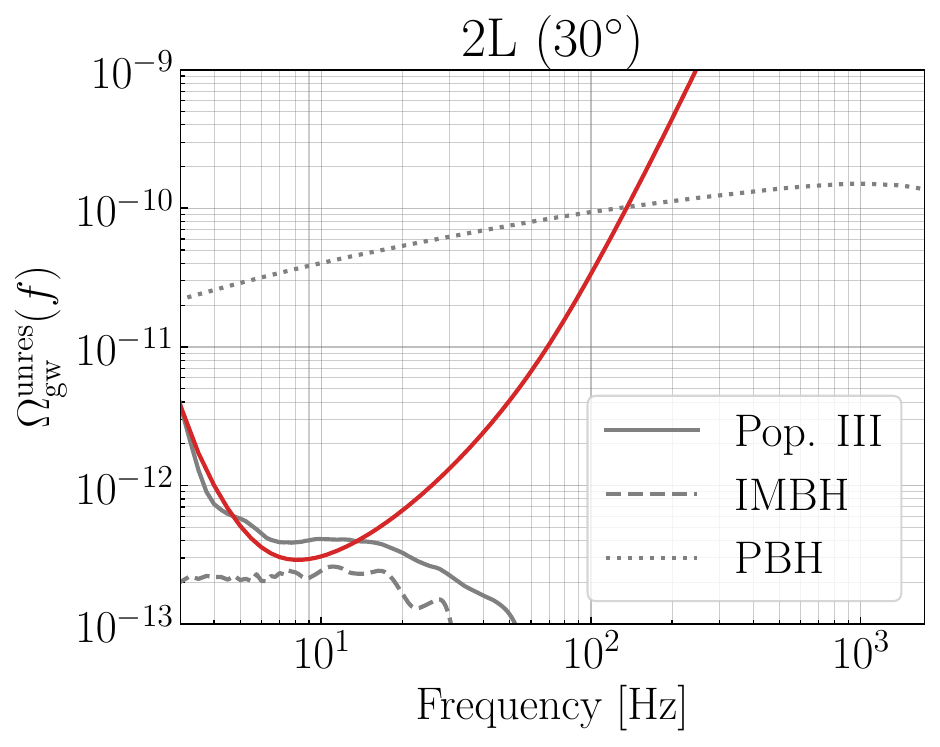}
\includegraphics[width=0.49\textwidth]{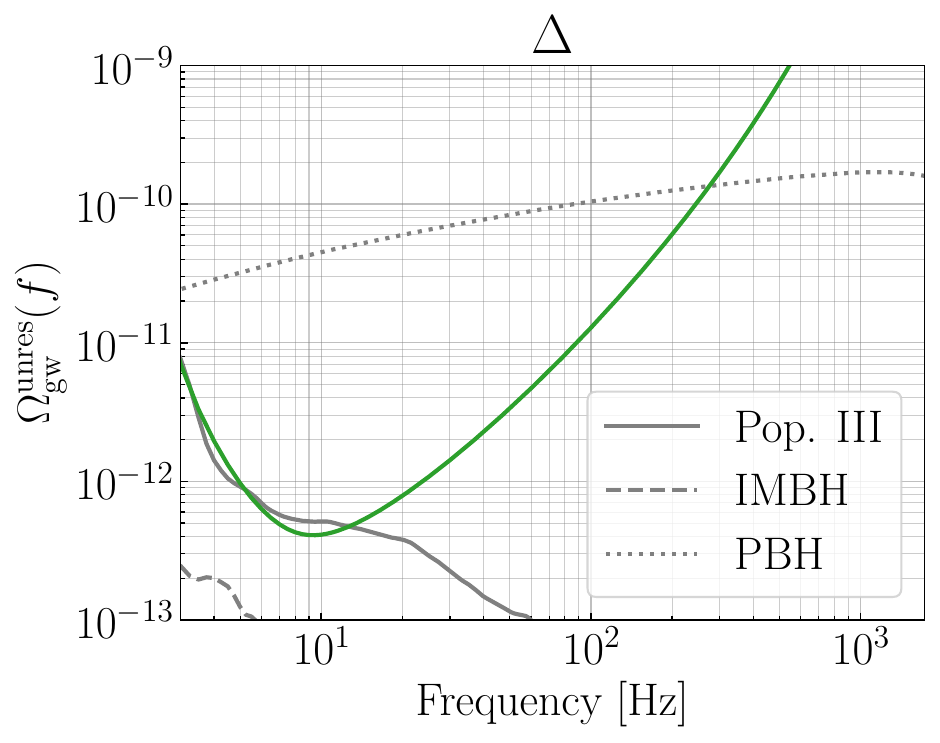}}
\captionof{figure}{\justifying Similar to Fig.~\ref{fig:Omega_after_subtraction_regular}, but focusing on the putative subpopulations of BBHs.}

\label{fig:Omega_after_subtraction_exotic}
\end{tcolorbox}
\end{minipage}
\hfill
\begin{minipage}{0.3\textwidth}
As in Fig.~\ref{fig:Omega_after_subtraction_regular}, we show the results for Pop. III, IMBH, and PBH foregrounds in Fig.~\ref{fig:Omega_after_subtraction_exotic}. 
We find that the unresolvable IMBH foreground is out of reach for all four geometrical configurations, while the PBH foreground would be detectable for all four cases. The Pop. III foreground could potentially be probed by all configurations except the 2L $(42.5^\circ)$ configuration, where the Pop. III foreground falls slightly below the PI curve. We stress that the existence of these three subpopulations remains uncertain, so the forecast shown here should be regarded as an order-of-magnitude estimate rather than a firm prediction. Nonetheless, if they do exist, the GW foreground they generate may be comparable to, or even stronger than, the BNS and NSBH foregrounds. A more thorough analysis based on a joint search would then be required. We leave this for future investigation.
\end{minipage}

\begin{tcolorbox}[colback=teal!5!white,colframe=red!75!black,title=\sc Geometry Comparison]
{ \begin{itemize}
    \item  {\em Sensitivities:} Based on geometry alone, the 2L $(-2.5^\circ)$, the 2L $(30^\circ)$, and the triangle configurations would all have a transformative impact on the searches for the stochastic GW background. The 2L $(42.5^\circ)$ configuration is $10\times$ less sensitive than the 2L $(-2.5^\circ)$ configuration, resulting in a significantly smaller science potential. The triangle and the 2L $(30^\circ)$ configurations are, respectively, about $3\times$ and $2\times$ less sensitive than the 2L $(-2.5^\circ)$ configuration, which maintains their substantial SGWB discovery potential while still supporting other science goals discussed in this document. However, correlated noise sources could significantly impact the triangle configuration sensitivity, and may require an extensive sensor network for monitoring and mitigation of these noise sources.
    \item  {\em Cosmological backgrounds:} A broad variety of cosmological background models, including inflationary models, phase transitions, and topological defects, will be accessible to ET 2L $(-2.5^\circ)$ and $(30^\circ)$ configurations, and to the triangle configuration depending on the mitigation of correlated noise. The 2L $(42.5^\circ)$ configuration will have a significantly smaller reach.
    \item  {\em CBC foregrounds:} All CBC foregrounds will be detectable with all ET configurations studied here, with the highest SNRs in the 2L $(-2.5^\circ)$ configuration and the lowest SNRs in the 2L $(42.5^\circ)$ configuration.
    \item  {\em Unresolved CBC foregrounds:} Unresolved CBC events will form a foreground for the cosmological GW background, for all ET configurations. This foreground could be significantly reduced using the joint analysis framework, to within a factor of $\sim 2$ of the detector sensitivity.
    \item  {\em Other unresolved astrophysical foregrounds:} All four ET configurations will be able to search for other unresolved astrophysical foregrounds, with the 2L $(-2.5^\circ)$ configuration being the most sensitive. In particular, the unresolved Pop. III foreground may be detectable with the 2L $(-2.5^\circ)$, 2L $(30^\circ)$, and triangle configurations, but not with the 2L $(42.5^\circ)$ configuration. The unresolved IMBH foreground is expected to be out of reach for all four configurations.     
\end{itemize}
}
\end{tcolorbox}

\subsection{Multimessenger astronomy with gravitational waves} \label{sec:mma}

\begin{tcolorbox}[colback=teal!5!white,colframe=yellow!75!black,title=\sc Science Target]
{\em Understand the astrophysics of neutron star mergers through joint gravitational-wave and electromagnetic observations, from the identification of their counterparts to their evolution across cosmic time.}
\end{tcolorbox}

\begin{figure*}
\begin{tcolorbox}[standard jigsaw,colframe=gray,colback=gray!10!white,opacityback=0.0,coltext=black, title=\small\sc Schematic overview of multimessenger observations of NS mergers]
\center{\includegraphics[width=1\textwidth]{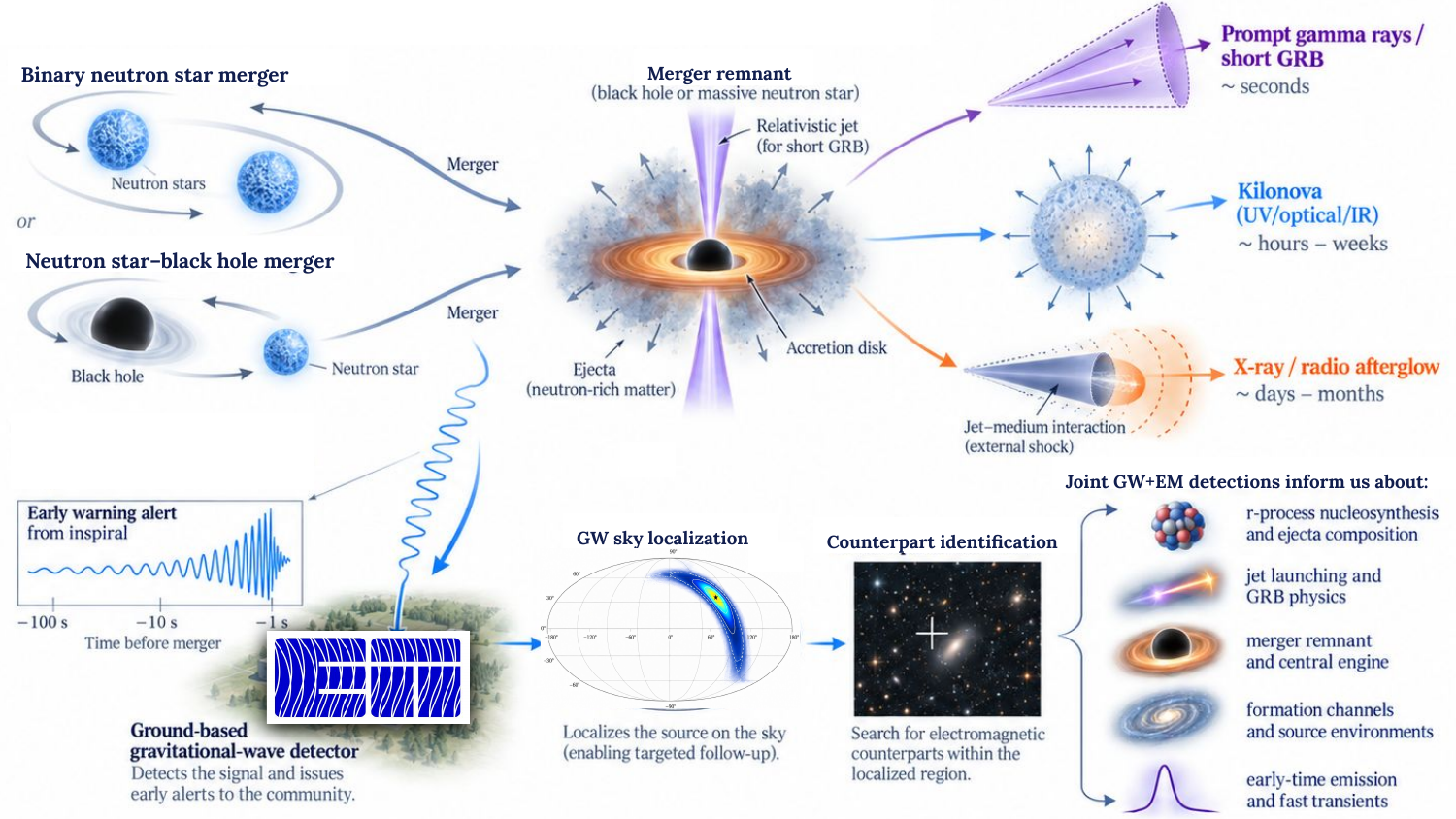}}
\caption{\justifying BNS and NSBH mergers produce GWs that can be detected and localized before or around merger, enabling targeted searches for electromagnetic counterparts such as short GRBs, kilonovae, and X-ray or radio afterglows. Combining the GW signal with these electromagnetic counterparts gives a much more complete picture of the merger. The GW signal tells us about the masses and geometry of the binary, while the electromagnetic emission reveals what happens to the matter after the collision: how much material is ejected, whether a relativistic jet is launched, what compact remnant is left behind, and how the outflow interacts with its surroundings. Together, these observations can reveal how heavy elements are synthesized, how short GRBs and their jets are produced, and how NSs and BHs form and evolve.} \label{fig:mma_schem}
\end{tcolorbox}
\vskip-15pt
\end{figure*}

Gravitational waves and electromagnetic radiation probe complementary aspects of NS mergers. The GW signal measures the dynamics of the compact binary and constrains binary properties, while the accompanying electromagnetic emission carries information about the material expelled during and after the merger and the physical processes governing the remnant. Observing both messengers from the same event makes it possible to connect the properties of the binary before merger to the phenomena that follow it, including kilonova emission, relativistic jets and short GRBs, and heavy-element nucleosynthesis. Hence, multimessenger observations are essential for building a coherent physical picture of NS mergers, from the inspiral through the merger and its electromagnetic aftermath.

\begin{tcolorbox}[standard jigsaw,colframe=ocre,colback=ocre!10!white,opacityback=0.0,coltext=black,title=\sc Key Science Goals]
\small
Multimessenger observations require GW detections to be translated into well-localized and timely targets for electromagnetic facilities. The key science goals span detailed studies of nearby mergers, joint observations across cosmic time, and rapid follow-up of transient emission.
\begin{itemize}[leftmargin=*]
\item {\bf Nearby counterparts and heavy-element nucleosynthesis:} Localize nearby NS mergers sufficiently well to identify their optical and infrared counterparts and establish their role as sites of $r$-process nucleosynthesis.
\item {\bf Multimessenger observations across cosmic time:} Localize distant NS mergers sufficiently well to enable joint observations with GRBs and study the merger population, relativistic outflows, and their environments across cosmic time.
\item {\bf Early alerts and rapid follow-up:} Provide sufficiently early and precise localizations to enable prompt electromagnetic observations of the merger and its rapidly evolving counterparts, including the early blue kilonova emission.
\end{itemize}
\end{tcolorbox}

\begin{figure*}
\begin{tcolorbox}[standard jigsaw,colframe=gray,colback=gray!10!white,opacityback=0.0,coltext=black, title=\small\sc Localization of NS mergers and their associated redshift and SNR]
\center{\includegraphics[width=1\textwidth]{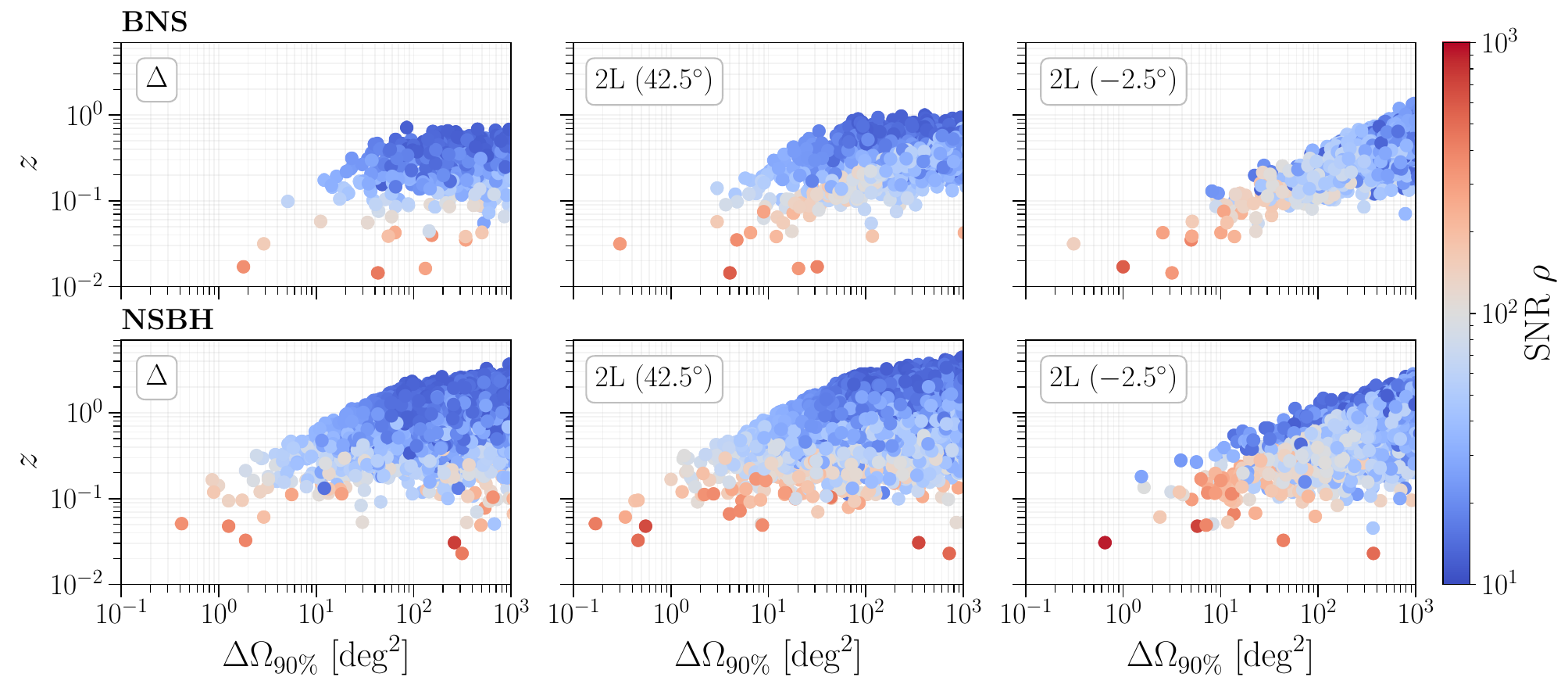}}
\caption{\justifying Redshift $z$ and 90\% credible sky-localization area, $\Delta\Omega_{90\%}$, for detected BNS (top) and NSBH (bottom) mergers in the three ET configurations. Each point represents a detected event and is colored by its SNR. The columns correspond to the triangular, misaligned 2L $(42.5^\circ)$, and the nearly aligned 2L $(-2.5^\circ)$ configurations, respectively.} \label{fig:mma_all}
\end{tcolorbox}
\vskip-15pt
\end{figure*}

Realizing the multimessenger science potential of NS mergers requires more than detecting their GW signals. The sources must also be localized sufficiently well for electromagnetic facilities to identify and follow their counterparts and, for rapidly evolving emission, sufficiently early for observations to begin around the time of merger. Hence, the sensitivity and geometry of ET will determine how many NS mergers can become useful multimessenger events, how far into the Universe such studies can be extended, and how rapidly electromagnetic follow-up can begin.

The sky-localization accuracy of a GW source depends on both the SNR of the signal and the detector geometry. Widely separated detectors provide longer baselines that improve localization, while their relative orientations help resolve the GW polarizations. For sufficiently long signals, the rotation of the Earth also modulates the detector response, providing additional localization information. This is particularly important for BNSs and low-mass NSBHs, which can remain in the ET band for hours. 

For both BNS and NSBH mergers, we use the 5-year catalog and report detections accordingly.
For BNS mergers, the three NS EOSs considered in this study produce very similar SNR and sky localization distributions (cf. Sec.~\ref{sec:population}). Since our focus here is on the impact of detector geometry on multimessenger performance, we show results only for the medium EOS.
For NSBH systems, we use parameter uncertainties obtained with waveforms that include higher-order modes but neglect tidal effects (see Appendix~\ref{app:methods}). In our simulations, tidal effects have a negligible impact on the sky localization. In contrast, higher-order modes can significantly improve the localization of asymmetric systems by providing additional angular information and helping to break parameter degeneracies. Higher-order multipoles can also enter the detector band before the dominant quadrupolar mode, making them particularly relevant for early localization.

Figure~\ref{fig:mma_all} illustrates how sensitivity and detector geometry jointly determine the localization of NS mergers. The longer arms of the 2L configurations yield systematically larger SNRs than the triangular configuration, allowing mergers to be detected out to larger redshifts. However, a larger SNR does not by itself guarantee better localization. The misaligned 2L $(42.5^\circ)$ configuration performs best overall because it combines the higher sensitivity of the 15-km detectors with a long geographical baseline and sufficiently different detector orientations to constrain both GW polarizations and break degeneracies among the extrinsic source parameters.

The nearly aligned 2L $(-2.5^\circ)$ configuration illustrates the importance of this geometrical information. Despite producing similarly large SNRs, the two detectors have limited ability to independently resolve the GW polarizations, leaving larger uncertainty in the inferred sky position. Nearby sources can still be localized well, aided by their high SNR and, for long-lived systems, the modulation of the detector response by the Earth's rotation. At $z\gtrsim1$, however, the loss of polarization information becomes increasingly important, and the triangular configuration can localize the sources more effectively despite its lower sensitivity.

The scientific importance of these differences depends on the counterpart being targeted and the timescale on which localization is required. In the following, we compare the configurations separately for nearby optical counterparts, multimessenger observations at cosmological distances, and pre-merger alerts for rapid electromagnetic follow-up.

\subsubsection{Nearby counterparts and heavy-element nucleosynthesis}
Kilonovae provide one of the most direct electromagnetic probes of the matter ejected in NS mergers. Neutron-rich material expelled during BNS and disrupting NSBH mergers undergoes rapid neutron capture, synthesizing heavy $r$-process nuclei. As these unstable nuclei decay, a fraction of their energy is thermalized in the expanding ejecta and reprocessed into ultraviolet, optical, and infrared emission, producing a rapidly evolving kilonova. The luminosity, color, and evolution of this transient carry information about the mass, velocity, and composition of the ejecta.

\noindent
\begin{minipage}{0.72\textwidth}
\vskip10pt
\begin{tcolorbox}[standard jigsaw,colframe=gray,colback=gray!10!white,opacityback=0.0,coltext=black, title=\small\sc Localization of nearby NS mergers]
\center{\includegraphics[width=1.0\textwidth]{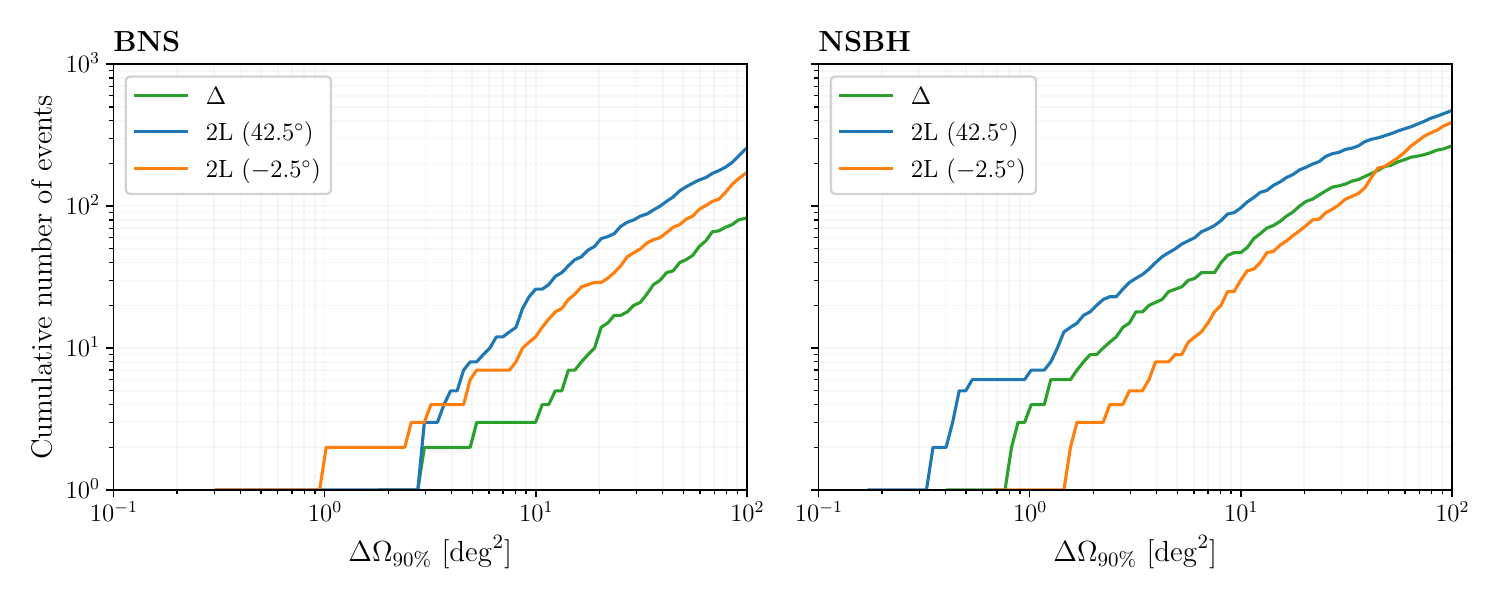}}
\captionof{figure}{\justifying Cumulative number of detected BNS (left) and NSBH (right) mergers at $z<0.4$ as a function of the 90\% credible sky-localization area, $\Delta\Omega_{90\%}$, for the triangular and the two 2L configurations. This redshift range approximately captures the volume within which optical counterparts can be targeted by current and next-generation facilities, making these events particularly relevant for kilonova observations and studies of heavy-element nucleosynthesis.} \label{fig:mma_z_0p4}
\end{tcolorbox}
\vskip10pt
\end{minipage}
\hfill
\begin{minipage}{0.24\textwidth}
For NSBH mergers, the production of an electromagnetic counterpart is conditional on the binary properties. The NS must be tidally disrupted before plunging into the BH, which requires the tidal disruption radius to lie outside the innermost stable circular orbit. 

\vskip10pt\hskip10pt
A massive BH tends to swallow the NS whole, leaving little material outside the horizon. Conversely, a rapidly spinning BH with its spin aligned with the orbital angular momentum has a smaller effective innermost stable circular orbit and can disrupt the NS at larger separations. 
\end{minipage}

The EOS also matters: a less compact NS, as generally predicted by a stiffer EOS, is more easily disrupted. Thus, the amount of matter available to power a kilonova is closely tied to the NS mass and EOS, the BH mass and spin, and the binary mass ratio.
Joint GW and kilonova observations make it possible to connect these binary properties directly to the ejecta. Building a population of such associations would help establish the contribution of NS mergers to the cosmic inventory of $r$-process elements and determine how the ejecta composition changes across the binary parameter space. For NSBHs in particular, the presence or absence of a counterpart can map the boundary between disruption and direct plunge, providing an independent probe of the BH mass and spin distribution and of the NS EOS.

Current studies indicate that optical and near-infrared facilities such as the Vera C.\ Rubin Observatory and the Nancy Grace Roman Space Telescope can detect kilonovae to redshifts of order $z\sim0.4$~\cite{Gupta:2023evt,Branchesi:2023mws}. Rubin has a field of view of $9.6\,{\rm deg}^2$, compared with $0.28\,{\rm deg}^2$ for Roman. However, Roman provides substantially greater depth and sensitivity in the near infrared, making the two facilities complementary. Motivated by this approximate observational reach, Fig.~\ref{fig:mma_z_0p4} shows the cumulative number of BNS and NSBH detections at $z<0.4$ as a function of their sky localization area. We restrict the comparison to $\Delta\Omega_{90\%}\leq100\,{\rm deg}^2$, corresponding to a region that Rubin could cover with of $\mathcal{O}(10)$ pointings.

The misaligned 2L $(42.5^\circ)$ configuration gives the largest set of well-localized nearby mergers for both source classes. It localizes 245 BNSs within $100\,{\rm deg}^2$ and 24 within $10\,{\rm deg}^2$, with one event localized to better than $1\,{\rm deg}^2$. For NSBHs, the corresponding numbers are 461, 92, and 7. The larger number of well-localized NSBHs reflects their larger total masses and, consequently, generally larger SNRs at the same distance. However, the systems that are easiest to detect and localize are not necessarily the most promising multimessenger sources. High-mass, strongly asymmetric NSBH binaries are also more likely to undergo direct plunge rather than tidal disruption.

Restricting the NSBH population to less asymmetric systems, for which an electromagnetic counterpart is more plausible, reduces the available sample substantially. Requiring $q\equiv m_{\rm NS}/m_{\rm BH}\geq1/4$, the misaligned 2L $(42.5^\circ)$ configuration yields 136 systems localized within $100\,{\rm deg}^2$, 20 within $10\,{\rm deg}^2$, and 2 within $1\,{\rm deg}^2$. Follow-up of the remaining, more asymmetric systems is nevertheless scientifically useful: a sufficiently deep non-detection can itself constrain the amount of ejecta and combinations of the BH spin, mass ratio, and NS compactness.

For BNS mergers, the nearly aligned 2L $(-2.5^\circ)$ configuration still benefits from the increased sensitivity of its longer arms. It localizes 175 events within $100\,{\rm deg}^2$ and 12 within $10\,{\rm deg}^2$, compared with 84 and 3 for the triangular configuration. At the much more restrictive $1\,{\rm deg}^2$ threshold, the counts are too small to establish a meaningful hierarchy, with 2 such detections for the 2L $(-2.5^\circ)$ configuration and none for the triangle. The comparison is clearer for NSBHs. The 2L $(-2.5^\circ)$ configuration benefits from its greater sensitivity and produces more events localized within $100\,{\rm deg}^2$, with 380 compared with 262 for the triangle. However, the triangle performs better when precise localization is required: the triangle localizes 47 NSBHs within $10\,{\rm deg}^2$, compared with 30 for the nearly aligned 2L configuration, and 4 versus 1 within $1\,{\rm deg}^2$. Thus, although the longer arms of the 2L $(-2.5^\circ)$ configuration increase the overall number of reasonably localized sources, the triangle's ability to resolve the GW polarizations makes it more effective for the best-localized NSBH events. The 2L $(42.5^\circ)$ configuration combines both advantages and performs substantially better, with 92 and 7 NSBHs localized within $10\,{\rm deg}^2$ and $1\,{\rm deg}^2$, respectively. Thus, the longer arms of an aligned 2L detector increase the number of detectable nearby systems, but substantial relative misalignment is required to fully exploit that sensitivity for precise sky localization.

The nearby and best-localized systems are also among the loudest events in Fig.~\ref{fig:mma_all}. As shown in Sec.~\ref{sec:population}, these high-SNR mergers will generally have their intrinsic parameters measured particularly well. A kilonova counterpart would then allow the observed ejecta properties to be associated with a well-characterized binary, providing precisely the joint information needed to connect merger dynamics, dense-matter physics, and $r$-process nucleosynthesis. Identifying the electromagnetic counterpart also fixes the source position and can identify its host galaxy, enabling an electromagnetic redshift measurement. Such events become bright standard sirens and will enable precise cosmological measurements (see Sec.~\ref{sec:cosmo}).

\subsubsection{Multimessenger observations across cosmic time}

Optical and infrared counterpart searches are largely restricted to the relatively nearby Universe, but high-energy transients from NS mergers can be detected to much greater distances. In particular, GRBs associated with relativistic jets can be observed at redshifts $z>2$, well beyond the typical reach of kilonova searches~\cite{Fong:2022mkv}. At these distances, the role of the GW localization is somewhat different: wide-field gamma-ray monitors can detect the burst independently, but a sufficiently constrained GW sky map is important for establishing a confident association and can also enable targeted searches for weak or sub-threshold gamma-ray emission. Joint detections at high redshift would extend multimessenger studies of NS mergers to earlier cosmic epochs, providing information on the evolution of the merger population and the environments in which these systems form.

Figure~\ref{fig:mma_z_2} shows the localization distributions in three redshift intervals, $0.5<z<1$, $1\leq z<2$, and $z\geq2$. As expected from Fig.~\ref{fig:mma_all}, the 2L $(42.5^\circ)$ configuration provides the best overall localization performance when comparatively small sky areas are required. For BNS mergers, it localizes 55 systems in the interval $0.5<z<1$ within $100\,{\rm deg}^2$, while no event at $1\leq z<2$ reaches the same localization accuracy. Relaxing the localization requirement to $1000\,{\rm deg}^2$ yields 1121 BNSs at $0.5<z<1$ and 170 at $1\leq z<2$, although none is retained at $z\geq2$. The triangular configuration is less effective, with only four BNSs at $0.5<z<1$ localized within $100\,{\rm deg}^2$, and 228 and 9 systems in the first two redshift intervals, respectively, within $1000\,{\rm deg}^2$.

For the nearly aligned 2L $(-2.5^\circ)$ configuration, no BNS at $0.5\leq z<1$ is localized within $100\,{\rm deg}^2$. At the broader $1000\,{\rm deg}^2$ threshold, however, its greater sensitivity yields 1040 systems at $0.5\leq z<1$, although only 12 at $1\leq z<2$. Thus, the misaligned geometry retains a clear advantage if useful intermediate-scale localization is required at $z\gtrsim1$. At still higher redshift, however, the larger detection reach of the 15-km interferometers becomes important. If the localization requirement is relaxed to $10^4\,{\rm deg}^2$, the 2L $(-2.5^\circ)$ configuration retains 637 BNS mergers at $z\geq2$, while the triangle retains none. Such sky areas are too large for conventional optical tiling, but remain useful for association with wide-field gamma-ray monitors such as Fermi-GBM.

Since prompt GRB emission is strongly beamed, only a fraction of these mergers will be favorably oriented for gamma-ray detection. As a simple proxy for this subset, we additionally select systems whose line of sight lies within $15^\circ$ of either direction of the orbital angular momentum (i.e., viewing angle $\Theta<15^{\circ}$). For BNS mergers localized within $1000\,{\rm deg}^2$, the misaligned 2L $(42.5^\circ)$ configuration retains 105 such systems at $0.5\leq z<1$, 22 at $1\leq z<2$, and none at $z\geq2$. The corresponding numbers are 17, 3, and 0 for the triangle, and 98, 3, and 0 for the 2L $(-2.5^\circ)$ configuration.

\noindent
\begin{minipage}{0.70\textwidth}
\begin{tcolorbox}[standard jigsaw,colframe=gray,colback=gray!10!white,opacityback=0.0,coltext=black, title=\small\sc Localization of NS mergers below and above redshift of 2]
\center{\includegraphics[width=1.0\textwidth]{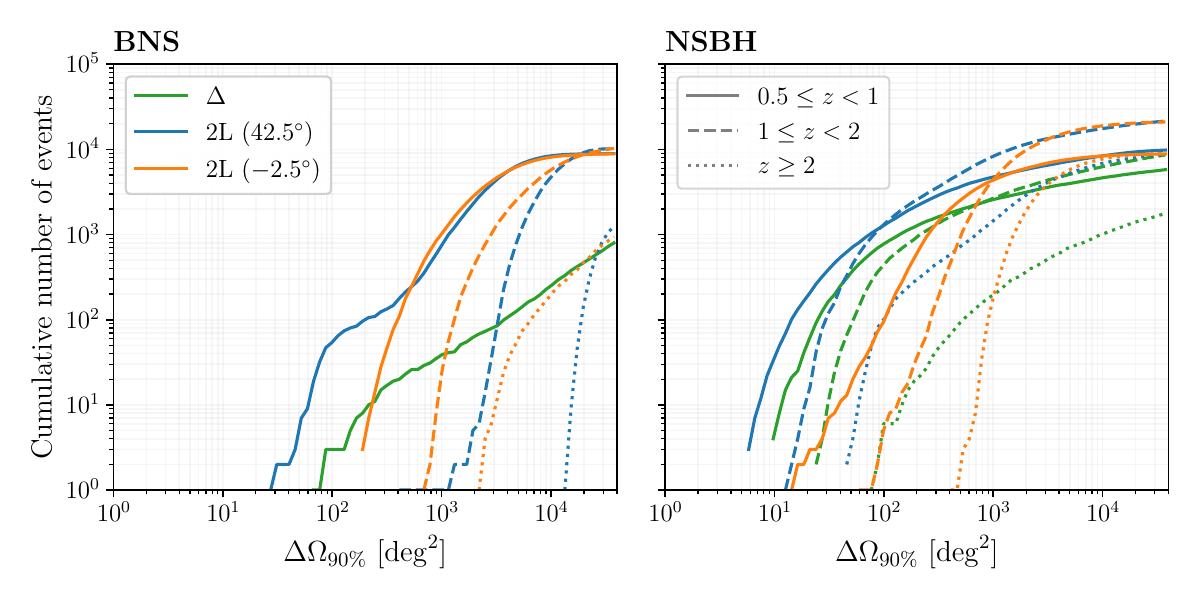}}
\captionof{figure}{\justifying Cumulative number of detected BNS (left) and NSBH (right) mergers as a function of the 90\% credible sky-localization area, $\Delta\Omega_{90\%}$, for the three ET configurations. The populations are separated into three redshift intervals, $0.5 \leq z<1$ (solid), $1\leq z<2$ (dashed), and $z\geq2$ (dotted), illustrating how the number of well-localized NS mergers evolves from the nearby Universe to cosmological distances.} 
\label{fig:mma_z_2}
\end{tcolorbox}
\end{minipage}
\hfill
\begin{minipage}{0.27\textwidth}
Thus, even after restricting to systems viewed close to the orbital axis, the misaligned geometry provides the largest population of usefully localized candidate BNS--GRB counterparts at $1\lesssim z\lesssim2$. None of the three configurations retains a BNS at $z\geq2$ within $1000\,{\rm deg}^2$ in this sample. If the localization requirement is relaxed to $10^4\,{\rm deg}^2$, however, the 2L $(-2.5^\circ)$ configuration retains 35 favorably oriented BNSs beyond $z=2$, while the triangle and 2L $(42.5^\circ)$ configurations retain none. Such very large sky areas can still be useful for association with all-sky gamma-ray monitors. The $15^\circ$ selection should be regarded only as a simple proxy for GRB visibility, since real jets may be structured and need not have a sharp opening angle. 
\end{minipage}

\vskip5pt
The advantage of the misaligned geometry is clearer for NSBH mergers, whose higher masses generally make them detectable with larger SNRs at a given redshift. At $z\geq2$, the misaligned 2L $(42.5^\circ)$ configuration localizes 91 NSBH mergers within $100\,{\rm deg}^2$, compared with five for the triangle and none for the aligned 2L configuration. Relaxing the localization requirement to $1000\,{\rm deg}^2$, the 2L $(42.5^\circ)$ configuration detects 1363 NSBHs beyond $z=2$.

As in the nearby sample, however, most of the tightly localized high-redshift systems are highly asymmetric and are less likely to tidally disrupt the NS. Requiring $q\geq1/4$ leaves no NSBH mergers at $z\geq2$ localized within $100\,{\rm deg}^2$ for any of the three main configurations. At the broader $1000\,{\rm deg}^2$ threshold, however, the 2L $(42.5^\circ)$ configuration retains 193 such systems beyond $z=2$, compared with 1 for the triangle and 41 for the 2L $(-2.5^\circ)$ configuration.

The inclination-selected sample remains sizable when the localization requirement is relaxed to $1000\,{\rm deg}^2$. For systems viewed within $15^\circ$ of either pole, the 2L $(42.5^\circ)$ configuration yields 370, 1095, and 346 NSBH mergers in the three redshift bins, respectively, including 127, 360, and 81 systems with $q\geq1/4$. The triangular configuration gives 308, 511, and 70 systems, of which 82, 73, and 1 satisfy the same mass-ratio cut, while the 2L $(-2.5^\circ)$ configuration yields 303, 619, and 45 systems, including 93, 203, and 15 with $q\geq1/4$. Thus, the high-redshift sample remains considerable even after requiring both a favorable viewing angle and a less asymmetric binary, with the misaligned geometry providing by far the largest population beyond $z=2$.

These results show that ET alone can provide useful localizations for searches for gamma-ray counterparts extending beyond $z=2$, particularly for NSBH mergers. The misaligned 2L $(42.5^\circ)$ geometry stands out when comparatively precise localization is required, while the longer arms of the nearly aligned 2L configuration can still retain large numbers of high-redshift sources once much broader localization areas are allowed. NSBH mergers provide the better-constrained high-redshift sample, including systems that are both favorably oriented and sufficiently symmetric for tidal disruption to remain plausible. Joint GW--gamma-ray observations of such systems would be particularly valuable because no NSBH merger has yet been securely associated with a GRB. They would also provide a direct way to test proposals that some merger-associated, apparently long-duration GRBs may originate from disrupting NSBH binaries~\cite{Zhu:2022kbt,Gottlieb:2023sja}.

A detected GRB can also be followed by longer-lived X-ray, optical, and radio afterglow emission. Because the afterglow can often be localized much more precisely than the prompt gamma-ray signal, it can substantially improve the source position and, in favorable cases, enable identification of the host galaxy and measurement of its redshift. Such systems would then become bright standard sirens, extending multimessenger cosmology to distances that are difficult to access with kilonova counterparts. The afterglow also probes physics that is complementary to both the prompt GRB and the GW signal. It is produced as the relativistic outflow interacts with the circumburst medium, so its temporal and spectral evolution constrains the density and structure of the binary environment, together with the energetics and geometry of the jet. Thus, joint GW--GRB--afterglow observations can connect the intrinsic properties of the binary not only to the relativistic outflow, but also to the environment in which the merger occurred.

\subsubsection{Early alerts and rapid follow-up}

Detecting an inspiral before coalescence is useful only if the source can also be localized well enough for electromagnetic facilities to act on the alert. Given both sufficient warning time and a useful sky area, telescopes can be pointed before the merger rather than responding after the transient has already begun. This is particularly valuable for capturing the earliest ultraviolet and optical emission from the blue kilonova, whose rapid evolution carries information about the fastest, relatively lanthanide-poor ejecta. Pre-merger observations could also probe more speculative electromagnetic precursors, such as resonant crust-shattering flares produced shortly before merger. More generally, early warning makes it possible to coordinate facilities across wavelengths and geographical locations.

\vskip10pt
\noindent
\begin{minipage}{0.73\textwidth}
\begin{tcolorbox}[standard jigsaw,colframe=gray,colback=gray!10!white,opacityback=0.0,coltext=black, title=\small\sc Pre-merger localization of NS mergers]
\center{\includegraphics[width=1.0\textwidth]{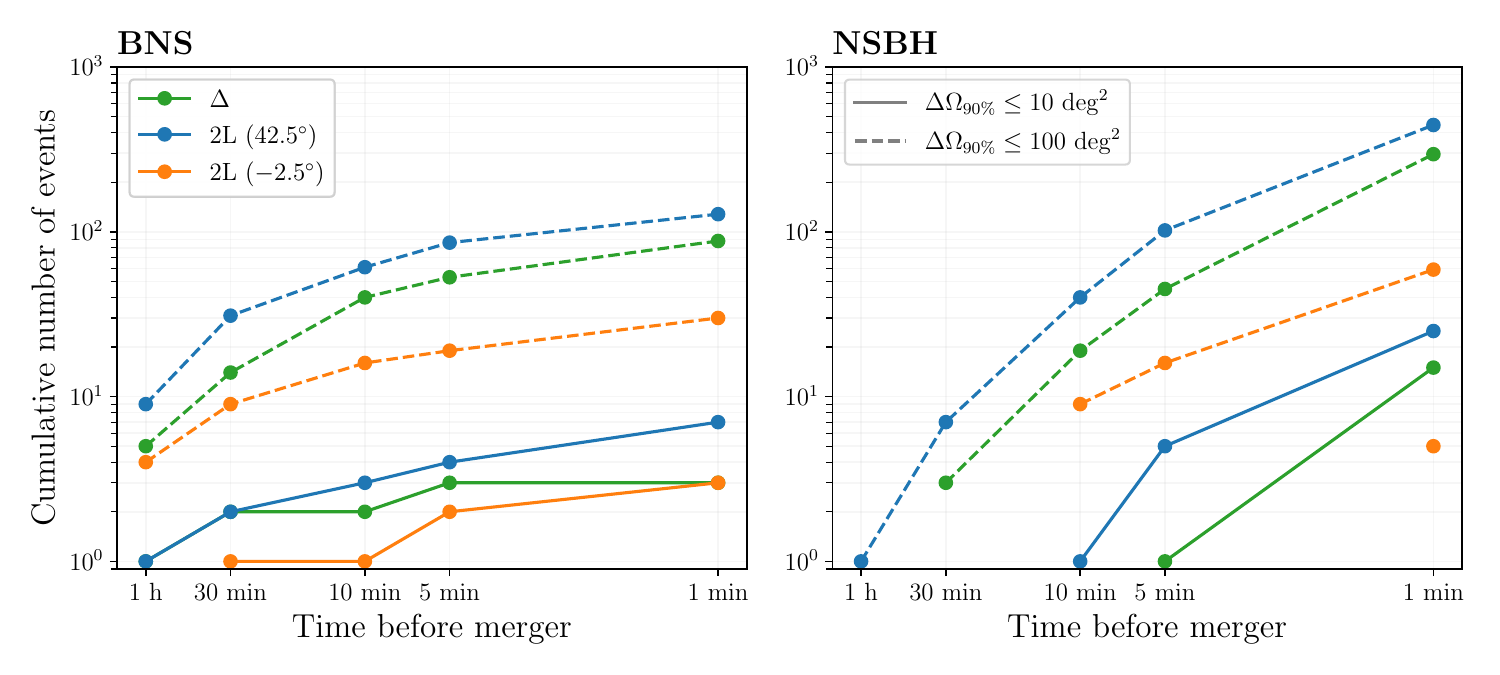}}
\captionof{figure}{\justifying Cumulative number of BNS (left) and NSBH (right) mergers that can be localized before coalescence for the three ET configurations. Solid and dashed curves show events constrained within $\Delta\Omega_{90\%}\leq10\,{\rm deg}^2$ and $\leq100\,{\rm deg}^2$, respectively, at times ranging from one hour to one minute before merger. The figure quantifies the number of sources for which ET could provide sufficiently early and precise sky localizations to enable rapid electromagnetic follow-up.} \label{fig:mma_premerger}
\end{tcolorbox}
\end{minipage}
\hfill
\begin{minipage}{0.24\textwidth}
We expect BNS mergers to be particularly favorable for early-warning localization. Their lower chirp masses cause them to evolve more slowly through the detector band, allowing them to remain observable at low frequencies for tens of minutes before merger and to accumulate measurable SNR well before coalescence. The long observation time also allows the changing detector response due to the Earth's rotation to contribute to the localization. NSBH systems are typically more massive and evolve through the band more rapidly, leaving less time to accumulate the SNR and localization information required for an early alert. 
\end{minipage}

\vskip10pt
\noindent
Their loudness can partly compensate for this, particularly close to merger, but useful pre-merger localizations are generally expected to emerge later than for BNSs.

Figure~\ref{fig:mma_premerger} shows the number of systems that can be localized within $100\,{\rm deg}^2$ and $10\,{\rm deg}^2$ at different times before coalescence. For BNS mergers, useful localizations can already be obtained an hour before merger. At $\Delta\Omega_{90\%}\leq100\,{\rm deg}^2$, the 2L $(42.5^\circ)$ configuration provides 9 such events one hour before merger, compared with 5 for the triangle and 4 for the 2L $(-2.5^\circ)$ configuration. The numbers rise to 31, 14, and 9 at 30 minutes, and to 61, 40, and 16 at 10 minutes before merger, respectively. By one minute before coalescence, the corresponding samples contain 128, 88, and 30 BNS mergers. Thus, although all three geometries can provide advance warning, the 2L $(42.5^\circ)$ configuration consistently delivers the largest sample of sources with an actionable sky localization.

Much tighter pre-merger localizations are naturally rarer. Within $10\,{\rm deg}^2$, only 1 BNS is localized an hour before merger by either the triangle or the 2L $(42.5^\circ)$ configurations, while none is obtained with the aligned configuration. By one minute before merger, these numbers increase to 3, 7, and 3, respectively. Such events are particularly valuable for optical follow-up: a localization of order $10\,{\rm deg}^2$ is comparable to the field of view of a wide-field facility such as Rubin, allowing observations to begin with very little tiling and potentially before any electromagnetic transient has appeared.

Pre-merger localization is more challenging for NSBH systems because their larger masses shorten the time spent in band. No NSBH is localized within $100\,{\rm deg}^2$ one hour before merger by either the triangle or 2L $(-2.5^\circ)$ configuration, while the misaligned configuration yields only 1. The number grows rapidly as the merger approaches. At 10 minutes before coalescence, the triangle, 2L $(42.5^\circ)$, and 2L $(-2.5^\circ)$ configurations localize 19, 40, and 9 NSBHs within $100\,{\rm deg}^2$, increasing to 296, 445, and 59 one minute before merger.

As before, only a subset of these NSBH systems is likely to undergo sufficient tidal disruption to produce an electromagnetic counterpart. Requiring $q>1/4$ leaves 10, 19, and 6 systems localized within $100\,{\rm deg}^2$ at 10 minutes before merger for the triangle, misaligned, and aligned configurations, respectively. At five minutes, these numbers increase to 21, 41, and 7, and by one minute to 69, 92, and 18. Even under this more restrictive selection, ET would provide a substantial sample of NSBH mergers for which electromagnetic facilities could be alerted before coalescence.

The $10\,{\rm deg}^2$ requirement is considerably more demanding for NSBHs. No configuration provides such a localization 30 minutes before merger. At 10 minutes, the 2L $(42.5^\circ)$ configuration yields 1 system with $q>1/4$, while the other configurations yield none. By one minute before merger, the triangle, 2L $(42.5^\circ)$, and 2L $(-2.5^\circ)$ configurations provide 3, 5, and 2 potentially disrupting systems, respectively. Although these samples are small, they represent particularly valuable events for which telescopes could already be observing a compact region of sky when the merger occurs.

These early alerts would enable more than simply a faster search for the kilonova. A coordinated network of gamma-ray, X-ray, ultraviolet, optical, infrared, radio, and neutrino facilities could observe the transition from inspiral to merger and follow the subsequent transient from its earliest stages. Capturing the onset and rapid color evolution of the kilonova would improve constraints on the ejecta components, while observations immediately around merger could search for prompt or precursor phenomena that would otherwise be easily missed.

Overall, Fig.~\ref{fig:mma_premerger} demonstrates that the scientific value of an early-warning network depends on both how early a source is detected and how rapidly its sky position becomes useful. The 2L $(42.5^\circ)$ configuration provides the strongest combination of these capabilities, delivering the largest number of BNS and NSBH mergers localized tens of minutes before coalescence and maintaining its advantage as tighter localization requirements are imposed. The triangular configuration generally performs second best, while the 2L $(-2.5^\circ)$ geometry is limited by its poorer ability to break extrinsic-parameter degeneracies despite the greater sensitivity of its longer arms.

Taken together, these results show that ET alone will already enable multimessenger studies across a wide range of distances and timescales, from nearby kilonovae and $r$-process nucleosynthesis to GRB counterparts at cosmological redshifts and pre-merger alerts for rapid follow-up. Across these science cases, the 2L $(42.5^\circ)$ configuration provides the most consistently favorable performance by combining the higher sensitivity of the longer arms with the geometrical information needed for precise localization. The 2L $(-2.5^\circ)$ configuration benefits from its larger reach and can retain large high-redshift events, but its poorer ability to break extrinsic-parameter degeneracies limits the number of tightly localized events, in some regimes allowing the triangular configuration to perform better. The results presented here are intentionally restricted to ET alone. In practice, an international network of GW observatories would further strengthen this science by improving sky coverage, extending the timing baselines, and breaking localization and polarization degeneracies. Even detectors less sensitive than ET can contribute valuable geometrical information to a coherent network, increasing both the number of actionable multimessenger events and the precision and warning time with which they can be localized.

\begin{tcolorbox}[colback=teal!5!white,colframe=red!75!black,title=\sc Geometry Comparison]
{
Multimessenger performance depends on both detector sensitivity and the ability of the network geometry to localize sources.
\begin{itemize}
    \item {\em Misaligned 2L performs best overall:} The 2L $(42.5^\circ)$ configuration provides the strongest multimessenger performance and is generally favored when the tightest localization is required.
    
    \item {\em Aligned 2L trades localization for reach:} The nearly aligned 2L $(-2.5^\circ)$ configuration benefits from the 15-km arms and can retain large high-redshift samples when broad sky areas are acceptable, but limited polarization resolution degrades precise localization and early-warning performance.
    
    \item {\em The triangle remains valuable for precise localization:} Despite its shorter arms and lower reach, the triangle can outperform the 2L $(-2.5^\circ)$ configuration when tight sky localization matters, particularly for well-localized NSBH mergers, although it does not match the overall yield of the 2L $(42.5^\circ)$ configuration.
\end{itemize}
}
\end{tcolorbox}

\subsection{Fundamental physics and tests of general relativity}

\begin{tcolorbox}[colback=teal!5!white,colframe=yellow!75!black,title=\sc Science Target]
    {\em 
    Uncover the fundamental nature of gravity and of compact objects, and characterize their environment.
    }
\end{tcolorbox}

GW observations access the strong-field, highly dynamical regime of gravity, providing laboratories for fundamental physics and opening new avenues for discovery~\cite{LIGOScientific:2016aoc}.
Signals from strongly self-gravitating objects, such as BHs and NSs, enable stringent tests of general relativity (GR) and of the nature of these objects~\cite{Will:2014kxa}. 
Compact-binary merger observations have already probed BH geometry, strengthened the case for horizons, constrained near-horizon modifications, and placed bounds on ultralight bosonic fields around BHs~\cite{LIGOScientific:2019fpa,LIGOScientific:2026oim,LIGOScientific:2018dkp}.
The improved sensitivity of next-generation detectors will enable precision GW astronomy and access to phenomena currently beyond reach~\cite{Kalogera:2021bya,ET:2025xjr,Evans:2021gyd}.
Key targets include the GW memory effect and possible signatures of new physics — a nonzero graviton mass, Lorentz or parity violations, extra spacetime dimensions, and additional GW polarizations. Precision measurements of tidal interactions and quasinormal-mode (QNM) spectra could also reveal exotic alternatives to BHs and NSs in compact binaries. GW observations will further offer complementary probes of dark matter and dark energy, testing whether the Universe appears the same in the gravitational and electromagnetic (EM) windows.

\begin{figure*}
  \begin{tcolorbox}[standard jigsaw,colframe=gray,colback=gray!10!white,opacityback=0.0,coltext=black, title=\small\sc Fundamental physics using gravitational waves]
        \centering
        \includegraphics[width=\linewidth]{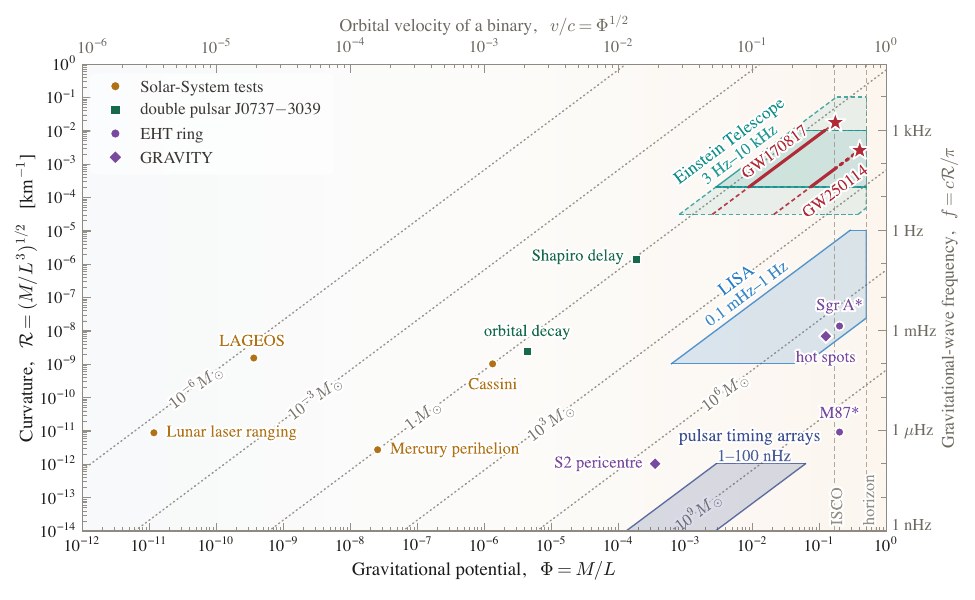}
        \caption{\small\justifying
        Landscape of gravitational tests in the potential--curvature plane. 
        Markers show representative tests using EM probes, while shaded regions indicate the parameter space accessible to GW probes such as pulsar timing arrays, LISA, and the ET. 
        ET will probe an expanded region of the parameter space at an improved sensitivity.
        Red tracks illustrate the evolution of selected GW sources, with the stars marking their mergers. 
        Here, $M$ and $L$ denote the characteristic mass and length scales associated with each observable.
        For binary systems, $M$ is the total mass and $L$ is the orbital separation, such that $M/L$ is related to $v^{2}/c^{2}$ through the virial theorem.
        The upper and right axes show the binary orbital velocity and GW frequency, respectively, and apply exclusively to GW probes.
        }
        \label{fig:fundamental_physics_schematic}
    \end{tcolorbox}
\end{figure*}

We now evaluate how effectively the three ET configurations address these scientific objectives, using a representative set of case studies. Compact-binary mergers offer a unique probe of gravity across a broad range of physical scales (see \cref{fig:fundamental_physics_schematic}). 
The long, early inspiral is particularly sensitive to environmental effects (gas and dark-matter halos) and to long-range interactions mediated by ultralight bosonic fields. The intermediate and late inspiral probe strong-field binary dynamics, nonlinear gravitational interactions, and the compact objects' properties. The merger and ringdown, finally, test the nature of the remnant, including its near-horizon structure.

\subsubsection{Probing the nature of compact objects and their gravitational interactions}

Traditional tests of binary dynamics rely either on Solar-System measurements, which probe the weak-field, slow-motion regime, where the gravitational potential and orbital velocity are both small, or on binary-pulsar observations, which involve strongly self-gravitating NSs but remain only mildly relativistic.~\cite{Will:2014kxa}. GWs from the late inspiral and merger of compact binaries probe a qualitatively different regime: separations comparable to the size of the objects, orbital velocities approaching the speed of light, and rapid, nonlinear spacetime evolution. The waveform encodes GR effects including spin–orbit and spin–spin interactions, GW tails, and nonlinear memory.

\begin{tcolorbox}[standard jigsaw,colframe=ocre,colback=ocre!10!white,opacityback=0.0,coltext=black,title=\sc Key Science Goals]
    \small
    GW observations provide a powerful means of addressing some of the most fundamental questions in physics.
    \begin{itemize}[leftmargin=*]
        \item {\bf Nature of gravity:} Does GR accurately describe binary dynamics? Are there additional light fields? What are the symmetries of nature? What are dark matter and dark energy?
        \item {\bf Nature of compact objects:} Do compact binaries contain exotic objects such as boson stars, gravastars, quark stars, or Planck stars? What are their gravitational signatures?
        \item {\bf Characteristics of the astrophysical environment:} Do compact binaries evolve within dense astrophysical environments, and to what extent does dark matter influence their evolution?
    \end{itemize}
\end{tcolorbox}

The GW phase evolves throughout the inspiral, accumulating over the tens to thousands of cycles observed within a detector's band. During the early and intermediate inspiral, it is well modeled by the post-Newtonian (PN) expansion --- a systematic series in $(v/c)$ that captures relativistic effects order by order~\cite{Blanchet:2013haa}. Coherent tracking of the phase across many cycles turns small modifications to the binary dynamics or GW emission into a measurable dephasing, making the phase a particularly sensitive probe of departures from GR.~\cite{Berti:2015itd}.

Broadly, tests of gravity follow two complementary strategies. Theory-specific tests compute the binary dynamics and emitted GWs within a given modification of GR and use observations to constrain its coupling constants --- as done for scalar–tensor theories, Einstein–dilaton–Gauss–Bonnet gravity, dynamical Chern–Simons gravity, Lorentz-violating frameworks, and massive-graviton theories~\cite{Berti:2015itd}. Because they exploit the detailed structure of the predicted signal, such analyses generally yield the strongest constraints on a given theory. Their reach, however, is limited to the specific models considered and, often, by the lack of complete and sufficiently accurate waveforms across the full inspiral, merger, and ringdown.

\begin{wrapfigure}{r}{0.5\linewidth}
    \begin{tcolorbox}[standard jigsaw,colframe=gray,colback=gray!10!white,opacityback=0.0,coltext=black, title=\small\sc Distribution of bounds on the positive-PN phase deviations]
        \centering
        \includegraphics[width=\linewidth]{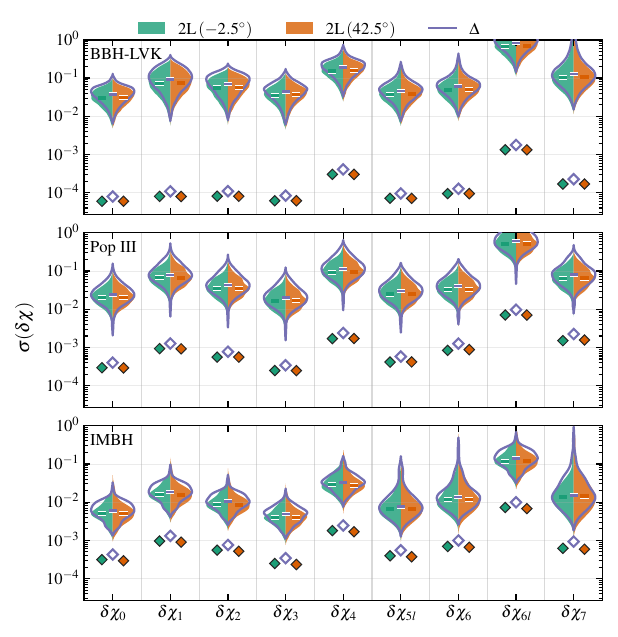}
        \caption{\small\justifying
        Distribution of 1$\sigma$ bounds (violin) and the combined bound (diamond) on the PN phase-deviation coefficients for the three ET geometries (colors) and the three source populations (panels: BBH-LVK, Pop.~III, and IMBH). 
        The results are grouped by coefficient, including $\delta\chi_n$ and $\delta\chi_{n(l)}$, and each violin also shows the median bound using horizontal dashes.
        }
        \label{fig:fti_combined}
    \end{tcolorbox}
\end{wrapfigure}

The complementary, theory-agnostic strategy introduces phenomenological modifications to quantities such as the amplitude, phase, dispersion relation, or polarization content, and constrains the corresponding parameters directly from the data. The resulting bounds are less targeted than those from theory-specific analyses but apply to a much broader range of possible deviations. A widely used implementation is the parameterized framework, in which deviations from GR are expressed as corrections at different orders in the PN expansion of the inspiral phase, i.e., terms of the form $\delta\chi_n (v/c)^n$ or $\delta\chi_{n(l)} (v/c)^n \log(v/c)$ (schematically, relative to the leading-order Newtonian phase) are added to the GW phase, and the free parameters $\delta\chi_{n}$ and $\delta\chi_{n(l)}$ are constrained~\cite{Arun:2006hn,Li:2011cg,Agathos:2013upa,Yunes:2009ke,Mehta:2022pcn}. The two strategies are complementary, since the predictions of any given theory can generally be mapped onto the phenomenological parameters.

Applied to growing GW catalogs, this framework has produced increasingly stringent tests of strong-field gravity~\cite{LIGOScientific:2026oim}. The sources that provide the strongest constraints depend on which part of the phase evolution is tested. 
The exceptionally loud GW250114~\cite{LIGOScientific:2026sit} yields high-order inspiral-to-ringdown constraints that are up to three times tighter than those obtained by combining many earlier events. 
The bounds on the 1PN coefficient are $\sigma(\delta\chi_{2}) \sim 0.07$ for GW250114 compared to $\sigma(\delta\chi_{2}) \sim 0.1$ for the combined events of the fourth Gravitational-Wave Transient Catalog (GWTC-4.0).
At low PN order, long-lived inspirals are more informative, making NS systems particularly important for testing dipole radiation. 
GW170817~\cite{LIGOScientific:2017vwq} and GW230529~\cite{LIGOScientific:2024elc} currently provide the strongest constraints in this sector, illustrating that different regions of the phase expansion are best probed by different source populations and event classes~\cite{LIGOScientific:2018dkp,Sanger:2024axs}.

In \cref{fig:fti_combined}, we show the distribution of 1$\sigma$ bounds on the PN deviation coefficients for the three ET configurations. In addition, we show the bounds obtained by combining all events in the population, i.e., the combined bound is given by $1/\sigma^2 = \sum_i 1/\sigma_i^2$, where $\sigma_i$ is the bound for the $i$-th event. 
For each BBH source population, we consider all sources detected with network inspiral ${\rm SNR}>12$.  
Each coefficient is constrained independently, marginalizing over the GR parameters. 
The range of the combined bounds across all the positive PN parameters for the BBH-LVK population is given in \cref{{tab:metrics}} (under ``Precision range of parameterized PN tests (positive PN orders)'').

All three ET configurations yield excellent constraints on every parameter, with the bounds from the median BBH-LVK event comparable to or stronger than those obtained by combining all events observed to date~\cite{LIGOScientific:2026oim}.
Nevertheless, the triangular configuration is consistently weaker by a median factor of ${\sim}1.3$ than both 2L configurations, which agree with each other to within a few percent. 
The weaker triangle bounds reflect its lower SNR, whereas the near-identical 2L results arise because the relative orientations of the detectors in the 2L configurations primarily impact the sky localization measurement, which is only weakly correlated with the GW phase evolution.
We find that the median event in the IMBH population constrains the positive PN coefficients more tightly than the other two populations, by a factor of a few to an order of magnitude, owing to its higher SNR (\cref{fig:snrs_ds1}). 
However, the number of suitable IMBH mergers is between 100 and 150 per year, depending on the geometry, compared to the BBH-LVK population, which contributes approximately 80,000 to 90,000 events per year.
Therefore, for the BBH-LVK population, the combined bounds are more than two orders of magnitude stronger than the median bounds and up to one order of magnitude stronger than those obtained for the other populations.

The leading radiative multipole in GR is the mass quadrupole since the equivalence principle plus conservation of mass-energy and momentum excludes monopole and dipole radiation, respectively. 
Therefore, the $-1$~PN dipole contribution differs fundamentally from the higher-order corrections considered above because it vanishes identically in GR. Hence, a measured nonzero dipole coefficient would provide a direct indication of physics beyond GR.

Dipole radiation arises naturally in theories with additional long-range interactions or new radiation channels. 

\begin{wrapfigure}{r}{0.49\linewidth}
    \begin{tcolorbox}[standard jigsaw,colframe=gray,colback=gray!10!white,opacityback=0.0,coltext=black, title=\small\sc Distribution of the dipole bound]
        \centering
        \includegraphics[width=\linewidth]{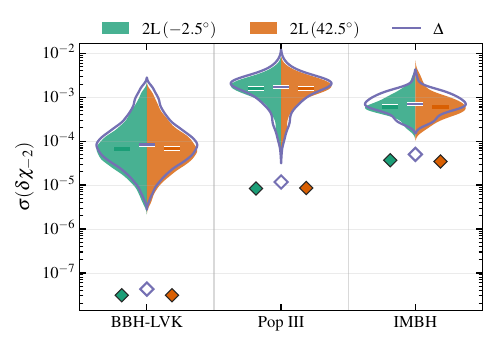}
        \caption{\small\justifying
        Distribution of the 1$\sigma$ bounds (violin) and the combined bound (diamond) on the $-1$~PN dipole coefficient $\delta\chi_{-2}$, shown for each source population (horizontal axis) and ET geometry (colors). 
        Each violin also shows the median bound using horizontal dashes.
        }
        \label{fig:dchiminus2}
    \end{tcolorbox}
\end{wrapfigure}

\noindent
In scalar–tensor gravity, compact objects can acquire scalar charges that depend on their internal structure and binding energy, so a binary whose components have different charge-to-mass ratios sources dipole radiation --- an energy-loss channel absent in GR. 
Dipole radiation also arises in Einstein--dilaton--Gauss--Bonnet gravity and in Lorentz-violating gravity, e.g., Einstein-aether and khronometric theories~\cite{Berti:2015itd}.
Because this contribution enters at $-1$~PN order, its relative importance increases at low orbital velocities, making the early inspiral especially sensitive to it. 
This is why the strongest bound on this term is from binary-pulsar observations~\cite{Kramer:2021jcw}, with the best GW bounds coming from the longest observed signals, GW170817 ($\sigma(\delta\chi_{-2}) \sim 2\times10^{-5}$) and the NSBH GW230529 ($\sigma(\delta\chi_{-2}) \sim 9\times10^{-5}$)~\cite{LIGOScientific:2018dkp,Sanger:2024axs}.

In \cref{fig:dchiminus2}, we show the distribution of the 1$\sigma$ bound for each ET geometry and source population under consideration, as well as the bounds obtained by combining all events from a population. As before, for each source, the dipole bound is marginalized over GR parameters. 
The combined bound for the BBH-LVK population is listed in \cref{{tab:metrics}} (under ``Precision of dipole test''). We observe that this population provides the most stringent bounds on this parameter due to the presence of low-mass binaries.  On the other hand, both the Pop.~III and IMBH populations primarily contain heavy BHs, thereby yielding weaker constraints. The tail in the Pop.~III distribution is due to the small fraction of relatively low-mass events in this population. We find that the combined bound for the BBH-LVK population in all three geometries is about three orders of magnitude better than the current best bound~\cite{LIGOScientific:2018dkp,LIGOScientific:2026oim}. 
However, as for the positive PN parameters, the triangular configuration is consistently weaker than either of the 2L configurations. 

GW constraints remain weaker than the binary-pulsar bound --- the projected combined constraint from the BBH-LVK population is about two orders of magnitude less stringent~\cite{Kramer:2021jcw} --- but the two probes are complementary. GWs test dipole radiation directly in the strong-field, highly relativistic regime, where orbital velocities approach the speed of light and the objects are close to merger. Such a regime is valuable because modifications of gravity may be suppressed in weak fields yet become significant in strong ones. In addition, in certain scalar–tensor theories BHs carry no scalar charge while NSs can develop substantial charge through their strong self-gravity, providing a further way to discriminate between such theories.

In GR, an isolated, stationary, asymptotically flat, and electrically neutral BH is described by the Kerr solution and fully characterized by its mass and angular momentum, as established by the \emph{no-hair theorem}~\cite{Carter:1971zc,Hansen:1974zz}. 
Its mass and current multipole moments are therefore uniquely determined by these two quantities, allowing GW observations to test whether the measured moments are consistent with the Kerr predictions~\cite{Krishnendu:2017shb,Krishnendu:2018nqa}.

\begin{wrapfigure}{r}{0.45\linewidth}
    \begin{tcolorbox}[standard jigsaw,colframe=gray,colback=gray!10!white,opacityback=0.0,coltext=black, halign title=left, title=\small\sc Distribution of the spin-induced quadrupole bound]
        \centering
        \includegraphics[width=\linewidth]{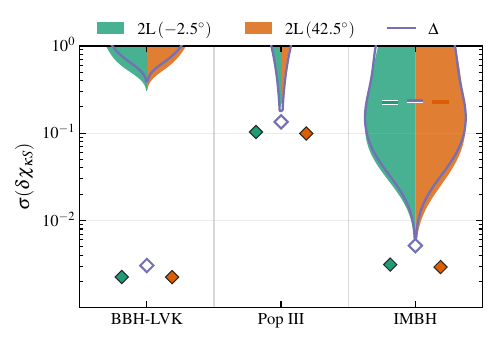}
        \caption{\small\justifying
        Distribution of the 1$\sigma$ bounds  (violin) and the combined bound (diamond) on the SIQM deviation coefficient
        $\delta\chi_{\kappa S}$, shown for each source population (horizontal axis) and ET geometry (colors). 
        }
        \label{fig:dchikappa}
    \end{tcolorbox}
\end{wrapfigure}

One such observable is the spin-induced quadrupole moment (SIQM), whose dimensionless coefficient equals unity for Kerr BHs in GR but can take different values for BH mimickers, such as boson stars, quark stars, and other exotic objects, depending on their internal structure.
In the PN expansion, the leading SIQM contribution enters through the spin--spin interaction and is therefore most pronounced for rapidly spinning compact objects.

Consistency with GR can be tested by parameterizing deviations through $\delta\chi_{\kappa S}$.
Owing to the large spin of its primary, GW241011 currently provides the most stringent constraint on this parameter with $\sigma(\delta\chi_{\kappa S}) \sim 0.2$~\cite{LIGOScientific:2025brd}.
However, most binaries observed by current detectors have relatively small spins, making measurements of this effect challenging.
Future GW detectors will improve these measurements in two key ways. 
First, their enhanced sensitivity will enable more precise spin measurements, helping to disentangle the SIQM contribution from other waveform effects. 
Second, their greater detection reach will probe astrophysical populations at higher redshifts, where binaries may have larger spins.

In \cref{fig:dchikappa}, we present the distributions of the $1\sigma$ bounds for each ET geometry and source population on the SIQM parameter $\delta\chi_{\kappa S}$, together with the corresponding bounds obtained by combining all events within each population.
The combined bound for the BBH-LVK population is listed in \cref{{tab:metrics}} (under ``Precision of SIQM test'').
The IMBH population provides the strongest individual-event constraints, with a large fraction of sources yielding fractional bounds below $100\%$ owing to their high spins (see \cref{sec:population}); by contrast, the other two populations are characterized by relatively low spins.
Nevertheless, the combined constraint for the BBH-LVK population is comparable to that of the IMBH population because its much larger number of binaries compensates for the weaker individual-event measurements. 
As before, the two 2L configurations yield nearly identical bounds, whereas the triangular configuration performs worse by a factor of ${\sim}1.3$.

The BH \emph{no-hair theorem} dictates the properties of the remnant as it settles into a stationary Kerr state~\cite{Carter:1971zc,Hansen:1974zz}.
The GW signal during the ringdown is a superposition of QNMs, damped sinusoids with complex frequencies, with oscillation frequency $\omega_{lmn}$ and damping time $\tau_{lmn}$, which are uniquely determined by the mass and spin of the remnant~\cite{Detweiler:1980gk}. 
Measuring the complex frequency of a single identified mode fixes these two parameters and thus predicts the complete Kerr QNM spectrum. 
The independent measurement of one or more additional modes, called BH spectroscopy, tests the consistency of those predictions and provides a stringent probe of the Kerr hypothesis and fundamental gravitational physics~\cite{Dreyer:2003bv,Berti:2006ew}.
Alternatively, tests incorporating the full inspiral, merger, and ringdown signal introduce theory-agnostic fractional deviations in the QNM frequencies $\delta\omega_{lmn}$ and damping times $\delta\tau_{lmn}$ within the ringdown portion of the waveform model~\cite{Brito:2018rfr,Ghosh:2021mrv}.
The reference Kerr QNM spectrum is then determined from the remnant properties predicted from the progenitor masses and spins using numerical-relativity fitting formulas.
Any deviation from the predicted QNM spectrum could signal modifications of the near-horizon geometry, due either to BH mimickers or to deviations from GR.

Analyses of sufficiently loud events from GWTC-4.0 constrained deviations in the dominant QNM frequency to the percent level and in its damping time to the tens-of-percent level, 
with both consistent with GR~\cite{LIGOScientific:2026oim}. 
The much louder event GW250114 has since provided tighter single-event constraints, reaching a few percent for the frequency ($\sigma(\delta\omega_{22}) \sim 0.04$) and around twenty percent for the damping time ($\sigma(\delta\tau_{22}) \sim 0.19$)~\cite{LIGOScientific:2025wao,LIGOScientific:2025rid}. 
Notably, a first constraint for the $(4,4)$ mode was found for the first time ($\sigma(\delta\omega_{44}) \sim 0.60$), bringing BH spectroscopy closer to a direct measurement rather than a single-mode consistency test. 

\begin{wrapfigure}{r}{0.55\linewidth}
    \begin{tcolorbox}[standard jigsaw,colframe=gray,colback=gray!10!white,opacityback=0.0,coltext=black, title=\small\sc Distribution of QNM bounds]
        \centering
        \includegraphics[width=\linewidth]{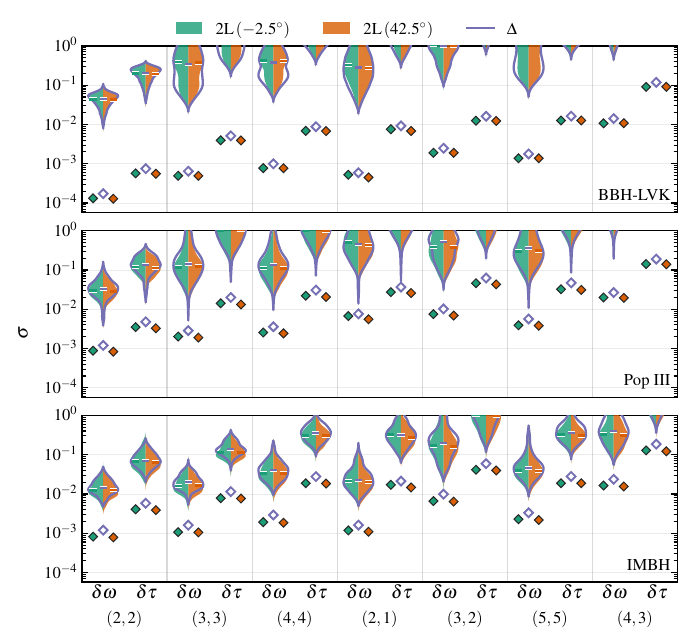}
        \caption{\small\justifying
        Distribution of 1$\sigma$ bounds (violin) and the combined bounds (diamond) on the QNM deviation parameters $\delta\omega_{\ell m}$ and $\delta\tau_{\ell m}$ for the three ET geometries (colors) and BH populations (panels) under consideration. 
        All modes are fundamental modes, with the overtone index $n=0$ omitted.
        }
        \label{fig:qnm_combined_by_catalog}
    \end{tcolorbox}
\end{wrapfigure}

In \cref{fig:qnm_combined_by_catalog}, we present the distribution of 1$\sigma$ constraints on the QNM deviation parameters $\delta\omega_{\ell m}$ and $\delta\tau_{\ell m}$ for the three ET configurations, together with the bounds obtained by combining all the events of a population.
The range of the combined bounds across oscillation frequencies and damping times for the different multipoles in the BBH-LVK population is listed in \cref{{tab:metrics}} (under ``Precision range of ringdown spectroscopy tests (oscillation frequency)'' and ``Precision range of ringdown spectroscopy tests (damping time)'', respectively).

\noindent
For each source population, the uncertainty is computed over all sources detected with both inspiral and merger-ringdown ${\rm SNR}>12$. 
We find that all three configurations enable the measurement of multiple oscillation frequencies and damping times, resulting in exquisite precision in the BH spectroscopy test. 
However, the triangular configuration is uniformly less constraining than either of the 2L configurations, 
by a median factor of $1.3$. 
Across all modes, geometries, and source populations, the frequency deviation is more tightly constrained than the damping-time deviation, with the improvement ranging from a factor of ${\sim}4$ for $(2,2)$ to ${\sim}13$ for $(2,1)$, as expected from previous studies~\cite{Berti:2005ys}. 
The $(2,2)$ mode provides the strongest overall constraint, whereas $(4,3)$ is the least well constrained, with bounds weaker by more than an order of magnitude. 
Events with a short inspiral and a prominent merger-ringdown are expected to give the best constraints, and we observe that the median events of the IMBH and Pop.~III outperform that of the BBH-LVK population owing to their larger BH masses. 
However, as with previous science objectives, the combined bound from the BBH-LVK population is better than those from the other populations by a factor of a few up to 10 for all deviation parameters owing to the sheer number of detected events. 
Finally, we find that the number of events with merger-ringdown SNR$>100$ is 614, 614, and 235 per year for the $\rm{2L} \, (-2.5^{\circ})$, $\rm{2L} \, (42.5^{\circ})$, and the triangular configurations, respectively.

In vacuum, Lorentz invariance requires all massless excitations to propagate at the speed of light. 
In the presence of a medium, however, Lorentz invariance is spontaneously broken, and the background selects a preferred frame, allowing different propagation speeds.
An analogous effective medium can arise in modified theories of gravity when additional degrees of freedom coupled to curvature acquire background configurations that select a preferred direction, such as the cosmological time direction~\cite{Baker:2017hug}.
Joint observations of GW and EM signals provide a direct means of constraining such violations of Lorentz invariance.
Although Lorentz symmetry in the Standard Model has been tested extensively in terrestrial experiments, constraints in the gravitational sector remain comparatively weak. 
The most stringent upper bound was obtained by comparing the arrival times of GW170817 and its GRB counterpart GRB 170817A, with the GRB detected $1.74 \pm 0.05$~s after the GW signal~\cite{LIGOScientific:2017zic}.

Future GW detectors are particularly well suited to this test because of their ability to observe sources at cosmological distances.
We find that all configurations considered here can detect joint GW--GRB events beyond $z=1$, with the assumed jet structure having little effect on the results.
Their relative performance, however, depends strongly on the imposed GW sky-localization requirement.
Assuming a structured GRB jet and no localization constraint, the three configurations perform similarly, with a median redshift, for the medium EOS, of ${\sim}1.3$ for the triangular configuration and ${\sim}1.4$ for both 2L configurations, and corresponding maximum redshifts of ${\sim}4.6$ and ${\sim}5.1$, respectively, and more than 1000 joint detections in each configuration. 
The differences among the GW detector configurations become pronounced when a modest sky-localization requirement of ${<}10^{4}\,\mathrm{deg}^{2}$ is imposed to control chance-coincidence association, reducing both the number of joint detections and their redshift reach.
The impact is greatest for the triangular configuration, which detects only $7$ events at $z{>}1$, compared with $489$ for $\rm{2L}\,(-2.5^{\circ})$ and $573$ for $\rm{2L}\,(42.5^{\circ})$.
The redshift reach also has a similar impact with $\rm{2L}\,(-2.5^{\circ})$, $\rm{2L}\,(42.5^{\circ})$, and the triangular configuration attaining redshifts of approximately $2.9$, $1.8$, and $1.2$, respectively.

We forecast constraints on the speed of gravity using both the median and the most distant joint GW--GRB detection of a BNS merger accessible to each detector configuration after imposing the sky-localization requirement. 
For the upper bound, we adopt an observed arrival-time difference of $1.74 \, (1 + z)$~s, using GW170817/GRB 170817A as a reference. 
For the lower bound, we allow the GRB to have been emitted $10$~s (in the source frame) after the GW signal~\cite{LIGOScientific:2017zic}. 
This delay should be regarded as a proxy and depends on the details of the GRB emission process.
For the median event, the resulting constraints on the fractional speed difference $\Delta v / v_{\rm{EM}}$ are 
$[-4.7\times10^{-17}, 9.9\times10^{-18}]$ for $\rm{2L}\,(-2.5^{\circ})$, 
$[-4.7\times10^{-17}, 9.9\times10^{-18}]$ for $\rm{2L}\,(42.5^{\circ})$, and 
$[-5.9\times10^{-17}, 1.2\times10^{-17}]$ for the triangular configuration.
For the most distant event, the corresponding constraints are 
$[-4.9\times10^{-17}, 1.0\times10^{-17}]$, 
$[-4.5\times10^{-17}, 9.5\times10^{-18}]$, and 
$[-4.5\times10^{-17}, 9.6\times10^{-18}]$, respectively.
These are more than an order of magnitude better than the current bounds, although both 2L configurations provide improved bounds over the triangular geometry by approximately a factor of $1.3$ for the median event. 
For $\rm{2L}\,(-2.5^{\circ})$, the most distant event yields a slightly weaker constraint than both its median event and the most distant event detected by the triangular configuration, because it lies beyond the redshift at which the angular diameter distance is maximal. 
Nevertheless, the substantially larger number of detections expected for the 2L configurations should lead to stronger combined constraints.

\subsubsection{Probing the astrophysical environment}

\begin{wrapfigure}{r}{0.48\linewidth}
    \begin{tcolorbox}[standard jigsaw,colframe=gray,colback=gray!10!white,opacityback=0.0,coltext=black, title=\small\sc Distribution of bounds on environmental effects]
        \centering
        \includegraphics[width=\linewidth]{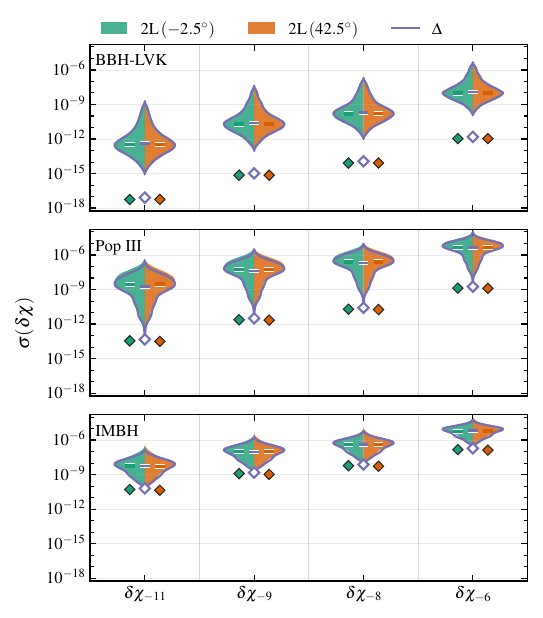}
        \caption{\small\justifying
        Distribution of $1\sigma$ bounds (violin) and the combined bounds (diamond) on the environmental phase coefficients $\delta\chi_n$ for the three ET geometries (color) and source populations (panels). 
        }
        \label{fig:fti_environmental}
    \end{tcolorbox}
\end{wrapfigure}

Compact binaries evolve in environments containing gas, stars, dark matter, or other matter fields. 
Interactions with these components can perturb the orbital dynamics and the emitted radiation, either through secular changes in the binding energy and the energy and angular-momentum fluxes or through bursts of GWs accompanied by rapid phase shifts. 
Both mechanisms leave characteristic imprints on the accumulated inspiral phase, so that GW measurements can constrain the environments in which compact binaries form and evolve.

ET's enhanced low-frequency sensitivity and its ability to track long-duration inspirals make it especially sensitive to these signatures. Environmental effects are typically strongest at large orbital separations, where the binary evolves slowly and accumulates many cycles, producing measurable dephasing relative to a vacuum-GR waveform. The phenomenological framework of the previous section captures such dephasings by mapping environmental effects onto specific negative-PN-order phase-deformation coefficients.

Coefficients at PN orders below -1 PN can first be constrained directly from the GW phase and later translated into limits on quantities such as the surrounding matter density, accretion rate, or other environmental parameters for a chosen model.
As representative examples, we consider here  dynamical friction ($-5.5$~PN, $\delta\chi_{-11}$), collisionless accretion ($-4.5$~PN, $\delta\chi_{-9}$), gravitational pull onto the binary components ($-3$~PN, $\delta\chi_{-6}$), and acceleration due to third-body interactions ($-4$~PN, $\delta\chi_{-8}$)~\cite{Cardoso:2019rou}.

In \cref{fig:fti_environmental}, we show both the distribution of $1\sigma$ bounds and the combined bound on the environmental phase coefficients $\delta\chi_n$ for the three ET geometries. 
The two 2L configurations agree to within $5\%$ for every coefficient, whereas the triangular configuration is consistently weaker, by a median factor of ${\sim}1.3$. 
The range of the combined bounds across all the environmental parameters for the BBH-LVK population is given in \cref{{tab:metrics}} (under ``Precision range of bounds on negative-PN-order coefficients'').
As with dipole radiation, the BBH-LVK population yields the strongest constraints, since the very-low-frequency part of the signal carries most of the information on negative-PN effects; long-duration inspirals and low-mass systems are therefore especially valuable. The contrast is sharper still for the environmental parameters: projected constraints on the median binary vary by up to four orders of magnitude across populations, and combined bounds improve by five orders of magnitude relative to the median for the BBH-LVK population.

\begin{tcolorbox}[colback=teal!5!white,colframe=red!75!black,title=\sc Geometry Comparison]
{
All three ET geometries will improve tests of GW generation and propagation by several orders of magnitude relative to current detectors. 
Their relative performance depends primarily on the detection reach and on the physical effect being tested.
\begin{itemize}
    \item \textit{The two 2L configurations perform similarly for generation effects and outperform the triangular configuration:} The two 2L configurations yield nearly identical constraints on effects associated with GW generation, including binary dynamics and the nature of compact objects, and give bounds approximately $25\%$ better than the triangular configuration.
    \item \textit{The two 2L configurations perform comparably for speed of gravity and outperform the triangular configuration:} The two 2L configurations yield one to two orders of magnitude more joint GW--GRB detections at cosmological distances than the triangular configuration, enabling stronger combined constraints.
\end{itemize}
}
\end{tcolorbox}

\subsection{Continuous and burst gravitational-wave sources}

\begin{tcolorbox}[colback=teal!5!white,colframe=yellow!75!black,title=\sc Science Target]
{\em Detect yet-elusive GW sources, in particular continuous waves emitted by isolated neutron stars, and burst signals from core-collapse supernov\ae{}.}
\end{tcolorbox}

\begin{figure}
    \begin{tcolorbox}[standard jigsaw,colframe=gray,colback=gray!10!white,opacityback=0.0,coltext=black, title=\small\sc Classes of GW signals]
        \centering
        \includegraphics[width=\linewidth]{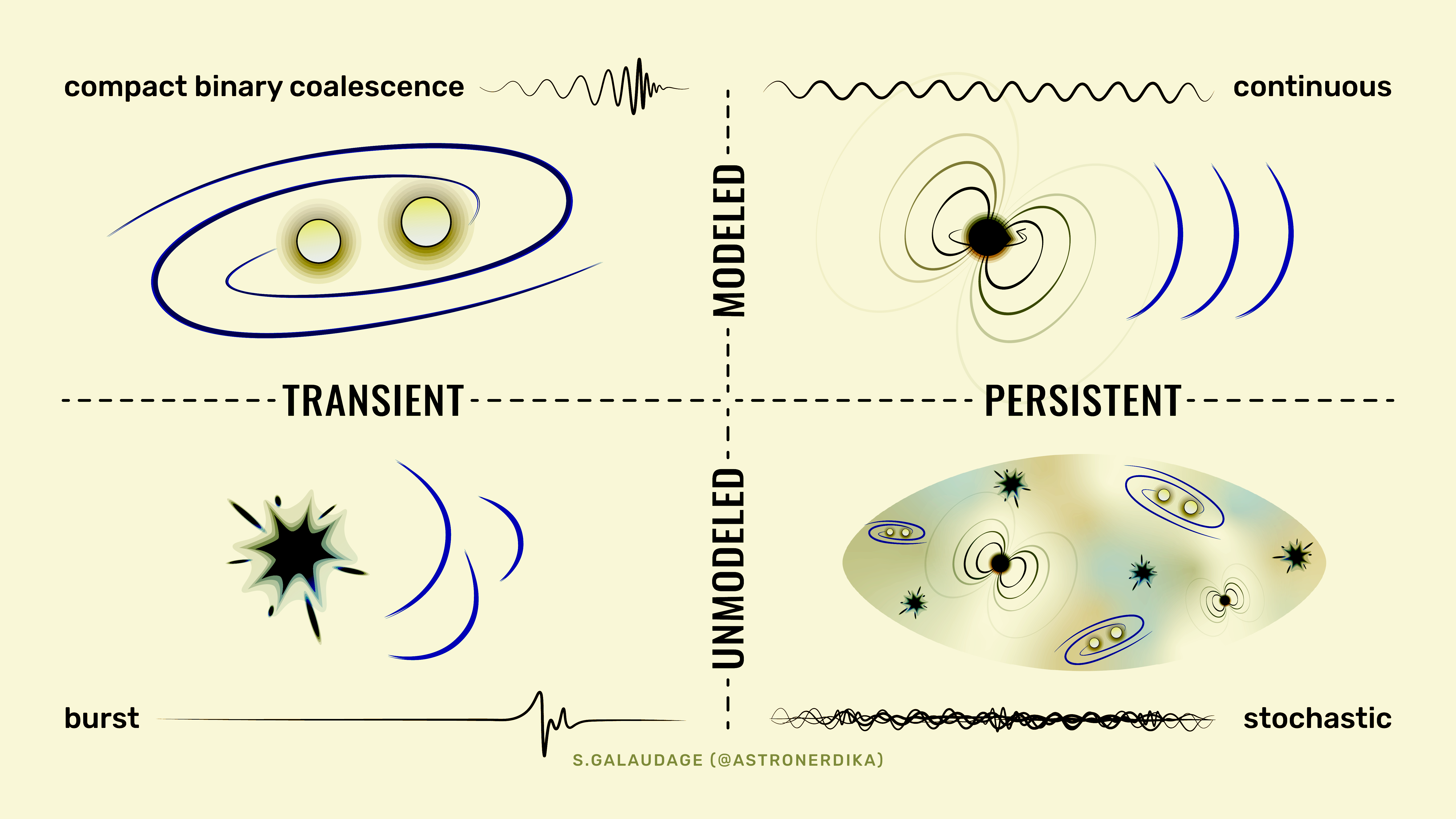}
        \caption{\justifying
        Four main classes of GW signals observable by ground-based detectors. While previous sections focused on short, well-modeled signals from compact binary coalescences and the persistent stochastic background, this section assesses the detectability of the remaining two classes: long, nearly monochromatic signals from spinning neutron stars, and short, partially modeled signals from stars at the end of their lives undergoing core collapse. Image taken from Ref.~\cite{galaudage_2022_20806149}.
        }
        \label{fig:gw-source-classes}
    \end{tcolorbox}
\end{figure}

\begin{tcolorbox}[standard jigsaw,colframe=ocre,colback=ocre!10!white,opacityback=0.0,coltext=black,title=\sc Key Science Goals]
\small
The science targets of ET extend beyond compact binaries and stochastic backgrounds. With its improved sensitivity, ET will be able to detect new classes of GW sources (Fig.~\ref{fig:gw-source-classes}).

\begin{itemize}[leftmargin=*]
\item {\bf Continuous waves from isolated neutron stars:} Can we detect the continuous GW emission from spinning neutron stars?
\item {\bf Bursts from core-collapse supernov\ae{}:} Can we detect the burst signal emitted by a star that collapses to form to form a neutron star or a black hole at the end of its life?
\end{itemize}
\end{tcolorbox}

\subsubsection{Continuous waves from neutron stars}

A qualitatively different class of GW sources is represented by long-lived, nearly periodic signals from spinning NSs. If an isolated NS is not perfectly axisymmetric with respect to its rotation axis, it carries a time-varying mass quadrupole and can emit continuous GWs~\cite{Maggiore:2007ulw}. For a triaxial star rotating about a principal axis, the dominant emission is expected at twice the stellar spin frequency, with an amplitude set by the moment of inertia, the rotation frequency, the distance, and the equatorial ellipticity. Other mechanisms, such as unstable oscillation modes, may also produce persistent emission~\cite{Andersson:1997xt,Owen:1998xg}, but in this section we focus on the standard ellipticity-driven scenario.
A detection of continuous waves would provide a unique probe of NS interiors~\cite{Riles:2022wwz,Wette:2023dom}. The inferred ellipticity would constrain the maximum elastic deformation sustainable by the crust, the role of internal magnetic fields, and possible exotic phases of dense matter. Even in the absence of a detection, improved upper limits are astrophysically informative: they test whether GW emission can account for a significant fraction of the observed spin-down of known pulsars, and progressively exclude regions of the parameter space allowed by NS structure models.
Search strategies for continuous waves depend strongly on how much is known about the source~\cite{Wette:2023dom}. Targeted searches use electromagnetic information on known pulsars, including sky position, spin frequency, spin-down, and, when relevant, binary parameters, allowing long coherent integrations and therefore maximal sensitivity. Narrow-band searches relax the assumption that the GW phase evolution is exactly tied to the electromagnetic ephemeris. Directed searches focus on promising sky regions, such as supernova remnants or the Galactic center, while all-sky broadband searches cover unknown NSs over a wide range of sky positions and spin parameters at much higher computational cost. Following this hierarchy, we consider figures of merit representative of targeted searches for known pulsars and broadband searches for previously unknown Galactic sources.
For the targeted analysis, we use the known pulsars in the \href{https://www.atnf.csiro.au/research/pulsar/psrcat/}{ATNF Pulsar Catalogue}~\cite{Manchester:2004bp} whose expected GW frequency lies in the detector band. For each object, the catalog provides the sky position, distance, rotational frequency, and frequency derivative, from which one can compute the spin-down limit: the GW amplitude obtained under the optimistic assumption that the entire observed loss of rotational kinetic energy is due to gravitational radiation. In the present analysis we additionally cap the signal amplitude using the latest LVK upper limits~\cite{LIGOScientific:2026qsb}, so that the assumed strain for each pulsar is not larger than what is already allowed observationally. We then evaluate the detectability of the catalog population for different assumptions on the ellipticity, ranging from the optimistic spin-down limit values to more conservative values motivated by long-term deformations sustainable by NS crusts.
As a complementary metric, we also estimate the minimum ellipticity detectable in a broadband all-sky search for a source at a representative Galactic distance, taken to be $8\,{\rm kpc}$. This quantity captures the sensitivity to unknown NSs in the Milky Way and depends on the detector noise curve, geometry, the observation time, and the semi-coherent search setup. Here we assume an observation period of one year, and a \texttt{FrequencyHough}---like pipeline~\cite{2014PhRvD..90d2002A}.
Together, the targeted and broadband metrics quantify two limiting regimes: the best-case sensitivity to known pulsars with accurately tracked phase evolution, and the more computationally demanding search for an unseen Galactic population.

\noindent
\begin{minipage}{0.55\textwidth}
\hskip10pt
For broadband searches, Fig.~\ref{fig:gw_broadb} shows the minimum detectable ellipticity for an isolated NS at a reference distance of $8\,{\rm kpc}$, representative of a source in the inner Galaxy. The sensitivity improves rapidly with frequency, reflecting the $h_0\propto \epsilon f_{\rm gw}^2$ scaling of the continuous-wave amplitude, together with the detector noise curve. 

\hskip10pt
At frequencies of a few tens of Hz, ellipticities of order $10^{-5}$--$10^{-6}$ become accessible, while at several hundred Hz the reach improves to $\epsilon \lesssim 10^{-7}$, approaching $10^{-8}$ at the highest frequencies considered. The $\Delta$ and 2L configurations give very similar results across the full band, with the 2L layout mildly more sensitive over most of the frequency range. The two curves converge at
the highest frequencies, where the noise curves for the different arm lengths are more similar to each other. Compared with current detectors, the sensitivity improves by a factor of $\gtrsim 10$ across the entire frequency range~\cite{LIGOScientific:2026plm}.

\hskip10pt
The targeted-pulsar results show the same qualitative picture. Assuming that each pulsar emits at the spin-down limit, which is an optimistic upper bound on the GW amplitude, the number of detectable sources is very large: $882$ for the $\Delta$ configuration and $1008$ for the 2L configuration. Note that, since we use an averaged antenna response, this analysis is insensitive to the relative orientation of the two L-shaped detectors. The same applies to the broadband-search metric. 
\end{minipage}
\hfill
\begin{minipage}{0.42\textwidth}
\begin{tcolorbox}[standard jigsaw,colframe=gray,colback=gray!10!white,opacityback=0.0,coltext=black, title=\small\sc Minimum detectable  ellipticity]
\center{\includegraphics[width=0.99\textwidth]{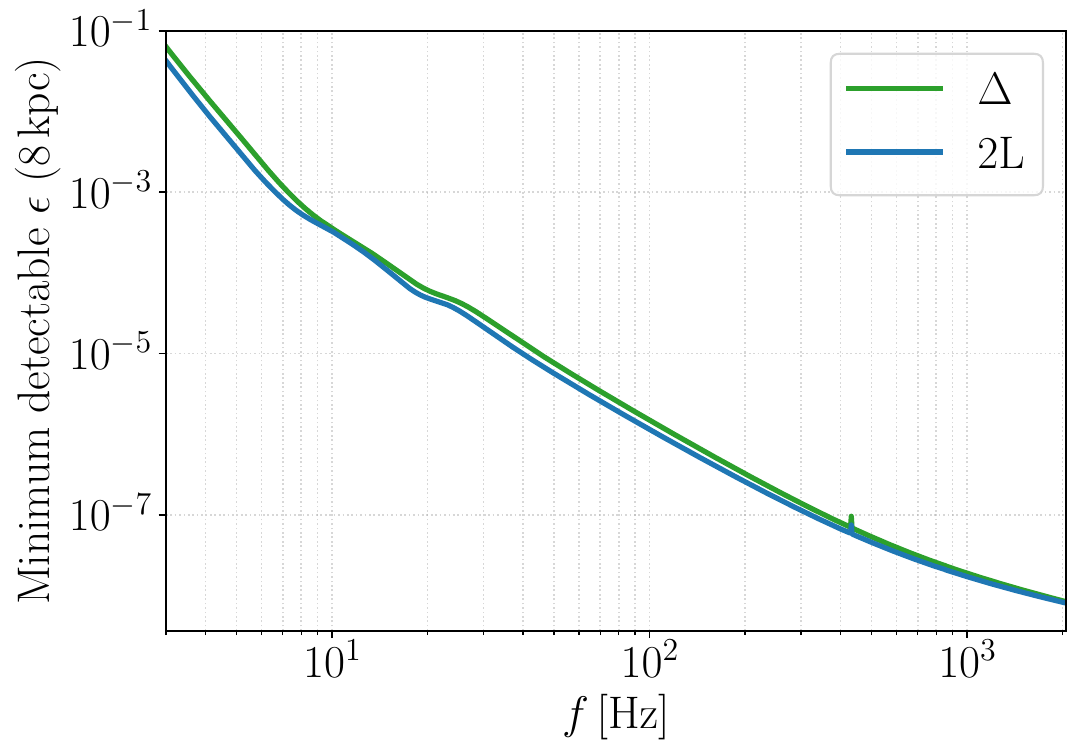}}
\captionof{figure}{Minimum detectable ellipticity as a function of frequency for an isolated pulsar at a distance of 8\,kpc for the $\Delta$ and 2L configurations. The curves assume a broadband search strategy.}
\label{fig:gw_broadb}
\end{tcolorbox}
\end{minipage}

Imposing the additional cap $\epsilon \leq 4\times10^{-6}$, motivated by the maximum deformation sustainable by an ideal NS crust, reduces these numbers to $204$ and $243$, respectively. For the more conservative long-term residual ellipticity cap, $\epsilon \leq 10^{-9}$, the number of detections drops substantially, but remains non-zero, with $13$ detectable pulsars for the $\Delta$ and $15$ for the 2L configuration. These numbers all assume an observation period of one year.

Overall, both metrics indicate that ET would provide a major improvement in continuous-wave sensitivity. The difference between the two geometries is modest compared with the astrophysical uncertainty in the ellipticity distribution, but the 2L configuration consistently gives a slightly larger number of detectable known pulsars and a lower minimum detectable ellipticity. This is mainly due to the longer arms. Even under conservative assumptions on the sustainable deformation, a small number of known pulsars could be within reach, while more optimistic scenarios would yield hundreds of detections.

\subsubsection{Bursts from core-collapse supernov\ae{}}

Core-collapse supernov\ae{} (CCSNe) are another major target for future GW detectors. When a massive star exhausts its nuclear fuel, the collapse of its iron core leads to the formation of a proto-NS or, in some cases, a BH. The dynamics of the collapse, bounce, accretion, convection, standing-accretion-shock instability, rotation, magnetic fields, and proto-NS oscillations can all generate gravitational radiation~\cite{Ott:2008wt,Muller:2026ofz}. Unlike CBCs, these signals are short, non-stationary, and inherently stochastic, with waveforms that depend sensitively on the progenitor structure and explosion mechanism.

A GW detection from a CCSN would give direct access to the central engine, which is hidden from electromagnetic observations~\cite{Powell:2023bex}. In combination with neutrinos and electromagnetic data, it would constrain the properties of the remnant and give information on the explosion mechanism~\cite{Nakamura:2016kkl,Warren:2019lgb}. Such observations would therefore be central to multimessenger studies of stellar death and compact-object formation.

The expected amplitudes are uncertain and span a broad range. Slowly rotating neutrino-driven explosions generally produce weaker, stochastic signals dominated by matter motions and proto-NS oscillations, while rapidly rotating collapse can produce stronger, more coherent emission around bounce~\cite{Dimmelmeier:2002bm}. BH-forming collapses may also generate distinctive high-frequency features associated with the final proto-NS evolution. Because of this diversity, burst searches are typically designed to be only weakly modeled, relying on excess power and time-frequency coherence across detectors rather than on a single template family~\cite{Astone:2018uge,KAGRA:2025yfg}.

Here we use a simpler approach: we estimate the detectability using a set of representative three-dimensional simulations. We consider six waveforms spanning progenitor masses between $9\,M_\odot$ and $23\,M_\odot$ from Ref.~\cite{Vartanyan:2023sxm}. These models sample a range of explosion dynamics and GW energetics, and therefore provide useful benchmarks for comparing detector configurations. For each waveform, we compute the optimal matched-filter SNR as a function of source distance and define the horizon distance as the distance at which the signal reaches ${\rm SNR}=8$. The results are obtained by randomizing over sky position and polarization angle.

\noindent
\begin{minipage}{0.52\textwidth}
\begin{tcolorbox}[standard jigsaw,colframe=gray,colback=gray!10!white,opacityback=0.0,coltext=black, title=\small\sc Horizon distance for selected CCSN simulations ]
\center{\includegraphics[width=0.99\textwidth]{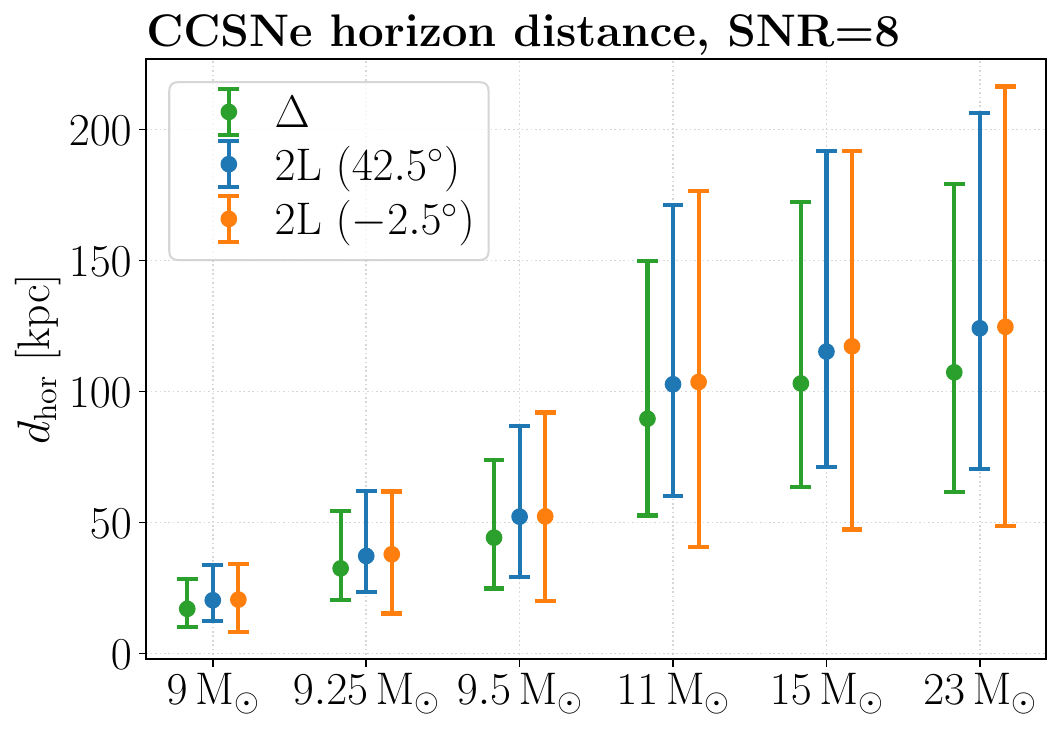}}
\captionof{figure}{Maximum distance out to which a CCSN signal from stars with different masses would be detected with an SNR of 8, for the three ET configurations. Markers show the median horizon distance and bars
the 90\% interval over the randomized extrinsic parameters. The calculation assumes matched filtering with an optimal template and random source extrinsic parameters.}
\label{fig:ccsne_hor}
\end{tcolorbox}
\end{minipage}
\hfill
\begin{minipage}{0.45\textwidth}

Because the expected horizons are at Galactic and Local-Group scales, cosmological effects can be neglected and the strain simply scales as $h\propto 1/d$. The SNR calculation is performed using the full detector response, without invoking the long-wavelength approximation, since CCSN signals can contain substantial power at high frequencies, where this approximation becomes inaccurate.

Figure~\ref{fig:ccsne_hor} reports the resulting horizon distances. As expected, the reach depends strongly on the progenitor model, reflecting the large intrinsic diversity of core-collapse GW emission. The weakest model, corresponding to the $9\,M_\odot$ progenitor, is detectable only out to $\sim 20$ kpc (and up to $\sim 30$ kpc for favorable sky positions and orientations), i.e., primarily within the Milky Way. More energetic models reach significantly farther: the $11\,M_\odot$,
$15\,M_\odot$, and $23\,M_\odot$ simulations have typical horizons of order $100$ kpc, with favorable sky locations extending beyond this, while the $9.25\,M_\odot$ and $9.5\,M_\odot$  models reach $\sim35$--$50$ kpc. This would include the Magellanic Clouds and, for the strongest orientations and models, other nearby satellite galaxies of the Milky Way (out to $\sim200$ kpc).

\end{minipage}

The geometry comparison reflects two competing effects. The 2L configurations benefit from their longer arms, which improve the strain sensitivity and generally lead to somewhat larger horizon distances than the $\Delta$ configuration. At the same time, CCSN waveforms contain a significant fraction of their power at high frequencies, where the finite-arm response becomes important. Since we do not use the long-wavelength approximation, this high-frequency response is included explicitly, and it partially penalizes the longer-arm 2L layouts. This helps explain why the differences between the configurations remain moderate despite the longer arms. Moreover, the 2L $(-2.5^\circ)$ configuration shows the largest spread in horizon distance across sky position and polarization. For favorable orientations, the aligned arms can combine to give a larger response and hence a higher SNR, producing the largest upper end of the horizon range; conversely, for unfavorable orientations both detectors are simultaneously insensitive to the same sky positions and polarizations, which lowers the lower end of the range. The differences are, however, modest compared with the spread induced by the supernova waveform itself and by the source orientation and sky position. 

These results should be interpreted as indicative rather than exhaustive. The selected waveforms do not span the full range of possible core-collapse dynamics. Moreover, realistic searches will generally rely on weakly modeled burst methods and may benefit substantially from electromagnetic or neutrino triggers. Nevertheless, the analysis demonstrates that ET would be sensitive to GWs from Galactic CCSNe across a broad range of models, and could extend the reach to nearby satellite galaxies for the more energetic explosions.

\begin{tcolorbox}[breakable,colback=teal!5!white,colframe=red!75!black,title=\sc Geometry Comparison]
{
ET could detect continuous waves from known pulsars even under conservative assumptions on the stellar ellipticity, and will provide strong sensitivity in broadband searches for unknown Galactic neutron stars. It will also be able to observe GW bursts from CCSNe across most of the Milky Way, with the more energetic models detectable beyond the Galaxy. 
\begin{itemize}
    \item {\em Both geometries can detect known pulsars:} Even for a conservative residual ellipticity, the $\Delta$ and 2L configurations can detect $13$ and $15$ pulsars respectively; for less conservative ellipticity assumptions, both geometries would detect hundreds of sources. 

    \item {\em Broadband continuous-wave sensitivity is very similar:} For an isolated neutron star at $8\,{\rm kpc}$, both layouts reach minimum detectable ellipticities of order $10^{-5}$--$10^{-6}$ at tens of Hz, improving to $\lesssim 10^{-7}$ at several hundred Hz and approaching $10^{-8}$ at the highest frequencies considered.

    \item {\em CCSN bursts are detectable beyond the Galaxy for energetic models:} Both geometries detect CCSNe across most of the Milky Way for all the waveforms considered (the weakest, $9\,M_\odot$ model is reached out to $\sim20$--$30$ kpc), while stronger explosions have horizons of order $100\,{\rm kpc}$, reaching the Magellanic Clouds and nearby satellite galaxies.

    \item {\em The 2L geometry is mildly favored but not decisively:} Its longer arms give a few more pulsar detections, lower detectable ellipticities, and mildly larger CCSN horizon distances, but the improvement remains small relative to uncertainties in the source physics and, for CCSNe, to the variation with waveform, sky position, and orientation.
\end{itemize}
}
\end{tcolorbox}

\newpage

\section{Summary of main results across the triangle and 2L geometries}
\label{summary-table}

\DefTblrTemplate{capcont}{default}{
  \UseTblrTemplate{caption-tag}{default}
  \UseTblrTemplate{caption-sep}{default}
  \InsertTblrText{entry}{\textit{Summary of main results across the three ET configurations considered.} (Continued)}
}
\begin{longtblr}[
    caption = {\textit{Summary of main results across the three ET configurations considered.}  ``X/yr'' denotes the number of detected events per year. The mass parameters in the table refer to source-frame masses. The BBH results under ``Compact-binary populations'' are reported jointly for BBH-LVK, Pop. III, and IMBH. The BNS results under ``Compact-binary populations'' are reported for the medium (SFHo) EOS. The fundamental physics results, except for the BNS-GRB row, are quoted for the BBH-LVK population only. Chirp masses and individual masses are in the source frame. Constraints on $H_0$ are obtained with BNS bright sirens, while $w_0$ and $w_a$ constraints are obtained with spectral sirens. Other cosmological constraints are reported in Table~\ref{tab:h0_summary}.
    },
    label = {tab:metrics}
]{width=0.99\linewidth,rowhead=1,colspec={X[5,l]X[3.5,c]X[3.5,c]X[3.5,c]}}
\toprule
Metric / criterion & 2L ($-2.5^\circ$) & 2L ($42.5^\circ$) & $\Delta$\\
\midrule
\SetCell[c=4]{l}    {{{\textbf{Compact-binary populations}}}} \\
BBH/yr                                                            & $86\,577$ & $90\,602$ & $81\,459$ \\
BBH/yr with SNR$>100$                                             & $3\,256$ & $3\,215$ & $1\,613$ \\
BBH/yr with $\Delta z/z<0.1$                                      & $7\,812$ & $15\,571$ & $6\,589$ \\
BBH/yr with $z>5$ and $\Delta z/z<0.1$                           & $129$ & $228$ & $79$ \\
BBH/yr with $\Delta {\cal M}_c/{\cal M}_c<10^{-2}$                & $288$ & $623$ & $220$ \\
BBH/yr with $\Delta m_1/m_1<0.1$                & $15\,599$ & $23\,760$ & $12\,016$ \\
BBH/yr with $\Delta m_2/m_2<0.1$                & $14\,361$ & $22\,437$ & $10\,930$ \\
BBH/yr with $\Delta d_L/d_L<10^{-2}$                              & $20$ & $42$ & $6$ \\
BBH/yr with $\Delta d_L/d_L<0.05$                              & $1\,285$ & $3\,113$ & $882$ \\
BBH/yr with $\Delta\Omega_{90\%}<10\,{\rm deg}^2$                 & $233$ & $1\,585$ & $746$ \\
BBH/yr with $\Delta\chi_{1,2}<10^{-2}$                            & $214, 73$ & $272, 74$ & $154, 53$ \\
BNS/yr                                                            & $12\,592$ & $12\,425$ & $7\,828$ \\
BNS/yr with SNR$>50$                                             & $108$ & $99$ & $48$ \\
BNS/yr with $z>1$ and $\Delta z/z<0.1$                           & $0$ & $3$ & $0$ \\
BNS/yr with $\Delta {\cal M}_c/{\cal M}_c<10^{-2}$                & $5$ & $20$ & $6$ \\
BNS/yr with $\Delta m_1/m_1<0.1$                & $34$ & $69$ & $26$ \\
BNS/yr with $\Delta m_2/m_2<0.1$                & $34$ & $74$ & $29$ \\
BNS/yr with $\Delta\Omega_{90\%}<100\,{\rm deg}^2$                & $35$ & $68$ & $21$ \\
BNS/yr with $\Delta\Omega_{90\%}<10\,{\rm deg}^2$                & $2$ & $5$ & $1$ \\
NSBH/yr                                                           & $10\,663$ & $11\,096$ & $7\,810$ \\
NSBH/yr with SNR$>50$                                            & $490$ & $502$ & $251$ \\
NSBH/yr with $z>1$ and $\Delta z/z<0.05$                          & $25$ & $54$ & $4$ \\
NSBH/yr with $\Delta {\cal M}_c/{\cal M}_c<10^{-2}$               & $35$ & $59$ & $18$ \\
NSBH/yr with $\Delta m_1/m_1<0.1$                & $1\,611$ & $2\,454$ & $1\,047$ \\
NSBH/yr with $\Delta m_2/m_2<0.1$                & $1\,615$ & $2\,480$ & $1\,063$ \\
NSBH/yr with $\Delta\Omega_{90\%}<100\,{\rm deg}^2$               & $113$ & $645$ & $312$ \\
NSBH/yr with $\Delta\Omega_{90\%}<10\,{\rm deg}^2$ & $6$ & $29$ & $12$ \\
Detection and inference horizon
    & $z_{\rm hor}=119$ & $z_{\rm hor}=119$ & $z_{\rm hor}=93$ \\
at $M=20\,M_\odot$ & $z_{\rm peak}=67$ & $z_{\rm peak}=66$ & $z_{\rm peak}=29$ \\
    & $z_{90\%}=33$    & $z_{90\%}=33$    & $z_{90\%}=11$ \\
\midrule
\SetCell[c=4]{l}{{{\textbf{Dense-matter physics}}}} \\
BNS/yr with $\Delta\tilde\Lambda<10$
    & soft: 2 & soft: 2 & soft: 1 \\
    & medium: 2 & medium: 2 & medium: 2 \\
    & stiff: 0 & stiff: 0 & stiff: 0 \\
BNS/yr with $\Delta\tilde\Lambda<50$
    & soft: 592 & soft: 668 & soft: 356 \\
    & medium: 293 & medium: 295 & medium: 149 \\
    & stiff: 92 & stiff: 107 & stiff: 51 \\
Population-stacked $\Delta R$ in km
    & soft: 0.150 & soft: 0.131 & soft: 0.141 \\
(using systems with ${\rm SNR}>100$) & medium: 0.083 & medium: 0.070 & medium: 0.073 \\
    & stiff: 0.032 & stiff: 0.027 & stiff: 0.028 \\
Postmerger SNR at 100\,Mpc
    & soft: $2.6^{+3.1}_{-1.9}$ & soft: $2.6^{+3.1}_{-1.5}$ & soft: $2.7^{+3.3}_{-1.6}$ \\
selected NR simulations, $q=1$ & stiff: $3.4^{+4.0}_{-2.5}$ & stiff: $3.4^{+4.0}_{-1.9}$ & stiff: $3.3^{+4.0}_{-1.9}$ \\
\midrule
\SetCell[c=4]{l}    {{{\textbf{Cosmology and the early Universe}}}} \\
Bright sirens: BNS/yr with $z < 0.4$, $\Delta\Omega_{90\%}<50\,{\rm deg}^2$ & $16$ & $26$ & $8$ \\
Golden (silver) dark sirens: BBH/yr, NSBH/yr with $\Delta d_L/d_L<0.1$, &  &  &  \\
$\Delta\Omega_{90\%}<0.1\,{\rm deg}^2$ (golden) & BBH: 0, NSBH: 0 & BBH: 0, NSBH: 0 & BBH: 0, NSBH: 0 \\  
$\Delta\Omega_{90\%}<1\,{\rm deg}^2$ (silver) & BBH:7, NSBH:0 & BBH:37, NSBH:1 & BBH:11, NSBH:1 \\
$\Delta H_0/H_0$ from $1\,$yr 68.3\% C.L. (target $<1\%$)                     & 2.9\% & 0.9\% & 2.4\% \\
Joint constraint on $w_0$ and $w_a$ from $1\,$yr 68.3\% C.L.    &  &  &  \\
$\Delta w_0 / w_0$ & 22.7\% & 22.8\% & 23.0\% \\
$\Delta w_a$ & 0.798 & 0.796 & 0.801 \\
PI sensitivity to SGWB $[\Omega_{\rm gw}]$ (population-independent, SNR = 5, $T_{\rm obs} = $1yr )                  & $7.3\times 10^{-13}$ & $8.3\times 10^{-12}$ & $2.0\times 10^{-12}$ \\
CBC foreground level after removing resolvable signals & $\sim10^{-11}$ & $\sim10^{-11}$ & $\sim10^{-11}$ \\
Discovery reach to primordial SGWB after performing joint analysis (SNR$=5,~T_\mathrm{obs}=1\mathrm{yr}$) & $\sim1.5\times 10^{-12}$ & $\sim1.7\times 10^{-11}$ & $\sim4\times 10^{-12}$ \\
SNR SGWB BBH ($T_{\rm obs} = $1yr) & 2300 & 200 & 880 \\
SNR SGWB BNS ($T_{\rm obs} = $1yr) & 51 & 4.5 & 20 \\
SNR SGWB NSBH ($T_{\rm obs} = $1yr) & 129 & 11 & 49 \\
SNR SGWB cosmic strings ($T_{\rm obs} = $1yr) & 747 & 65 & 266 \\
SNR SGWB FOPTs ($T_{\rm obs} = $1yr) & 49 & 4.3 & 17 \\
SNR SGWB axion inflation ($T_{\rm obs} = $1yr) & 1.96 & 0.17 & 0.70 \\
Degradation in SGWB sensitivity due to correlated noise (magnetic, Newtonian, medium mitigation)               & $\leq 2$ & $\leq 2$ & $\leq 25$ \\
\midrule
\SetCell[c=4]{l}    {{{\textbf{Fundamental physics and tests of general relativity}}}} \\
Precision range of parameterized PN tests (positive PN orders)           & $5.8\times10^{-5} \text{--} 1.3\times10^{-3}$ & $5.7\times10^{-5} \text{--} 1.3\times10^{-3}$ & $7.7\times10^{-5} \text{--} 1.7\times10^{-3}$ \\
Precision of dipole test           & $3.1\times10^{-8}$ & $3.1\times10^{-8}$ & $4.3\times10^{-8}$ \\
Precision of SIQM test           & $1.8\times10^{-3}$ & $1.8\times10^{-3}$ & $2.6\times10^{-3}$ \\
Precision range of bounds on negative-PN-order coefficients              & $5.6\times 10^{-18} \text{--} 1.1\times 10^{-12}$ & $5.6\times 10^{-18} \text{--} 1.1\times 10^{-12}$ & $8.2\times 10^{-18} \text{--} 1.5\times 10^{-12}$ \\
Precision range of ringdown spectroscopy tests (oscillation frequency)       & $1.3\times10^{-4} \text{--} 8.2\times10^{-3}$ & $1.2\times10^{-4} \text{--} 8.1\times10^{-3}$ & $1.7\times10^{-4} \text{--} 1.1\times10^{-2}$ \\
Precision range of ringdown spectroscopy tests (damping time)       & $5.6\times10^{-4} \text{--} 6.5\times10^{-2}$ & $5.4\times10^{-4} \text{--} 6.4\times10^{-2}$ & $7.3\times10^{-4} \text{--} 8.8\times10^{-2}$ \\
BBH/yr with merger-ringdown SNR$>100$                                   & $614$ & $614$ & $235$ \\
Joint BNS--GRB/yr with $\Delta\Omega_{90\%}<10^4\,{\rm deg}^2$ for $z>1$ (structured jet)    & 489 & 573 & 7 \\
\midrule
\SetCell[c=4]{l}    {{{\textbf{Multimessenger astronomy}}}} \\
BNS/yr with $\Delta\Omega_{90\%}<100\,{\rm deg}^2$ for $z>2$      & 0 & 0 & 0 \\
BNS/yr with $\Delta\Omega_{90\%}<100\,{\rm deg}^2$ for $z<2$  & 35 & 68 & 21 \\
BNS/yr with $z<0.4$ and & & & \\
$\Delta\Omega_{90\%}<10\,{\rm deg}^2$  & 2 & 5 & 1 \\
$\Delta\Omega_{90\%}<1\,{\rm deg}^2$  & 0 & 0 & 0 \\
BNS pre-merger/yr with $\Delta\Omega_{90\%}<100,\,10,\,1\,{\rm deg}^2$ at & & & \\
$1\,$h & 1, 0, 0 & 2, 0, 0 & 1, 0, 0 \\
$30\,$min & 2, 0, 0 & 6, 0, 0 & 3, 0, 0 \\
$10\,$min & 3, 0, 0 & 12, 1, 0 & 8, 0, 0 \\
$5\,$min & 4, 0, 0 & 17, 1, 0 & 11, 1, 0 \\
$1\,$min & 6, 1, 0 & 26, 1, 0 & 18, 1, 0 \\
NSBH/yr with $\Delta\Omega_{90\%}<100\,{\rm deg}^2$ for $z>2$ & 0 & 18 & 1 \\
NSBH/yr with $\Delta\Omega_{90\%}<100\,{\rm deg}^2$ for $z<2$ & 113 & 627 & 311 \\
NSBH/yr with $z<0.4$ and & & & \\
$\Delta\Omega_{90\%}<10\,{\rm deg}^2$  & 6 & 18 & 9 \\
$\Delta\Omega_{90\%}<1\,{\rm deg}^2$  & 0 & 1 & 1 \\
NSBH pre-merger/yr with $\Delta\Omega_{90\%}<100,\,10,\,1\,{\rm deg}^2$ at & & & \\
$1\,$h & 0, 0, 0 & 0, 0, 0 & 0, 0, 0 \\
$30\,$min & 0, 0, 0 & 1, 0, 0 & 1, 0, 0 \\
$10\,$min & 2, 0, 0 & 8, 0, 0 & 4, 0, 0 \\
$5\,$min & 3, 0, 0 & 20, 1, 0 & 9, 0, 0 \\
$1\,$min & 12, 1, 0 & 89, 5, 0 & 59, 3, 0 \\
Subset of with viewing angle $\Theta<\pi/12$:                &  &  &  \\
BNS/yr with $\Delta\Omega_{90\%}<100\,{\rm deg}^2$ & 2 & 5 & 1 \\
NSBH/yr with $\Delta\Omega_{90\%}<100\,{\rm deg}^2$ & 13 & 106 & 53 \\
Joint GW--EM detection rate/yr with reference EM facilities          &  &  &  \\
Rubin ($\Delta\Omega_{90\%}<100\,{\rm deg}^2,\ z<0.4$) & BNS: 35, NSBH: 76 & BNS: 49, NSBH: 92 & BNS: 17, NSBH: 52 \\
Roman ($\Delta\Omega_{90\%}<3\,{\rm deg}^2,\ z<0.4$) & BNS: 1, NSBH: 1 & BNS: 1, NSBH: 5 & BNS: 0, NSBH: 3 \\
\midrule
\SetCell[c=4]{l}    {{{\textbf{Continuous and burst sources}}}} \\
PSR detections from targeted searches (1 yr of data) for conservative $\epsilon$ ($\epsilon \leq 10^{-9}$) & $15$ & $15$ & $13$ \\
Minimum detectable PSR ellipticity, all-sky broadband (target $\lesssim 10^{-8}$) at $d=8$\,kpc and $f=1$\,kHz & $1.72\times10^{-8}$ & $1.72\times10^{-8}$ & $1.90\times10^{-8}$ \\
Maximum distance for CCSN detection $M_\star = 23M_\odot$    & $125^{+92}_{-76}$\,kpc & $124^{+82}_{-54}$\,kpc & $107^{+72}_{-46}$\,kpc \\
\end{longtblr}

\newpage

\section{Methodology and limitations of the analyses}
\label{app:methods}

\subsection{Methodology}

\paragraph*{\bf Forecasting approach.} 
Assessing the science capabilities of a future observatory requires forecasting
the properties of populations that have not yet been observed and the precision
with which their parameters could be measured. For the very large populations
expected with ET, the only computationally practical method currently available
for carrying out such forecasts over the full source population is the Fisher
information matrix (FIM)~\cite{Cutler:1994ys,Vallisneri:2007ev,Iacovelli:2022bbs}. 
The latter has been widely employed in studies of this kind 
for ground-based and space-based GW interferometers, see e.g.~\cite{Cutler:1997ta,Dupletsa:2022scg,Iacovelli:2022bbs,Borhanian:2022czq}. The FIM provides a local, quadratic approximation to the log likelihood around the true source parameters; its inverse, when it exists and is well conditioned, approximates the covariance matrix of the inferred parameters~\cite{Cutler:1994ys,Vallisneri:2007ev}. This makes it possible to evaluate parameter uncertainties and
correlations for the several hundred thousand sources required for a comprehensive comparison of detector geometries. Since the validity of this approximation is not guaranteed for every ET source, we complement the population-scale FIM analysis with full Bayesian inference on a representative subset of signals. The combination allows us both to survey the full science reach of each geometry and to assess where the approximations entering those forecasts are reliable.

\paragraph*{\bf Simulated datasets and analysis strategy.}

In the ``Science Golden Channels''  document, we proposed to organize the analysis around three datasets of increasing realism~\cite{ET-GCR}. Dataset 1 consists of simulated astrophysical populations corresponding to 1 yr of BBH observations and 5 yr of BNS and NSBH observations, the longer period for NS systems being used to obtain adequate statistics for the relevant science metrics. Source-parameter uncertainties are evaluated using the FIM. For BBHs, uncertainties in population hyperparameters are also computed using a population Fisher analysis~\cite{Gair:2022fsj, DeRenzis:2024dvx}. Population catalogs are generated using \texttt{gwfast.population}~\cite{DeRenzis:2024dvx}; optimal SNRs and event-level Fisher matrices are calculated using \texttt{gwfast}~\cite{Iacovelli:2022bbs}/\texttt{GWFish}~\cite{Dupletsa:2022scg}, and population FIM calculations use \texttt{gwfast.population}. Dataset 1 provides the statistics required for the principal geometry comparison presented in this report.

Dataset 2 was intended to address data-analysis effects that cannot be represented by the Fisher calculation. It was to consist of at least one day of simulated strain data containing stationary Gaussian noise consistent with the official ET sensitivity curves, astrophysical signals drawn from the same populations as Dataset 1, and instrumental and environmental glitches. Glitches would be mitigated using \texttt{BayesWave}  ~\cite{Cornish:2014kda}, for the 2L configuration and the null stream together with \texttt{BayesWave} for the triangle. The cleaned data would then be analyzed with \texttt{Bilby}~\cite{Ashton:2018jfp,Romero-Shaw:2020owr} to obtain Bayesian posterior distributions for the source parameters. Time-dependent ET antenna patterns would be included and likelihood heterodyning would be used, when possible, to reduce the computational cost. For science cases not adequately represented in one day of observations, Dataset 2 would be supplemented with selected signals drawn from Dataset 1. Importantly, because \texttt{BayesWave} cannot be employed below 20 Hz, the Dataset-2 analysis would have been limited to signals above 20 Hz.

Dataset 3 was designed to share the same signal populations, Gaussian noise, and Bayesian analysis as Dataset 2, but to contain no glitches. It therefore serves two purposes. A comparison of Datasets 2 and 3 (when restricting to signals above 20 Hz) would isolate the impact of glitches and their mitigation on the science metrics, while a comparison of Datasets 1 and 3 provides a direct validation of the FIM approximation against full Bayesian inference. The latter comparison is performed at the level of individual events and distributions of measurement uncertainties. Particular attention is paid to whether any discrepancy between FIM and Bayesian inference depends systematically on detector geometry, rather than merely affecting the absolute forecast for all configurations.

Given the time available for the geometry comparison, we prioritized Dataset 1, which we completed in full, together with a subset of Dataset 3, used to cross-validate Dataset 1 and to identify potential pitfalls of the FIM approximation. A full treatment of Dataset 2 lay beyond the timescale of this report and will be pursued as an independent study, to be published in a peer-reviewed journal. Dataset 3 will be completed within the next three months and results will be summarized in an Addendum.

\paragraph*{\bf Source populations and waveform models.}
The source populations are chosen to reflect current astrophysical knowledge.
For BBHs we adopt the \textsc{Broken Power-Law + 2 Peaks} source-frame mass
distribution and the Gaussian precessing-spin model inferred in the latest LVK
analysis \cite{LIGOScientific:2026ctl}. BNS source-frame masses are drawn from
a uniform distribution, with small aligned spins,
$|\chi_{1,2}|\leq0.05$. NSBH systems follow the \textsc{NSBH-pop} model
\cite{Biscoveanu:2022iue,LIGOScientific:2024elc}; because the present LVK NSBH
mass distribution is inferred from only a few observations, we allow its BH
mass power law to extend naturally to higher masses rather than imposing the
best-fit upper cutoff. Merger rates follow a Madau--Dickinson form, with an
additional power-law delay-time distribution for BNSs and NSBHs and a minimum
delay of 40\,Myr. We also consider selected populations of high-redshift
Pop~III BBHs, IMBHs, and primordial black holes, for which ET's low-frequency
sensitivity is particularly important. For NS matter we consider
representative soft, medium, and stiff equations of state (APR4, SFHo, and H4, respectively). Unless
otherwise specified, remaining parameters are drawn uniformly and Planck18
cosmology is assumed \cite{Planck:2018vyg}.

Waveforms are generated with
\texttt{IMRPhenomXPHM}~\cite{Pratten:2020ceb} for BBHs and
\texttt{IMRPhenomXAS\_NRTidalv3}~\cite{Pratten:2020fqn,Abac:2023saz}
for BNSs. For NSBHs we use
\texttt{IMRPhenomXHM}~\cite{Garcia-Quiros:2020qpx} for detectability
and extrinsic-parameter forecasts and
\texttt{IMRPhenomNSBH}~\cite{Matas:2020wab} for
tidal-deformability measurements. Tests of general relativity use the
Flexible Theory Independent (\texttt{FTI}) framework
\cite{Mehta:2022pcn} for parameterized PN tests and \texttt{pSEOBNR}
\cite{Brito:2018rfr,Ghosh:2021mrv,Pompili:2025cdc} for ringdown tests.
Following previous ET studies
\cite{Branchesi:2023mws,ET:2025xjr}, we adopt detection thresholds of
SNR 12 for BBHs and NSBHs and 8 for BNSs. Dataset~1 analyses use a
lower frequency cutoff of 3\,Hz.

\paragraph*{\bf Detector duty cycle.}
For the population-scale FIM calculations, we retain the
simple duty-cycle prescription adopted in previous ET forecasts, namely we consider an uncorrelated duty cycle of 85\% for each of the two L detectors in both the 2L configurations, 2L\,($-2.5^{\circ}$) and 2L\,($42.5^{\circ}$), and for each of the three detectors in the triangle.

\subsection{Limitations}

\paragraph*{\bf Validity of the FIM approximation.}
The scale of this geometry-comparison study makes the FIM indispensable, but
also requires care in interpreting its predictions. The FIM is obtained by
expanding the log likelihood to quadratic order about the true parameters.
The resulting likelihood is therefore locally Gaussian, and the inverse
Fisher matrix approximates the covariance of the inferred parameters ~\cite{Balasubramanian:1995bm,Vallisneri:2007ev}. 
This approximation becomes asymptotically reliable at high SNR, although the SNR
required in practice depends on the dimensionality of the parameter space,
parameter degeneracies, multimodalities, waveform structure, and the parameters being measured~\cite{Vallisneri:2007ev, Rodriguez:2013mla}.

These asymptotic conditions are not uniformly satisfied by the cosmological
population accessible to ET. Its extraordinary reach implies that many
detected sources will lie close enough to the detection threshold that the
likelihood need not be well described by a quadratic expansion. Moreover,
compact-binary inference involves a high-dimensional parameter space with
strong correlations among distance, inclination, polarization, sky position,
masses, and spins. Consequently, Fisher forecasts should not be interpreted
as exact event-by-event credible intervals for the entire detected
population~\cite{Vallisneri:2007ev}. Nevertheless, for the majority of detected sources, we expect the
approximation to provide a useful estimate of measurement precision, and it
should be substantially more reliable for the nearby, high-SNR sources that
dominate many of ET's precision-science applications.

The distinction is particularly important for parameters with strong
degeneracies or non-Gaussian likelihoods. The FIM can either underestimate or
overestimate measurement uncertainties when the quadratic approximation
fails. It cannot describe multiple separated modes of a likelihood and does
not by itself enforce physical parameter boundaries or informative priors~\cite{Vallisneri:2007ev}.
These effects can be especially relevant to extrinsic parameters, including
distance, inclination, polarization, and sky position. Priors and physical
boundaries can be imposed by sampling the Gaussian likelihood implied by the
FIM but this does not restore non-Gaussian structure or
multimodality absent from the quadratic approximation~\cite{Dupletsa:2024gfl, Santoliquido:2025lot}. Nonetheless, their inclusion can improve the accuracy of the FIM and the associated measurement uncertainties. We do not use this methodology here.

\paragraph*{\bf Network geometry and FIM conditioning.}
An additional issue arises from the detector networks considered here. The
triangle is a single-site observatory and 2L consists of only two
geographically separated sites. In the absence of a third non-collocated
site, some combinations of extrinsic parameters can be only weakly
constrained, particularly for short-duration signals for which Earth's
rotation provides little additional baseline. The resulting FIM can
be poorly conditioned, making its numerical inversion and hence the inferred
covariance matrix unreliable. Appropriate numerical conditioning and
diagnostics are therefore essential \cite{Borhanian:2020ypi, Dupletsa:2022scg, Iacovelli:2022mbg}. 
Full Bayesian inference does not require
inversion of a FIM: it evaluates the likelihood over parameter space
and can consequently represent extended degeneracies, curved correlations,
physical boundaries, and multiple posterior modes. It thus provides the
appropriate benchmark against which the FIM
forecasts should be tested.

\paragraph*{\bf Validation of FIM Dataset 1 against Bayesian inference from Dataset 3.} For the purpose of this report, the most important question is not whether FIM and Bayesian inference give identical uncertainties for every source, but whether departures from the FIM approximation introduce a
\emph{geometry-dependent bias} in the comparison between the triangle and 2L.
Previous comparisons of FIM and Bayesian parameter estimation indicate that discrepancies can occur, particularly for distance and sky localization, but overall do not show evidence that they systematically favor one of these detector geometries over the other (e.g., see~\cite{Santoliquido:2025aiq, Pandey:2024mlo}). This point is important because localization enters directly or indirectly into several of the science metrics considered here, including multimessenger astronomy, dark-siren measurements of $H_0$, and dense-matter EOS studies.

We use Dataset 3 to test this assumption. For a subset of events corresponding to 1 day for BBHs and 5 days for NSBHs, we perform a full Bayesian inference of the posterior distribution using \texttt{Bilby}~\cite{Ashton:2018jfp} in combination with nested sampling. For BBH sources, we analyze the signals embedded in Gaussian noise starting from a minimum frequency of 5\,Hz (3\, Hz are used in the FIM analysis in the main text), employing the same waveform approximant. We choose standard priors~\cite{Romero-Shaw:2020owr}, namely uniform in detector-frame component masses, uniform in comoving volume and source-frame time for the luminosity distance, uniform in cosine (sine) inclination and spin tilts (declination), uniform for the remaining angles and merger time, and uniform in spin magnitudes. For NSBHs we adopt similar prior distributions and start the analysis from a minimum frequency of 3\,Hz, exploiting relative binning to make the analysis of these long signals computationally feasible. Since using this technique with signals embedded in Gaussian noise can lead to issues in the sampling procedure, for this analysis we restrict to zero-noise injections. 

Notice that, while for the selected BBHs Earth motion during the observation has a negligible impact of the analysis, for NSBHs this is not the case. This impact is not accounted for in the Bayesian analysis, as no reviewed implementation is available at the time of writing. Accounting for this effect would result in an improvement in the localization constraints. Additionally, we do not include any duty cycle for the analysis of Dataset 3. It should also be noted that the FIM results have been recomputed accordingly for the Dataset 3 events, matching the starting frequency used in \texttt{Bilby} and dropping Earth's rotation. FIM analysis of Dataset 3 works in the zero-noise approximation for both BBHs and NSBHs. Finally, we again stress that no prior is applied to the FIMs in this analysis.

Of the $\approx 310$ BBHs and $\approx 340$ NSBHs in the two catalogs, we analyzed 176 and 212 events, respectively, in both configurations, 2L ($42.5^\circ$) and $\Delta$. This sample, although limited, provides a representative benchmark for assessing the validity of the FIM approximation and, crucially, whether it biases the comparison between configurations.

\begin{figure}
\begin{tcolorbox}[standard jigsaw,colframe=gray,colback=gray!10!white,
opacityback=0.0,coltext=black,
title=\small\sc Relative ratio of the FIM likelihood and Bayesian posterior distribution for selected parameters for BBHs]
\centering
\includegraphics[width=\textwidth]{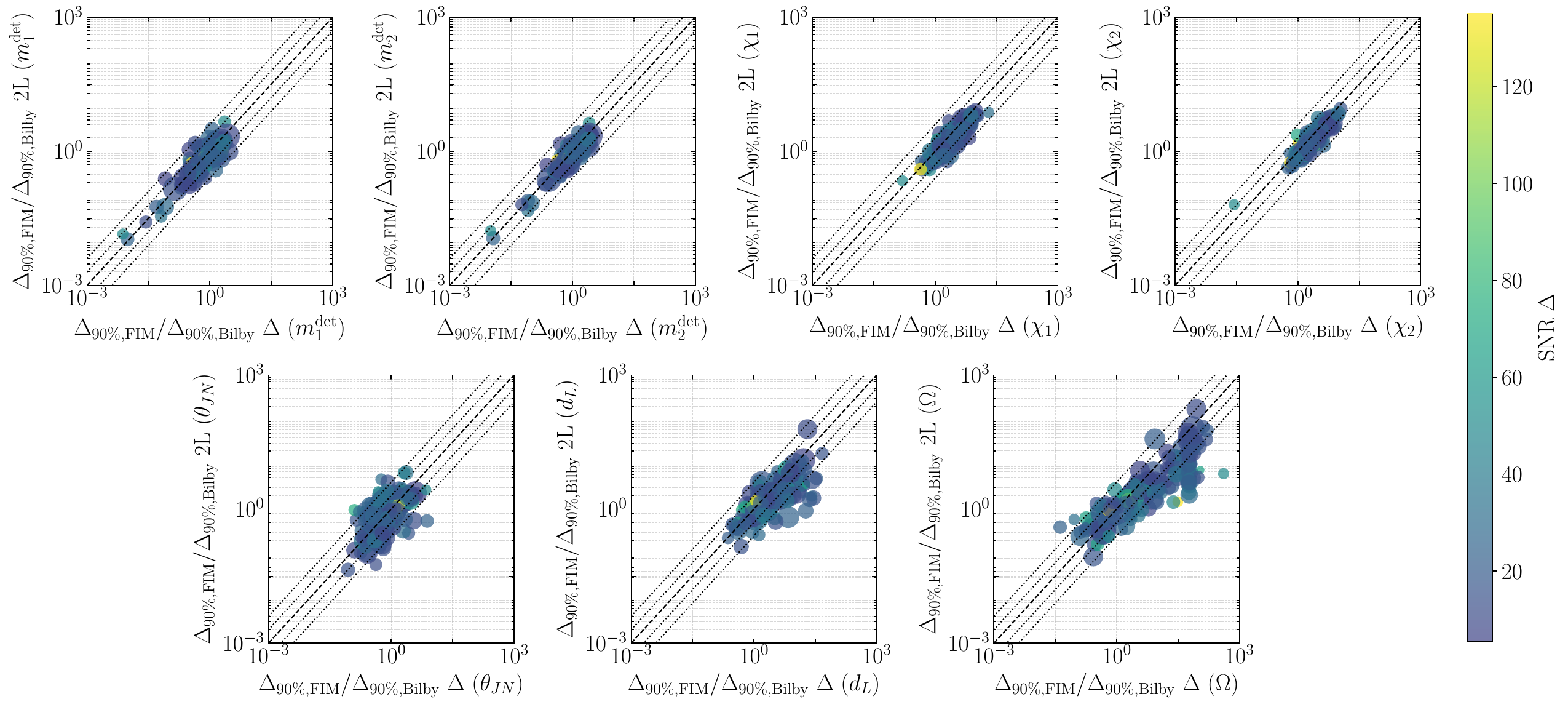}
\caption{\justifying We report the ratio between the uncertainty predicted for BBHs by the FIM likelihood and the posterior obtained with Bayesian analysis (performed with \texttt{Bilby}) at 90\% credible level for the two detector configurations considered, 2L ($42.5^\circ$) on the $y$-axis and $\Delta$ on the $x$-axis. The color of the points reports their SNR in the $\Delta$ configuration, while their size is proportional to the luminosity distance of the source. In each panel, the dashed lines highlight the diagonal $y=x$ and the dotted lines show $y=\alpha x$ with $\alpha=\{4,2,1/2,1/4\}$.}
\label{fig:app_methodology_ratio_FIMBil_BBH}
\end{tcolorbox}
\vskip-15pt
\end{figure}

\begin{figure}
\begin{tcolorbox}[standard jigsaw,colframe=gray,colback=gray!10!white,
opacityback=0.0,coltext=black,
title=\small\sc Relative ratio of the FIM likelihood and Bayesian posterior distribution for selected parameters for NSBHs]
\centering
\includegraphics[width=0.74\textwidth]{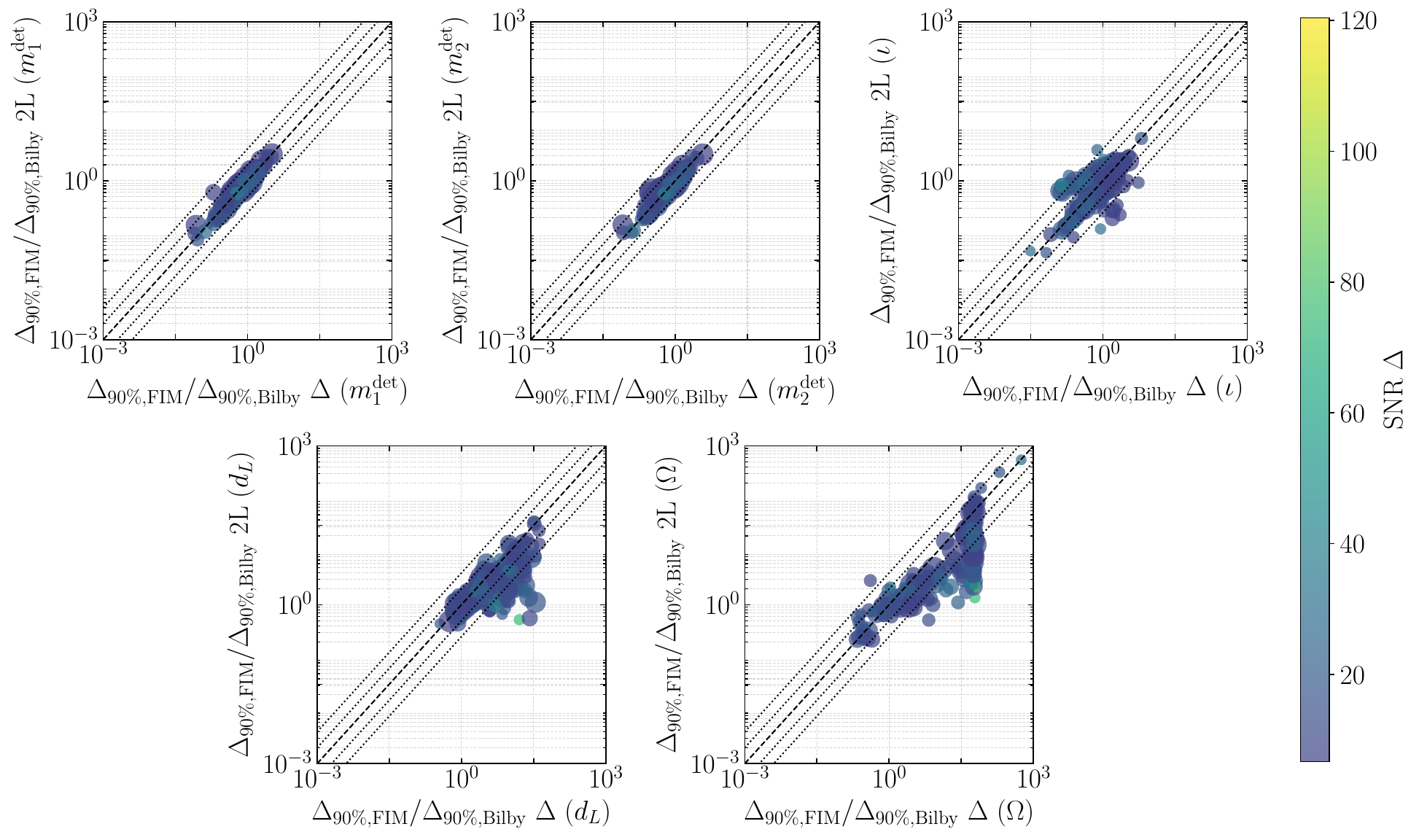}
\caption{\justifying Same as in Fig.~\ref{fig:app_methodology_ratio_FIMBil_BBH} for NSBH systems.}
\label{fig:app_methodology_ratio_FIMBil_NSBH}
\end{tcolorbox}
\vskip-15pt
\end{figure}

The results of this comparison are presented in Figs.~\ref{fig:app_methodology_ratio_FIMBil_BBH} and~\ref{fig:app_methodology_ratio_FIMBil_NSBH}. For the BBH injections, we see that the FIM formalism gives statistical uncertainties within factors of ${\cal O}({\rm few})$ of the Bayesian posteriors for the detector-frame masses and spins, and without any strong preference for one configuration over the other. The similarity of the two results improves for higher SNR systems, as expected. This points to the fact that the FIM formalism is a good approximation for intrinsic parameters.
For the inclination angle $\theta_{\rm JN}$, the results show a similar trend, albeit exhibiting a larger spread. This can be traced to the fact that the FIM is not able to capture multimodal posteriors. For the luminosity distance, the ratios are generally higher than unity, especially for low-SNR systems. This is due to correlations between this parameter and the other extrinsic parameters, whose multimodality is not captured by the FIM, and to the absence of physical prior boundaries. Nonetheless, more than 50\% of the points are within a factor of 2 of the posterior width found with Bayesian inference, and 80\% within a factor of 3 (4) for the 2L ($\Delta$) configuration. For the sky localization the situation is instead more complex. Here the FIM is not able to capture multimodalities in the relevant parameters present for a large fraction of the considered systems, leading to severe under- or overestimation of the sky area. This is exacerbated by the absence of priors in the FIM analysis, which can lead to extremely broad FIM error estimates. For the 2L configuration, we find most of the results to lie within a factor of 10 of the posterior found with Bayesian inference, while for the triangle the spread is even larger. For NSBH systems we observe a similar trend. Agreement between the FIM and Bayesian results for the extrinsic parameters is expected to greatly improve when adding more detectors in different locations, which help cure the multimodal structure of the posterior~\cite{Pandey:2024mlo}.

We next turn our attention to potential biases induced by the FIM in the analysis of the two configurations. As can be seen from Figs.~\ref{fig:app_methodology_ratio_FIMBil_BBH} and~\ref{fig:app_methodology_ratio_FIMBil_NSBH}, we do not find a relevant bias in the statistical uncertainties predicted by the FIM formalism for the mass and spin parameters. For extrinsic parameters, however, the situation is different. In this case, the FIM tends to overestimate the uncertainty on the luminosity distance for the triangular configuration. This can be partly traced to the fact that the 2L configuration always has a higher SNR, a regime in which the FIM is known to perform better. %
For the sky localization, we observe a mild preference for the 2L configuration in the BBH results, and a strong preference in the NSBH ones. The preference mostly impact events with broad sky localization in the Fisher analysis. The observed vertical feature in the NSBH populations corresponds to events of this kind, which are more common in the $\Delta$ configuration.

Finally, in Table~\ref{tab:metrics_bilbycomparison} we report the number of sources among the ones analyzed with Bayesian inference that respect the same criteria as in Table~\ref{summary-table} for compact binaries. The quoted numbers on the left come from the Bayesian analysis, while those on the right come from the corresponding FIM analysis of the same events. %
Both for BBH and NSBH systems we find good agreement for the events passing the cut on the redshift uncertainty, with a mild preference for the 2L configuration. Concerning the source-frame masses, instead, we find the FIM results to overestimate the precision. This can be traced to the correlated structure of the posterior in those parameters, which the FIM is not able to capture, and leads to a broader distribution within the Gaussian approximation. This is more severe for NSBH systems.
Nonetheless, we find the Bayesian results, albeit with a limited statistics, to confirm the hierarchy in the configurations present in the FIM results. %
We notice, however, numbers closer to Bayesian inference for the 2L configuration concerning the mass cuts. Moreover, we find the localization results with Bayesian analysis for both BBHs and NSBHs to show a superiority of the 2L configuration even larger than that suggested by the FIM approach for this selection of events.

\paragraph*{\bf Summary.} The comparison with full Bayesian inference, therefore, supports the use of the FIM for the principal conclusions of this study. Although the FIM can misestimate uncertainties for individual events---most notably for extrinsic parameters with multimodal or non-Gaussian posteriors---it performs well for intrinsic parameters and, crucially, does not introduce a significant bias in the relative comparison of the detector geometries for the principal science drivers used in this study. Where differences do arise, e.g., for luminosity distance and sky localization, the Bayesian analysis generally preserves the hierarchy found with the FIM for the two configurations. Thus, while absolute FIM uncertainties and event counts should be interpreted with caution, especially for localization and redshift-dependent quantities, the comparative conclusions drawn from the FIM analysis are broadly robust.

\DefTblrTemplate{capcont}{default}{
  \UseTblrTemplate{caption-tag}{default}
  \UseTblrTemplate{caption-sep}{default}
  \InsertTblrText{entry}{Results for the two ET configurations, comparing full parameter estimation (left, Bayesian posterior sampling) with the FIM approximation (right). (Continued)}
}
\begin{longtblr}[
    caption = {Results for the two ET configurations, comparing full parameter estimation (left, Bayesian posterior sampling) with the FIM approximation (right). Results are shown for the analyzed subset of Dataset 3: 176 BBHs and 212 NSBHs. Zero entries indicate that no event in the subset meets the corresponding criterion.
    },
    label = {tab:metrics_bilbycomparison}
]{width=0.99\linewidth,rowhead=1,colspec={X[5,l]X[3.5,c]X[3.5,c]}}
\toprule
Metric / criterion & 2L ($42.5^\circ$) & $\Delta$\\
\midrule
\SetCell[c=3]{l}    {{{\textbf{Compact-binary populations}}}} \\
BBH with $\Delta z/z<0.1$                                      & 36\,/\,34 & 21\,/15  \\
BBH with $z>5$ and $\Delta z/z<0.1$                           & 0\,/\,1 & 0\,/\,0  \\
BBH with $\Delta {\cal M}_c/{\cal M}_c<10^{-2}$                & 0\,/\,0 & 0\,/\,0  \\
BBH with $\Delta m_1/m_1<0.1$                & 59\,/\,42 & 43\,/\,19 \\
BBH with $\Delta m_2/m_2<0.1$                & 48\,/\,36 & 39\,/\,15 \\
BBH with $\Delta d_L/d_L<10^{-2}$                              & 0\,/\,0 & 0\,/\,0 \\
BBH with $\Delta d_L/d_L<0.05$                              & 4\,/\,3 & 1\,/\,1 \\
BBH with $\Delta\Omega_{90\%}<100\,{\rm deg}^2$                 & 20\,/\,28 & 4\,/\,16 \\
BBH with $\Delta\Omega_{90\%}<10\,{\rm deg}^2$                 & 1\,/\,1 & 0\,/\,0 \\
BBH with $\Delta\chi_{1,2}<10^{-2}$                            & 0\,/\,0 & 0\,/\,0 \\
NSBH with$\Delta z/z<0.1$ & 16\,/\,$16$ & 10\,/\,$7$ \\
NSBH with $z>1$ and $\Delta z/z<0.05$                          & 0\,/\,$1$ & 0\,/\,$0$ \\
NSBH with $\Delta {\cal M}_c/{\cal M}_c<10^{-2}$               & 1\,/\,1 & 0\,/\,$0$ \\
NSBH with $\Delta m_1/m_1<0.1$ & 87\,/\,41 & 70\,/\,20 \\
NSBH with $\Delta m_2/m_2<0.1$ & 84\,/\,41 & 70\,/\,22 \\
NSBH with $\Delta\Omega_{90\%}<100\,{\rm deg}^2$               & 9\,/\,12 & 3\,/\,8 \\
NSBH with $\Delta\Omega_{90\%}<10\,{\rm deg}^2$ & 0\,/\,0 & 0\,/\,0 \\
\bottomrule
\end{longtblr}

\acknowledgments

The authors thank INFN-Tier 1 at CNAF and the Max Planck Institute for Gravitational Physics for providing the computational resources to carry out some of the analyses presented in this work.
\\
We thank Daniel Holz, Raffaella Margutti, Alexander Nitz, Jocelyn Read, Gabriele Vajente, and Nico Yunes for their careful review of, and helpful comments on, the definition of the Golden Channels used in this report. We also thank Sophie Bini, Giacomo Bruno, Francesco Cireddu, Tito Dal Canton, Justin Janquart, Michela Mapelli, Anuradha Samajdar, Filippo Santoliquido, Jacopo Tissino, Isaac Wong, and Thibeau Wouters for helpful discussions.

\appendix 

\section{Detector locations for the geometry-comparison study}
\label{detectors}

\begin{figure*}
\begin{center}
    \includegraphics[width=0.49\linewidth]{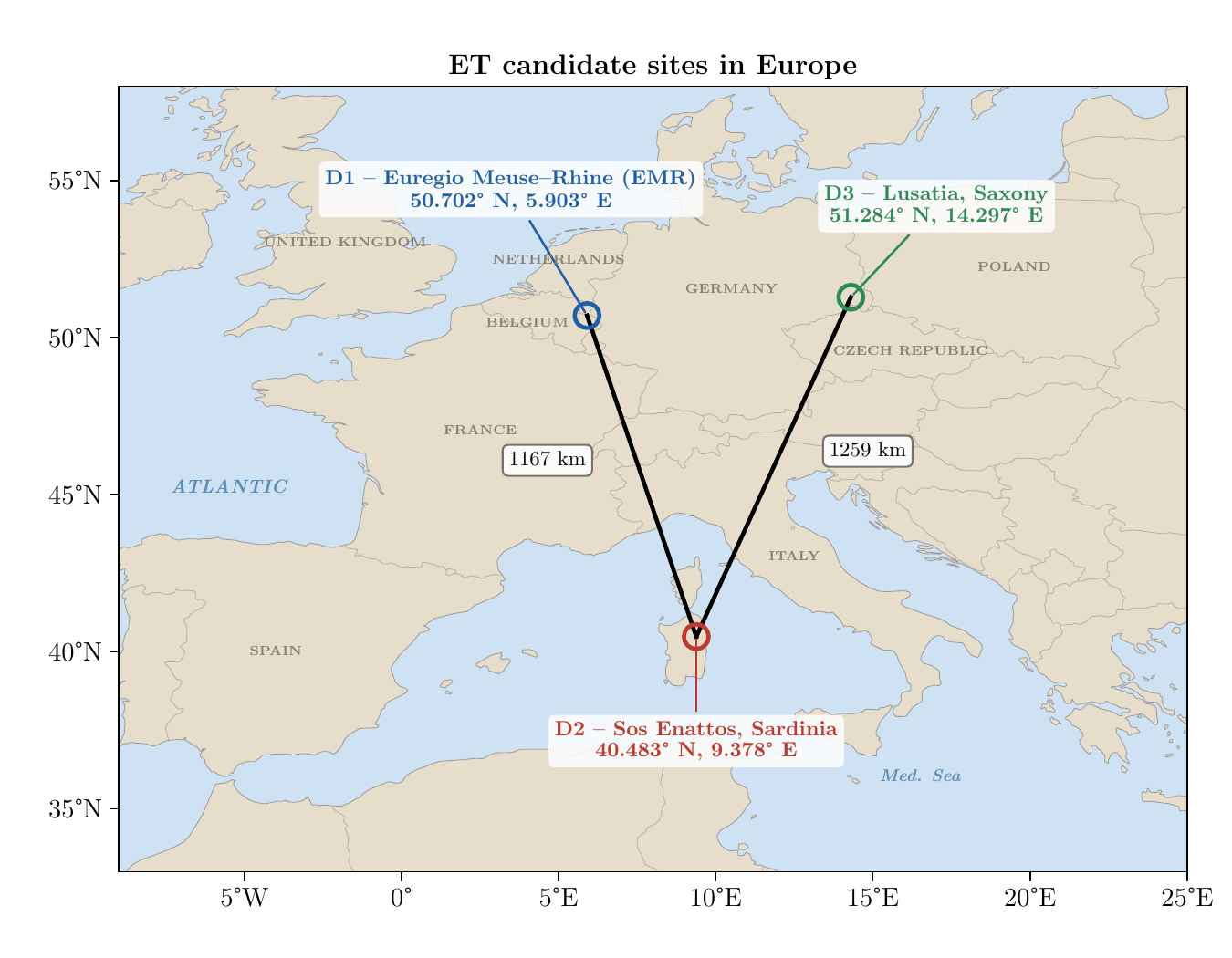}
     \includegraphics[width=0.49\linewidth]{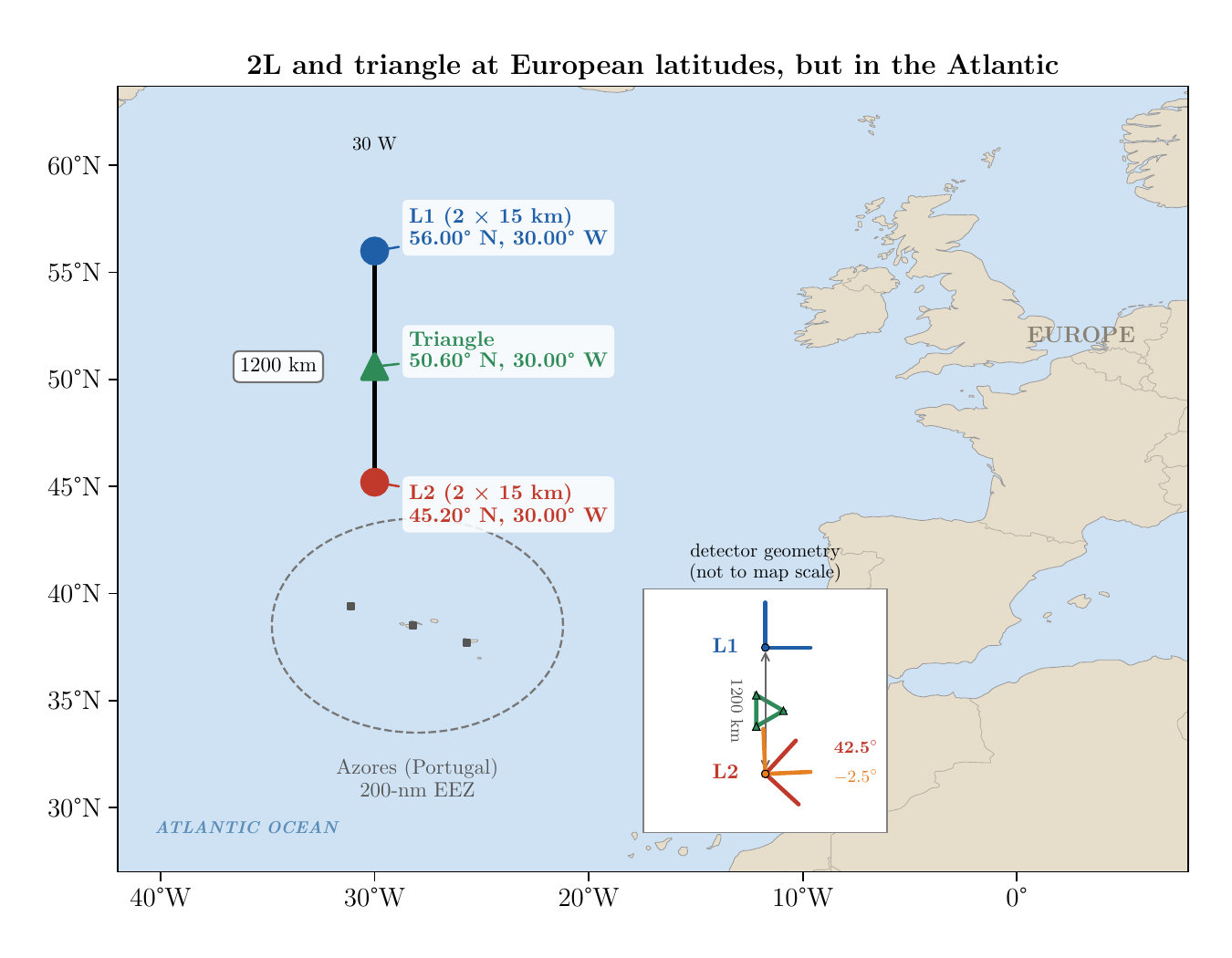}
\end{center}
    \caption{\justifying Candidate ET sites (left) and the detector locations adopted in this study (right), drawn at the same scale. (left panel) The three European candidate detector sites, each marked by a circle:
  D1, Euregio Meuse--Rhine (EMR; $50.702^\circ$N, $5.903^\circ$E; approximate); D2, Sos Enattos,
  Sardinia ($40.483^\circ$N, $9.378^\circ$E; approximate); and D3, Lusatia, Saxony
  ($51.284^\circ$N, $14.297^\circ$E; approximate).
  (right panel)~We use an ocean site for the 2L--$15\,$km and $\Delta$--$10\,$km
  geometries, at a North-Atlantic location on the $30^\circ$W meridian but
  translated north to European latitudes (L1 at $56.0^\circ$N, triangle at
  $50.6^\circ$N, L2 at $45.2^\circ$N). The two L-shaped detectors (L1, L2) are
  $1200\,$km apart and the $\Delta$--$10\,$km triangular detector sits at their
  midpoint; each L has $90^\circ$ between its two $15\,$km arms, with the two
  L's rotated by $42.5^\circ$ or $-2.5^\circ$  relative to each other (inset, oriented along the
  true baseline). The arms are negligibly small against the $1200\,$km baseline
  and are not drawn to map scale. All three detectors lie in the high seas: the
  configuration is shifted north along the meridian so that L2 clears Portugal's
  Azores Exclusive Economic Zone (dashed ellipse shows the approximate $200$ nautical-mile EEZ; small squares mark the Azores islands).}
    \label{fig:2LT_sites}
\end{figure*}

\begin{figure*}
\begin{center}
    \includegraphics[width=0.49\linewidth]{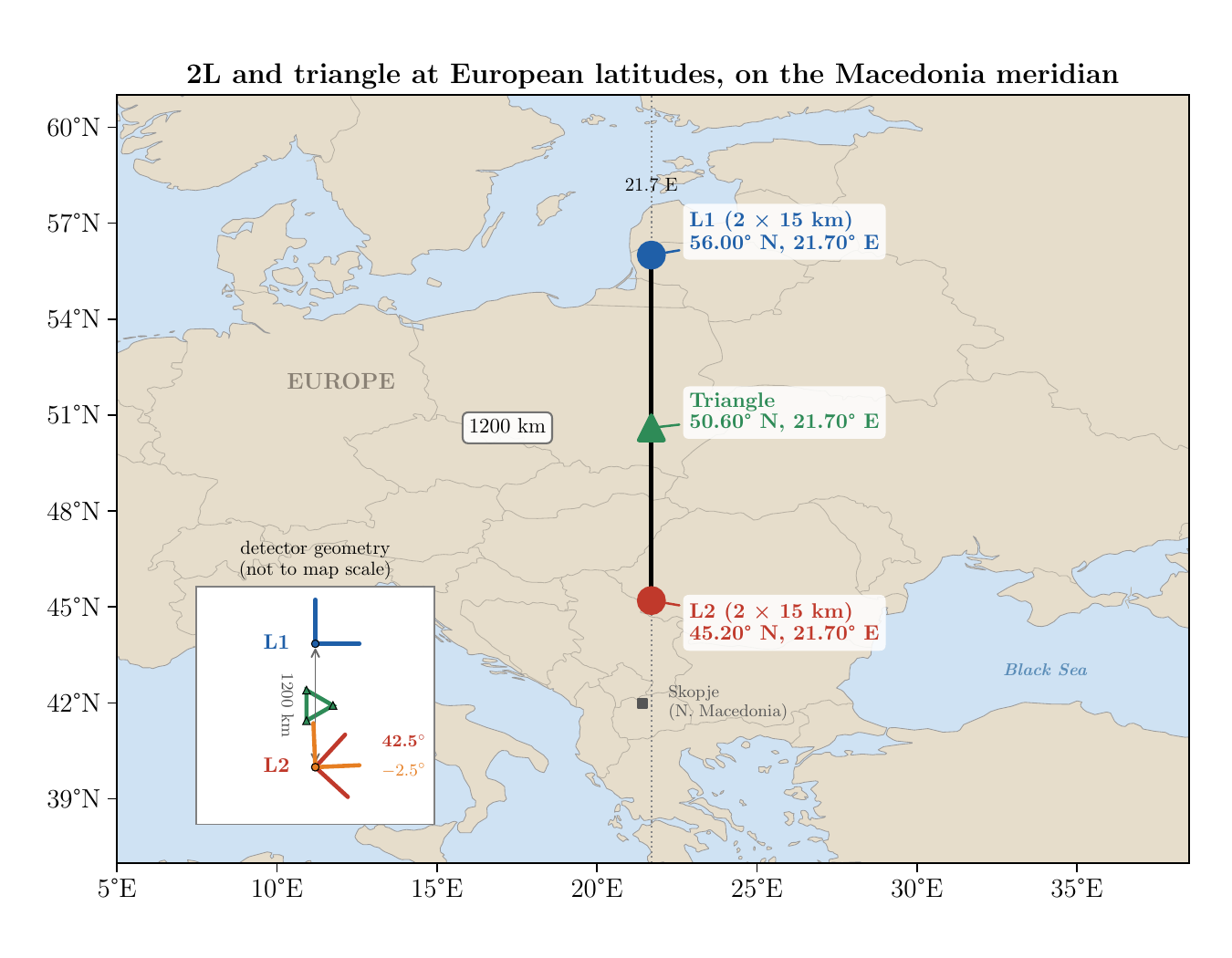}
\end{center}
    \caption{\justifying Alternative on-shore siting of the 2L--$15\,$km and
  $\Delta$--$10\,$km geometries. The three detectors keep the same latitudes as
  the North-Atlantic ocean site of Fig.~\ref{fig:2LT_sites} (L1 at $56.0^\circ$N,
  triangle at $50.6^\circ$N, L2 at $45.2^\circ$N) but are rigidly translated in
  longitude onto the meridian of North Macedonia ($21.7^\circ$E, near Skopje;
  marked). The two L-shaped detectors (L1, L2) remain $1200\,$km apart with the
  $\Delta$--$10\,$km triangular detector at their midpoint; each L has $90^\circ$
  between its two $15\,$km arms, and the two L's are shown for both relative
  orientations, $42.5^\circ$ and $-2.5^\circ$ (inset, oriented along the true
  baseline). The arms are negligibly small against the $1200\,$km baseline and are
  not drawn to map scale.}
    \label{fig:2LT_sites_macedonia}
\end{figure*}

Here we show the locations of the ET configurations employed for the geometry comparison. The three European candidate ET sites are illustrated in the left panel of Fig.~\ref{fig:2LT_sites}. To minimize dependence on specific site choices while preserving European latitudes, we locate both the 2L and the $\Delta$ configurations in the Atlantic Ocean, outside any Exclusive Economic Zone (EEZ), as shown in the right panel of Fig.~\ref{fig:2LT_sites}. We have verified that our conclusions are insensitive to the longitude choice: placing the detectors on land at the same latitude, but at a different longitude (e.g., along the $21.7^\circ$E meridian through North Macedonia; Fig.~\ref{fig:2LT_sites_macedonia}) yields equivalent results.

\section{Exploring the intermediate misaligned \texorpdfstring{$\mathrm{2L}\ (30^{\circ})$}{2L (30 deg)} configuration}
\label{app:2L30}

Our main analysis compares the nearly aligned $\mathrm{2L}\ (-2.5^{\circ})$ and strongly misaligned $\mathrm{2L}\ (42.5^{\circ})$ configurations, which emphasize different aspects of the ET science case. The nearly aligned orientation is the most favorable for stochastic background searches, whereas $\mathrm{2L}\ (42.5^{\circ})$ is the least favorable, but provides substantially better parameter estimation and sky localization for compact binary sources. Here, we also consider an intermediate relative orientation, $\mathrm{2L}\ (30^{\circ})$, to test whether much of the compact binary science advantage of the strongly misaligned geometry persists without requiring the orientation that is least favorable for the stochastic background.

\begin{figure*}
\begin{tcolorbox}[standard jigsaw,colframe=gray,colback=gray!10!white,
opacityback=0.0,coltext=black,
title=\small\sc Signal-to-noise ratio comparison for the intermediate $\mathrm{2L}$ orientation]
\centering
\includegraphics[width=0.32\textwidth]{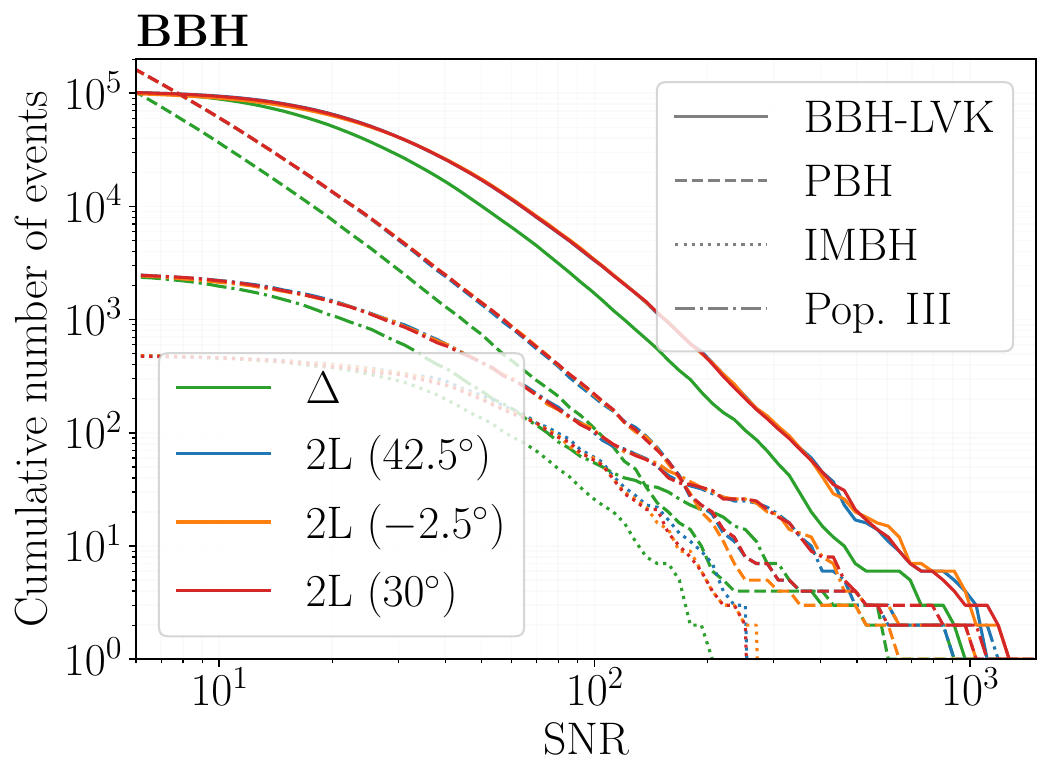}
\hfill
\includegraphics[width=0.32\textwidth]{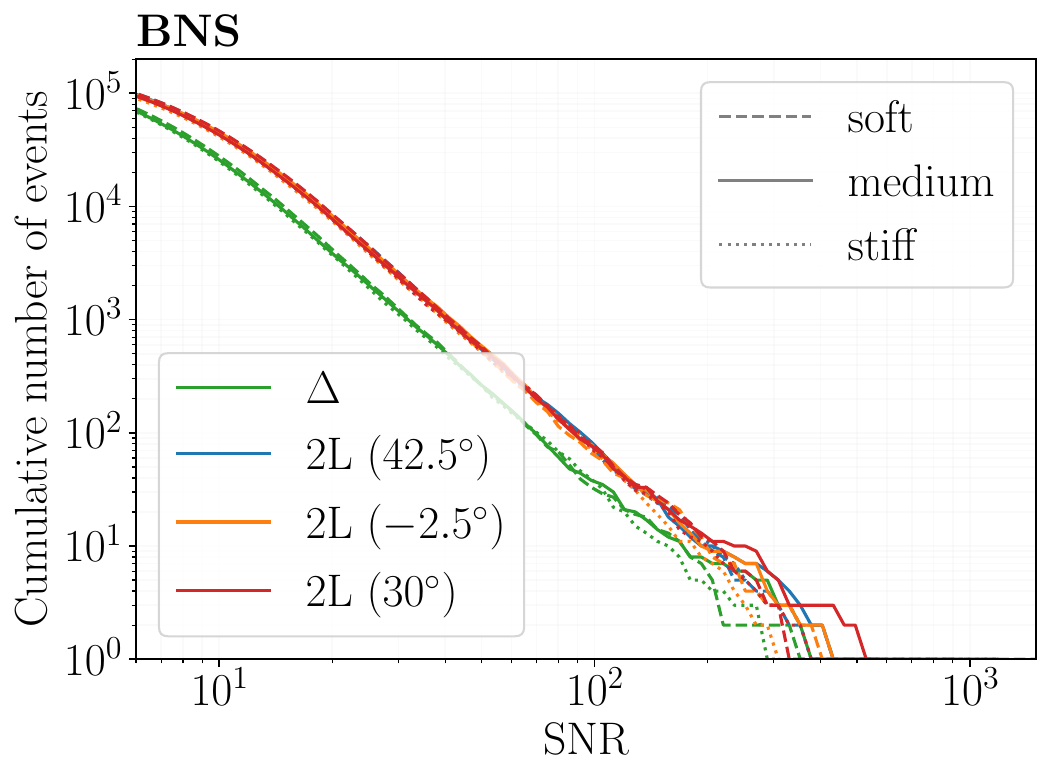}
\hfill
\includegraphics[width=0.32\textwidth]{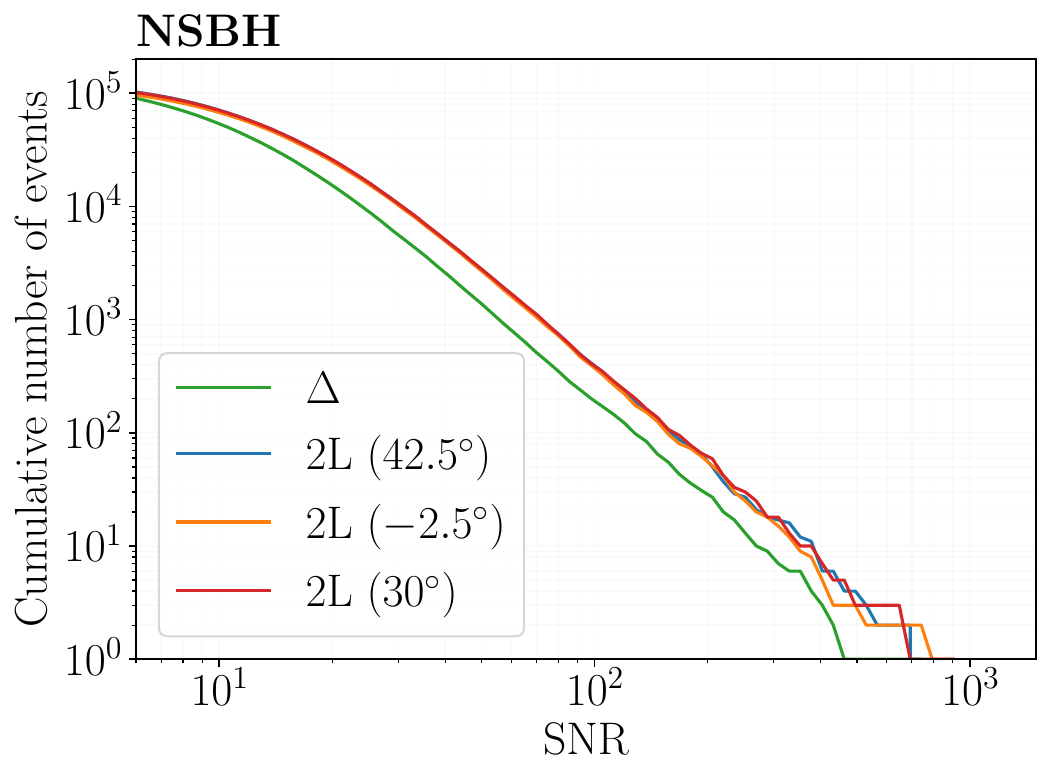}
\caption{\justifying Cumulative network SNR distributions for BBH (left), BNS (middle), and NSBH (right) populations for the triangular configuration and the three $\mathrm{2L}$ relative orientations. The $\mathrm{2L}\ (30^{\circ})$ and $\mathrm{2L}\ (42.5^{\circ})$ distributions are nearly indistinguishable, showing that the intermediate orientation preserves essentially the same compact binary detection reach.}
\label{fig:app_2L30_snr}
\end{tcolorbox}
\vskip-15pt
\end{figure*}

Figure~\ref{fig:app_2L30_snr} shows that changing the relative orientation of the two 15-km interferometers, while keeping them substantially misaligned, has little effect on the overall SNR distribution. For example, in the 1-year populations, the numbers of events with ${\rm SNR}\geq100$ are 3056 and 3067 for BBHs, 13 and 13 for BNSs, and 72 and 71 for NSBHs in the $\mathrm{2L}\ (42.5^{\circ})$ and $\mathrm{2L}\ (30^{\circ})$ configurations, respectively. The nearly aligned $\mathrm{2L}\ (-2.5^{\circ})$ configuration gives similarly large high-SNR populations. 

The reconstruction of intrinsic parameters and luminosity distance for  $\mathrm{2L}\ (30^{\circ})$ also remains close to the performance of $\mathrm{2L}\ (42.5^{\circ})$. In particular, 2\,817 (22\,642) [254] BBHs have $\Delta d_L/d_L<5\%$ ($\Delta m_1/m_1<10\%$) [$\Delta \chi_1<1\%$] at $\mathrm{2L}\ (30^{\circ})$, compared with 3\,133 (23\,760) [272] at $\mathrm{2L}\ (42.5^{\circ})$ and 1\,285 (15\,599) [214] at $\mathrm{2L}\ (-2.5^{\circ})$.

\begin{figure*}
\begin{tcolorbox}[standard jigsaw,colframe=gray,colback=gray!10!white,
opacityback=0.0,coltext=black,
title=\small\sc Sky localization comparison for the intermediate $\mathrm{2L}$ orientation]
\centering
\includegraphics[width=0.32\textwidth]{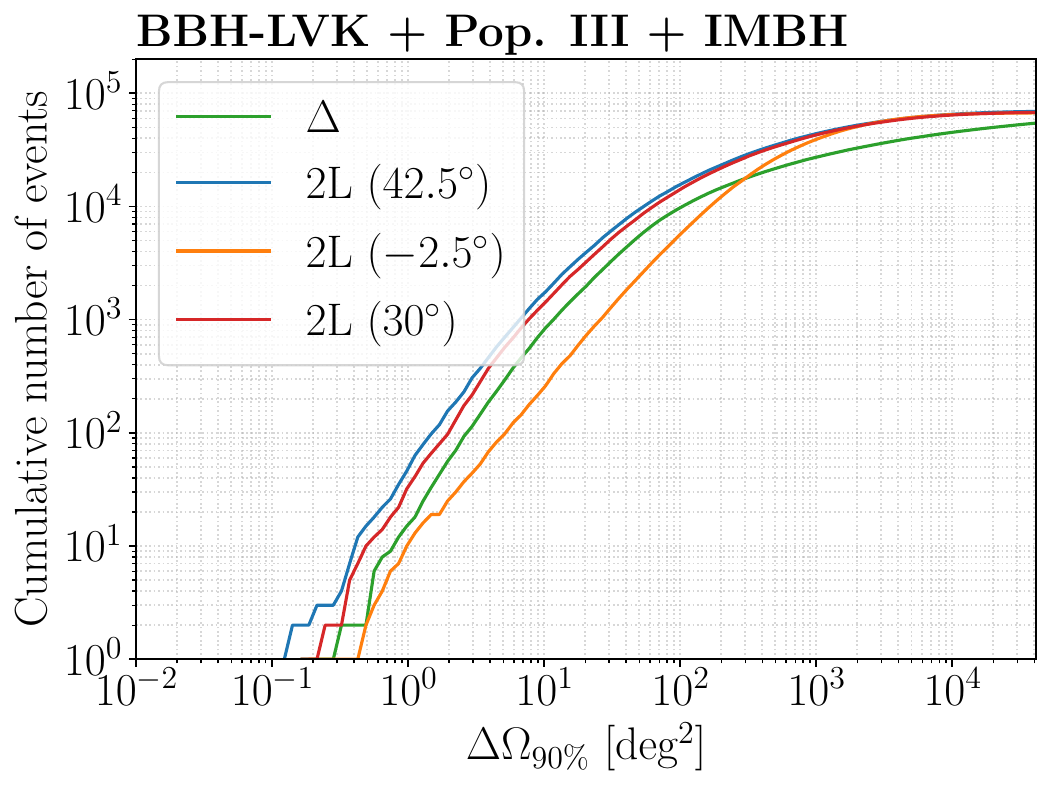}
\hfill
\includegraphics[width=0.32\textwidth]{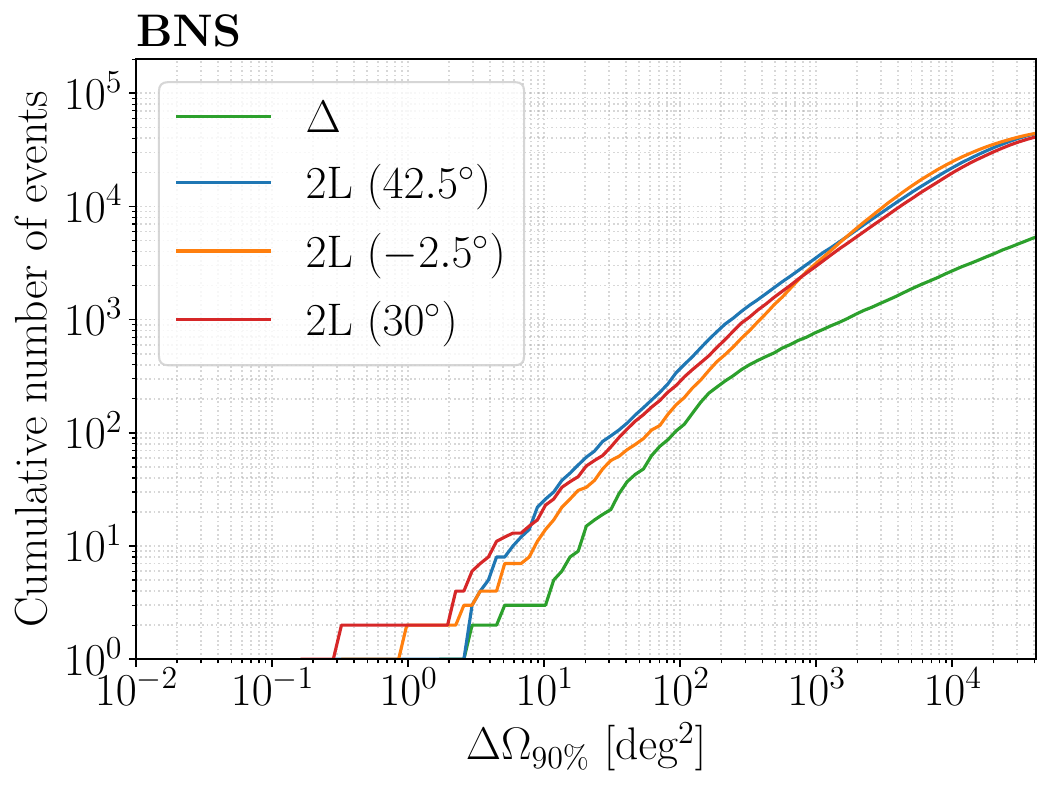}
\hfill
\includegraphics[width=0.32\textwidth]{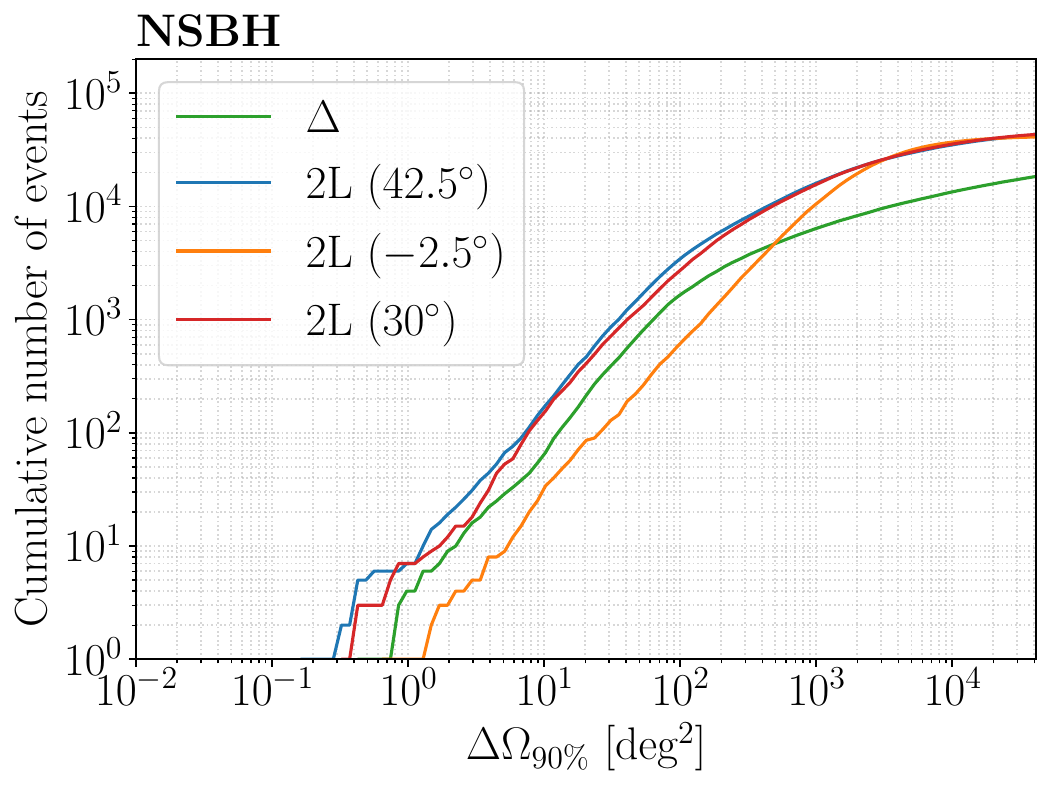}
\caption{\justifying Cumulative 90\% credible sky-localization distributions for BBH (left), BNS (middle), and NSBH (right) mergers. The $\mathrm{2L}\ (30^{\circ})$ configuration closely follows $\mathrm{2L}\ (42.5^{\circ})$ and retains most of its localization advantage over the nearly aligned $\mathrm{2L}\ (-2.5^{\circ})$ configuration.}
\label{fig:app_2L30_skyloc}
\end{tcolorbox}
\vskip-15pt
\end{figure*}

The same conclusion holds for sky localization, although $\mathrm{2L}\ (42.5^{\circ})$ retains an advantage for the most demanding localization thresholds. In the 1-year populations, $\mathrm{2L}\ (42.5^{\circ})$ and $\mathrm{2L}\ (30^{\circ})$ localize 14\,664 and 13\,183 BBHs, 68 and 52 BNSs, and 645 and 508 NSBHs within $100\,{\rm deg}^2$, respectively. At $10\,{\rm deg}^2$, the corresponding numbers are 1530 and 1224 for BBHs, 5 and 4 for BNSs, and 29 and 27 for NSBHs. Both misaligned configurations remain substantially better for precise localization than $\mathrm{2L}\ (-2.5^{\circ})$, which yields only 222 BBHs and 6 NSBHs within $10\,{\rm deg}^2$ in one year.

This similarity carries directly into the multimessenger forecasts, for which, following Sec.~\ref{sec:mma}, we use 5-year populations of BNS and NSBH mergers. For nearby mergers ($z<0.4$), $\mathrm{2L}\ (42.5^{\circ})$ localizes 245 BNSs and 461 NSBHs within $100\,{\rm deg}^2$, compared with 195 and 466 for $\mathrm{2L}\ (30^{\circ})$. Within $10\,{\rm deg}^2$, the corresponding numbers are 24 and 18 for BNSs and 92 and 93 for NSBHs. Thus, the two orientations provide essentially the same NSBH counterpart sample and broadly similar BNS performance. At higher redshift, the distinction becomes more pronounced for the tightest localization requirements. For example, at $z\geq2$ the number of NSBHs localized within $100\,{\rm deg}^2$ decreases from 91 for $\mathrm{2L}\ (42.5^{\circ})$ to 17 for $\mathrm{2L}\ (30^{\circ})$. Once the localization requirement is relaxed to $1000\,{\rm deg}^2$, however, the two configurations again become very similar, retaining 1363 and 1270 NSBHs, respectively. Thus, $\mathrm{2L}\ (30^{\circ})$ preserves the broad multimessenger science reach of the strongly misaligned configuration, while $\mathrm{2L}\ (42.5^{\circ})$ remains preferable when the tightest possible high-redshift localization is required.

The cosmological forecasts show the same pattern. For BNS bright sirens with the medium EOS, $\mathrm{2L}\ (42.5^{\circ})$ yields 128 events in 5 years and constrains $H_0$ to $0.40\%$ and $\Omega_m$ to $52\%$, while $\mathrm{2L}\ (30^{\circ})$ yields 108 events and corresponding uncertainties of $0.65\%$ and $68\%$. The degradation is noticeable but does not remove the strong advantage of a misaligned $\mathrm{2L}$ geometry over the nearly aligned configuration. For NSBH bright sirens, the difference is much smaller. With the $z\leq0.4$ selection, $\mathrm{2L}\ (42.5^{\circ})$ and $\mathrm{2L}\ (30^{\circ})$ yield 313 and 316 events and constrain $H_0$ to $0.35\%$ and $0.36\%$, and $\Omega_m$ to $7.6\%$ and $8.4\%$, respectively. Restricting the sample to $z\leq0.2$ gives essentially identical $H_0$ constraints of $0.49\%$ for both configurations, with $\Omega_m$ uncertainties of $16.7\%$ and $16.1\%$. Thus, the NSBH bright siren science case is effectively unchanged by reducing the relative orientation from $42.5^\circ$ to $30^\circ$.

The same conclusion extends to the high-redshift NSBH--GRB forecasts. The $\mathrm{2L}\ (42.5^{\circ})$ and $\mathrm{2L}\ (30^{\circ})$ configurations yield 661 and 608 usable associations and constrain $w_0$ to approximately $49\%$ and $44\%$, and $\Omega_m$ to approximately $43\%$ and $41\%$, respectively. The modest reversal in this particular realization is small compared with the overall statistical uncertainty and does not indicate a meaningful preference for the $30^\circ$ orientation. Likewise, the tests-of-GR forecasts show no appreciable difference between $\mathrm{2L}\ (30^{\circ})$ and $\mathrm{2L}\ (42.5^{\circ})$, so the corresponding science conclusions remain unchanged.

Taken together, these results show that $\mathrm{2L}\ (30^{\circ})$ retains most of the compact binary science advantages of $\mathrm{2L}\ (42.5^{\circ})$. The SNR distributions are essentially identical, the localization performance remains close for most source classes and sky-area thresholds, and the resulting multimessenger, cosmological, and tests-of-GR forecasts are generally unchanged at the level relevant for the science case. The main cost is a reduction in the most demanding BNS and high-redshift localization measurements, for which $\mathrm{2L}\ (42.5^{\circ})$ remains superior. Hence, the intermediate $\mathrm{2L}\ (30^{\circ})$ configuration provides a useful compromise when stochastic background sensitivity is considered together with the compact binary science enabled by detector misalignment.
\newpage

\section{Correlated instrumental noise}
\label{sec:NoiseModels}

Correlated noise impacts all GW analyses. It alters the effective signal-to-noise ratio (SNR) of transient GW signals, and it causes excess noise in observations of a stochastic GW background, which rely on correlations between detectors. Noise correlations are caused by the geophysical environment of the detectors and are therefore typically stronger between collocated than distant detectors. Given the dependence of environmental noise on site conditions and partially unknown properties of environmental fields, it is impractical to account for correlated noise in all of the plots presented this study. This section is dedicated to a comparison of the impact of correlated noise on GW analyses in the two ET configurations. These findings must be accounted for to correct the ideal sensitivities assumed in the scientific comparison of the configurations.

\subsection{Impact on transient gravitational-wave signals}
If one analyzes signals in data from a network of detectors (which includes the three-detector triangle configuration), it is important to consider noise correlations between detectors, since otherwise the results can be biased. Noise correlations impact signal SNRs and consequently all signal analyses \cite{Prince:2002hp,Wong2022}. 

\begin{figure*}
\begin{tcolorbox}[standard jigsaw,colframe=gray,colback=gray!10!white,opacityback=0.0,coltext=black, title=\small\sc SNR variation in the presence of correlated environmental noise]
\center{
\includegraphics[width=0.45\textwidth]{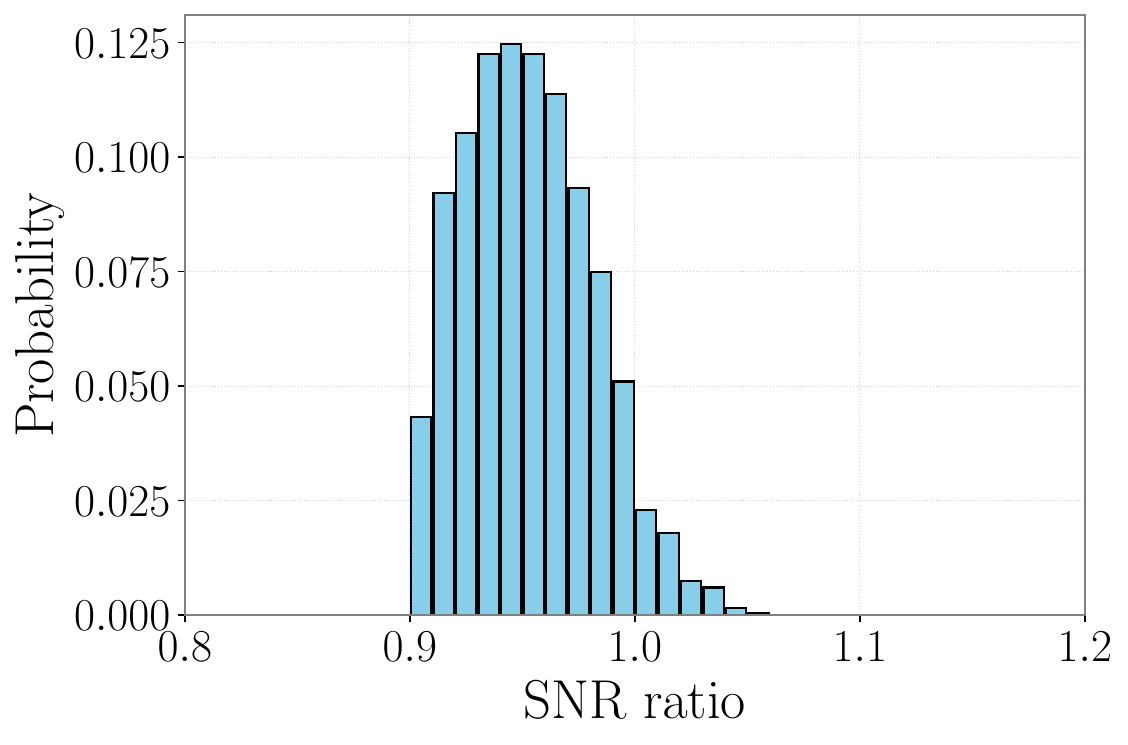}
}
\caption{\justifying Distribution of SNRs of transient GW signals in the presence of correlated noise normalized by the SNRs with uncorrelated noise in the case of the triangle configuration. Coherence values between the three detectors are drawn from a random distribution consistent with an environmental origin. In realistic scenarios, this effect applies to signals below about 10\,Hz where environmental noise can dominate detector noise and be correlated.}
\label{fig:snr_correlated}
\end{tcolorbox}
\vskip-15pt
\end{figure*}

While correlated noise can generally have a positive or negative impact on signal analyses \cite{Wong2022}, in a realistic scenario it is always produced by the geophysical environment of the detector, and in this case, the impact is almost always detrimental \cite{Harms2026}. Figure \ref{fig:snr_correlated} shows the change of SNR in transient signals due to correlated noise in the triangle. For this plot, (frequency-independent) coherence values were drawn randomly from a distribution of possible coherence values consistent with a geophysical origin. The minimum possible SNR ratio is 0.81, which corresponds to equal coherence of -0.5 between all three detector pairs. The impact of correlated noise on SNRs is mild. In addition, since the coherence values are here assumed to be frequency independent, the shown effect on SNRs is representative only of frequencies where environmental noise can dominate and be correlated, i.e., below about 10\,Hz. The SNRs and analyses of higher-frequency signals are unaffected, at least according to the environmental-noise models that we can produce today.
 
\subsection{Impact on stochastic gravitational-wave backgrounds}
The most sensitive method to detect a stochastic GW background is to correlate data between two or more detectors under the assumption that the noise in the detectors is independent. To estimate the sensitivity of future detector networks to stochastic backgrounds, it is common practice to use the statistical errors of correlation measurements between detectors \cite{Christensen1992,Allen1999}, an approach also used in Sec.~\ref{sec:backgrounds} of this report. 

The geophysical environment produces correlated instrument noise, which creates excess noise in stochastic searches. An important example is the stochastic search using data from the two collocated LIGO detectors at the Hanford site. Its sensitivity was limited by correlated environmental noise despite all efforts to mitigate the noise \cite{Fotopoulos2008}.

The geophysical environment of ET, even deep underground, will also produce correlated instrument noise \cite{Amann2020,Janssens2022,Janssens2024} leading to excess noise in its stochastic searches \cite{ET:2025xjr,Harms2026}. While we cannot today model all correlated noise sources of ET, two important ones have been studied extensively. One is the so-called Newtonian noise (NN) produced by terrestrial gravity fluctuations from seismic fields \cite{Harms2019,Bader2022,Harms2022}, the other is the magnetic noise from fluctuations of Earth's magnetic field \cite{Thrane2013,Cirone2018,Janssens2025}. 

\begin{figure*}
\begin{tcolorbox}[standard jigsaw,colframe=gray,colback=gray!10!white,opacityback=0.0,coltext=black, title=\small\sc Impact of correlated noise on stochastic GW searches]
\center{
\includegraphics[width=0.4\textwidth]{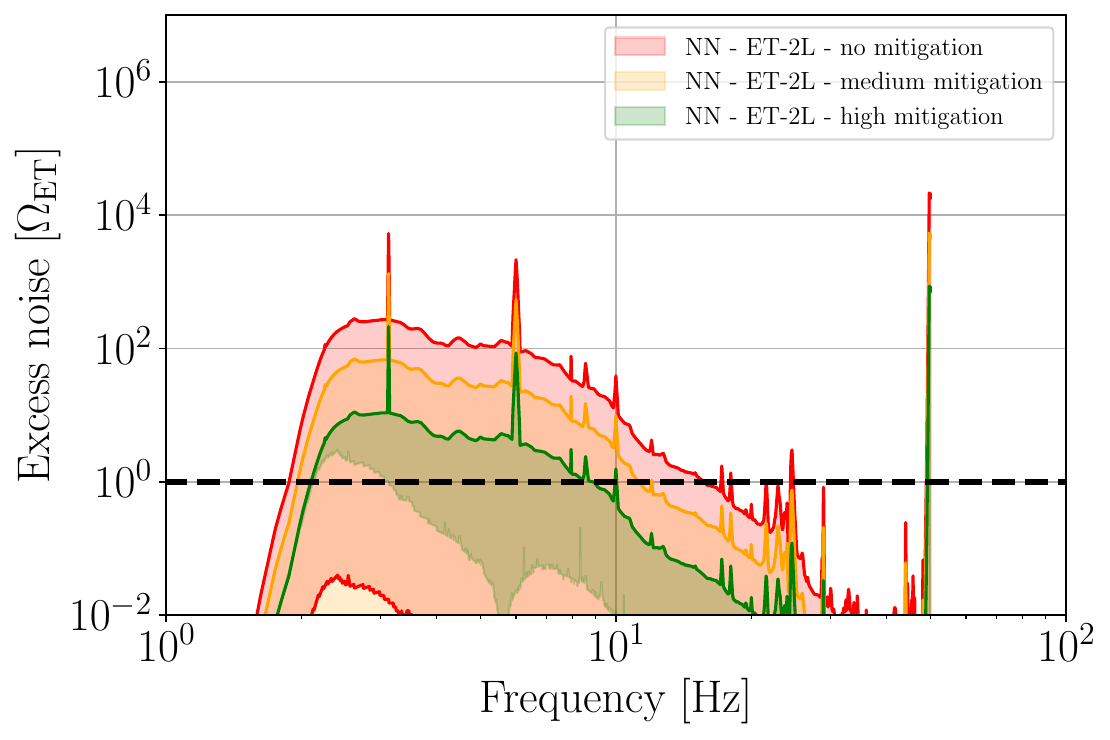}
\includegraphics[width=0.4\textwidth]{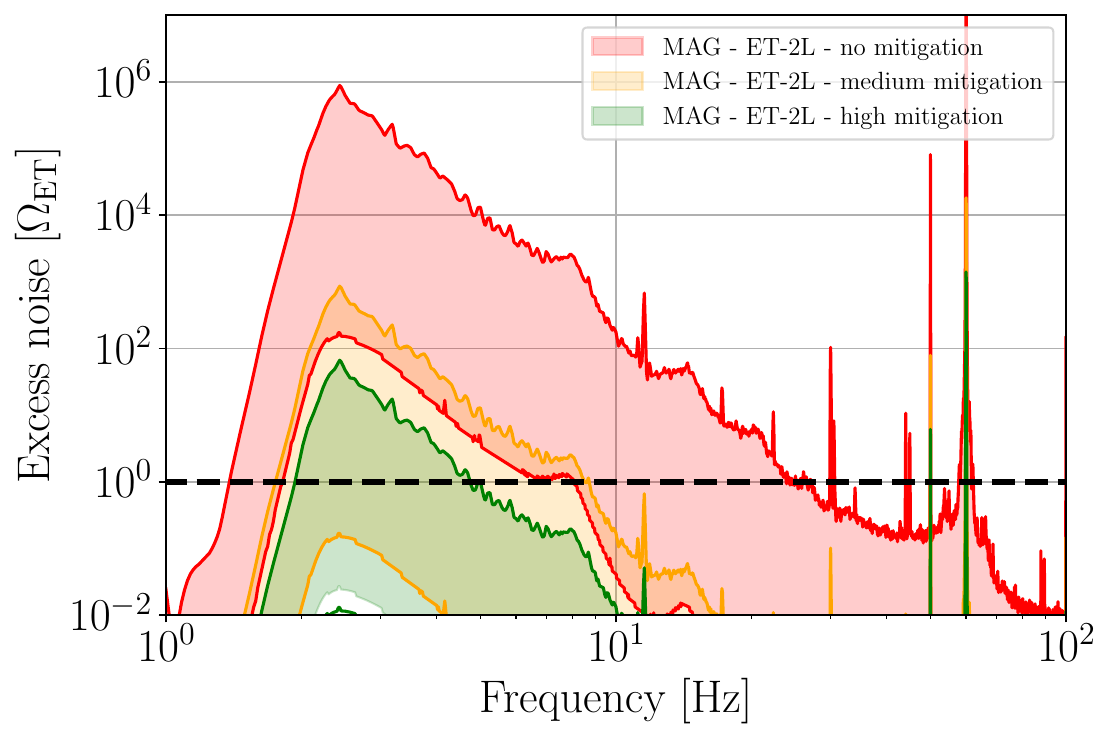}\\[0.1cm]
\hspace*{3pt}\includegraphics[width=0.4\textwidth]{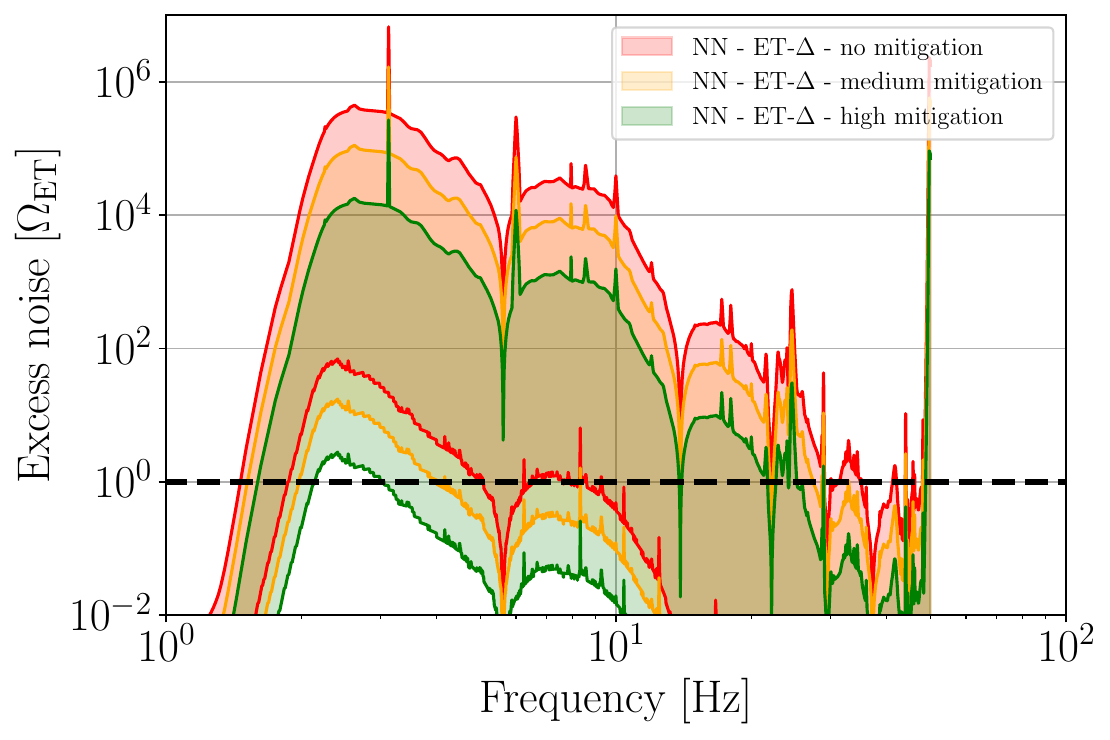}
\includegraphics[width=0.4\textwidth]{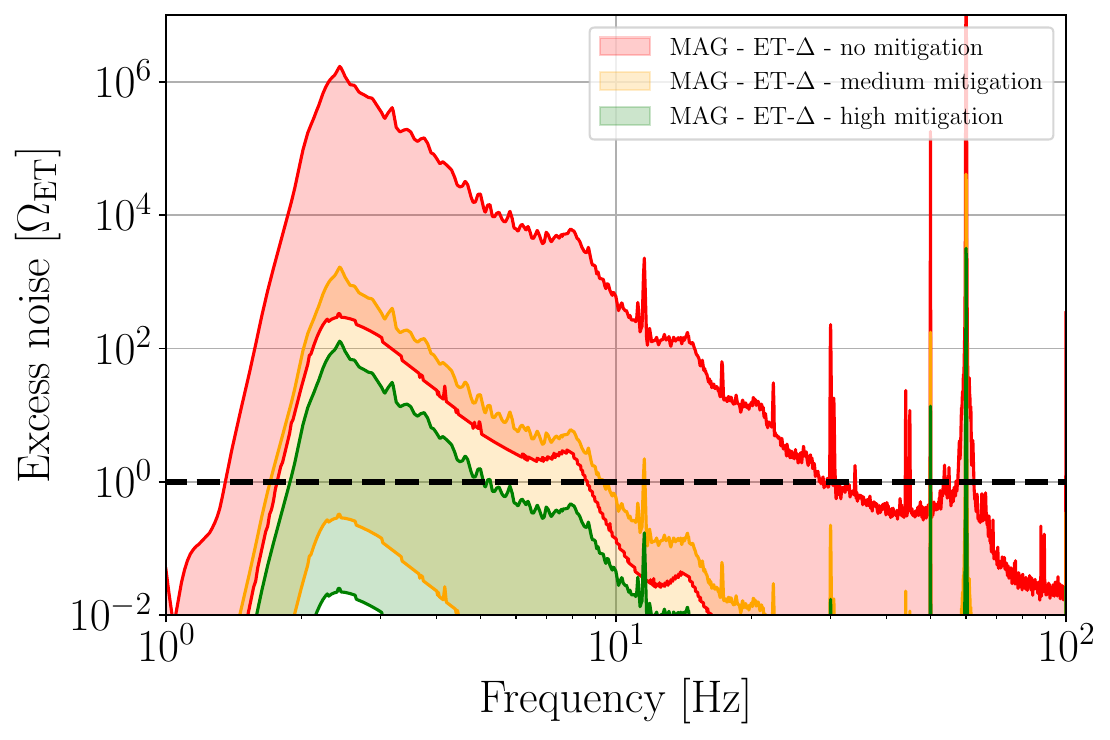}
}
\caption{\justifying Excess noise in stochastic GW searches produced by correlated Newtonian and magnetic noise normalized by the sensitivity without environmental noise. Upper row: 2L configuration. Lower row: triangle configuration. Left: Newtonian noise. Right: magnetic noise. The colored bands represent the plausible range of site-dependent environmental noise with red representing no mitigation, yellow medium mitigation, and green high mitigation.}
\label{fig:excess_omega}
\end{tcolorbox}
\vskip-15pt
\end{figure*}

Figure \ref{fig:excess_omega} shows the excess noise in searches for an isotropic GW background normalized by the sensitivity without correlated noise. The upper row is for the 2L configuration, and the lower row for the triangular configuration. Each colored band represents the range of variation of seismic and magnetic fluctuations observed underground over time and across candidate sites. Temporal variations of noise are accounted for by using the 10th and 90th percentiles of noise spectra observed over a year to construct the noise bands. 

If these noises enter the stochastic searches without mitigation, then the red bands represent the range of excess noise. The normalized magnetic noise is similar in both configurations. This is because the natural magnetic fluctuations at these low frequencies come from the so-called Schumann resonances, which are globally coherent \cite{Coughlin2016, Coughlin:2018tjc}. Magnetic noise can reduce ET's sensitivity to stochastic backgrounds by up to 6 orders of magnitude depending on site quality.

Newtonian noise affects stochastic searches differently in the two configurations. In the frequency band of interest, NN is uncorrelated between sites of the 2L configuration. It still affects the 2L stochastic search because it can increase the average instrument-noise spectrum shown in Fig.~\ref{fig:sensitivity}. The corresponding result is shown in the upper-left panel with excess noise up to a factor of 100 -- 1000 above the statistical error (neglecting the narrow noise peaks). In the triangle configuration, Newtonian noise can cause excess noise up to 5 -- 6 orders of magnitude above the statistical error as shown in the lower-left panel.

Figure \ref{fig:excess_omega} shows that noise-mitigation technology is needed to reach ET's ultimate sensitivity to stochastic backgrounds represented by the statistical error (horizontal dashed line). The yellow and green bands are calculated by applying a frequency-independent noise-mitigation factor to the red bands. This is common practice in the ET Instrument Science Board in the absence of a mature design of the mitigation technologies \cite{ISB:2026}. For the magnetic noise, we apply mitigation factors of 32 (medium) and 115 (high), and for Newtonian noise 2 (medium) and 5 (high); all factors need to be squared when applied to noise in units of $\Omega_{\rm gw}$. These factors are current best estimates based on preliminary design studies \cite{ISB:2026} and should only be taken as an indication of what mitigation can be achieved.

Newtonian-noise sensitivity limitations cannot be avoided in the case of the triangle configuration even if the detector is located at the quietest known site and mitigation is applied. In the case of the 2L configuration, Newtonian noise does not limit the sensitivity if the seismic noise levels at the sites fall within the lower half of the noise band. 

The sensitivities in Sec.~\ref{sec:backgrounds} represent the statistical error, and depending on configuration (2L vs triangle), on site noise and on mitigation, they need to be corrected by multiplying by the excess-noise factors shown in Fig.~\ref{fig:excess_omega} with a minimum correction factor of 1 representing the ideal case when the search is limited by statistical errors.

\newpage

\bibliography{references}

\end{document}


\title{The Scientific Added Value of Einstein Telescope Geometries in a Network with Cosmic Explorer }

\maketitle

\section{Impact of a Cosmic Explorer detector in the network}
\label{app:ce_network}

\begin{figure*}
\begin{tcolorbox}[standard jigsaw,colframe=gray,colback=gray!10!white,
opacityback=0.0,coltext=black,
title=\small\sc Signal-to-noise ratio comparison for the three configurations in a network with CE]
\centering
\includegraphics[width=0.32\textwidth]{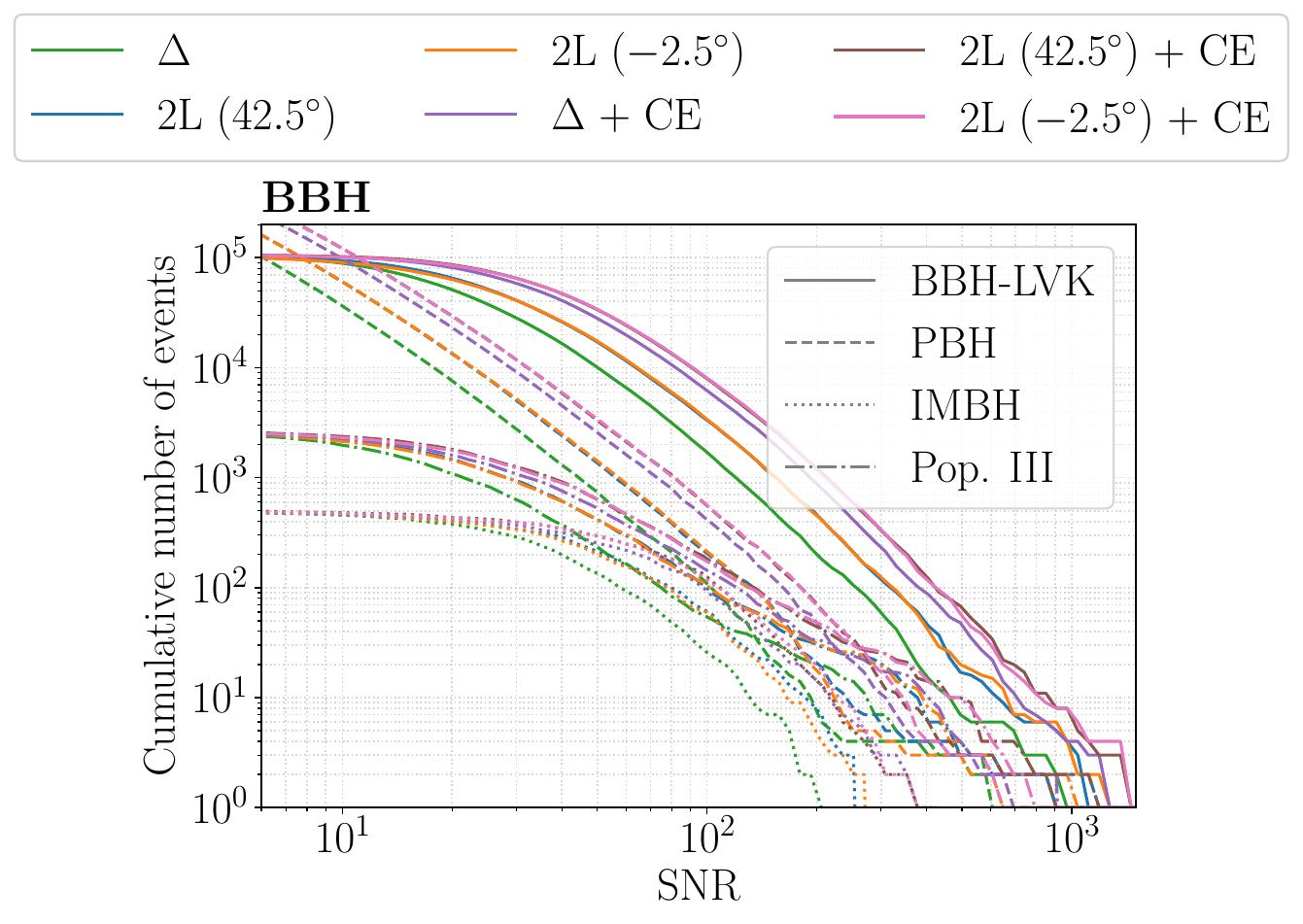}
\hfill
\includegraphics[width=0.32\textwidth]{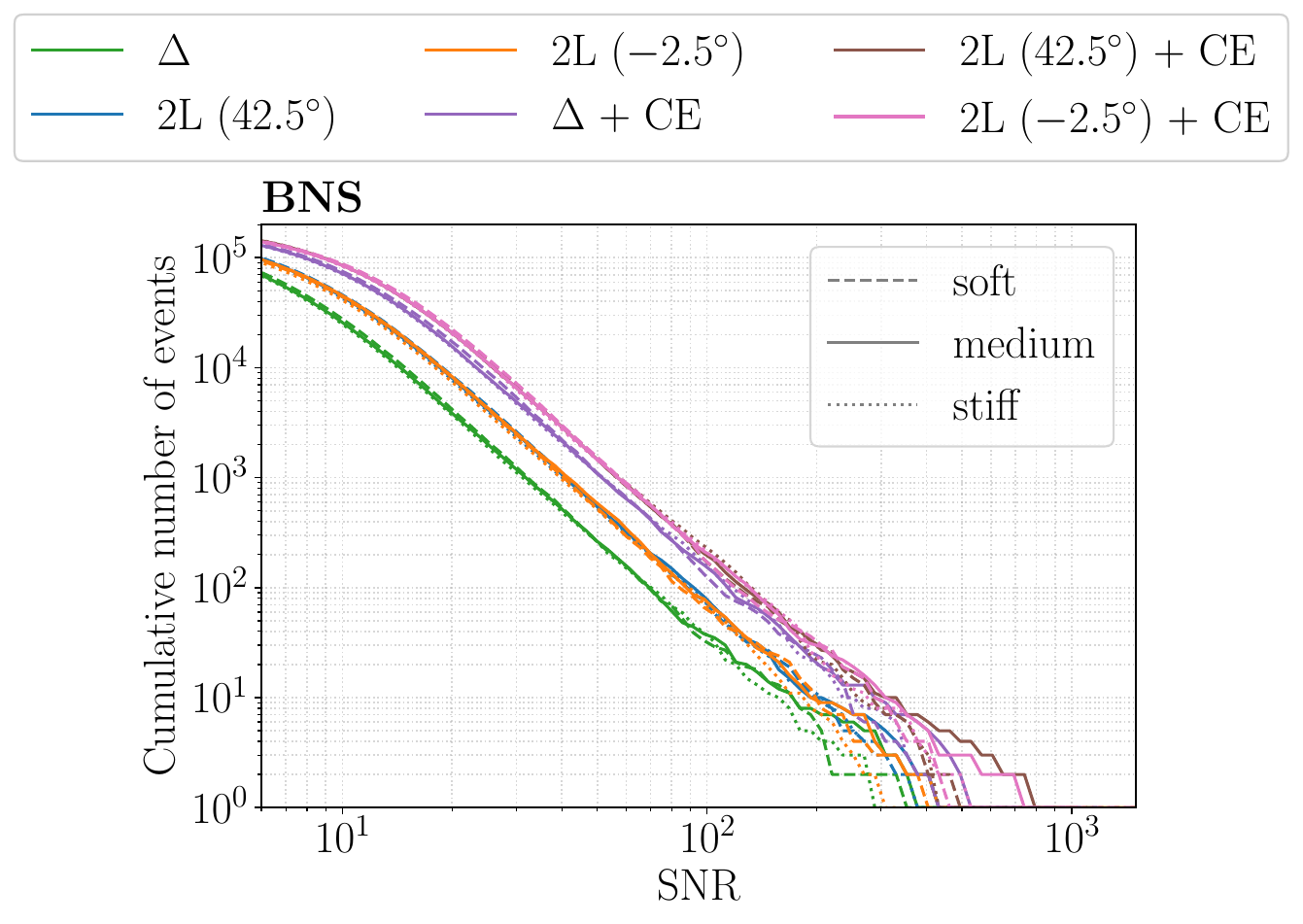}
\hfill
\includegraphics[width=0.32\textwidth]{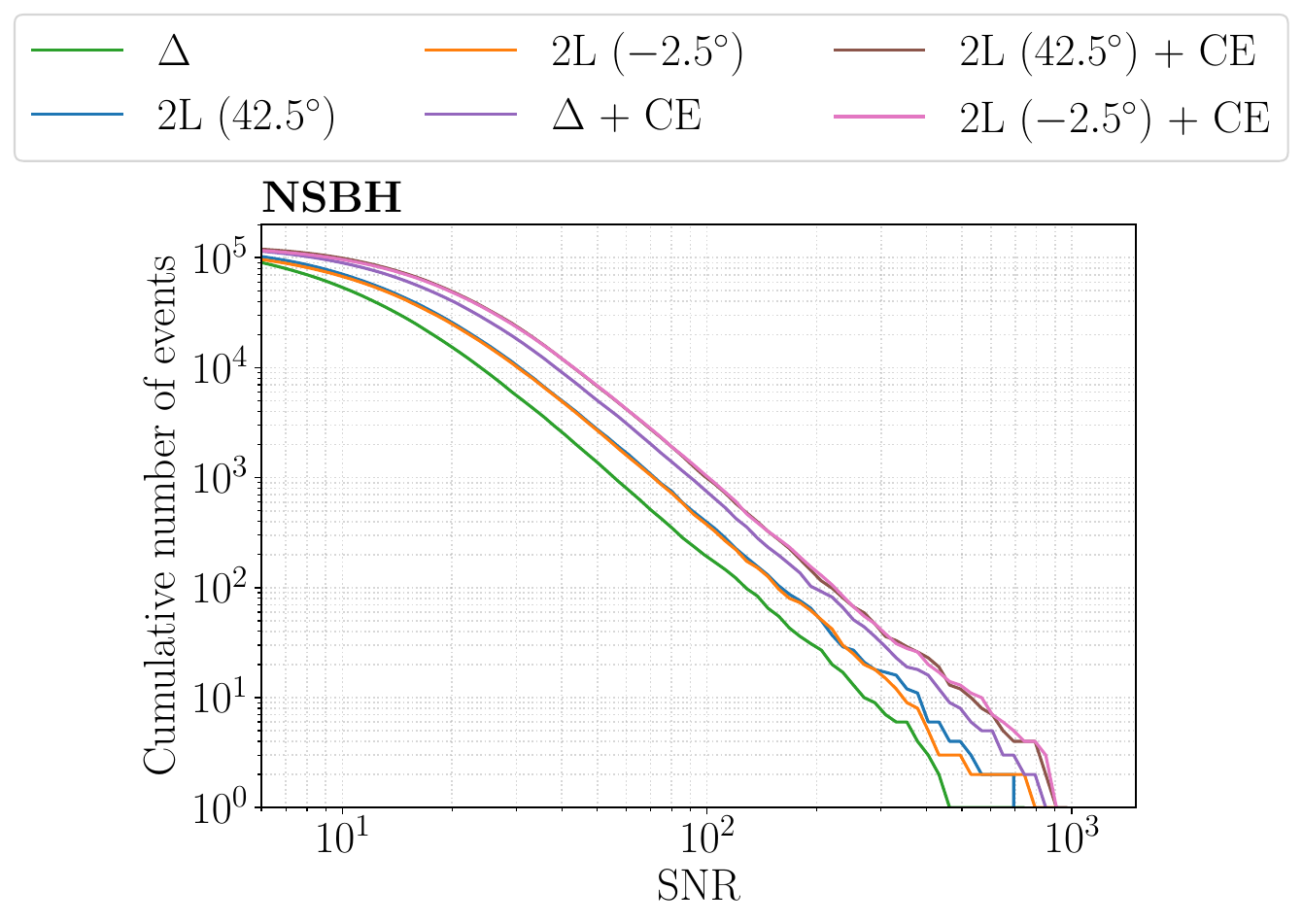}
\caption{\justifying Cumulative network SNR distributions for BBH (left), BNS (middle), and NSBH (right) populations for the triangular configuration and the two 2L relative orientations ($-2.5^{\circ}$ and $42.5^{\circ}$), for ET alone and for ET in a network with CE.}
\label{fig:app_CE_snr}
\end{tcolorbox}
\vskip-15pt
\end{figure*}

The results presented in the ``Scientific Added Value'' chapter (hereafter, the main chapter) consider the Einstein Telescope (ET) as a stand-alone observatory, comparing three configurations: a triangle ($\Delta$) of 10-km interferometers, and two L-shaped detectors with 15-km arms (2L) at separate sites, either nearly aligned (2L $(-2.5^\circ)$) or misaligned (2L $(42.5^\circ)$). In practice, ET is likely to operate within a global network, potentially including a Cosmic Explorer (CE) detector in the United States. We therefore assess here the impact of adding one CE detector to each ET configuration. The results quoted below use the same source populations of binary black holes (BBHs), binary neutron stars (BNSs), and neutron-star-black-hole (NSBH) binaries, and the same detection thresholds as the main chapter. As in the main analysis, BBH numbers refer to one year of observation, whereas BNS and NSBH numbers quoted in the text refer to the five-year catalogs; the corresponding per-year values are reported in Table~\ref{tab:metrics}.
For CE, we select a location in the Pacific Ocean, near the north-west of the US mainland, without any specific alignment with respect to ET, and consider a 40\,km-arms detector with the power spectral density (PSD) model of Ref.~\cite{Srivastava:2022slt}. The coordinates are the ones provided for the CE-A detector in Table 2 in~\cite{Gupta:2023lga}. 

The addition of CE substantially increases the overall compact-binary science yield of the network. For the BBH population, the number of detections with $\mathrm{SNR}>12$ increases from $81\,459$, $90\,602$, and $86\,577$ for the triangle, 2L $(42.5^\circ)$, and 2L $(-2.5^\circ)$ configurations, respectively, to $100\,469$, $102\,395$, and $101\,022$ when CE is added. The number of exceptionally loud BBHs with $\mathrm{SNR}>100$ increases more strongly: from $1\,613$, $3\,215$, and $3\,256$ to $5\,895$, $7\,590$, and $7\,714$, respectively. The same trend is found for BNSs and NSBHs. For example, for the medium-EOS BNS population, the number of events with $\mathrm{SNR}>8$ increases from $39\,138$, $62\,125$, and $62\,960$ to $94\,009$, $107\,943$, and $106\,976$ for the triangle, 2L $(42.5^\circ)$, and 2L $(-2.5^\circ)$ configurations, respectively. Similarly, the number of NSBH detections with $\mathrm{SNR}>12$ rises from $39\,052$, $55\,482$, and $53\,316$ to $75\,230$, $85\,313$, and $83\,321$. Note that throughout this section, the numbers provided for BBH sources refer to 1 year of observations, whereas for BNSs and NSBHs the reference is to 5 years of observations. Cumulative SNR distribution for the three populations are shown in Fig.~\ref{fig:app_CE_snr}.

\begin{figure*}
\begin{tcolorbox}[standard jigsaw,colframe=gray,colback=gray!10!white,opacityback=0.0,coltext=black, title=\small\sc Distributions for the statistical uncertainties on the masses and spins for BBHs for the three configurations in a network with CE]
\center{\includegraphics[width=0.5\textwidth]{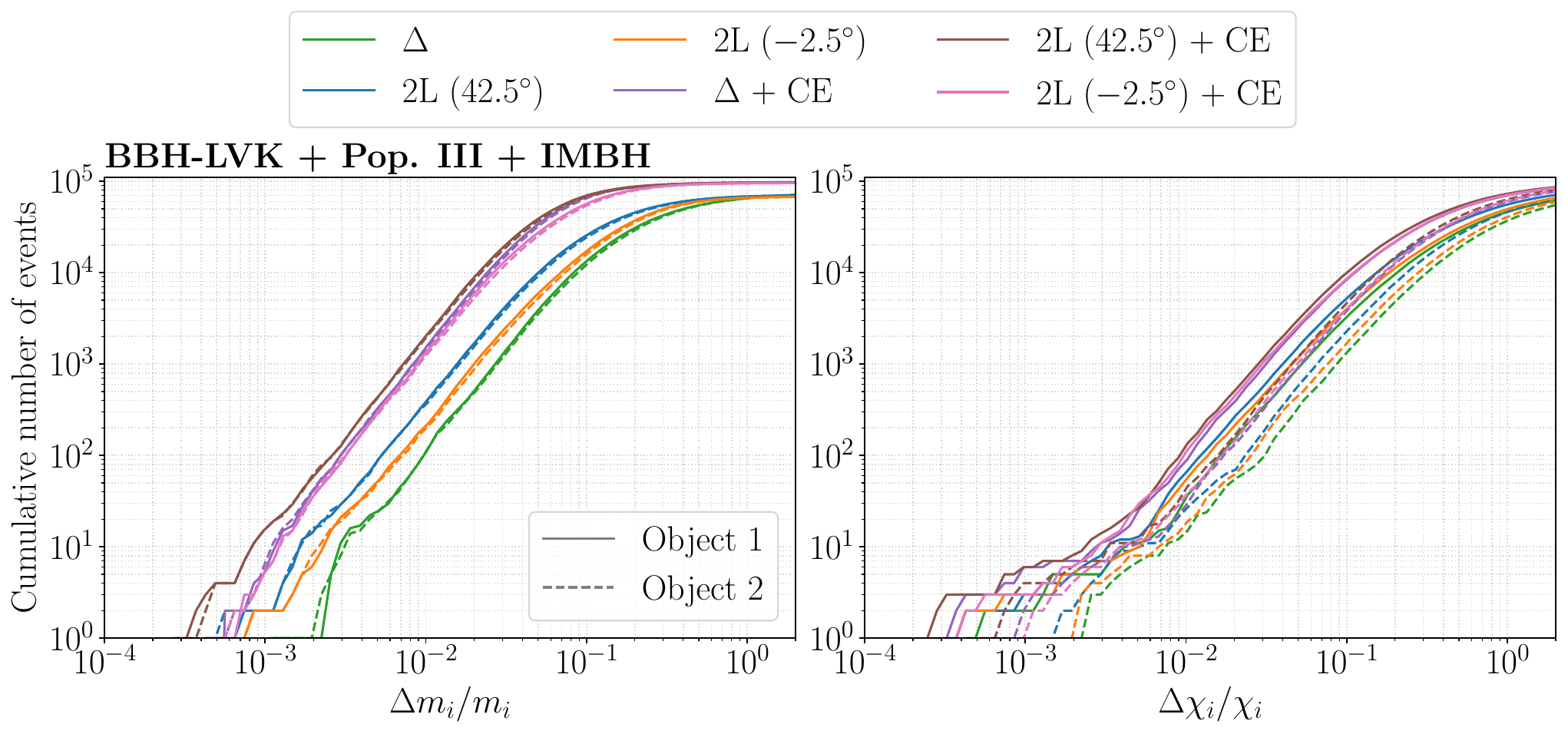} \includegraphics[width=0.32\textwidth]{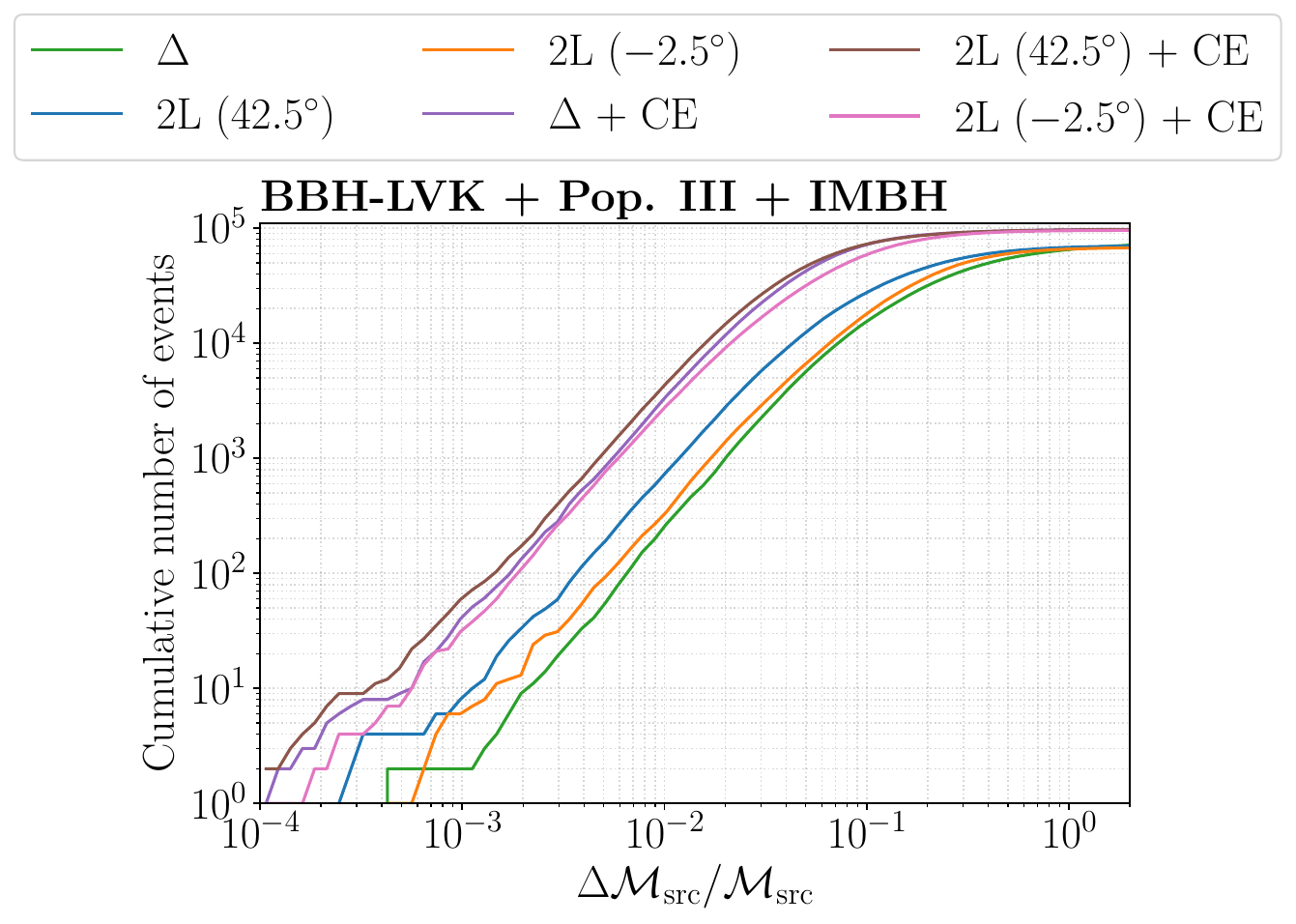}}
\captionof{figure}{\justifying Cumulative distributions of the relative uncertainty on the individual source-frame masses, spin magnitudes, and source-frame chirp masses of the astrophysical BBH populations used in this work. Results are shown for ET alone and for ET in a network with CE.}
\label{fig:errs_ds1_bbh_intr_app_CE}
\end{tcolorbox}
\vskip-15pt
\end{figure*}

\begin{figure}
\begin{tcolorbox}[standard jigsaw,colframe=gray,colback=gray!10!white,opacityback=0.0,coltext=black, title=\small\sc Distributions for the statistical uncertainties on the luminosity distance and sky localization for BBHs for the three configurations in a network with CE.]
\center{
\includegraphics[width=0.32\textwidth]{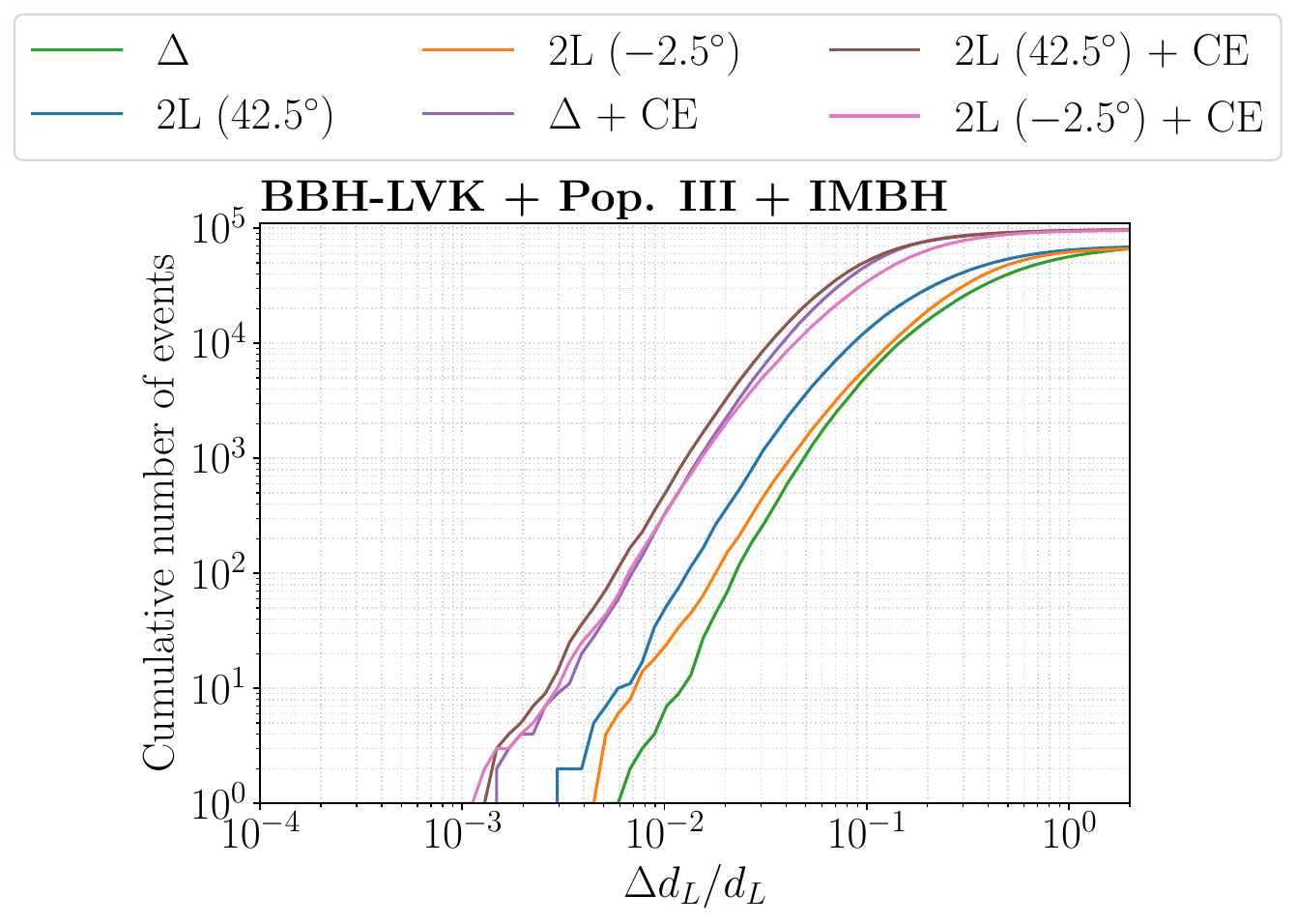} 
\includegraphics[width=0.32\textwidth]{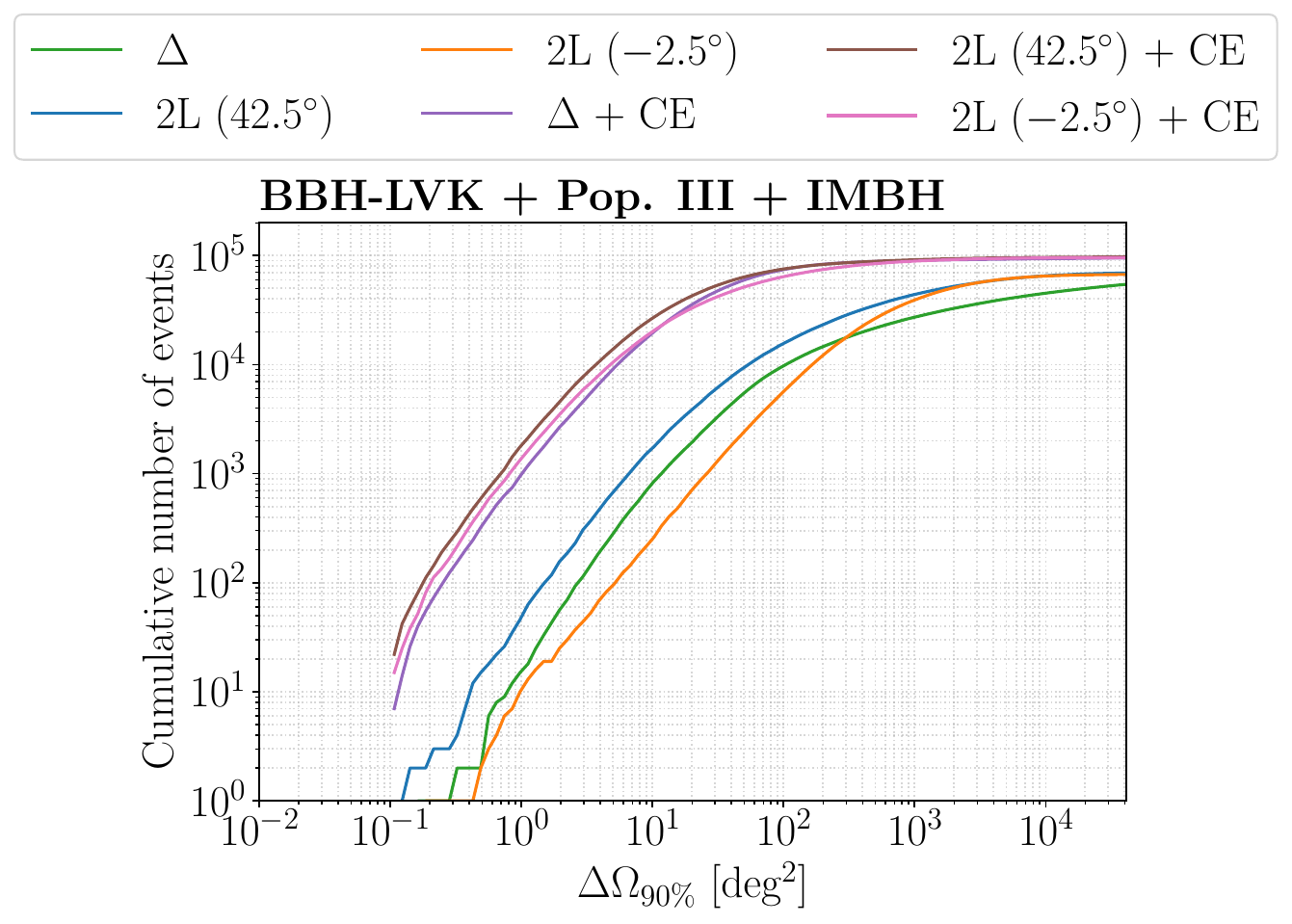}}
\captionof{figure}{\justifying Cumulative distributions of the relative uncertainty on the luminosity distance and 90\% sky localization of the astrophysical BBH populations used in this work. Results are shown for ET alone and for ET in a network with CE.}
\label{fig:errs_ds1_bbh_extr_app_CE}
\end{tcolorbox}
\end{figure}

Despite these large absolute gains, the relative detection performance of the ET geometries is unchanged. The two 15-km 2L configurations continue to yield more detections than the triangle, owing to their greater strain sensitivity, while the two 2L orientations remain very similar in total detection yield. The intermediate 2L$(30^\circ)$ configuration follows the same pattern, yielding $102\,176$ BBH detections, $106\,373$ BNS detections, and $84\,455$ NSBH detections when combined with CE.

\begin{figure*}
\begin{tcolorbox}[standard jigsaw,colframe=gray,colback=gray!10!white,opacityback=0.0,coltext=black, title=\small\sc Distributions for the statistical uncertainties on the source parameters for BNSs for the three configurations in a network with CE]
\center{\includegraphics[width=0.32\textwidth]{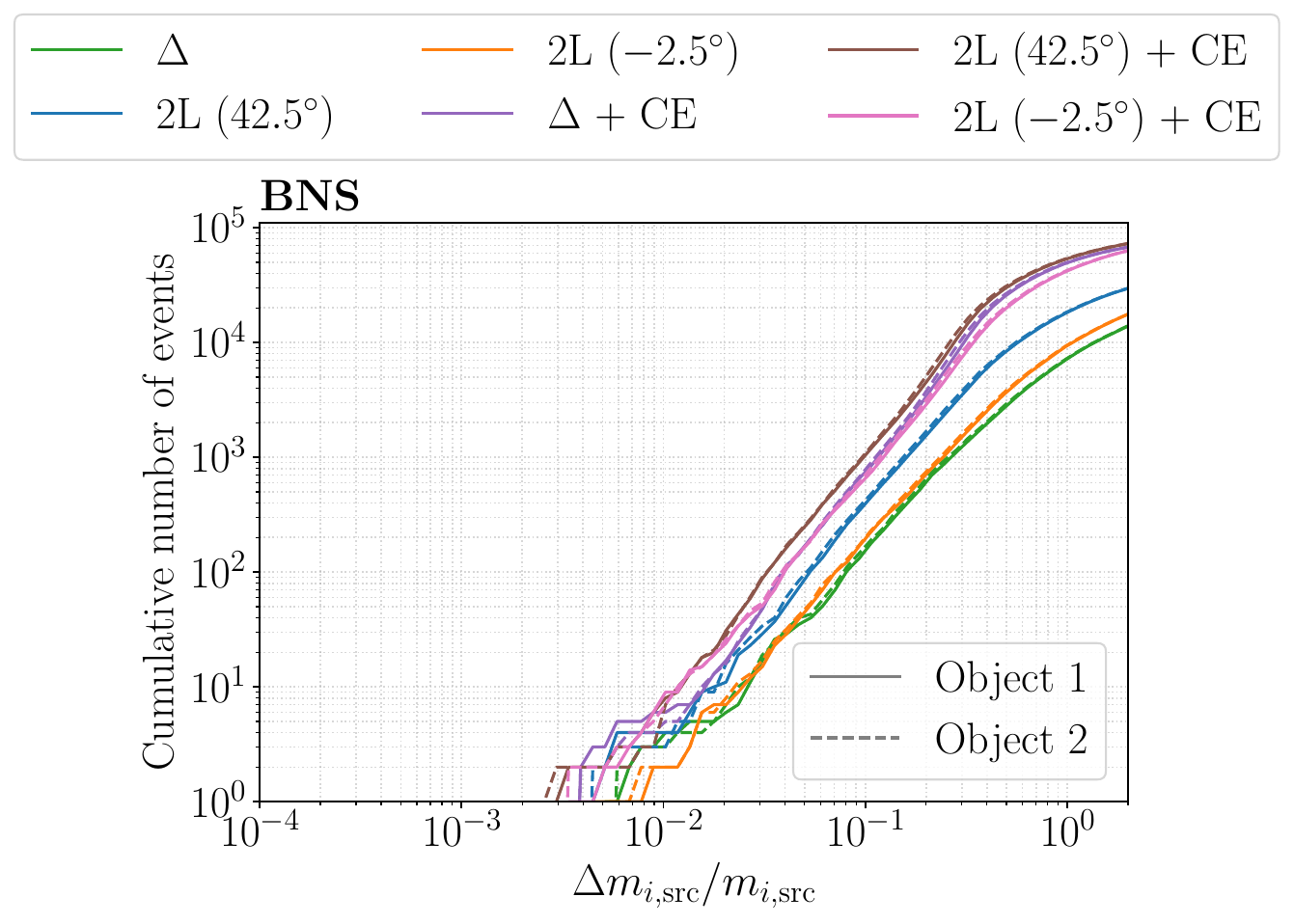}\hfill\includegraphics[width=0.32\textwidth]{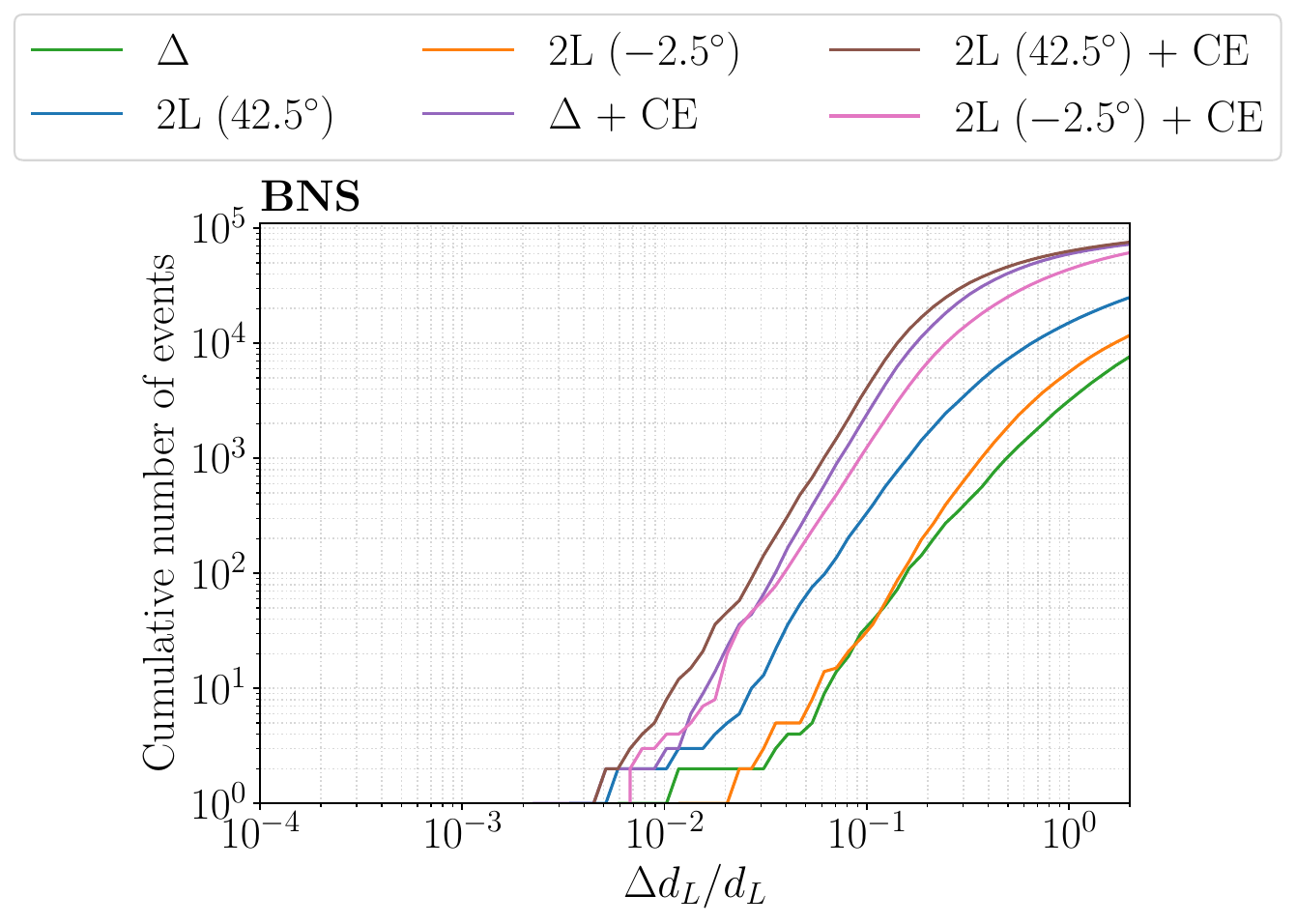}\hfill\includegraphics[width=0.32\textwidth]{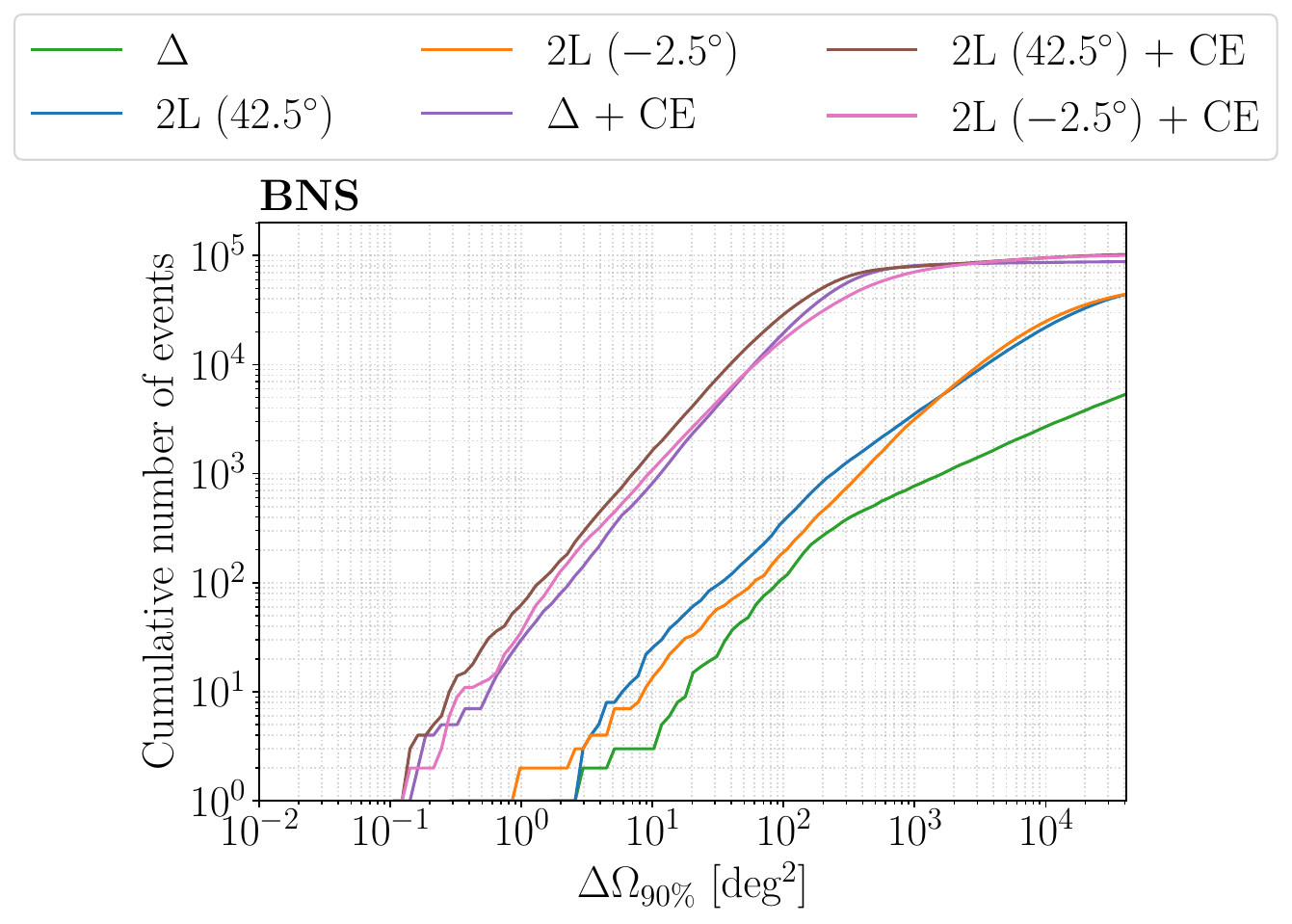}}
\captionof{figure}{\justifying Cumulative distributions of the relative uncertainty on the individual source-frame masses, luminosity distance, and 90\% sky localization of the BNS population used in this work. We only show the results for the medium EOS. Results are shown for ET alone and for ET in a network with CE.}
\label{fig:errs_ds1_bns_app_CE}
\end{tcolorbox}
\vskip-15pt
\end{figure*}

\begin{figure*}
\begin{tcolorbox}[standard jigsaw,colframe=gray,colback=gray!10!white,opacityback=0.0,coltext=black, title=\small\sc Distributions for the statistical uncertainties on the source parameters for NSBHs for the three configurations in a network with CE]
\center{\includegraphics[width=0.32\textwidth]{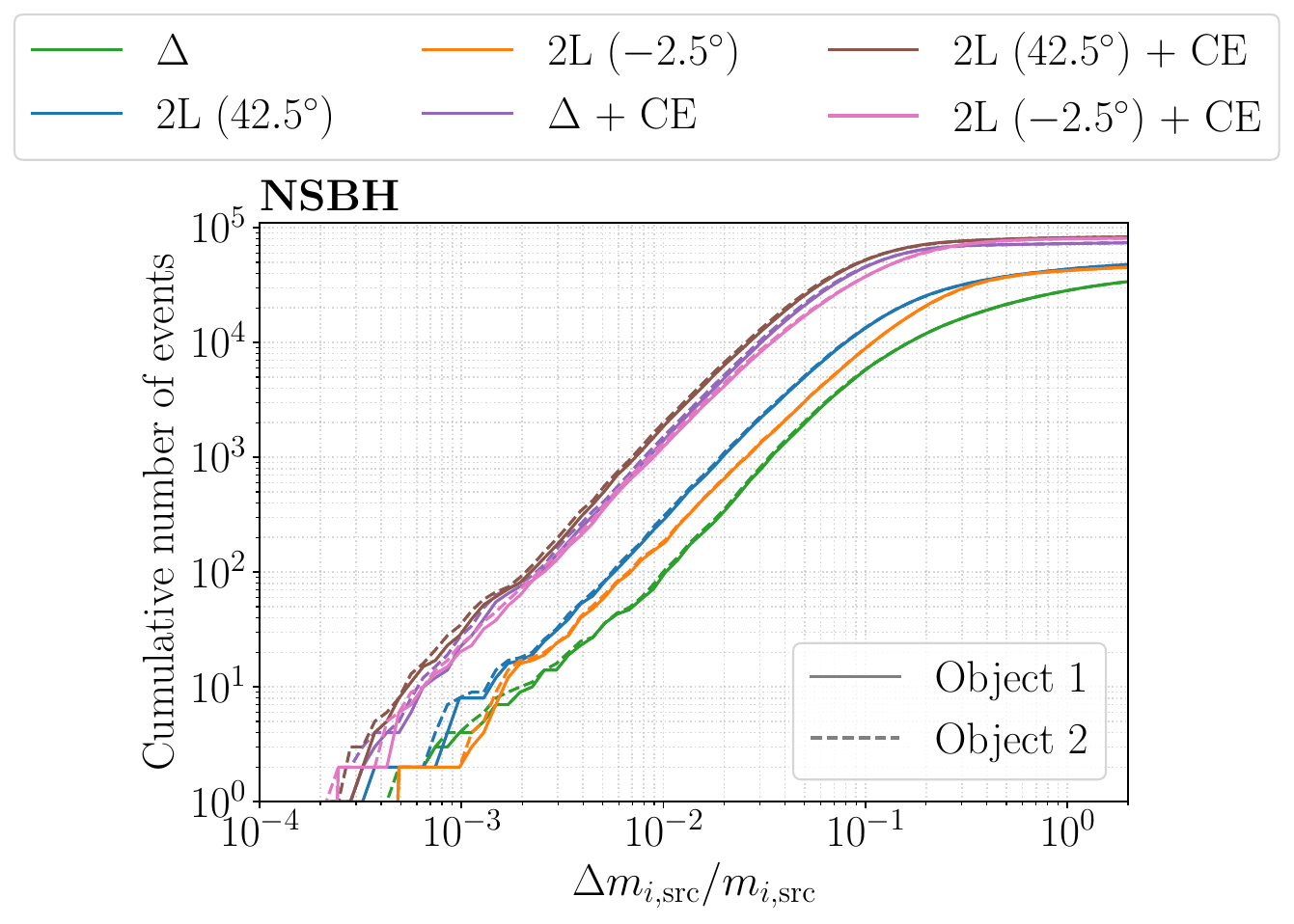}\hfill\includegraphics[width=0.32\textwidth]{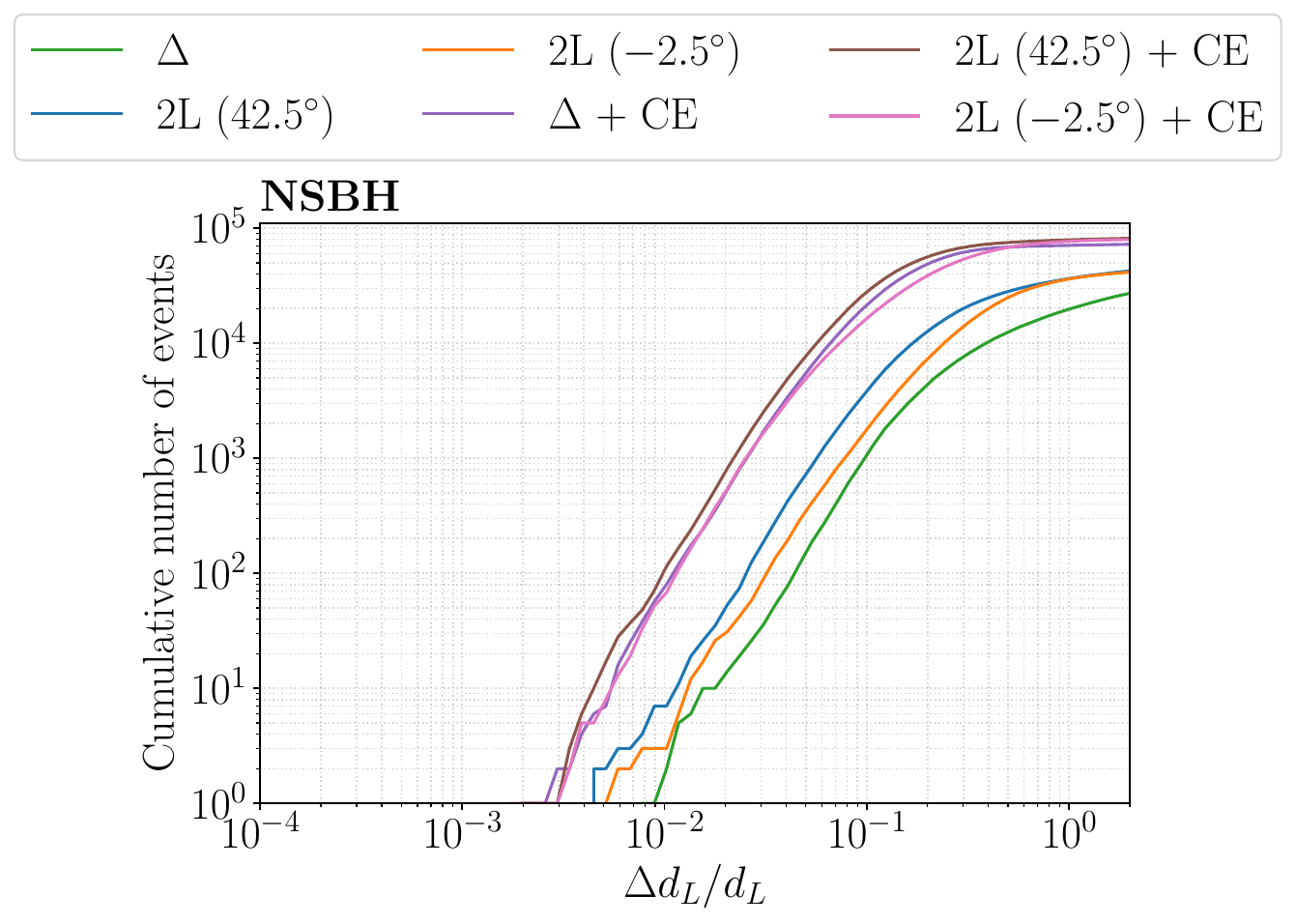}\hfill\includegraphics[width=0.32\textwidth]{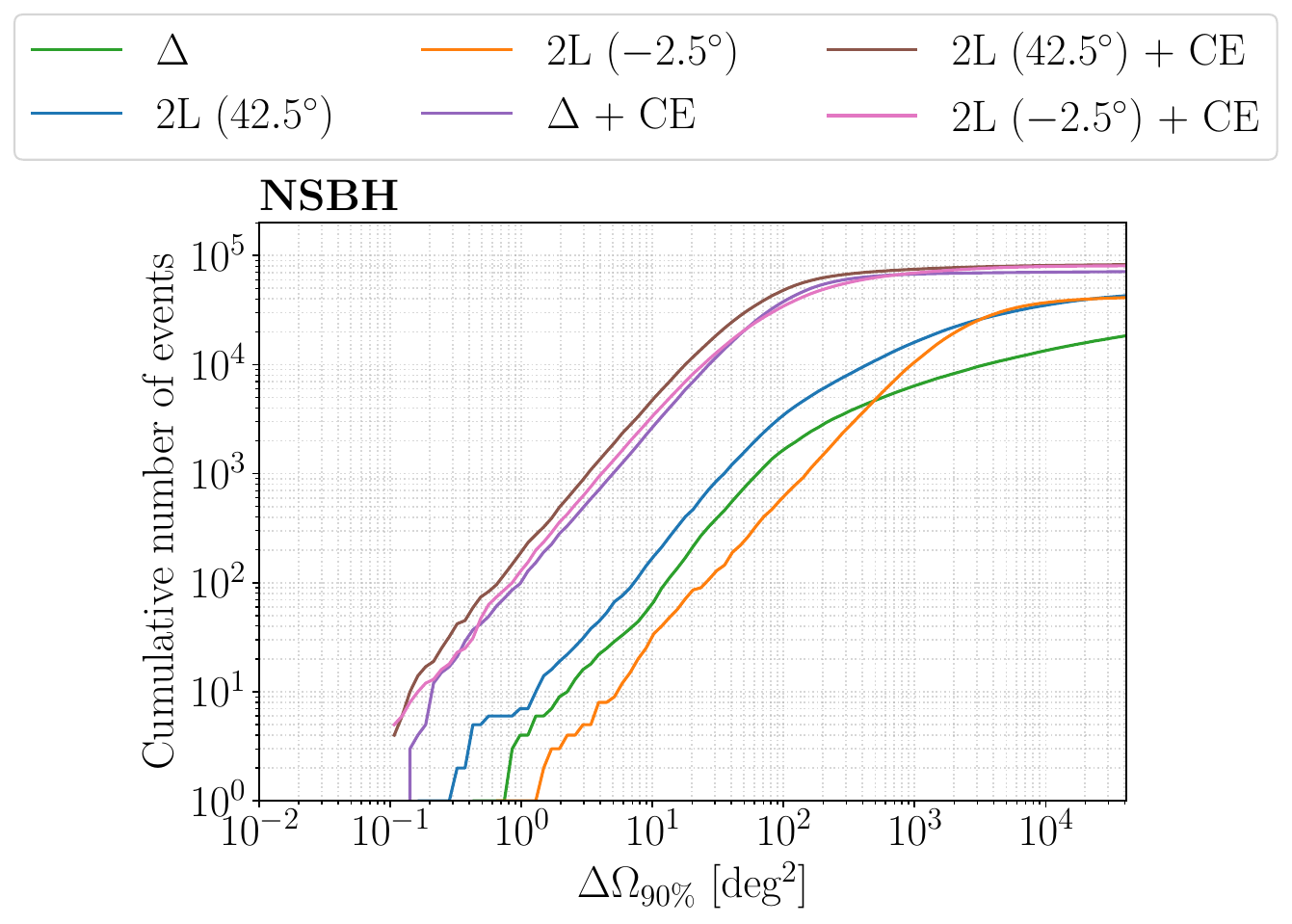}}
\captionof{figure}{\justifying Cumulative distributions of the relative uncertainty on the individual source-frame masses, luminosity distance, and 90\% sky localization of the NSBH population used in this work. We show the results obtained when employing a waveform including higher-order modes and no tidal effects. Results are shown for ET alone and for ET in a network with CE.}
\label{fig:errs_ds1_nsbh_app_CE}
\end{tcolorbox}
\vskip-15pt
\end{figure*}

The most important effect of CE is the dramatic improvement in the reconstruction of extrinsic parameters. The additional intercontinental baseline and independent antenna response substantially improve sky localization, luminosity-distance inference, and consequently redshift determination (Figs.~\ref{fig:errs_ds1_bbh_extr_app_CE}-\ref{fig:errs_ds1_nsbh_app_CE}). For BBHs, the number of events with fractional luminosity-distance uncertainty $\Delta d_L/d_L<5\%$ increases from $882$, $3\,113$, and $1\,285$ for ET alone to $15\,047$, $19\,239$, and $11\,068$ with CE for the triangle, 2L $(42.5^\circ)$, and 2L $(-2.5^\circ)$ configurations, respectively. At the more stringent $1\%$ distance threshold, the corresponding numbers increase from $6$, $42$, and $20$ to $264$, $403$, and $269$. The number of BBHs localized within $10\,\mathrm{deg}^2$ at 90\% credibility similarly grows from $746$, $1\,585$, and $233$ to $18\,148$, $25\,222$, and $19\,014$.

The improvement is even more striking for neutron-star systems. For medium-EOS BNSs, the number localized within $100\,\mathrm{deg}^2$ rises from $104$, $340$, and $177$ with ET alone to $17\,959$, $26\,935$, and $15\,908$ with CE. Within $10\,\mathrm{deg}^2$, the corresponding numbers are $747$, $1\,490$, and $990$, compared with only $3$, $24$, and $12$ without CE. For NSBHs, CE increases the number localized within $100\,\mathrm{deg}^2$ from $1\,559$, $3\,225$, and $564$ to $36\,089$, $46\,337$, and $32\,949$, while the number localized within $10\,\mathrm{deg}^2$ becomes $2\,415$, $4\,323$, and $3\,074$, respectively.

The addition of CE reduces the relative differences between the ET configurations for extrinsic-parameter reconstruction, because the CE--ET baseline already provides powerful timing, localization, and polarization information. Nevertheless, we confirm that the misaligned 2L $(42.5^\circ)$ configuration provides the best overall reconstruction of distance, redshift, and sky position. The relative performance between 2L $(-2.5^\circ)$+CE and triangle--CE exhibits a reverse trend with respect to not adding CE in these cases. With the addition of CE, the 2L $(-2.5^\circ)$+CE network outperforms triangle--CE, contrary to the case without CE. As a consequence, when CE is included the triangle--CE has the most brights sirens events (comparable to 2L $(42.5^\circ)$--CE, considering also the duty cycle) and gives the tightest constraints the Hubble constant parameter. The 2L$(30^\circ)$ configuration remains close to the $42.5^\circ$ case: for example, it yields $18\,238$ BBHs with $\Delta d_L/d_L<5\%$ and $25\,138$ localized within $10\,\mathrm{deg}^2$.

The improved distance measurement also has a major impact on source-frame quantities. For BBHs, the number of events with source-frame chirp-mass uncertainty below $1\%$ increases from $220$, $623$, and $288$ to $2\,913$, $3\,793$, and $2\,431$ for the triangle, 2L $(42.5^\circ)$, and 2L $(-2.5^\circ)$ configurations. Likewise, the numbers with source-frame component masses measured to better than 10\% become $64\,598$ and $62\,199$ for the two components in the triangle--CE network, $67\,066$ and $64\,699$ for 2L $(42.5^\circ)$+CE, and $53\,724$ and $51\,050$ for 2L $(-2.5^\circ)$+CE. This improvement should not be interpreted primarily as a direct improvement in intrinsic-parameter reconstruction. Detector-frame chirp masses are already measured at the percent level for nearly all detected BBHs, both with and without CE. Rather, the large improvement in source-frame masses is driven mainly by the much more precise luminosity-distance and redshift inference enabled by the expanded network.

The intrinsic parameters that are determined predominantly from the waveform phase evolution are comparatively less affected. For instance, the fraction of BBHs with detector-frame chirp-mass uncertainty below $1\%$ remains close to $98\%$ for all ET configurations, with or without CE. Spin measurements improve owing to the larger number of loud sources, but the differences among ET geometries remain modest compared with those for distance and localization. Similarly, CE increases the number of BNSs with accurately measured tidal deformability, but does not qualitatively change the comparison between ET configurations. For the medium EOS, the number of systems with $\Delta\tilde{\Lambda}<50$ increases from $746$, $1\,475$, and $1\,465$ for ET alone to $2\,892$, $3\,839$, and $3\,814$ with CE for the triangle, 2L $(42.5^\circ)$, and 2L $(-2.5^\circ)$ configurations, respectively.

Overall, the inclusion of one CE detector greatly enhances the absolute CBC science reach of the network. It produces substantially larger detection samples, many more high-SNR events, and especially large gains in luminosity-distance and sky localization measurement. These gains propagate into source-frame mass measurements and strengthen the prospects for multimessenger astronomy and cosmology. However, the conclusions of the main chapter regarding the comparison of ET geometries remain valid in this broader network context. For CBC science, the misaligned 2L configuration continues to provide the strongest extrinsic-parameter reconstruction, the triangle remains more effective than the nearly aligned 2L configuration when distance inference and localization
at the $\sim100\,{\rm deg}^2$ level are important, and the 15-km 2L layouts retain their advantage in overall detection yield.

Moreover, the additional detections and improved source characterization strengthen population-level inference. In particular, the improved distance and redshift reconstruction reduces the uncertainty with which the source-frame mass spectrum and merger-rate evolution can be recovered. Nevertheless, population constraints are already extremely tight for ET alone, owing to the very large compact-binary catalogs detected in all configurations. We thus expect that the inclusion of CE therefore would improve the absolute precision of the reconstructed population parameters, without introducing an appreciable difference between the ET geometries or modifying the conclusions of the main
chapter.

\begin{figure*}
    \begin{tcolorbox}[standard jigsaw,colframe=gray,colback=gray!10!white,opacityback=0.0,coltext=black, title=\small\sc Horizon distance for selected CCSN simulations ]
\center{\includegraphics[width=0.32\textwidth]{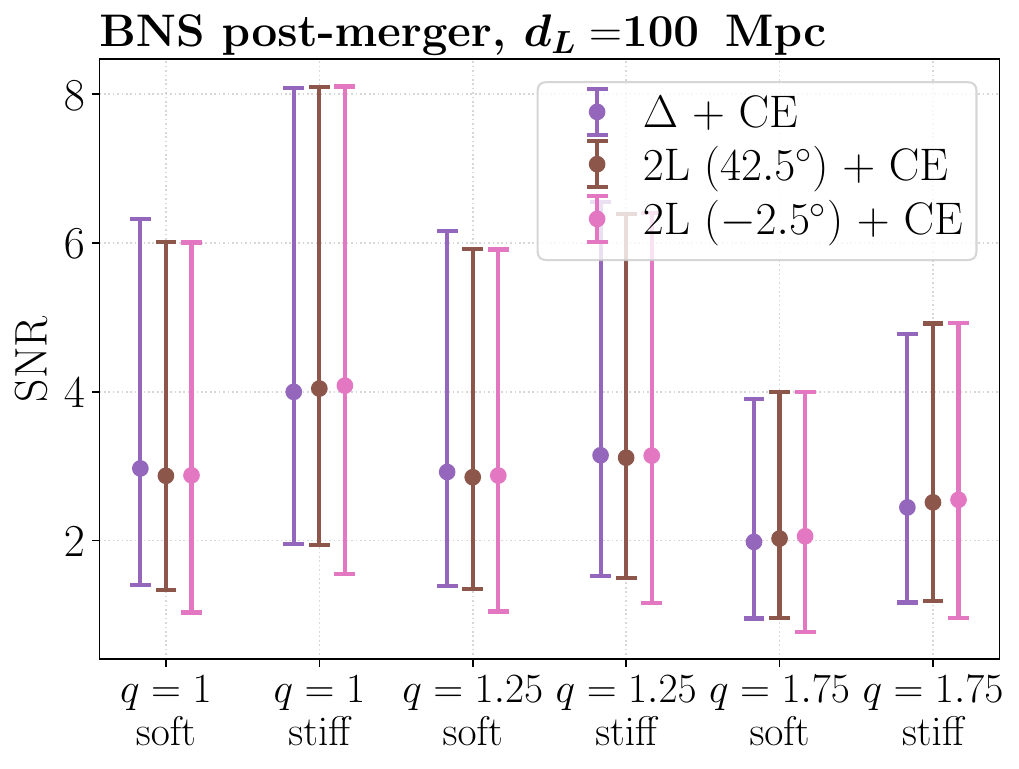}\includegraphics[width=0.32\textwidth]{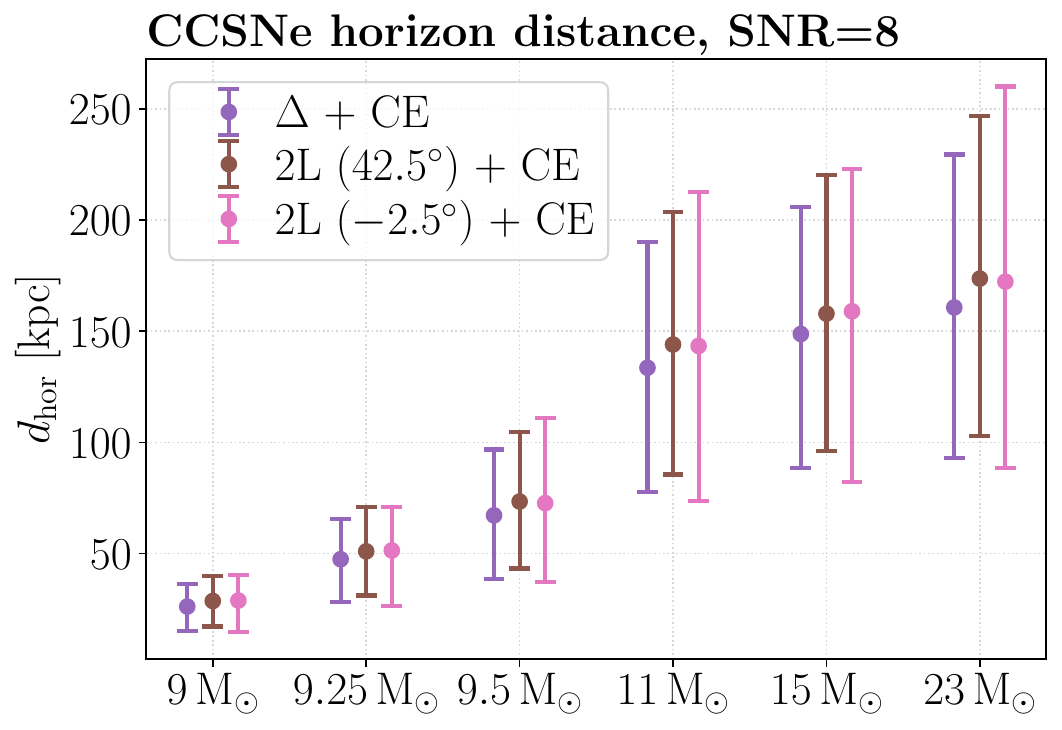}}
\captionof{figure}{\justifying \emph{Left}: Distribution of the SNR in the post-merger phase of selected BNS signals with different mass ratios and EOS, obtained by randomizing over different sky positions, inclinations and polarization angles. \emph{Right}: Maximum distance out to which a CCSN signal from stars with different masses would be detected with an SNR of 8. The calculation assumes matched filtering through an optimal template and random source extrinsic parameters. We consider a global network of ET and CE (the corresponding ET-only results are given in the main chapter).}
\label{fig:bnspm_ccsne_hor_app_CE}
\end{tcolorbox}
\end{figure*}

The addition of CE also improves the prospects for the other source classes considered in this study, without changing the qualitative comparison between ET configurations. For BNS post-merger signals (Fig.~\ref{fig:bnspm_ccsne_hor_app_CE}, left), CE modestly increases the network SNR through its additional sensitivity and independent detector response. This increases the number of potentially informative post-merger observations, although the comparison among the ET geometries remains substantially unchanged. In particular, the modest differences found for ET alone---which are driven by the high-frequency response of the individual interferometers and the breakdown of the long-wavelength approximation in the post-merger band---are retained, but become somewhat smaller in the network with CE.

A similar conclusion applies to CCSNe (Fig.~\ref{fig:bnspm_ccsne_hor_app_CE}, left). The CE detector increases the network sensitivity to CCSN burst signals and therefore extends the horizon distance for the representative waveform set considered here. However, the spread associated with the progenitor model, source sky position, and polarization remains more important than the residual differences among ET geometries. The relative performance of the triangle and the two 2L configurations is therefore unchanged, while the addition of CE reduces the practical importance of the geometry-dependent differences.

For targeted continuous-wave searches for known pulsars, the improved network sensitivity increases the number of detectable sources for all assumptions on the stellar ellipticity. For the three ellipticity scenarios considered in the main chapter, the numbers of detectable pulsars increase from $882$, $204$, and $13$ for the triangle to $990$, $313$, and $29$ when CE is included. The corresponding numbers for the 2L configuration increase from $1008$, $243$, and $15$ to $1085$, $332$, and $34$. Thus, CE increases the continuous-wave discovery potential in both optimistic and conservative ellipticity scenarios, while preserving the mild advantage of the longer-arm 2L geometry. As in the ET-only analysis, these differences remain small compared with the astrophysical uncertainty in the ellipticity distribution of Galactic neutron stars.

\begin{figure*}
\begin{tcolorbox}[standard jigsaw,colframe=gray,colback=gray!10!white,opacityback=0.0,coltext=black, title=\small\sc Polar Histograms Comparing the Science Reach of the Two Geometries]
    \includegraphics[width=0.43\textwidth]{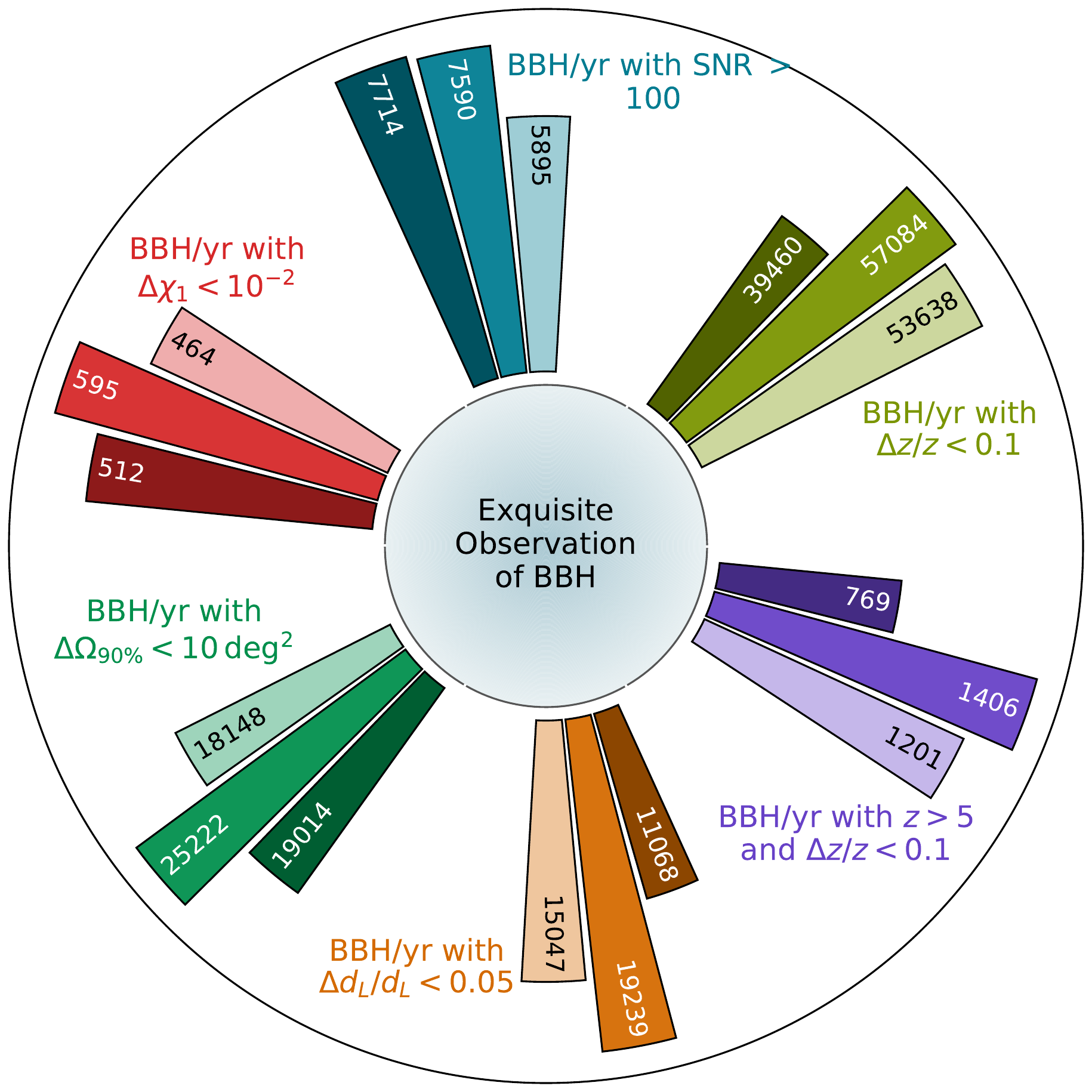} \hspace{0.5cm}
    \includegraphics[width=0.43\textwidth]{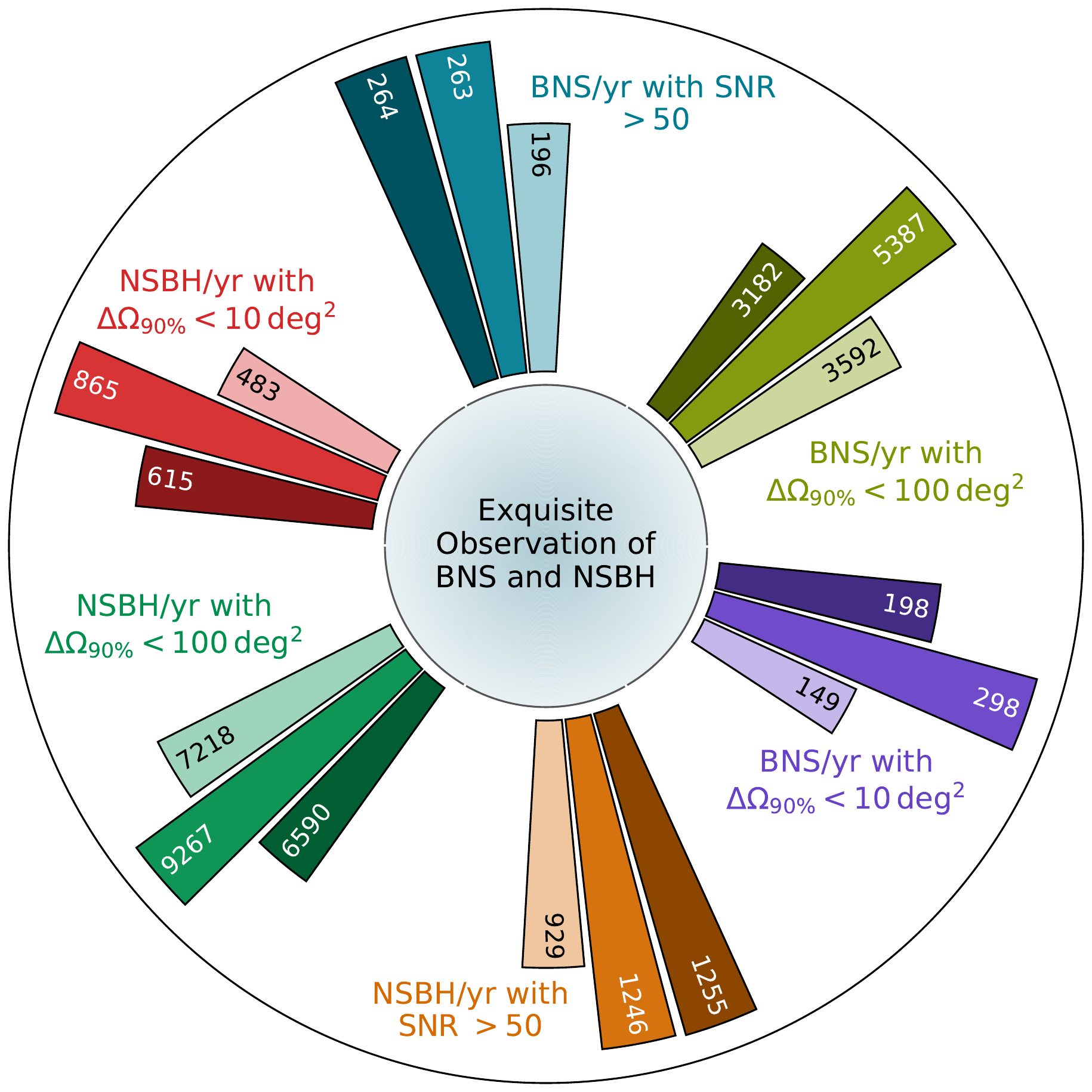}\\[8pt]
    \vskip-10pt
    \includegraphics[width=0.40\textwidth]{src/legend.pdf}\\[2pt]
    \vskip-7pt
    \includegraphics[width=0.43\textwidth]{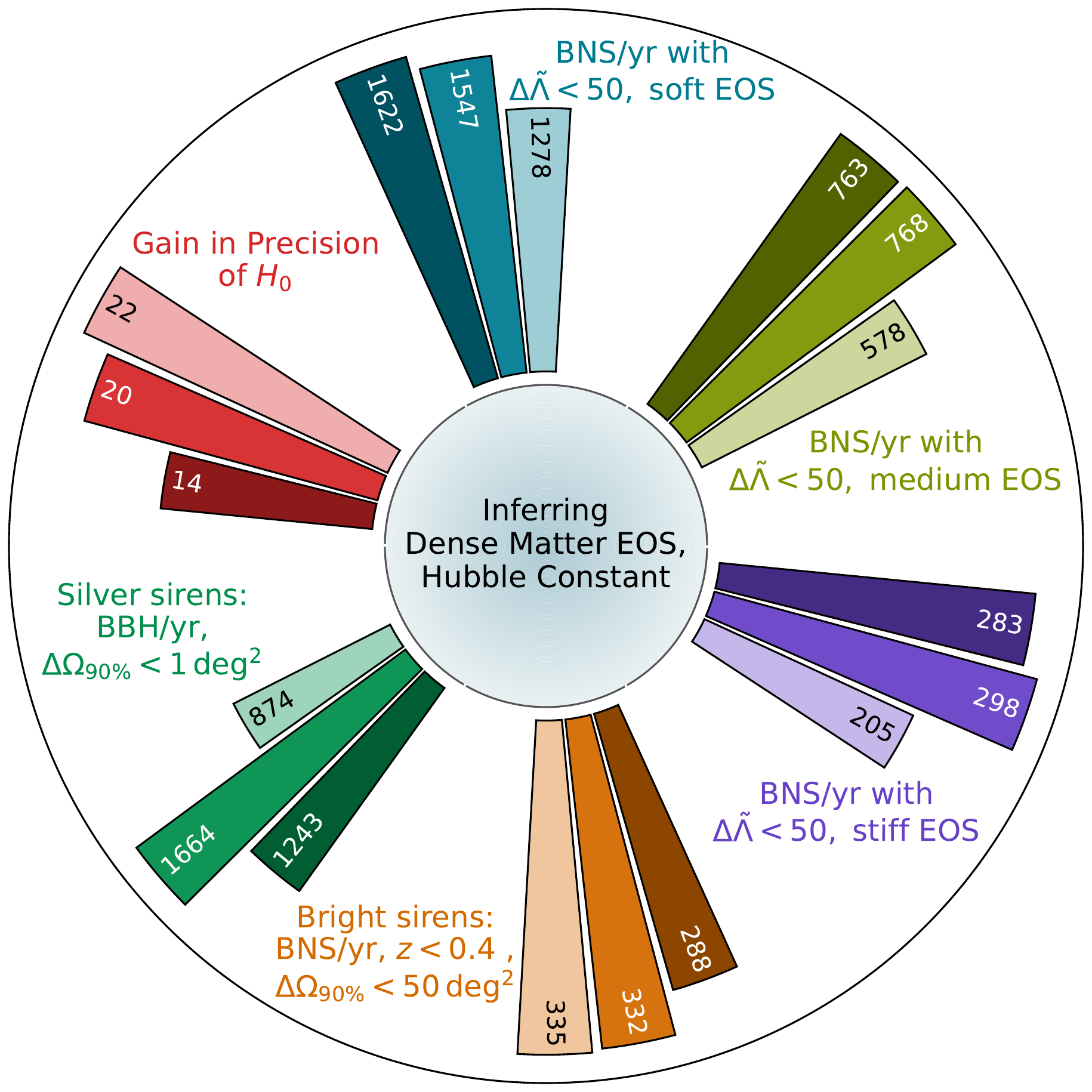} \hspace{0.5cm}
    \includegraphics[width=0.43\textwidth]{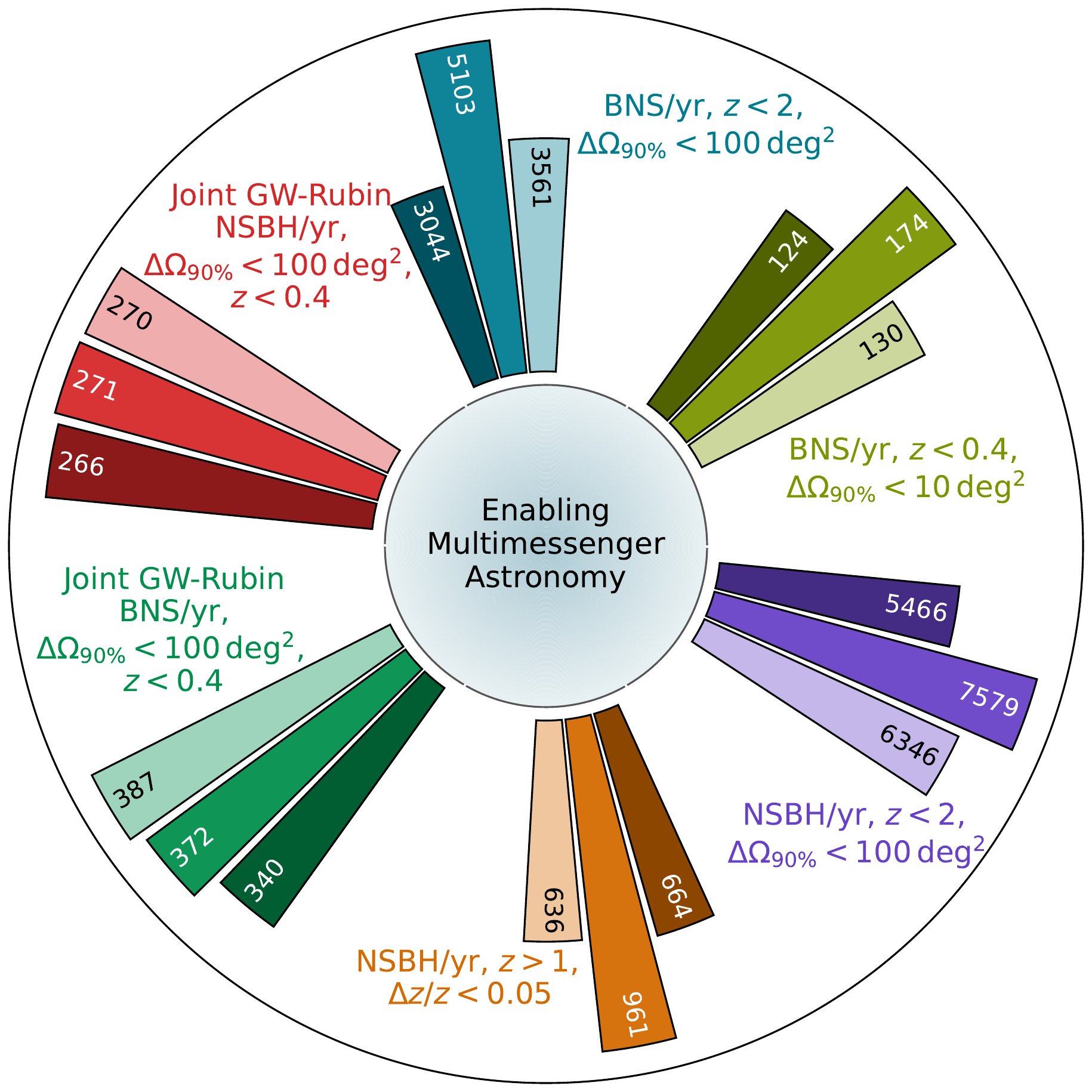}
    \caption{\justifying \textbf{
    Comparison of selected science metrics for the ET geometries in a network with CE.}
Each network includes the same Cosmic Explorer detector and differs only in the
ET configuration. Within each wedge, the three bars show, from dark to light,
CE combined with the nearly aligned 2L ($-2.5^\circ$), the misaligned
2L ($42.5^\circ$), and the triangle. Bar lengths are normalized 
separately within each wedge and therefore compare the
three networks for a given science metric, not different metrics with one
another; the number on each bar gives the corresponding absolute value. 
The four panels illustrate performance for exquisite
observations of BBHs (top left), BNSs and NSBHs (top right), dense-matter
physics (bottom left), and multimessenger astronomy (bottom right).
For BBHs, the metrics include the annual numbers of exceptionally loud events,
sources with precise redshift or luminosity-distance measurements, high-redshift
systems, well-localized sources, and events with accurately measured component
spins. For BNSs and NSBHs, they similarly quantify loud and well-localized
events. The ability to infer the physics of dense-matter
is illustrated by the annual number of BNSs for which the
mass-weighted tidal deformability $\tilde\Lambda$ can be measured to better
than 50\%, for representative soft, medium, and stiff equations of state.
The multimessenger metrics give annual numbers of BNS and NSBH mergers satisfying
the indicated redshift, localization, and distance-precision requirements,
including events jointly accessible to GW observations and the Rubin
Observatory.  Bright sirens are mergers with an identified electromagnetic counterpart, which supplies
the redshift; silver sirens are dark sirens (no counterpart) localized within $1 \deg^2$ 
with $\Delta d_L/d_L < 0.1$, for which a statistical host-galaxy association is possible. 
Throughout, ``X/yr'' denotes the number of events per year; SNR is the
signal-to-noise ratio; $z$ is redshift; $d_L$ is luminosity distance;
$\Delta\Omega_{90\%}$ is the 90\%-credible sky-localization area; and
$\chi_{1}$ is the dimensionless spin of the primary component. Quantities of the form
$\Delta X/X$ denote fractional measurement uncertainties.}

\label{fig:polar1}
\end{tcolorbox}
\end{figure*}

\DefTblrTemplate{capcont}{default}{
  \UseTblrTemplate{caption-tag}{default}
  \UseTblrTemplate{caption-sep}{default}
  \InsertTblrText{entry}{\textit{Summary of main results across the three ET configurations in combination with CE.} (Continued)}
}
\begin{longtblr}[
    caption = {\textit{Summary of main results across the three ET configurations in combination with CE.}  ``X/yr'' denotes the number of detected events per year. The mass parameters in the table refer to source-frame masses. The BBH results under ``Compact-binary populations'' are reported jointly for BBH-LVK, Pop. III, and IMBH. The BNS results under ``Compact-binary populations'' are reported for the medium (SFHo) EOS. Chirp masses and individual masses are in the source frame. Constraints on $H_0$ are obtained with BNS bright sirens.},
    label = {tab:metrics}
]{width=0.99\linewidth,rowhead=1,colspec={X[5,l]X[3.5,c]X[3.5,c]X[3.5,c]}}
\toprule
Science metric &
2L ($-2.5^\circ$)+CE &
2L ($42.5^\circ$)+CE &
$\Delta$+CE \\
\midrule
\SetCell[c=4]{l}{\textbf{Compact-binary populations}} \\
BBH/yr & $101\,022$ & $102\,395$ & $100\,469$ \\
BBH/yr with SNR$>100$ & $7\,714$ & $7\,590$ & $5\,895$ \\
BBH/yr with $\Delta z/z<0.1$ & $39\,460$ & $57\,084$ & $53\,638$ \\
BBH/yr with $z>5$ and $\Delta z/z<0.1$ & $769$ & $1\,406$ & $1\,201$ \\
BBH/yr with $\Delta {\cal M}_c/{\cal M}_c<10^{-2}$ & $2\,431$ & $3\,793$ & $2\,913$ \\
BBH/yr with $\Delta m_1/m_1<0.1$ & $53\,724$ & $67\,066$ & $64\,598$ \\
BBH/yr with $\Delta m_2/m_2<0.1$ & $51\,050$ & $64\,699$ & $62\,199$ \\
BBH/yr with $\Delta d_L/d_L<10^{-2}$ & $269$ & $403$ & $264$ \\
BBH/yr with $\Delta d_L/d_L<0.05$ & $11\,068$ & $19\,239$ & $15\,047$ \\
BBH/yr with $\Delta\Omega_{90\%}<10\,{\rm deg}^2$ & $19\,014$ & $25\,222$ & $18\,148$ \\
BBH/yr with $\Delta\chi_{1,2}<10^{-2}$ & $512, 149$ & $595, 169$ & $464, 145$ \\
BNS/yr & $21\,395$ & $21\,589$ & $18\,802$ \\
BNS/yr with SNR$>50$ & $264$ & $263$ & $196$ \\
BNS/yr with $z>1$ and $\Delta z/z<0.1$ & $36$ & $274$ & $69$ \\
BNS/yr with $\Delta {\cal M}_c/{\cal M}_c<10^{-2}$ & $45$ & $91$ & $65$ \\
BNS/yr with $\Delta m_1/m_1<0.1$ & $116$ & $181$ & $131$ \\
BNS/yr with $\Delta m_2/m_2<0.1$ & $121$ & $187$ & $135$ \\
BNS/yr with $\Delta\Omega_{90\%}<100\,{\rm deg}^2$ & $3182$ & $5387$ & $3592$ \\
BNS/yr with $\Delta\Omega_{90\%}<10\,{\rm deg}^2$ & $198$ & $298$ & $149$ \\
NSBH/yr & $16\,664$ & $17\,063$ & $15\,046$ \\
NSBH/yr with SNR$>50$ & $1\,255$ & $1\,246$ & $929$ \\
NSBH/yr with $z>1$ and $\Delta z/z<0.05$ & $664$ & $961$ & $636$ \\
NSBH/yr with $\Delta {\cal M}_c/{\cal M}_c<10^{-2}$ & $265$ & $410$ & $312$ \\
NSBH/yr with $\Delta m_1/m_1<0.1$ & $7\,022$ & $9\,932$ & $8\,584$ \\
NSBH/yr with $\Delta m_2/m_2<0.1$ & $7\,060$ & $10\,027$ & $8\,691$ \\
NSBH/yr with $\Delta\Omega_{90\%}<100\,{\rm deg}^2$ & $6\,590$ & $9\,267$ & $7\,218$ \\
NSBH/yr with $\Delta\Omega_{90\%}<10\,{\rm deg}^2$ & $615$ & $865$ & $483$ \\
\midrule
\SetCell[c=4]{l}{\textbf{Dense matter}} \\
BNS/yr with $\Delta\tilde\Lambda<10$ & soft: $8$ & soft: $6$ & soft: $5$ \\
 & medium: $5$ & medium: $5$ & medium: $4$ \\
 & stiff: $1$ & stiff: $2$ & stiff: $1$ \\
BNS/yr with $\Delta\tilde\Lambda<50$ & soft: $1\,622$ & soft: $1\,547$ & soft: $1\,278$ \\
 & medium: $763$ & medium: $768$ & medium: $578$ \\
 & stiff: $283$ & stiff: $298$ & stiff: $205$ \\
Post-merger SNR at 100\,Mpc
    & soft: $2.9^{+3.1}_{-1.8}$ & soft: $2.9^{+3.1}_{-1.5}$ & soft: $3.0^{+3.3}_{-1.6}$ \\
selected NR simulations, $q=1$
    & stiff: $4.1^{+4.0}_{-2.5}$ & stiff: $4.0^{+4.1}_{-2.1}$ & stiff: $4.0^{+4.1}_{-2.0}$ \\
\midrule
\SetCell[c=4]{l}{\textbf{Cosmology and early Universe}} \\
Bright sirens: BNS/yr with $z<0.4$, $\Delta\Omega_{90\%}<50\,{\rm deg}^2$
& $288$ & $332$ & $335$ \\
Golden (silver) dark sirens: BBH/yr, NSBH/yr with $\Delta d_L/d_L<0.1$, &  &  &  \\
$\Delta\Omega_{90\%}<0.1\,{\rm deg}^2$ (golden) & BBH:38 , NSBH:1  & BBH:51 , NSBH:1  & BBH: 20, NSBH:1  \\  
$\Delta\Omega_{90\%}<1\,{\rm deg}^2$ (silver) & BBH:1243, NSBH:24 & BBH:1664, NSBH:36 & BBH:874, NSBH:20 \\
$\Delta H_0/H_0$ & 0.74\% & 0.51\% & 0.45\% \\
\midrule
\SetCell[c=4]{l}{\textbf{Multimessenger astronomy}} \\
BNS/yr with $\Delta\Omega_{90\%}<100\,{\rm deg}^2$ for $z>2$     & 138 & 284 & 31 \\
BNS/yr with $\Delta\Omega_{90\%}<100\,{\rm deg}^2$ for $z<2$  & 3044 & 5103 & 3561 \\
BNS/yr with $z<0.4$ and & & & \\
$\Delta\Omega_{90\%}<10\,{\rm deg}^2$  & 124 & 174 & 130 \\
$\Delta\Omega_{90\%}<1\,{\rm deg}^2$  & 7 & 12 & 5 \\
BNS pre-merger/yr with $\Delta\Omega_{90\%}<100,\,10,\,1\,{\rm deg}^2$ at & & & \\
$1\,$h & 1, 0, 0 & 3, 0, 0 & 2, 0, 0 \\
$30\,$min & 5, 0, 0 & 10, 1, 0 & 6, 0, 0 \\
$10\,$min & 20, 2, 0 & 48, 4, 0 & 40, 2, 0 \\
$5\,$min & 37, 2, 0 & 94, 6, 0 & 88, 5, 0 \\
$1\,$min & 94, 5, 0 & 305, 16, 1 & 287, 13, 0 \\
NSBH/yr with $\Delta\Omega_{90\%}<100\,{\rm deg}^2$ for $z>2$ & 1124 & 1688 & 871 \\
NSBH/yr with $\Delta\Omega_{90\%}<100\,{\rm deg}^2$ for $z<2$ & 5466 & 7579 & 6346 \\
NSBH/yr with $z<0.4$ and & & & \\
$\Delta\Omega_{90\%}<10\,{\rm deg}^2$  & 153 & 201 & 179 \\
$\Delta\Omega_{90\%}<1\,{\rm deg}^2$  & 24 & 35 & 20 \\
NSBH pre-merger/yr with $\Delta\Omega_{90\%}<100,\,10,\,1\,{\rm deg}^2$ at & & & \\
$1\,$h & 0, 0, 0 & 0, 0, 0 & 0, 0, 0 \\
$30\,$min & 0, 0, 0 & 1, 0, 0 & 1, 0, 0 \\
$10\,$min & 5, 0, 0 & 19, 1, 0 & 14, 1, 0 \\
$5\,$min & 20, 1, 0 & 64, 4, 0 & 52, 3, 0 \\
$1\,$min & 159, 11, 0 & 450, 29, 1 & 403, 25, 2 \\
Subset of with viewing angle $\Theta<\pi/12$:                &  &  &  \\
BNS/yr with $\Delta\Omega_{90\%}<100\,{\rm deg}^2$ & 203 & 398 & 308 \\
NSBH/yr with $\Delta\Omega_{90\%}<100\,{\rm deg}^2$ & 521 & 608 & 522 \\
Joint GW--EM detection rate/yr with reference EM facilities          &  &  &  \\
Rubin ($\Delta\Omega_{90\%}<100\,{\rm deg}^2,\ z<0.4$) & BNS: 340, NSBH: 266 & BNS: 372, NSBH: 271  & BNS: 387, NSBH: 270  \\
Roman ($\Delta\Omega_{90\%}<3\,{\rm deg}^2,\ z<0.4$) & BNS: 41, NSBH: 74  & BNS: 53, NSBH: 111  & BNS: 26, NSBH: 79  \\
\midrule
\SetCell[c=4]{l}{\textbf{Continuous and burst sources}} \\
PSR detections/yr from targeted searches for conservative $\epsilon$
& $34$ & $34$ & $29$ \\
Maximum distance for CCSN detection $M_\star = 23\,M_\odot$ & $172^{+88}_{-84}$\,kpc & $174^{+73}_{-71}$\,kpc & $161^{+69}_{-68}$\,kpc \\
\end{longtblr}
\newpage

\newpage

\bibliography{references}